\documentclass[12pt,a4paper]{report}

\usepackage{packages/scienceThesis}
\usepackage{verbatim}
\usepackage{amsfonts}
\usepackage{amsmath}
\usepackage{amssymb}
\usepackage{amsthm}
\usepackage{bibunits}
\usepackage[acronym,toc,nonumberlist]{glossaries}

\theoremstyle{definition}

\theoremstyle{definition}

\theoremstyle{definition}

\theoremstyle{definition}

\theoremstyle{definition}

\theoremstyle{definition}

\usepackage{fancyhdr}
\makeatletter
\renewcommand{\chaptermark}[1]{\markboth{\small\textsc{\@chapapp}\ \thechapter:\ \sc{#1}}{}}
\makeatother
\usepackage{sectsty}
\chapterfont{\large\sc\centering}
\chaptertitlefont{\sc\centering}
\subsubsectionfont{\centering}

\usepackage[colorlinks]{hyperref}
\hypersetup{
    bookmarksnumbered,
    pdfstartview={FitH},
    citecolor={black},
    linkcolor={black},
    urlcolor={black},
    pdfpagemode={UseOutlines}
}
\usepackage{siunitx}
\usepackage{url}
\usepackage{subfig}
\usepackage{nicefrac}
\usepackage{algorithm}
\usepackage{algorithmic}    
\usepackage{breakcites}
\usepackage{arydshln}
\usepackage{subfiles}
\usepackage{rotating}
\usepackage{multirow}
\usepackage{rotfloat}
\usepackage[bottom]{footmisc}
\usepackage{mathtools}
\usepackage{enumerate}
\usepackage[gen]{eurosym}
\usepackage{array}
\usepackage{pdfpages}

\DeclareMathOperator\erf{erf}

\begin{document}

\setlinespacing{oneandhalf}

\title{A Real-Time, GPU-Based, Non-Imaging Back-End for Radio Telescopes}
\author{Alessio Magro}
\date{27th September 2013}
\supervisor{Dr. Kristian Zarb Adami}
\cosupervisor{Dr. John Abela} 
\department{Physics}
\degree{Ph.D}

\cosupervisor{Dr. John Abela}
\nomen{acronyms.tex}
\acknowledgements{ack.tex}
\publications{publications.tex}
\abstract{abstract.tex}

\include{} 


\chapter{Introduction}

Radio Astronomy is a relatively modern field, which started with Karl G. Jansky in 1931. He worked as a radio engineer with Bell Telephone Laboratories in New Jersey. While studying radio frequency interference (RFI) from thunderstorms, he noticed a steady hiss type of static whose origin could not be attributed to weather systems. Rotating the
antenna revealed that the static changed gradually, with a period of about 24 hours. He concluded that this static originated from the Milky Way \cite{Jansky1933}. This field has grown exponentially since Jansky's time and has led to a substantial increase in astronomical knowledge, coupled with the construction of large radio instruments. The advent of radio interferometry enabled individual dishes and receiving elements to be combined into huge observatories, even spanning multiple continents, culminating in the current design and eventual deployment of the Square Kilometre Array, which will pose several data processing and transport challenges, some of which will be tackled in this thesis.

One of the more exotic areas of study in this field is the discovery and observation of several classes of radio transients, which are celestial objects that do not emit a steady amount of electromagnetic radiation. Their discovery is an excellent example of serendipity in astronomy. Pulsars were discovered in 1967 by Jocelyn Bell and Antony Hewish \cite{Hewish1968} during a low-frequency survey of extragalactic sources that scintillate in the interplanetary plasma, while Gamma Ray Bursts (GRBs) were first identified in the late 1960s and early 1970s by the Vela satellite system which was launched to verify Soviet adherence to the nuclear test ban treaty \cite{Klebesadel1973}.
Both these examples demonstrate the necessity for current and future telescopes to explore uncharted regions of parameter space, in particular the need for higher sensitivity, finer time and spectral resolution and faster survey speed.

\section{Telescope Sensitivity}

In radio astronomy, radiation is generally measured as units of flux density, $\mathcal{F}$, which is the power per unit frequency interval that passes through a surface of unit area. Therefore the power density received by a radio receiver is
\begin{equation}
\label{powerEquation}
 P = \frac{\mathcal{F}A_{\text{eff}}}{2}
\end{equation}
where the factor of 2 reduction is due to the fact that we can only observe in the horizontal and vertical polarisation, and $A_{\text{eff}}$ is the effective cross section of the antenna, which for wavelength $\lambda$ and antenna gain G is defined as:
\begin{equation}
 A_{\text{eff}} = \frac{\lambda^2 G}{4\pi}
\end{equation}
For the small fluxes encountered in radio astronomy it is convenient to define 
the unit Jansky (Jy), where 1 Jy = 10$^{-26}$ W Hz$^{-1}$ m$^{-2}$. For 
comparison, the Sun, which is the brightest natural radio emitter in the sky, 
has a flux density of about 4 MJy when observed at 10 GHz.

A radio receiver measures the power density $P$ picked up by an antenna, which can be compared with the thermal noise produced by a bandwidth limited resistor of a given temperature $T$:
\begin{equation}
\label{antennaPowerEquation}
 P = k_B\Delta vT
\end{equation}
where $k_B$ is the Boltzmann constant, equal to 1.38 $\times$ 10$^{-23}$ Ws/K \cite{Johnson1928}.  The system temperature of the telescope, $T_{\text{sys}}$, is a measure of the instantaneous noise contribution to a measurement from the telescope receiver chain, the sky, which is at a constant temperature itself, and any other sources of noise in the telescope. Using equation \ref{antennaPowerEquation}, the system temperature can be defined as
\begin{equation}
  T_{\text{sys}} = \frac{P}{k_B\Delta v}
\end{equation}

Using this definition, the noise of a measurement from a telescope is then determined by the ideal radiometer
equation (see \cite[chapter 3]{BurkeGraham2010}:
\begin{equation}
 \Delta T = \frac{T_{sys}}{\sqrt{\Delta v \Delta t}}
\end{equation}

The sensitivity to a point source can be expressed by considering the flux 
collecting area of a telescope $A_{\text{eff}}$, which yields an expression for 
the noise on a measurement of flux density $\Delta S$ from a point source
\begin{equation}
\label{radiometerEquation}
 \Delta S = \frac{2k_BT_{\text{sys}}}{A_{\text{eff}}\sqrt{\Delta v \Delta t}}
\end{equation}
This shows the critical dependence of point source sensitivity to 
$A_{\text{eff}}/T_{\text{sys}}$, and thus this ratio is often quoted when 
specifying the top level requirements of a telescope. The sensitivity of a 
telescope to a point source can always be improved by building a more directive 
antenna (or antenna array) with larger effective area. If $\beta$ is the minimum 
S/N for a source to be considered detected, then the minimum detectable flux 
density, $S_{\text{min}}$, is
\begin{equation}
\label{sminEquation}
  S_{\text{min}} = \frac{2k_B\beta}{A_{\text{eff}}}\frac{T_{\text{sys}}}{\sqrt{N_p\Delta v \Delta t}}
\end{equation}
where $N_p$ is the number of orthogonal polarisations averaged, and can have a 
value of either 1 or 2. In the case of pulsars, the data obtained from the 
receiver can be folded at the pulse period to enhance the S/N of the pulse, and 
$S_{\text{min}}$ will then be dependent on the pulse period, duty cycle and 
integration time (interested readers may refer to \cite{Bhat1998} for an 
analytic analysis of how $S_{\text{min}}$ is affected).

The intensity, or brightness temperature, $I$, is also used as a measure of radiation. This is the power per unit frequency interval passing through a surface of unit area and from a cone of unit solid angle. The solid angle $\Omega$ measures the fraction of the entire sky which is covered by the source, the whole area having a solid angle of $4\pi$. The unit of intensity is W Hz$^{-1}$ m$^2$ sr$^{-1}$, where sr stands for steridian, meaning ``per unit angle''. Flux and intensity are related by
\begin{equation}
 \mathcal{F} = I\Omega
\end{equation}
The intensity along a line of sight in a vacuum does not change, irrespective of the distance from the source. It only changes if there are additional light sources, or if radiation is absorbed by intervening material. However, the flux received from a source decreases with the square of the distance.

\section{Effects of the Interstellar Medium}
\label{ISMSection}

Before signals from astrophysical sources reach the Earth they pass through the interstellar medium (ISM) and, for the case of signals originating from extragalactic sources, the intergalactic medium (IGM). This is a cold, ionised plasma which causes these signals to experience a frequency-dependent index of refraction as they propagate through it. Following \cite{LorimerKramer2005}, the refractive index is
\begin{equation}
\label{refrEquation}
 \mu = \sqrt{1 - \frac{f_p}{f}^2}
\end{equation}
where $f$ is the observing frequency and the plasma frequency
\begin{equation}
 f_p = \sqrt{\frac{e^2n_e}{\pi m_e}} \simeq 8.5\; \text{kHz}\left(\frac{n_e}{\text{cm}^{-3}}\right)^{\frac{1}{2}}
\end{equation}
where $n_e$ is the electron number density, and $e$ and $m$ are the charge and mass of an electron respectively. For the ISM, $n_e \simeq 0.003$ cm$^{-3}$ and $f_p \simeq 1.5$ kHz.

\subsection{Dispersion}

From equation \ref{refrEquation}, $\mu < 1$, therefore the group velocity of a propagating wave $v_g = c\mu$ is less than the speed of light $c$. Therefore, the propagation of a radio signal along a path of length $d$ from a source to Earth will be delayed in time with respect to a signal of infinite frequency by an amount
\begin{equation}
   t = \left( \intop^d_0\frac{\text{d}l}{v_g}\right) - \frac{d}{c}
\end{equation}
Substituting $v_g = c\mu$ and noting $f_p \ll f$ to approximate $\mu$, we get
\begin{equation}
 t = \frac{1}{c} \intop^d_0 \left[ 1 + \frac{f_p^2}{2f^2} \right]\text{d}l - \frac{d}{c} = \frac{e^2}{2\pi m_e c} \frac{\intop^d_0 n_e \text{d}l}{f^2} \equiv \mathcal{D} \times \frac{\text{DM}}{f^2}
\end{equation}
where the dispersion measure
\begin{equation}
\label{dmEquation}
 \text{DM} = \intop^d_0 n_e \text{d}l
\end{equation}
is generally expressed in pc cm$^{-3}$, and the dispersion constant (\cite{LorimerKramer2005})
\begin{equation}
 \mathcal{D} = \frac{e^2}{2\pi m_e c} = (4.148808 \pm 0.000003) \times 10^3\; \text{MHz}^2\; \text{pc}^{-1}\; \text{cm}^3\; \text{s}
\end{equation}
where the uncertainty in $\mathcal{D}$ is determined by the uncertainties in $e$ and $m_e$.

\begin{figure}[t!]
  \begin{center}
  \includegraphics[width=300pt]{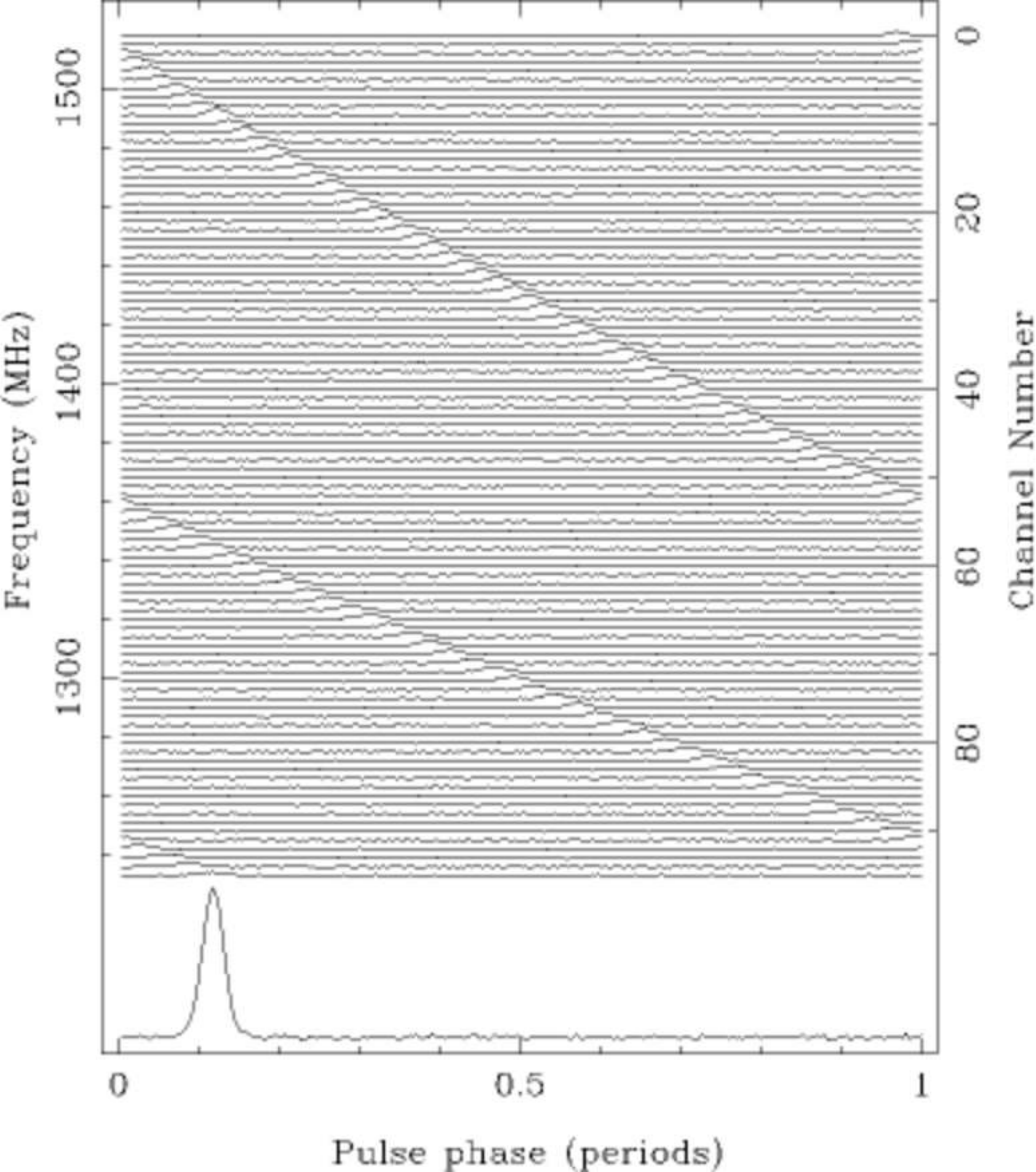}
  \end{center}
  \caption[Pulse dispersion]{Pulse dispersion shown in the Parkes observation of the 128 ms pulsar B1356-60, with a dispersion measure of 295 cm$^{-3}$ pc. Source: \cite[figure 1.8]{LorimerKramer2005}}
\label{dispersionFigure}
\end{figure}

The above process has the effect of delaying the signals in a frequency dependent manner, whereby signals at lower frequencies are delayed more than those at higher frequencies. We can calculate the time delay $\Delta t$ between two signals $f_1$ and $f_2$ with the dispersion relationship equation:
\begin{equation}
\label{dispRelationshipEquation}
 \Delta t = k_{\text{DM}} \cdot \mbox{DM} \cdot \left( f^{-2}_1 - f^{-2}_2 \right)
\end{equation}
where $k_{\text{DM}}$ = $\mathcal{D} \times 10^3$. In practice the 
approximation
\begin{equation}
 k_{\text{DM}} = \frac{1}{2.41 \times 10^{-7}} \times 10^3
\end{equation}
is used. Figure \ref{dispersionFigure} demonstrates 
the dispersion effect on a pulse from the 128 ms pulsar B1356-60, observed with 
the Parkes Radio Telescope, having a DM of 295 pc cm$^{-3}$. The DM value for 
any source can be be determined by measuring the phase as a function of 
frequency, which allows the distance to the object to be calculated by 
numerically integrating equation \ref{dmEquation}, assuming one has a model for 
the Galactic electron density distribution.

When observing radio sources this dispersive effect needs to be corrected for, in order to avoid excessive smearing
in the time series and the consequent loss in signal-to-noise (S/N). This can be 
performed by using a dedispersion algorithm, some of which are discussed in 
detail in sections \ref{dispRemovalSection} and \ref{coherentSection}. These 
techniques require the source's DM value to correct for dispersion, however, 
when searching for new sources, such as in blind transient surveys, the distance 
to any potential discovered source is unknown, and so this procedure needs to be 
repeated for many trial dispersion measures. This is a computationally expensive 
task, where a simple approach involves summing all the frequency channels across 
the observing band ($\sim$10$^3$ frequency channels) for a number of trial DMs 
($\sim$10$^4$, typically) for every time sample, where modern sampling intervals 
are of the order of $\mu s$. This task is even more challenging when multiple 
beams need to be processed (for interferometric arrays or
multiple-feed dishes).

\subsection{Scattering}

\begin{figure}[t!]
  \begin{center}
  \includegraphics[width=300pt]{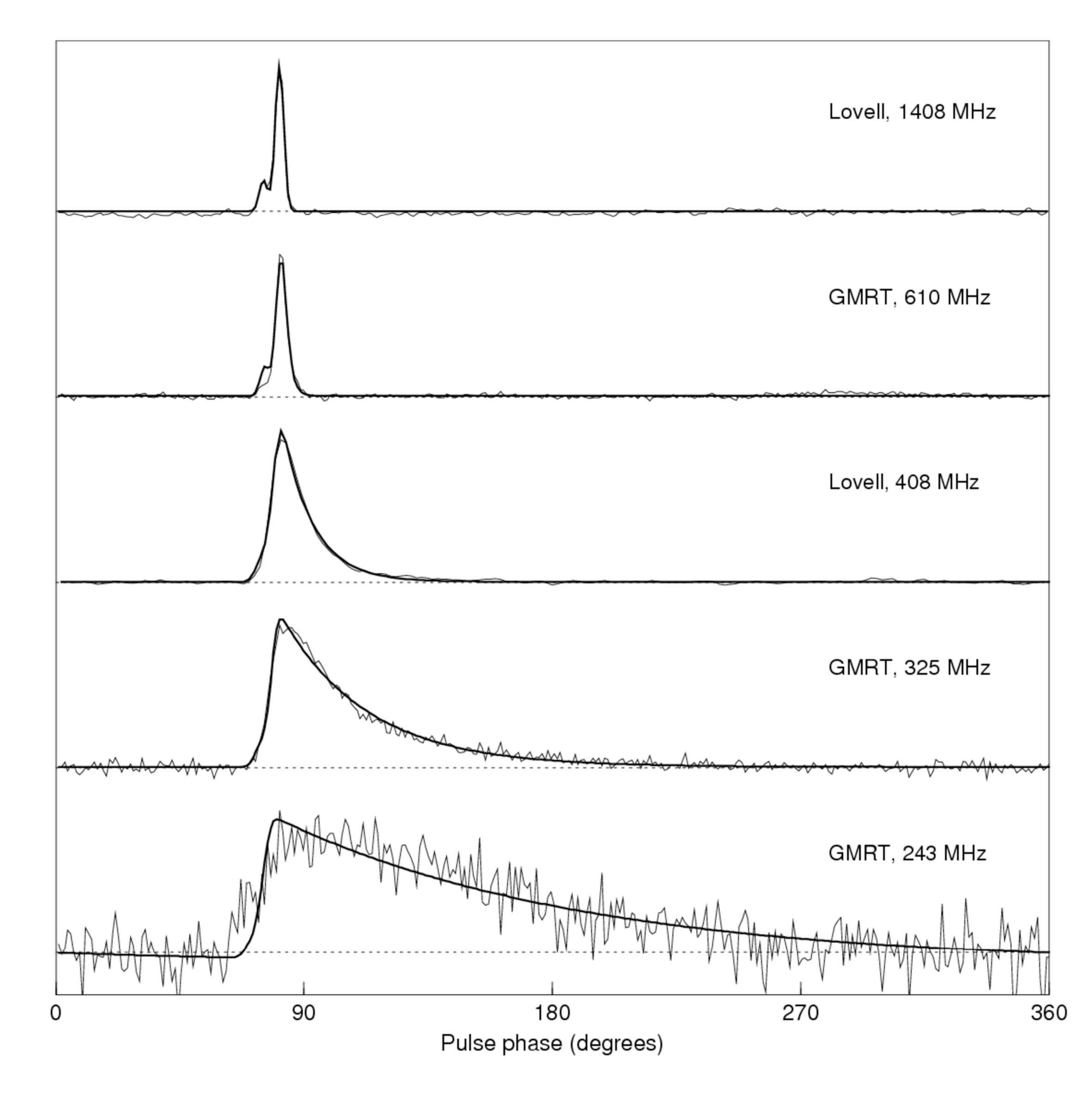}
  \end{center}
  \caption[Pulse scattering]{Pulse profiles for PSR B1831-03 observed at five different frequencies with the Lovell telescope and the GMRT, clearly showing the increasing effect of scattering at lower frequencies. The solid lines show exponential fits to the data. Source: \cite[figure 1.11]{LorimerKramer2005}}
\label{scatteringFigure}
\end{figure}

The electron density in the ISM is not homogeneous but shows variations on a wide range of scales. This results in temporal variations in dispersion measure, which also distort and scatter the pulse shape. Multi-path propagation temporally broaden narrow pulses emitted from a radio source and are observed as a one-sided exponential function with a scattering timescale $\tau_s$. More complicated pulse shapes appear as a convolution of the pulse with the exponential. This effect is depicted in figure \ref{scatteringFigure}, where pulse profiles of PSR B1831-03, observed at five different frequencies with the Lovell telescope and the GMRT, show the increasing effects of scattering at lower frequencies. Several authors have investigated a possible empirical relationship between the scattering timescale and the dispersion measure. We reproduce the empirical fit by \cite{Bhat2004} here:
\begin{equation}
 \label{scatteringRelationship}
 \mbox{log }\tau_s = -\mbox{6.46} + \mbox{0.154 log(DM)} + \mbox{1.07(log(DM))}^2 - \mbox{3.86 log}f
\end{equation}
where $\tau_s$ is in ms and $f$ in GHz. The scattering effect limits the sensitivity of transient surveys, which causes higher DM pulses to be stretched by a substantial amount along the time series, significantly reducing the S/N of detected pulses, especially for low frequency telescopes.

This relationship applies to sources in the Milky Way. When observing extragalactic source $\tau_s$ will be smaller, even for large DM values, although the distribution of the intergalactic medium (IGM) is not yet well understood. Recently \cite{Lorimer2013} rescaled this relationship for extragalactic fast transients, based on measured properties of the bursts discovered by \cite{Lorimer2007}, \cite{Keane2012} and \cite{Thornton2013}, where the leading term (-6.5) is replaced by -9.5. They however state that since only one measurement of the scattering timescale has been made so far, it is likely that this is an upper limit to the average amount of scattering as a function of DM.

It should be noted that the observed $\tau_s$ for a given DM can deviate from the predicted value by up to two orders of magnitude. Following \cite{Hassall2013}, assuming the unscattered pulse is a step function with intrinsic width $\tau_0$ and a peak flux density $S_{v0}$, then its intrinsic influence is $\mathcal{F}=S_{v0\tau 0}$. As the pulse is broadened by scattering, the fluence at a given frequency $F_v$ is given by:
\begin{equation}
 \mathcal{F}_v = S_v \sqrt{\tau_0^2 + \left( \intop_0^\infty e^{-t/\tau_s}\text{d}t \right)^2} = S_v\sqrt{\tau_0^2 + \tau_s^2}
\end{equation}
where $S_v$ is the observed peak flux density at frequency $v$. Assuming that the fluence is conserved by scattering, the peak flux density is given by
\begin{equation}
 S_v = \frac{\mathcal{F}}{\sqrt{\tau_0^2 + \tau_s^2}}
\end{equation}

\subsection{Scintillation}

Relative motion between the source, the scattering medium and the observer leads to the phenomenon of interstellar
scintillation (ISS), which manifests itself as intensity variations on various timescales. This effect is
very similar to what causes stars to twinkle in the night sky. ISS is not taken into consideration in the work presented in this thesis, however interested readers can refer to \cite[section 4.2]{LorimerKramer2005} for a more thorough description of this phenomenon.

\section{Radio Transients}
\label{transientsSection}

The exploration of the transient universe is an exciting and rapidly growing area in radio astronomy. Known transient phenomena range in time scales from sub-nanosecond to years, yielding a rich diversity  in their underlying physical processes and acting as probes to physical and astrophysical phenomena in extreme environments, where their density, temperature, pressure, velocity, gravitational and magnetic fields far surpass the capabilities of any Earth-based laboratory. These transient phenomena are thought to be likely locations of cataclysmic or dynamic events, thus providing enormous potential to uncover a wide range of new physics. 

Most of the objects have been discovered through a limited number of surveys and serendipitous discoveries, meaning that the time domain sky has, up till recent times, been sparsely explored, suggesting that there is much to be found and studied. Over the next decade, the ``new generation of radio telescopes'', with improved sensitivity, wider fields of view, higher survey speed and flexible digital signal processing backends will be able to explore radio transient parameter space more comprehensively and systematically. This parameter space, also referred to as the transient phase space, is extensive. Transients have been detected, and are predicted for, all radio wavelengths, across a wide range of timescales, and may originate from nearly all astrophysical environments including the solar system, star-forming regions, the Galactic centre, and beyond our host galaxy.

\begin{figure}[t!]
 \centering
 \includegraphics[width=400pt]{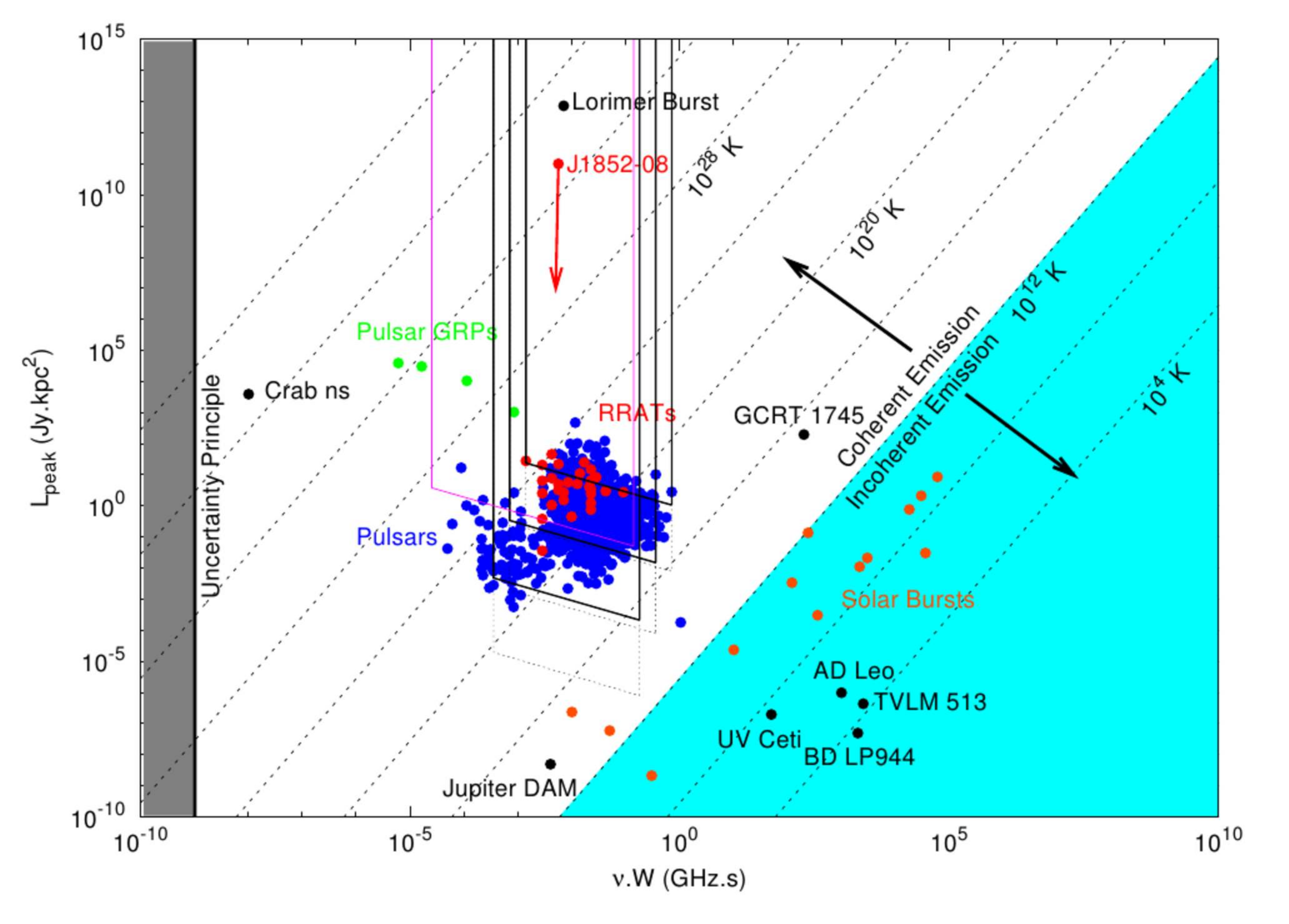}
 \caption[Transient phase space]{The transient 'phase space' with known sources identified, 
          plotting the pseudo-luminosity $L=SD^2$ versus $vW$, where $S$ is flux density, $D$ is distance, $v$ is 
	  observing frequency and $W$ is pulse width, with a representative number of sources
	  plotted, including pulsars \cite{Hobbs2003}, RRATs \cite{Laughlin2006}, pulsar giant radio pulses \cite{Cognard1996,Romani2001}, flare stars \cite{Bastian1994,Richards2003,Osten2008}, auroral radio emission from the Sun and planets \cite{Dulk1985,Zarka1998}, GCRT 1745-3009 \cite{Hyman2006} and the Lorimer burst \cite{Lorimer2007}. Source: \cite{Keane2011}}
  \label{phaseSpaceFigure}
\end{figure}

The transient population is generally divided into two types, although the exact cutoff currently relies on the snapshot time of current radio instruments, which is related to the shortest time at which an image can be generated. Transients can be classified as either ``fast'' or ``slow''. Fast transients have a duration of less than $\sim$1 s and arise from coherent emission. These include pulsar and neutron star phenomena, solar bursts, flare stars and annihilating black holes. Due to the nature of these signals, real-time data processing is a challenge, and severe effects caused by the Inter Stellar Medium (ISM), especially at low frequencies, where the field of view is highest, and large bandwidths, need to be corrected for. Slow transients arise from incoherent (synchrotron) emission, and can be defined as those transients with time scales longer than the time it takes to image the relevant region of the sky.
Known source classes include novae, Active Galactic Nuclei (AGN) and Gamma-ray bursts (GRB). Figure \ref{phaseSpaceFigure} provides an illustration of the transient phase space, as depicted by \cite{Keane2011}, with known sources identified.

\subsection{Classes of Radio Transients}

\begin{figure}[t!]
  \begin{center}
  \includegraphics[width=420pt]{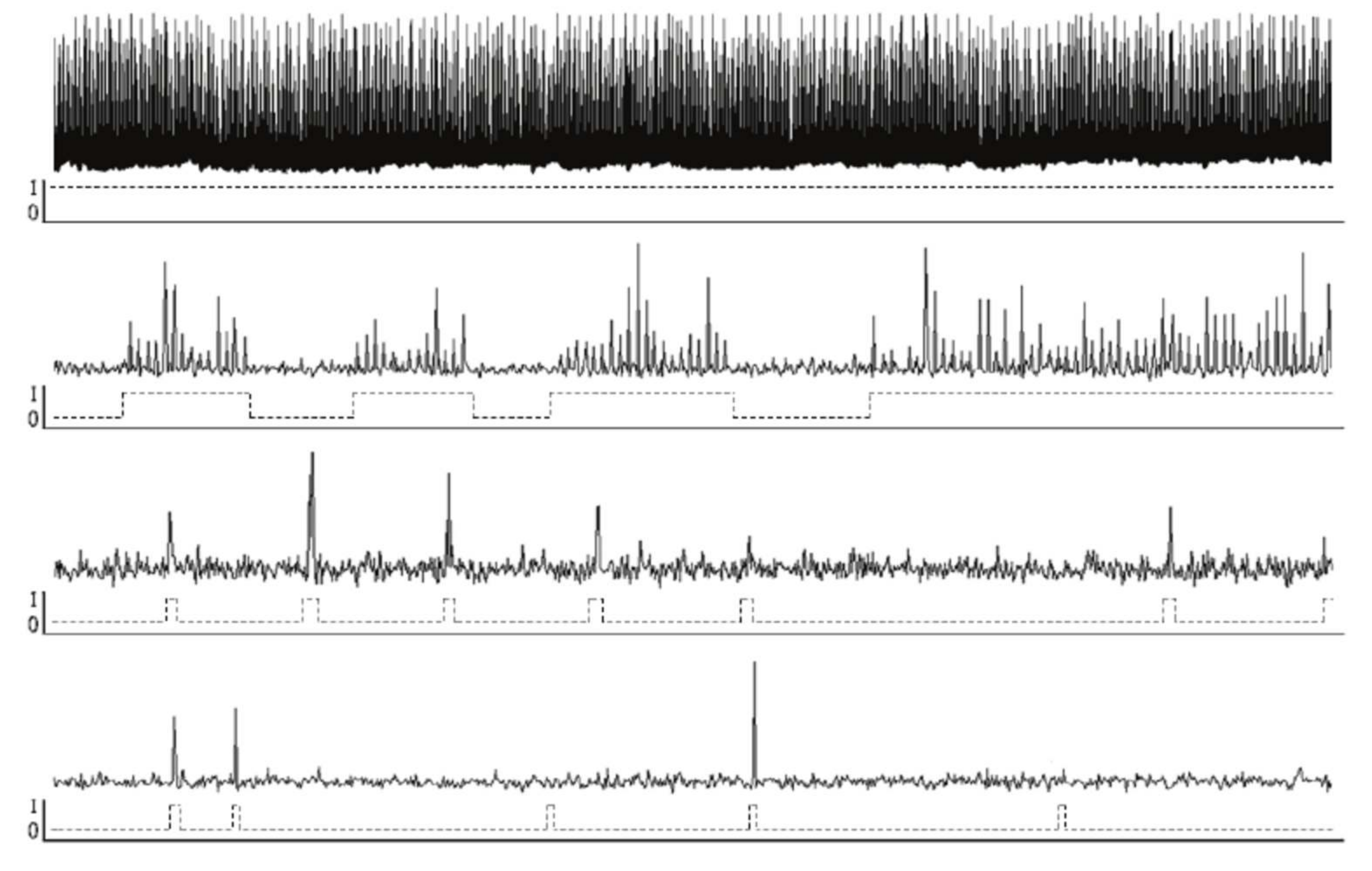}
  \end{center}
  \caption[Classes of rotating neutron stars]{A series of pulsars exhibiting a 
range of emission activity timescales. Top to bottom: Vela, PSRs J1646-6831, 
J1647-36, J1226-32. All panels are of equal duration. The binary scale below 
each timeline shows an estimated representation of the null/emission state. 
Source: \cite[figure 2]{Burke2012}}
\label{rratFigure}
\end{figure}

Classes of transients are diverse, ranging from nearby stars to objects at cosmological distances, and touching upon nearly every aspect of astronomy, astrophysics and astrobiology. These classes include, but are not limited to:

\;(i) {\it Neutron Stars}. Neutron stars are the most populous member of the transient radio phase space, and are thus the most well studied. They are highly compact stars with radii of $\sim$10 km and masses of $\sim$1.4 $M_{\odot}$, with extremely strong gravity and magnetic fields, 10$^{12}$ G being typical (see \cite{Shapiro1983}). Pulsars are those rapidly rotating neutron stars which emit an apparently steady, narrow beam of emission. This emission seems to originate from fixed regions on/above the neutron star surface so that, modulated by the star's rotation. the beams can sweep across our line of sight and be detected at Earth. Thus observers see a pulse of emission per rotation. Rotating Radio Transients (RRATs) \cite{Laughlin2006} are those rapidly rotating neutron stars whose emission does not seem to be steady and they are usually seen to be ``off''. These are generally single, dispersed bursts of duration 2 - 30 ms, repeating with the same DM and having a recurrence time of 4 min - 3 hours. In addition to their pulsed emission, neutron stars can be transient in other respects. There are pulsars known in eclipting binaries, where the orbit plane of a binary pulsar lies so close to the line of sight that the components undergo mutual eclipses, as well as ``nulling'' pulsars, which turn on and off for weeks at a time. Figure \ref{rratFigure} shows a series of pulsars exhibiting a range of emission activity timescales.

\;(ii) {\it Pulsar Giant Pulses}. While all pulsars show pulse-to-pulse intensity variations, some pulsars emit so-called giant pulses, with strengths 100 or even 1000 times the mean pulse intensity. The Crab was the first pulsar found to exhibit this phenomenon, and giant pulses have since been detected from numerous other pulsars, for example  \cite{Cognard1996,Romani2001,Johnston2003}. Pulses with flux densities of order 10$^3$ Jy at 5 GHz and with durations of only 2 ns have been detected from the Crab \cite{Hankins2003}. These ``nano-giant'' pulses imply brightness temperatures of 10$^{38}$ K, which are considered to be amongst the most luminous emissions from any astronomical object. In addition to being probes of particle acceleration in pulsar magnetosphere, giant pulses may serve as probes of the local intergalactic medium \cite{McLaughlin2003}.

\begin{figure}[t]
  \centering 
  \subfloat[Lorimer Burst]{\includegraphics[width=190pt]{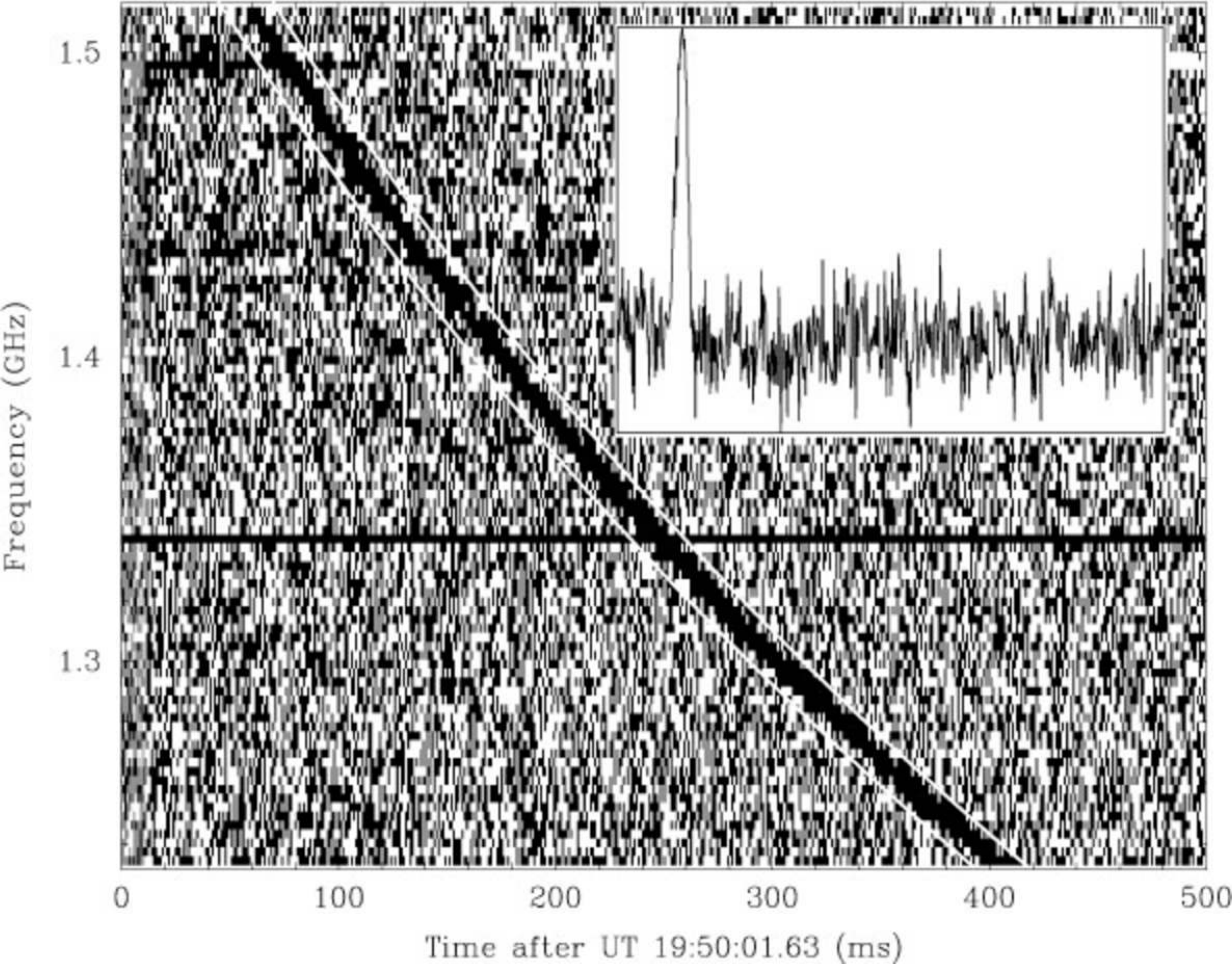}}
  \hspace{8mm}
  \subfloat[J1852-08]{\includegraphics[width=200pt]{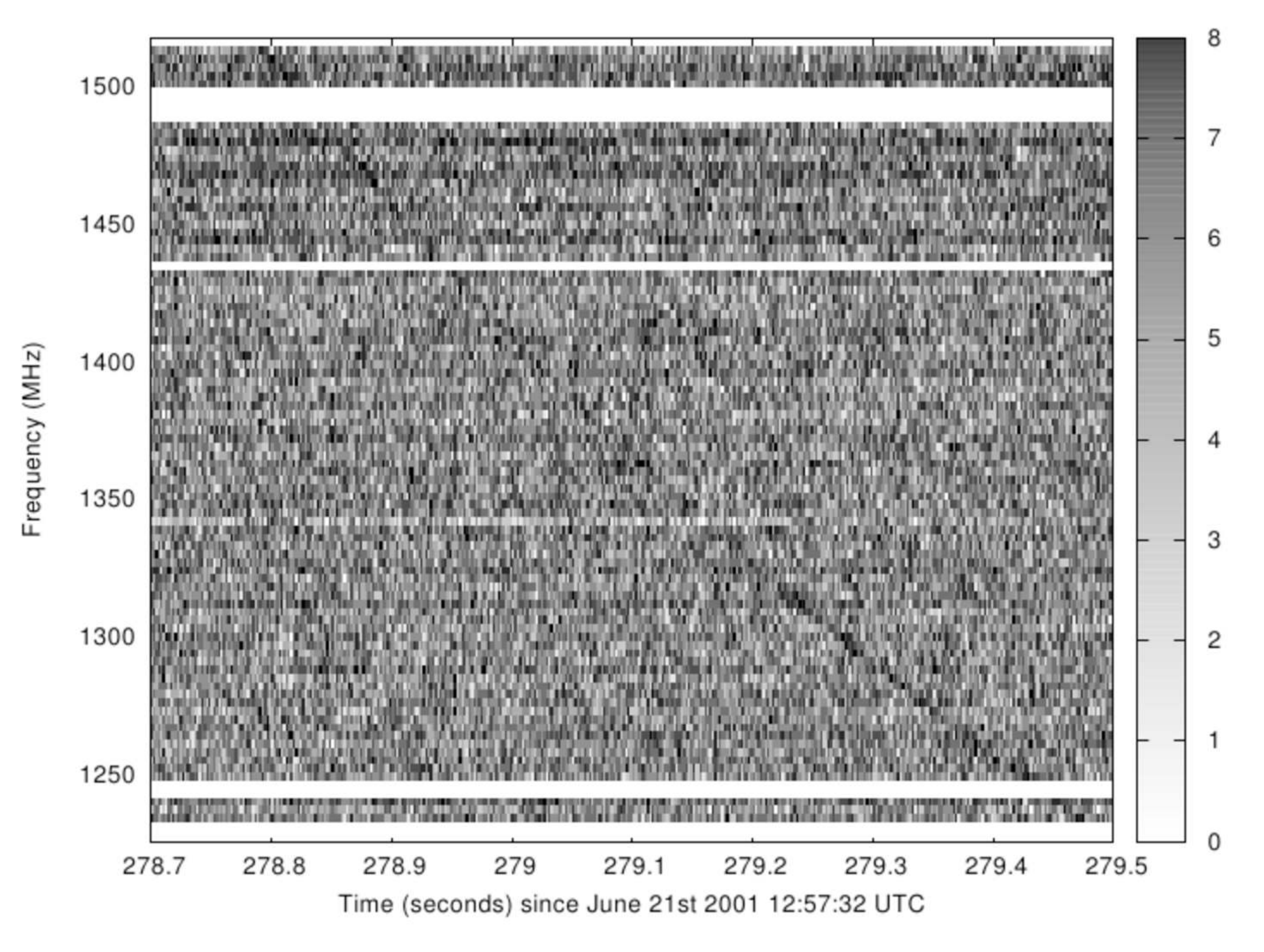}}
  \caption[Candidate extragalactic bursts]{Waterfall plots for two candidate extragalactic bursts.
           (Left) The Lorimer Burst, a 30Jy dedispersed burst of duration less than 
	   5ms located 3$^{\circ}$ south of the centre of the Small Magellanic Cloud (SMC). 
	   It was found by \cite{Lorimer2007} in archival survey data . (Right) A 400 mJy pulse from J1852-08, 
	   discovered in the Parkes Multi-beam Pulsar Survey (PMPS) by \cite{Keane2012} with a DM of 746 
	   cm$^3$ pc and pulse width of 7.8ms.}
  \label{extragalacticBursts}
\end{figure}

\;(iii) {\it Fast Radio Bursts}. In 2007, the detection of a 30 Jy, 5 ms duration, highly dispersed (DM of 375 cm$^{-3}$ pc) burst, detected in 3 independent beams at Parkes, was reported by \cite{Lorimer2007} (figure \ref{extragalacticBursts}a), thought to be of extragalactic origin. A second burst was detected by \cite{Keane2011} (figure \ref{extragalacticBursts}b), with a DM of 745 cm$^{-3}$ pc. Recently, \cite{Thornton2013} reported an additional 4 millisecond-duration radio transients, all more than 40$^\circ$ from the Galactic plane, also of extragalactic origin. Currently no temporally coincident X- or $\gamma$ ray signatures were identified in association with these bursts. These recent discoveries have spurred an increased interest in conducting surveys for highly dispersed, single pulses signals, with recent studies suggesting that the low frequency component of the SKA$_1$ could find an extragalactic burst every hour \cite{Hassall2013}.

\;(iv) {\it Flare Stars, Brown Dwarfs and Extrasolar Planets}. Active stars and star systems have been known to produce radio flares attributed to particle acceleration from magnetic field activity \cite{Gudel2002}. Flares from late-type stars and brown dwarfs have also been discovered \cite{Berger2001,Hallinan2007}, in some cases with periodicities indicative of rotation. Finally, Jupiter is bright below 40 MHz, and many stars with ``hot Jupiters'' show signatures of magnetic star-planet interactions \cite{Shkolnik2005}, suggesting that extrasolar planets may also be radio sources \cite{Zarka2007}, indicating that searches for exoplanets can be conducted with next generation radio telescopes.

\;(v) {\it Radio Supernovae and GRBs}. Frequenct observation of a large area of 
sky, as made possible by recent and future radio telescopes, can be used to 
find GRBs and supernovae that emit in the radio regime, as well as to follow up 
on such transients detected at other wavelengths. Multi-wavelength, multi-epoch 
observations (eg \cite{Cenko2006}) can provide information on progenitors, the 
surrounding medium and models of GRB energetics and beaming.

\;(vii) {\it Annihilating Black Holes}. Annihilating black holes are predicted to produce radio bursts \cite{Rees1977}. Advances in $\gamma$-ray detectors has renewed interest in possible high-energy signatures from primordial black holes \cite{Dingus2002,Linton2006}. Observations at the extremes of the electromagnetic spectrum are complementary as radio observations attempt to detect the pulse from an individual black hole, while high-energy observations generally search for the integrated emission.

\;(viii) {\it Gravitational wave events.} The progenitors for gravitational waves may generate associated electromagnetic signals or pulses. For example, the in-spiral of a binary neutron star system may produce electromagnetic pulses, both at high energies and in the radio due to the interaction of the magnetosphere of the neutron stars \cite{Hansen2001}. More generally, the combined detections of both electromagnetic and gravitational wave signals may be required to produce localisations and understanding of the gravitational wave emitters \cite{Bloom2009}.

\;(ix) {\it Extraterrestrial transmitters}. While none are known, searches for extraterrestial intelligence (SETI) have found non-repeating signals that are otherwise consistent with expected signals from an ET transmitter. \cite{Cordes1997} show how ET signals could appear transient, even if intrinsically steady. More recently, several searches were conducted towards Kepler's field of view, which to date has confirmed 151 planets, with more than 3,000 candidates still being analysed.

\section{Transient Detection Metrics}

\begin{figure}[t!]
  \begin{center}
  \includegraphics[width=430pt]{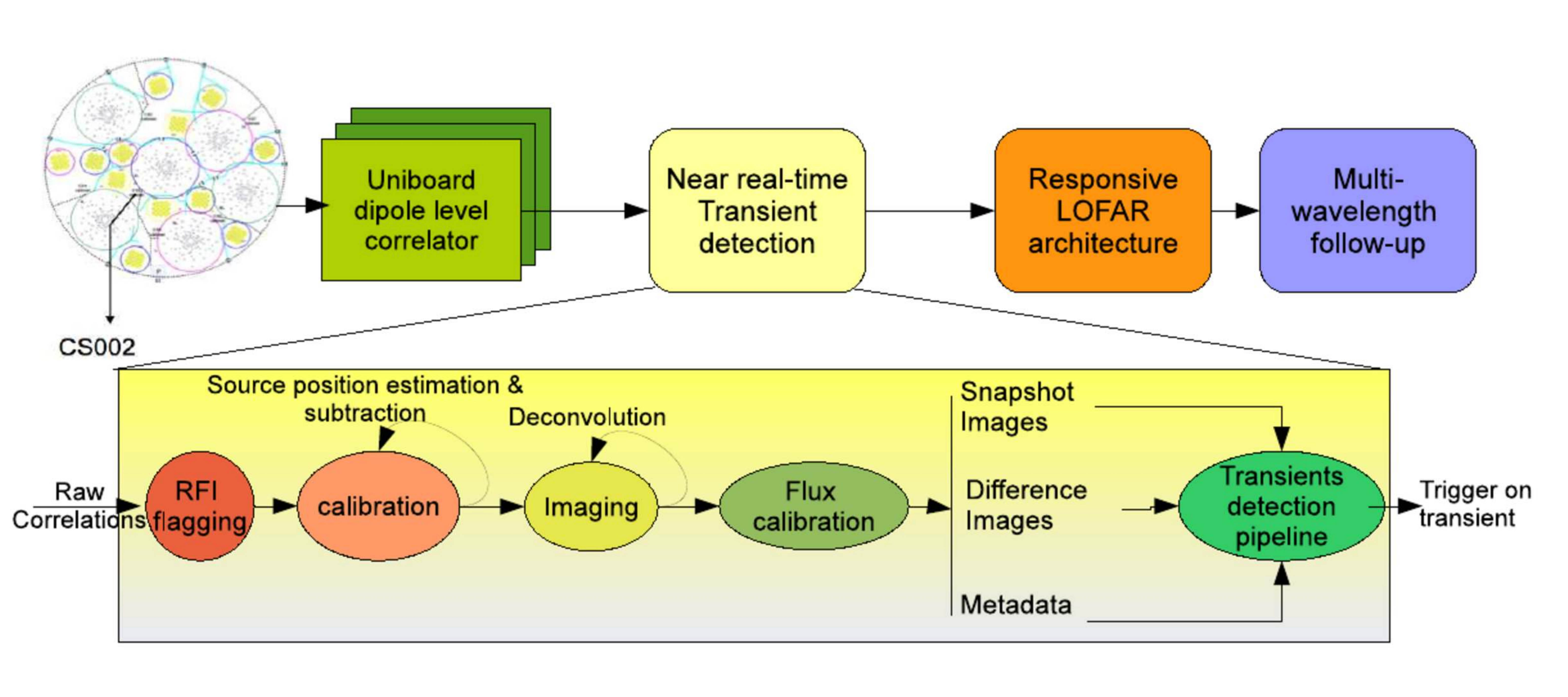}
  \end{center}
  \caption[AARTFAAC pipeline]{The main components of the AARTFAAC ASM. The correlator outputs are first passed through an RFI excision stage which generates appropriate RFI masks. The correlations are then calibrated and imaged with low latency, with the output images passing through a flux calibration stage and then through the Transients Pipeline. This carries out source extraction, source association, and the generation of light curves from existing observations for transient detection. Source: \cite[figure 2]{Prasad2012}.}
\label{aartfaacFigure}
\end{figure}

Phased-arrays with very wide fields of view can essentially image the entire sky 
in a time $T_c$, which is the fastest possible correlator dump time. Figure 
\ref{aartfaacFigure} shows a schematic overview of the Amsterdam-ASTRON Radio 
Transient Facility And Analysis Center (AARTFAAC) \cite{Prasad2012}, which aims 
to implement a near real-time, 24x7 All-Sky Monitor (ASM) for LOFAR which will 
be capable of monitoring low frequency radio transients over most of the sky 
locally visible to LOFAR, at timescales ranging from seconds to several days. 
Developments in millisecond imaging, such as \cite{Law2011}, are constantly 
reducing the snapshot time for these type of surveys. 

Fast transients are those which cannot be well sampled through imaging surveys, 
unless they are very frequent and there is tolerance for a low completeness 
coefficient. Rare, fast transients are better sampled through staring 
observations of large solid angles. We follow \cite{Cordes2009} in defining a 
survey metric based on the amount of volume surveyed by a radio instrument. 
Surveys that involve time-variable sources need to take into account the event 
rates and durations of transient sources, as well as sensitivity requirements. 
Slow transients are those for which the sky may be sampled by raster scanning, 
where the telescope beam with beam size $\Omega_i$ scans a patch of sky of total 
solid angle $\Omega_s$ in a time $T_s$. This scan is then repeated for the 
duration of the survey. The dwell time per sky position, that is the time spent 
pointing to the same patch of sky, is $\tau$ = $(\Omega_i/\Omega_s)T_s$.

For blind searching, the rate of sky coverage $\dot{\Omega}$ (deg$^2$ s$^{-1}$) needs to be maximised, while also achieving the desired search depth, characterised by the maximum detection distance $D_{\text{max}}$. The yield of a survey, or the number of events detected per unit time, involves the product of source number density $n_s$ and search volume $V_{\text{max}}=\frac{1}{3}\Omega_sD^2_{\text{max}}$. If the survey of the solid angle $\Omega_s$ is conducted in a time $T_s$, the resulting search volume yields a combination of parameters similar to that obtained by calculating the survey speed (SS), which is $\Omega_i/\tau$. These two approaches lead to the figure of merit for steady sources (FoMSS), which can be defined as (see \cite[appendix a]{Cordes2009}):
\begin{equation}
 \mbox{FoMSS}=B\left( \frac{N_{\text{FoV}}\Omega_{FoV}}{N_{\text{sa}}} \right) \left( \frac{f_cA_{\text{eff}}}{mT_{\text{sys}}} \right)^2
\end{equation}
where $B$ is the bandwidth, $N_{\text{FoV}}$ is the number of fields of view (or pixels) for each antenna, $\Omega_{\text{FoV}}$ is the solid angle for each FoV, $N_{\text{sa}}$ is the number of subarrays into which the array is divided, which are assumed to be equal in size and pointed in non-overlapping directions, $f_c$ is the fraction of the total effective area $A_\text{eff}$ usable in the survey, $m$ is the threshold S/N in the survey for which an event is considered to be detected and $T_{\text{sys}}$ is the system temperature. 

FoMSS is relevant to surveys of sources that are homogeneously distributed within the spatial domain, that are steady, standard candles. A more general metric which applies to transients is defined by \cite{Cordes2009}:
\begin{equation}
  \mbox{FoMTS}  =  \mbox{FoMSS} \times \mathcal{K}(\eta W, \tau / W)
\end{equation}
where $\mathcal{K}(\eta W, \tau / W)$ accounts for the fact that sources with burst time $W$ may only be on for a fraction of the dwell time $\tau$, and the burst event time follows a Poisson distribution with rate $\eta$. 

\subsection{Transient Event Rates}
\label{fomSection}

The detection rate for fast transients has also been well studied by 
\cite{Macquart2010} and \cite{Colegate2011}. Here we present a brief overview of 
their results, the interested reader can refer to the original documents. 

The main goal of transient surveys is to maximise the number of events detected, which primarily depends on the search strategy employed on the telescope. Given each search strategy has a different processing cost, we can define a new FoM, the event rate per unit cost $\mathcal{R}_{\text{cost}^{-1}}$, which can be a more comprehensive FoM than survey speed. Since signal and search processing costs are architecture specific, the ``event rate per beamformed and searched'', referred to as $\mathcal{R}_{\text{beam}^{-1}}$, can be used to generalise the problem, and is based on the rate of transient events detectable in a volume of sky. It is assumed that $\mathcal{R}_{\text{cost}^{-1}} \propto \mathcal{R}_{\text{beam}^{-1}}$, where cost increases linearly with number of beams processed.

For fast transient searches, one volume of sky is considered as likely as another to contain transient events, and each time a volume is revisited there is new detection potential. This enables the deployment of digital backends for commensal surveys (sometimes called ``piggy-back'' surveys), where the survey does not point the telescope beams itself but rather branches off the data stream generated by other observations from the signal processing chain and performs a transient search on it. Based on \cite{Macquart2010}, \cite{Colegate2011} define the event rate as 
\begin{equation}
 \mathcal{R}_v = \rho \frac{\Omega_p}{4\pi}V_{\text{max}}\; \mbox{events s}^{-1}
\end{equation}
where $V_{\text{max}}$ is the maximum volume out to which an object is detectable and $\Omega_p$ is the processed
FoV, which is the product of the number of beams formed ($N_{\text{beam}}$) and the FoV of each beam. Thus the event
rate per beam is
\begin{equation}
 \mathcal{R}_{\text{beam}^{-1}} = \frac{\mathcal{R}_v}{N_{\text{beam}}}
\end{equation}
An extragalactic survey is described as a search for a homogeneously distributed population of isotropically emitting
fast transients of fixed intrinsic luminosity. For such a population, the event rate is
\begin{equation}
\mathcal{R}_v = \frac{1}{3}\rho\Omega_p\left(\frac{W_i}{W} \right)^{\frac{3}{4}} \left( \frac{\mathcal{L}_i}{4\pi S_{min}} \right)^{\frac{3}{2}}\; \mbox{events s}^{-1}
\label{eventRateEquation}
\end{equation}
where $\rho$ (events s$^{-1}$ pc$^{-3}$) is the event rate density, $\mathcal{L}_i$ (Jy pc$^2$) is the intrinsic
luminosity of the population, $W_i$ is the intrinsic width, $W$ is the observed pulse width and $S_{\text{min}}$ is the minimum detectable flux density of the telescope, defined in equation \ref{sminEquation}, with integration time $\tau=W_i$. The $(W_i/W)^{\nicefrac{3}{4}}$ term approximates the loss in S/N due to pulse broadening.

\subsection{Signal Combination Modes}
\label{signalCombinationSection}

Incoherent addition of an array of $N_0$ elements increases the sensitivity by a 
factor of $\sqrt{N_0}$ over a single element while retaining its full FoV. 
Forming $N_{b-0}$ station beams linearly increases the FoV. To even further 
increase the FoV, $N_{\text{sa}}$ subarrays can be incoherently combined, where 
subarrays are pointed in a different directions, increasing the FoV by a factor 
$N_{\text{sa}}$, but only increasing the sensitivity by a factor 
$\sqrt{N_{0/\text{sa}}}$ over a single element, where $N_{0/\text{sa}}$ is the 
number of elements per subarray. Fly's eye pertains to the case where 
$N_{0/\text{sa}} = 1$.

\begin{figure}[t!]
  \begin{center}
  \includegraphics[width=430pt]{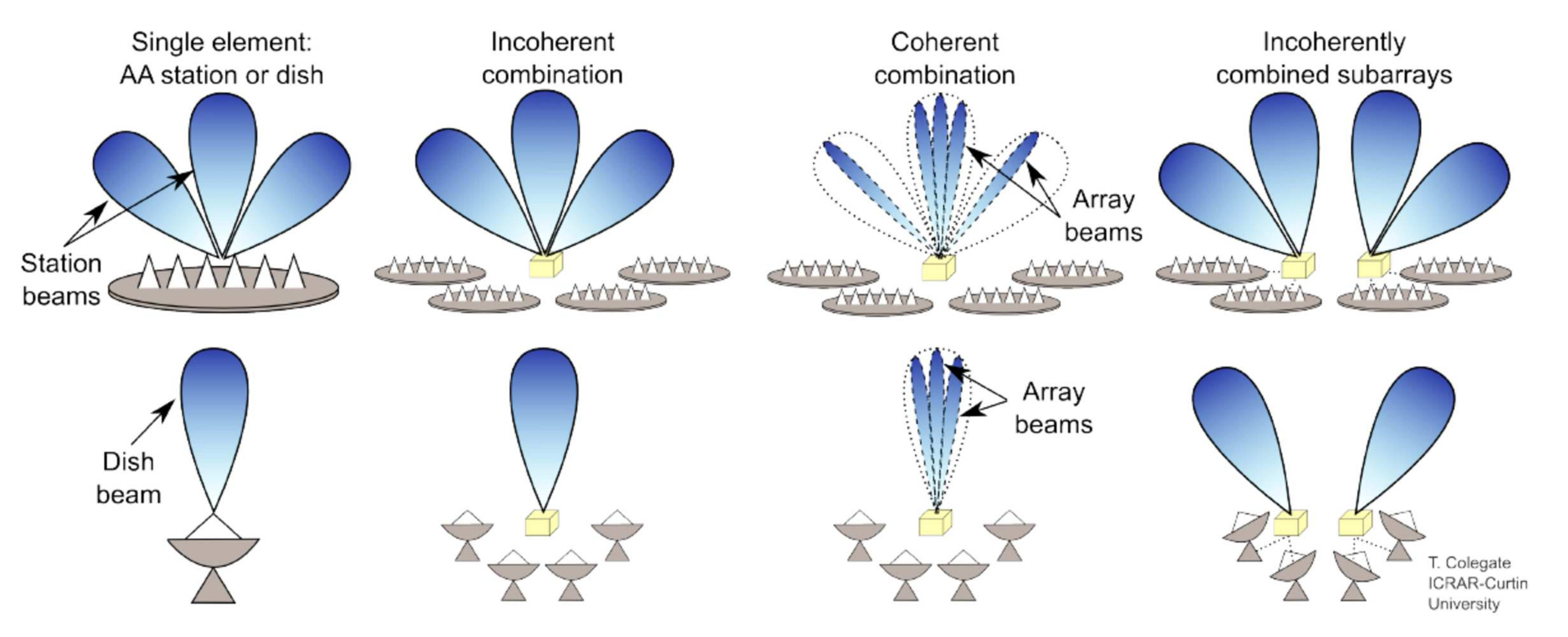}
  \end{center}
  \caption[Signal combination modes]{Signal combination modes, resultant beam patterns and beam terminology for dishes and aperture array. Beams sizes are not to scale. Source: \cite[figure 2]{Colegate2011}}
\label{signalCombinationFigure}
\end{figure}

Sensitivity of the coherent combination of an array of $N_0$ elements is higher than incoherent combination and subarraying as it increases proportional to $N_0$. However the FoV of the array beam, $\Omega_{\text{arr}}$, is much smaller and is proportional to $D_{\text{arr}}^{-2}$, where $D_{\text{arr}}$ is the diameter of the array of elements being combined. The FoV can be linearly increased by forming $N_{b-\text{arr}}$ array beams. Applying these relationships to equation \ref{eventRateEquation} gives the total event rate for each signal combination mode (see \cite[section 4]{Colegate2011}):
\begin{eqnarray}
 \mathcal{R}_v & = & \frac{1}{3}\rho\left(\frac{W_iN_p\Delta v \tau}{W} \right)^{\frac{3}{4}} \left( \frac{\mathcal{L}_i A_{\text{eff}}}{4\pi \sigma 2 k_B T_{\text{sys}}} \right)^{\frac{3}{2}} \mathcal{M} \\
 \mathcal{M}   & = & \begin{cases}
                      N_{b-0} \Omega_0 N_0^{\frac{3}{4}} & \mbox{Incoherent combination} \\
                      N_{b-\text{arr}} \Omega_{\text{arr}} N_0^{\frac{3}{2}} & \mbox{Coherent combination} \\
                      N_{\text{sa}}^{\frac{1}{4}} N_{b-0} \Omega_0 N_0^{\frac{3}{4}} & \mbox{Subarraying}
                     \end{cases}
\end{eqnarray}
where $N_p$ is the number of polarisations summed, $\Delta v$ is the processed bandwidth, $\tau$ is the post-detection integration time and $\sigma$ is the S/N ratio required for event detection. The coherent combination mode is more effective if the elements are closely spaced, thus having a higher filling factor, resulting in a larger array beam FoV. Figure \ref{signalCombinationFigure} shows a schematic illustration of the various combination modes mentioned in this section.

\section{Single Pulse Searches for Fast Transients}
\label{fastPipelineSection}

\begin{figure}[t!]
  \begin{center}
  \includegraphics[width=400pt]{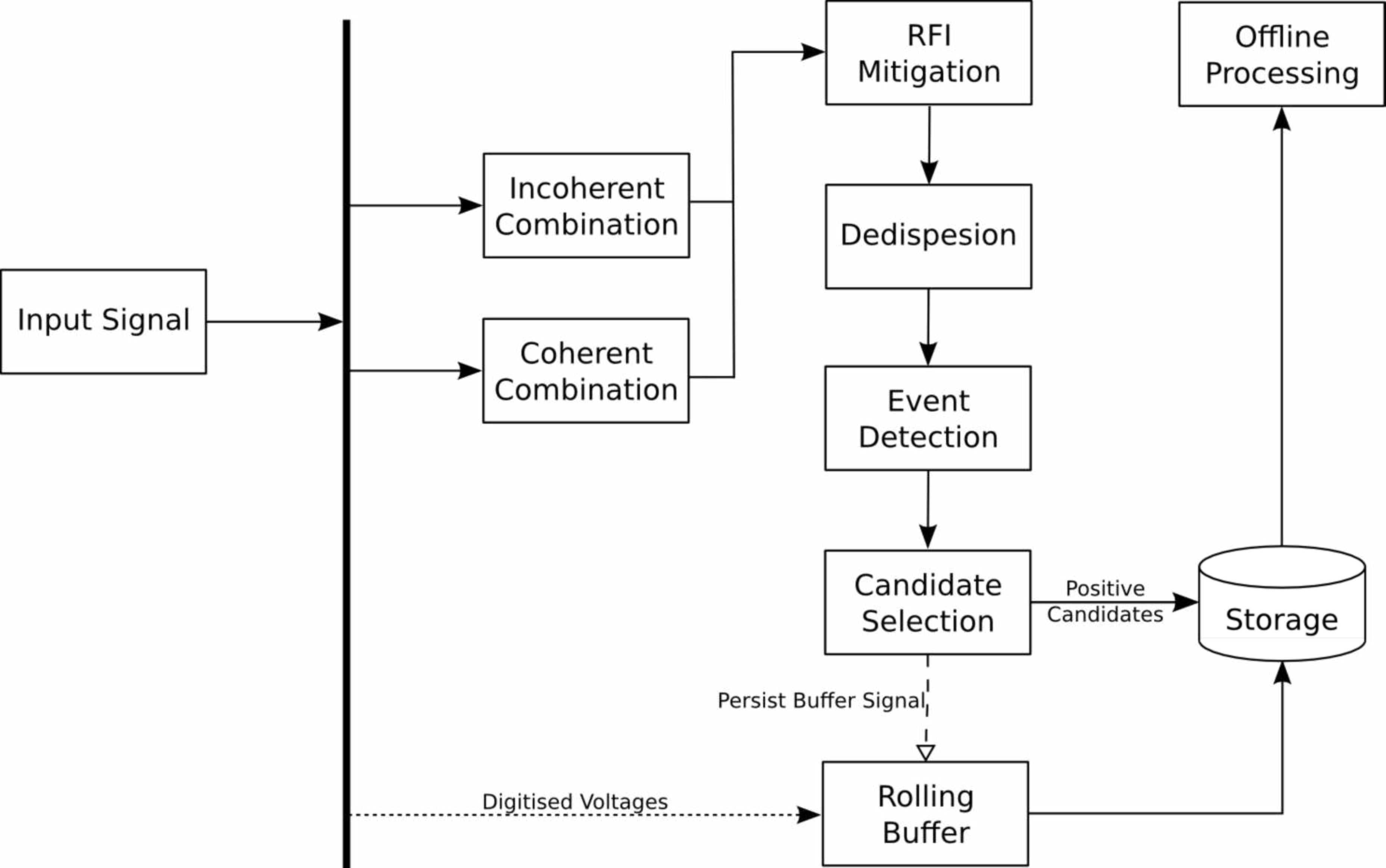}
  \end{center}
  \caption[Generic fast transient detection pipeline]{Generic fast transient detection pipeline.}
\label{transietPipelineFigure}
\end{figure}

Since the discovery of RRATs, interest in single radio pulse searches has increased dramatically. Single pulse detection extensions have been incorporated into current pulsar surveys, and several archival pulsar data have been reanalysed for detecting such events. Due to the large data volumes generated by these searches, especially in planned surveys for future radio telescopes, too large to store cost-effectively, such searches have to be conducted in real-time, on a data stream which is a continuous observation of the sky. Such systems are referred to as pipelines, whereby the input antenna voltages pass through a chain of processing elements, each performing a specific task.
In order to be able to further process positive candidate events offline, after the observation has been conducted, a rolling buffer is required, which stores a small period of the data as it is being observed. This buffer is sometimes referred to as a transient buffer board (TBB). The specific implementation of such a pipeline depends on the target or expected source population, as well as the predicted cost and performance factors of components in the processing pipeline.  Figure \ref{transietPipelineFigure} shows a generic fast transients pipeline, the components of which are described below.

{\it (i) Signal reception}: Radio signals collected by radio dishes or aperture 
arrays are digitised and subsequently quantised, to reduce the required 
transmission data rate. This processing can be performed on a digital backend or 
conventional compute cluster. Data transfer between this backend and hardware 
running the transients pipeline generally occurs via high-throughput network 
links, at which point the packetised data are decoded, assembled into 
processable buffers and prepared for analysis.

{\it (ii) Signal Combination}: Signals from dish or station beams can be combined coherently, incoherently  or not at all, as discussed in the previous section. Incoherent (phase insensitive) combination sums the detected signals from antennas pointing in the same direction. Coherent combination forms phased or tied array beams, where voltages measured at each antenna are phase aligned towards a specific direction in the sky. Smaller groups of antennas (subarrays) can be incoherently combined and each subarray pointed in a different direction.

{\it (iii) RFI Mitigation}: Radio Frequency Interference (RFI) mitigation is one of the major problems in transient surveys, especially for radio telescope which are close to urban centres. Discriminating between astronomical and terrestrial signals is a challenge, and several techniques have been investigated for this purpose. This step is typically performed prior to dedispersion, through clipping and thresholding algorithms, or in the case where multiple beams are being processed, during event detection with coincidence matching techniques.

{\it (iv) Dedispersion}: As discussed in section \ref{ISMSection}, astronomical signals pass through a cosmic medium of unknown dispersion measure, therefore detection needs to be trialled for many DMs. This is generally the costliest step in a single-pulse transient pipeline, computationally (for pulsar surveys, the dedispersion cost can be dwarfed by searches for binary systems). The DM range to be trialled depends on several factors, including the sky location being observed, the observation parameters, as well as the available computational resources.

{\it (v) Event Detection}: An event detection algorithm needs to be applied to each dedispersed time series. This generally takes the form of thresholding and matched filtering techniques, and is discussed in detail by \cite{Cordes2003}. This step can generate a large number of detections, especially in observations which are dominated by terrestrial RFI.

{\it (vi) Candidate Selection}: The large number of detections from the event detection stage need to be grouped together, such that detections caused by the same event can be processed collectively. These clusters of detections can then be classified as either caused by RFI or by an astrophysical event. Several techniques can be employed to perform this classification, including clustering, patter-matching and machine learning techniques, and is an area of ongoing research.

{\it (vi) Storage}: The rolling buffer stores a small period of data, and on receipt of a trigger from the candidate selection stage, will dump the voltage data to storage for offline processing, which could include RFI filtering information, analysis of the candidate detection stage as well as a correlation of the dish or station beams for source localisation and imaging.

\subsection{Single Pulse Searches S/N}
\label{singlePulseSection}
After correcting for dispersion, each dedispersed time series must be searched individually for pulses with amplitudes above some S/N threshold. This search is generally performed using matched filtering techniques. Given a time series of predominantly Gaussian noise of known mean and standard deviation, these techniques search for individual events that deviate by several standard deviations from the mean. Considering a rectangular pulse of amplitude $S_{\text{peak}}$ and width $W$, and for the optimal case when $W$ is equal to the sampling time $t_{\text{samp}}$, \cite{Cordes2003} show that the S/N ratio of the pulse
\begin{equation}
\label{pulseSNREquation}
 S/N = \frac{S_{\text{peak}}W}{S_{\text{sys}}}\sqrt{\frac{N_p\Delta v}{W}}
\end{equation}
where $S_{\text{sys}}$ is the system equivalent flux density (SEFD), $N_p$ is the number of polarisations summed
and $\Delta v$ is the receiver bandwidth. For a fixed pulse area it follows that S/N $\propto1/\sqrt{W}$, 
thus narrower pulses are easier to detect than broader ones, however, a low-amplitude, broad pulse is more easily detectable than a sharp narrow pulse if its area is sufficiently large. In general, $W$ will not usually be a good match to $t_{\text{samp}}$ and the S/N will be less than expected from equation \ref{pulseSNREquation}. In order to match optimal detections more closely the time series is smoothed by successively adding groups of adjacent samples and searching for statistically significant events. If the true pulse shape and width are unknown, the smoothing approach is a straight-forward and efficient approximation to optimal detection.

A sensible choice for S/N threshold should be made to avoid recording too many candidate pulses that most likely
are caused by random noise fluctuations. For the ideal case of Gaussian noise with zero mean and unit standard
deviation, \cite{Cordes2003} show that the number of events expected to occur by chance above some threshold
$S/N_{\text{min}}$ is
\begin{equation}
 n\left(>S/N_{\text{min}} \right) \sim 2n_{\text{samp}}\int^\infty_{\text{SNR}_{\text{min}}}{\text{exp}(-x^2)dx}
\end{equation}
where $n_{\text{samp}}$ is the number of samples in the time series. Requiring that n $<$ 3 usually leads to S/N$_{\text{min}}$ = 4. In practice, however, radio frequency interference (RFI) usually increases the number of false detections so that a more practical S/N threshold is 5-6 \cite{LorimerKramer2005}. 

\subsection{Transient Surveys}

Despite the scientific potential, the transient and time variable sky is a relatively unexplored region of parameter space. The restricted survey speeds of older generation radio facilities has meant that detections of radio transients lag far behind that which is possible with, for example, X- and $\gamma$-ray instruments. Four dominant methods have so far been used to detect radio transients: (i) dedicated surveys, (ii) multi-wavelength triggered detections, (iii) archival studies and (iv) serendipitous and commensal detections.

\begin{table}[t!]
  \centering
  \footnotesize
  \begin{tabular}{ | l  c  c  c  c  c  c  |}
    \hline
     \multirow{2}{*}{Experiment$^a$} & Telescope   & $v_{\text{centre}}$ & $\Delta v$ & $B_{\text{max}}$ & $\mathcal{R}_{\text{beam}^{-1}}$ & Max. beams \\
               & and status  & (MHz)        & (MHz)      & (km)$^b$          & (normalised)$^c$              & available  \\ 
    \hline
    \hline
    Archival searches$^d$ & Parkes & N/A & - & - & - & - \\
    \hline
    Fly's eye radio  & ATA & 1420 &  210 & N/A & 10$^{-3}$ (fly) & 42 \\
    transient search$^e$                      & completed & & & & & \\
    \hline
    \multirow{2}{*}{HTRU Survey}$^f$ & Parkes  & 1352 & 340 & N/A & 10$^{-2}$ & 13 \\
                     & (operational)   & & & & & \\
    \hline
     \multirow{2}{*}{PALFA Survey$^g$} & Arecibo  & 1440 & 100 & N/A & 10$^{-2}$ & 7 \\
                               & (operational)  & & & & & \\
    \hline
    \multirow{2}{*}{V-FASTR$^h$} & VLBA & 1400 & 64 & 6000 & 10$^{-2}$ (inc.) & 1 \\
           & (operational) & & & & & \\ 
    \hline
    LOFAR & LOFAR & 120 & 32 & $<$100 & 10$^{-1}$ (inc.) & 1$^*$ \\
    Transients KSP$^i$      & (in progress) & & & & 10$^{-4}$ (coh.) & thousands$^*$ \\
    \hline
     \multirow{2}{*}{CRAFT Survey$^j$}  & ASKAP     & 1400 & 300 & 6 & 10$^{-2}$ (inc.) & 36 \\
                  & (planned) &      &     &   & 10$^{-6}$ (coh.) & N/A \\
    \hline
     \multirow{2}{*}{SKA$_1$ AA-low}            & & 260 & 380 & 200 & 1 (inc.) & hundreds$^*$ \\
                            & &     &     &     & 10$^{-1}$ (coh) & thousands$^*$ \\
    \hline
     \multirow{2}{*}{SKA$_1$ LB dishes}   & & 725 & 550 & 200 & 1 (inc.) & 1 \\
                            & &     &     &     & 10$^{-2}$  (coh.) & thousands$^*$ \\
    \hline
  \end{tabular}

  \caption[Transient Surveys]{Radio searches of the high time resolution universe. Table reproduced and adapted from \cite[table 1]{Colegate2011}
  \newline $^a$ Only experiments within SKA1 frequencies (70 MHz - 3 GHz) are listed, and surveys which are insensitive to single pulses are excluded. N/A is not applicable or information not available
  \newline $^d$ Maximum baseline, for event localisation and triggering
  \newline $^c$ Order of magnitude estimation, normalised to the incoherent combination of SKA1 low band dishes. For radio telescope arrays, the calculation is for fly's eye (fly), incoherent combination (inc.) or coherent combination (coh.). A flat spectrum and no scatter broadening is assumed.
  \newline $^d$ \cite{Lorimer2007,Burke2010,Keane2011}
  \newline $^e$ \cite{Siemion2012}
  \newline $^f$ High Time Resolution Universe Pulsar Survey \cite{Keith2010}
  \newline $^g$ Pulsar Arecibo L-Band Feed Array (ALFA) Survey. \cite{Cordes2006,Deneva2009}
  \newline $^h$ \cite{Wayth2011}
  \newline $^i$ Commensal Real-Time ASKAP (CRAFT) Survey. \cite{Hessels2008,Leeuwen2010}
  \newline $^j$ \cite{Macquart2010}
  \newline $^*$ Limited by the available beamformer processing and data transport
  }
  \label{surveysTable}
\end{table}

Major facilities are currently under construction, and older facilities are undergoing dramatic upgrades, driven in large part by the desire to achieve the full Square Kilometre Array (SKA) in the next decade (\cite{Carilli2004}). \cite{Colegate2011} list some current and ongoing radio searches of the high time resolution universe, together with the event rate per beam, using the figures of merit discussed in section \ref{fomSection}. The event rate per beam have been normalised to the incoherent combination mode of SKA$_1$ low band dishes, based on the specification listed in \cite{Dewdney2010}. Their table is reproduced in table \ref{surveysTable}.

Several large collaborations are conducting or planning surveys for fast radio 
transients. This includes groups using LOFAR 
\cite{Stappers2011,Fender2012,Hessels2008}, the VLBA: V-FASTR \cite{Wayth2011}; 
Parkes: High Time Resolution Universe Survey \cite{Keith2010}; ASKAP: CRAFT 
\cite{Macquart2010}; MeerKAT: TRAPUM (PIs: Stappers and Kramer), Arecibo: PALFA 
\cite{Cordes2006,Deneva2009} and the GBT: 
GBNCC (PI: Scott Ransom). Although highly sensitive in 
their own right, these surveys are envisaged as precursors to those that will 
ultimately be conducted with the SKA \cite{Cordes2009}. A crucial element to the 
detection of rare events is the large FoV afforded by the new technologies 
employed by these surveys. The Parkes telescope utilises multibeam technology, 
the MWA and LOFAR use aperture array technology which can, in principle, detect 
objects over a large fraction of the visible sky, while APERITIF and ASKAP 
employ focal plane aperture array technology to achieve a FoV of 8 and 
30$^{\circ}$ respectively. 

\section{Thesis Outline}

This thesis works towards building a real-time, GPU-based, non-imaging backend for radio telescopes, emphasising
on online fast transient discovery. This poses several challenges due to the considerable amount of computational resources required and high data rates involved. We propose GPUs as ideal candidates to tackle various aspects of the 
several stages within this pipeline, and present a viable prototype which we demonstrate to be scalable to large-N,
wide bandwidth telescopes, and propose a tentative architecture for several components of SKA$_1$.

Chapter \ref{dispersionChapter} starts with a brief overview of many-core technologies, with special emphasis on GPUs. A generic GPU framework is then presented, which parallelises input, processing and output stages across
multiple GPUs and CPUs. The challenges posed by dedispersion are then discussed, and a GPU implementation for incoherent (direct method) as well as coherent dedispersion is presented, together with performance and speedup benchmarks. 

In chapter \ref{pipelineChapter} we give a brief overview on the current state of the art transient detection pipelines, and present our own. The overall software architecture is first discussed, after which
we describe in detail the various processing stages, including: bandpass correction, channel thresholding, spectrum 
thresholding, dedispersion, median filtering, detrending, candidate selection through clustering and cluster
parametrisation, as well as quantisation and data persistence. We then present various figures of merit
for some of these stages, particularly the classification stage. We also discuss several performance
benchmarks which were performed on the pipeline, and compare this to similar state of the art pipelines.

In chapter \ref{medicinaChapter} we introduce the BEST-II SKA pathfinder and then describe the digital backend
deployed by a team based in Oxford, to which we attached our transient detection pipeline. Real-time 
additions to the pipeline are discussed, and then we move on to list several test observation data products, 
including observations of PSR B0329+54.

In chapter \ref{beamformingChapter} we discuss the applicability of GPUs for beamforming aperture arrays and 
describe our multi-beam coherent beamformer. Several performance benchmarks are then presented, highlighting the 
major bottleneck for such implementations. Real-time benchmarks are also presented, aimed to simulate the BEST-II
digital backend and highlight a use case where transient detection pipelines and beamforming kernels can be 
integrated, enabling dynamic observations with real-time beam observation adjustment possibilities.

Chapter \ref{skaChapter} starts by briefly exploring the digital signal 
processing challenges posed by SKA$_1$ and propose a speculative design for 
station-level, GPU-based beamforming and transient detections for SKA$_1$-low, 
providing estimates for hardware requirements and cost. The same procedure is 
then applied to beamforming and dedispersion stages of the SKA$_1$-mid 
non-imaging pipeline.

Chapter \ref{conclusionChapter} draws some conclusion on the work presented in this thesis and presents some possible
extensions and future work.


\chapter{Dispersion Removal Techniques and Implementation}
\label{dispersionChapter}

The need for higher computational power, required by online transient detection 
systems, has been the driving force of several projects in recent year. 
Conventional systems are either composed of several interconnected 
servers/workstations (compute clusters) or employ custom hardware for 
specialised processing. Field Programmable Gate Arrays (FPGAs) have played an 
important role in this, especially with the development of custom programmable 
boards such as CASPER's FPGA-based ROACH board which reduces the amount of time 
required to test prototype digital designs for use in radio astronomy. Graphics 
Processing Units (GPUs) have gained popularity and support in recent years, and 
their application to radio astronomy has been extensively investigated (see 
\cite{Barsdell2010} for a theoretical analysis of several algorithms which would 
benefit a performance gain when implemented on GPUs). Several existing 
instruments, including pulsar timing experiments (for example, see 
\cite{Allal2009,Ransom2009}), transient detectors \cite{Serylak2012}, 
spectrometers \cite{Kondo2010} and array correlators \cite{Clarke2013} make use 
of GPUs to speed up processing.

Most dispersion removal techniques are inherently parallel over multiple dimensions and so would potentially
gain a performance boost when run on GPUs. In this chapter we investigate this for two dedispersion algorithms:
direct (incoherent) dedispersion and coherent dedispersion. Both techniques were implemented and optimised
for the latest GPU architecture developed by NVIDIA. Where applicable, performance benchmarks are also
presented compared to existing implementations.

\section{Incoherent Dispersion Removal Techniques}
\label{dispRemovalSection}

The effect of dispersion on a short-duration pulse is to smear it out in time. In order to determine the emitted pulse shape, or to detect the pulse with maximum S/N, the effect of dispersion must be corrected for through a process known as dedispersion. Two methods are generally employed. {\it Incoherent Dedispersion} works on channelised, detected data, whereupon a range of trial DMs are searched by summing across frequency channels, after delaying each channel according to the trial DM of interest. {\it Coherent Dedispersion} operates on raw telescope voltages and involves convolving the telescope voltages with the impulse response corresponding to the inverse of the ISM. This is generally used for pulsar monitoring, when the DM is approximately known, as the inverse filtering preserves the emitted pulse shape more faithfully than incoherent dedispersion. In this section we describe some of the incoherent dedispersion techniques which are in use, whilst in section \ref{coherentSection} we describe coherent dedispersion in more detail.

\subsection{Direct Dedispersion}
\label{directDispSecion}

The simplest way to correct for the dispersion effect is to split the incoming frequency band into a number of
narrow, independent frequency channels, record the signal in each band separately and then apply an appropriate
time delay to each channel, such that the received pulses arrive at the output of each channel at the same time. These
delays are calculated by using the dispersion relationship (equation \ref{dispRelationshipEquation}), where
$f_2$ is a channel with a frequency $f_{\text{chan}}$ and $f_1$ is the reference frequency $f_{\text{ref}}$ (usually the top frequency). These delayed channel frequencies can then be summed to produce a dedispersed time series.

The direct dispersion removal technique, also known as brute-force dedispersion, sums the frequency channels
along a quadratic dispersion trail for each input time sample. This process is repeated for every DM value
to be processed, and thus generates a set of dedispersed and summed time series for input dataset $A$ as follows:
\begin{equation}
\label{bruteEquation}
 D_{d,t} = \sum\limits_{c}^{N_c} A_{f,t+\Delta t(d,c)}
\end{equation}
where subscripts $d$, $t$ and $c$ represent the DM value, time sample and frequency channel respectively,
$N_c$ is the total number of frequency channels and $D_{d,t}$ represents the dedispersed trail ($d$,$t$).
$\Delta t(d,v)$ is a discretised version of equation \ref{dispRelationshipEquation}, and gives the time delay relative to the start of the band in integer time samples for a given DM and frequency channel:
\begin{eqnarray}
\label{deltaTEq}
 \Delta T(c) & \equiv & \frac{k_{\text{DM}}}{\Delta t} \left( (v_0 + c\Delta v)^{-2} - (v_0)^{-2} \right) \\
 \Delta t(d,c) & \equiv & \text{round}(\text{DM}(d)\Delta T(c))
\end{eqnarray}
where $\Delta t$ is the sampling time, $v_0$ is the frequency at the top of the band in MHz and $\Delta v$ is the width of a frequency channel. DM$(d)$ represents the range of DM trials to be computed. This technique has a complexity of $\mathcal{O}(N_tN_c N_{\text{DM}})$, where $N_{\text{DM}}$ is the total number of dispersion measure trials and $N_t$ is the number of time samples. 

When dedispersing for large DMs the dispersed signal is significantly smeared within a single frequency channel. This occurs when the gradient of the dispersion curve on the time-frequency grid is less then unity (referred to as the 'diagonal DM'). When this effect is significant it becomes inefficient to dedisperse the time series at its full sampling rate. One option is to reduce the sampling interval by a factor of 2 when the DM exceeds the diagonal, and then repeat the procedure at two times the diagonal, four times the diagonal, and so on. We refer to this process as ``time binning'', and it results  in an overall reduction in computational cost, however it also introduces a degradation in pulse S/N if the intrinsic pulse width is comparable to that of the dispersion smear.

Transient surveys generally require dedispersion over a large number of DM 
trials, of order $\mathcal{O}(10^4)$, requiring considerable computational 
resources. Methods for speeding up these processes using accelerator 
technologies, such as GPUs, will be discussed shortly. 
\cite{Addario2010,Clarke2011,ClarkeN2013} describe an FPGA-based implementation 
of a transient detection system, based on direct dedispersion, for the CRAFT 
survey, and examine the computational and S/N performance of their system. In 
chapter \ref{pipelineChapter} we'll compare their system to our GPU-based 
transient detection pipeline. An alternative to speeding up implementations is 
to increase computational efficiency. 

\subsection{Tree Dedispersion}
\label{treeDedispSection}

\begin{figure}[t!]
  \begin{center}
  \includegraphics[width=200pt]{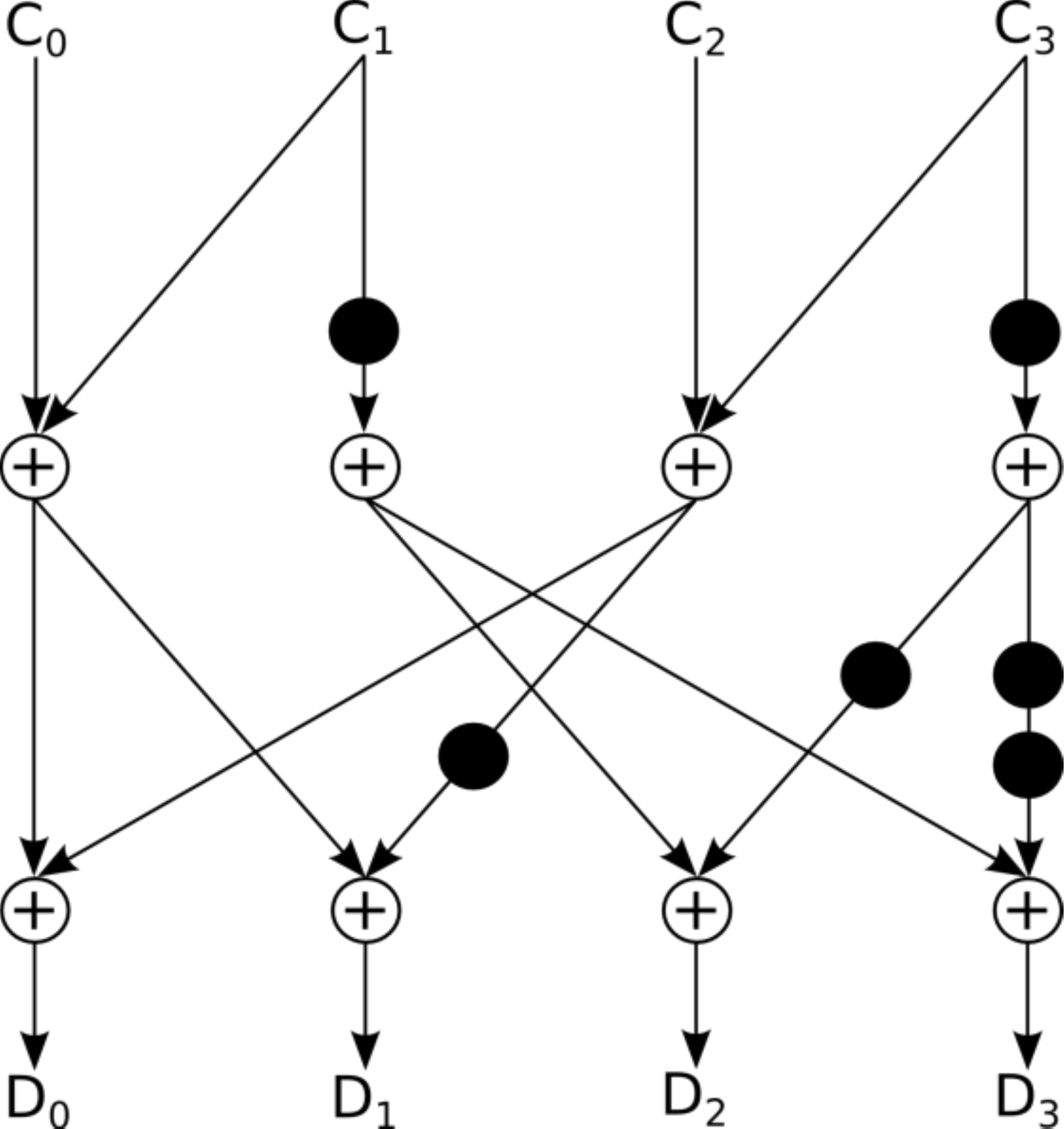}
  \end{center}
  \caption[Tree dedispersion algorithm]{Visualisation of the tree dedispersion algorithm. $N$ channels are input to the system, and $N$ dedispersed time series are then generated and output. Arrows represent data flow, whilst solid circles indicate unit time delays.}
  \label{taylorTreeFigure}
\end{figure}

General brute-force dedispersion algorithms involve many redundant operations, in that they add the same sample multiple times for different DMs. The Taylor tree method \cite{Taylor1974} attempts to reduce the complexity of the dedispersion computation from $\mathcal{O}(N_tN_cN_{\text{DM}})$ to $\mathcal{O}(N_tN_c\log_2(N_c))$. A Taylor tree consists of a network of delay and sum elements interconnecting $N$ inputs to $N$ outputs, a 4-channel version of which is shown in figure \ref{taylorTreeFigure}. The basic algorithm generates the dedispersed time series as follows:
\begin{eqnarray}
D_{d',t}' & =  & \sum_c^{N_c}T_{c,t+\Delta t'(d',c)} \\
\Delta t'(d',c) & = & \text{round} \left( d'\frac{c}{N_c - 1} \right)
\end{eqnarray}
where $d$ has a value $0 \leq d < N_c$. Due to the regularisation scheme employed by the algorithm, the delay function $\Delta t'(d',c)$ becomes a linear function of $c$ that ranges from 0 to exactly $d'$ across the band, resulting in a DM range of
\begin{equation}
\label{treeDMRange}
 \text{DM}(d') = \frac{d'}{\Delta T(N_c - 1)}
\end{equation}

One of the shortcomings of Taylor trees is that they implement linear approximations to dispersion, not proportional to the inverse-square of the frequency, which are less accurate for lower frequencies and wider bandwidths.  Additionally, the computed dispersion measures are constrained to those given by equation \ref{treeDMRange}, and the number of input frequency channels, $N_c$, must be a power of two. The last constraint is not a significant one, since zero-padded frequency channels can be inserted between existing channels, thus spreading the signal out in frequency, with the additional effect of alleviating somewhat the first constraint. Alternative methods can be employed to work around these limitations.

\subsubsection{Piecewise linear tree}

The piecewise linear tree method \cite{Manchester1996} approximates the dispersion curve using piecewise linear segments, where the input data is divided into $N_{\text{sub}}$ subbands of length
\begin{equation}
 N_c' = \frac{N_c}{N_{\text{sub}}}
\end{equation}
where the $n^{\text{th}}$ subband starts at frequency channel $nN_c'$. This method involves two stages of computation, the first of which involves applying the basic tree dedispersion algorithm to each subband independently
\begin{equation}
 S_{n,d',t} = \sum_{c'}^{N_c'}T_{c_n+c',t+\Delta t'(d',c_n+c')}
\end{equation}
This stage approximates the quadratic dispersion trail with a linear one. The linear DM in the $n^{\text{th}}$ subband that approximates the true DM indexed by $d$ is computed as follows:
\begin{eqnarray}
 d_n'(d) & = & \Delta t(d,c_{n+1}) - \Delta t(d,c_n) \\
  & = & \text{round}(\text{DM}(d) [\Delta T(c_{n+1}) - \Delta T(c_n)])
\end{eqnarray}
During the second stage, the dedispersed subbands are then combined to approximate the result of equation \ref{bruteEquation}
\begin{eqnarray}
 D_{d,t} & \approx & \sum_n^{N_{\text{sub}}} S_{n,d_n'(d),t+\Delta t_n''(d)} \\
 \Delta t_n''(d) & = & \text{round}\left( \text{DM}(d) \sum_m^n \Delta T(c_{m+1}) - \Delta T(c_m) \right)
\end{eqnarray}

It should be noted that the this method introduces additional smearing into the 
dedispersed time series as a result of approximating the quadratic dispersion 
curve with a piecewise linear one. \cite[appendix B]{Barsdell2012} derive an 
analytic upper limit for this smearing.

\subsubsection{Frequency-padded tree}

An alternative approach is to linearise the input data by changing the frequency coordinates, whereby $\Delta T(c)$ is stretched to a linear function $\Delta T'(c') \propto c'$, such that
\begin{equation}
\label{freqPaddingEquation}
 c' = \text{round}\left( \frac{v_0}{2\Delta v} \left[ 1 - \left( 1 + \frac{\Delta v}{v_0}c \right)^{-2} \right] \right)
\end{equation}

Evaluating $c=N_c$ gives the total number of frequency channels in the linearised coordinates, which determine the additional computational overhead introduced by the procedure. This number must be rounded up to the nearest power of 2 before the tree dedispersion algorithm can be applied. This linearisation is generally implemented by padding the frequency dimension with blank channels such that the real channels are spaced according to equation \ref{freqPaddingEquation}.

\subsubsection{Computing Larger DMs}

The basic tree dedispersion algorithm computes exactly the DMs specified by equation \ref{treeDMRange}, however it is often necessary to search a much larger DM range. This can be achieved by transforming the input data and repeating the dedispersion computation for each transformed series. The following operations can be used to compute an arbitrary range of DMs:
\begin{enumerate}[i]
 \item Apply the tree algorithm, or any variant mentioned above, to obtain DMs from zero to DM$_{\text{diag}}$ (the diagonal DM)
 \item Apply a time delay across the entire band, such that $\Delta t = c$
 \item Apply the dedispersion algorithm to obtain DMs from DM$_{\text{diag}}$ to  2DM$_{\text{diag}}$
 \item Repeat from step (ii) to obtain DMs up to 2$^n$DM$_{\text{diag}}$
\end{enumerate}

An alternative approach is to downfactor, or ``bin'', the input time series. This method provides a performance benefit, however at the cost of a minor reduction in the S/N for pulses of intrinsic width near the DM smearing time. This procedure is applied as follows:

\begin{enumerate}[i]
 \item Apply the dedispersion algorithm to obtain DMs zero to DM$_{\text{diag}}$
 \item Apply a time delay across the band
 \item Apply the dedispersion algorithm to obtain DMs from DM$_{\text{diag}}$ to  2DM$_{\text{diag}}$
 \item Downfactor the time series by a factor of 2 by summing adjacent samples
 \item Apply the dedispersion algorithm to obtain DMs from 2DM$_{\text{diag}}$ to  4DM$_{\text{diag}}$
 \item Repeat from step (iv) to obtain DMs up to $2^n$DM$_{\text{diag}}$
\end{enumerate}

\subsection{Subband Dedispersion}

``Subband'' dedispersion, a technique implemented by \cite{Ransom2001,Magro2011,Barsdell2012}, rather than exploiting a regularisation of the dedispersion algorithm, takes a simple approximation approach, involving two steps. In the first step the set of trial DMs is approximated by a reduced set of $N_{\text{DMnom}}=N_{\text{DM}}/N^{'}_{\text{DM}}$ 'nominal' DMs, each separated by $N^{'}_{\text{DM}}$ trial dispersion measures. The direct method is applied to subbands of $N^{'}_c$ channels to compute a dedispersed time series for each nominal DM and subband. In the second step the DM trials near each nominal value are computed by applying the direct algorithm to the reduced filterbank data, formed by the time series for the subbands at each nominal DM. These data have a reduced frequency resolution $N_{\text{SB}}=N_c/N^{'}_{c}$ channels across the band. The two steps thus operate at reduced dispersion measure and frequency resolution respectively, resulting in an overall reduction in the computational cost, at the expense of inducing some additional smearing into the dedispersed time series. See \cite[appendix B]{Barsdell2012} for an analytical upper-bound of this error.

\subsection{Alternative Dedispersion Methods}

Alternative methods have also been investigated which minimise the computational cost for dedispersion.
\cite{Fridman2010} propose the use of the Hough transform to reduce the number of trials for incoherent dedispersion, which measures the slopes of the the transient's tracks on the time-frequency plane and thus gives estimates of the required DM range. They also state that the cumulative sum method, which is occasionally used for RFI detection, can be used to provide an estimate for transient duration matching. \cite{Bannister2011} propose two new methods, the Chirpolator and Chimageator, which exploit the observation that when a linear chirp received by one antenna is multiplied by a delayed linear chirp received at another antenna, the result is a fixed-frequency tone whose frequency is proportional to the geometric delay. \cite{Law2012} introduce the bispectrum method, which is the product of visibilities around a closed-loop of baselines of an interferometer. The bispectrum is calibration-independent, resistant to interference and computationally efficient, and can be easily integrated into correlators for real-time transient detection. Note that the latter two methods require an interferometer, and thus are unsuitable for dishes with single feeds.

\section{Many-core Architectures}
\label{manycoreSection}

The number of transistors on a chip continues to climb with successive generations of processor technology
(Moore's Law, which is the observation that the number of transistors on integrated circuits doubles approximately every two years), however the power available to the chip is decreasing. This has led to a ``power wall'' and has
shifted focus of computer architecture from raw performance to performance per watt. Complex cores running at high
frequencies were replaced with multiple, simpler and lower power cores within a chip. Current generation
multiprocessors currently offer CPUs with a modest number of cores. This concept can be extended further to the
idea of many-core architectures, where a chip can contain hundreds or thousands of simple, low-frequency and
low-power cores sharing on-chip resources.

This concept is not entirely new. GPU computing is the use of GPUs, together with a host system, to accelerate
general-purpose scientific and engineering applications. These devices were originally designed to rapidly
manipulate and alter memory to accelerate the creation of images in a frame buffer, intended for output to
a display. The process of shading is inherently parallel, and so GPUs evolved to include multiple shader
pipelines by increasing the number of cores on the device, vastly improving computation throughput,
With the development of high-level APIs, including CUDA\footnote{\;https://developer.nvidia.com/category/zone/cuda-zone} and OpenCl\footnote{\;http://www.khronos.org/opencl/},
this raw performance became available to general-purpose computation on the GPU. In this chapter we will
focus on NVIDIA GPUs, however similar arguments can be applied to GPUs from other vendors. These have gone
through a number of architectural changes, all of which affect the execution behaviour of CUDA kernels in
some way, including improvements in data access coalescing, introduction of cache memory, an increase in the
number of cores and the way they are partitioned, block-level and thread-level instructions, and scheduling
schemes, amongst others.

\subsection{GPU Architecture}

\begin{figure}[t!]
  \begin{center}
  \includegraphics[width=360pt]{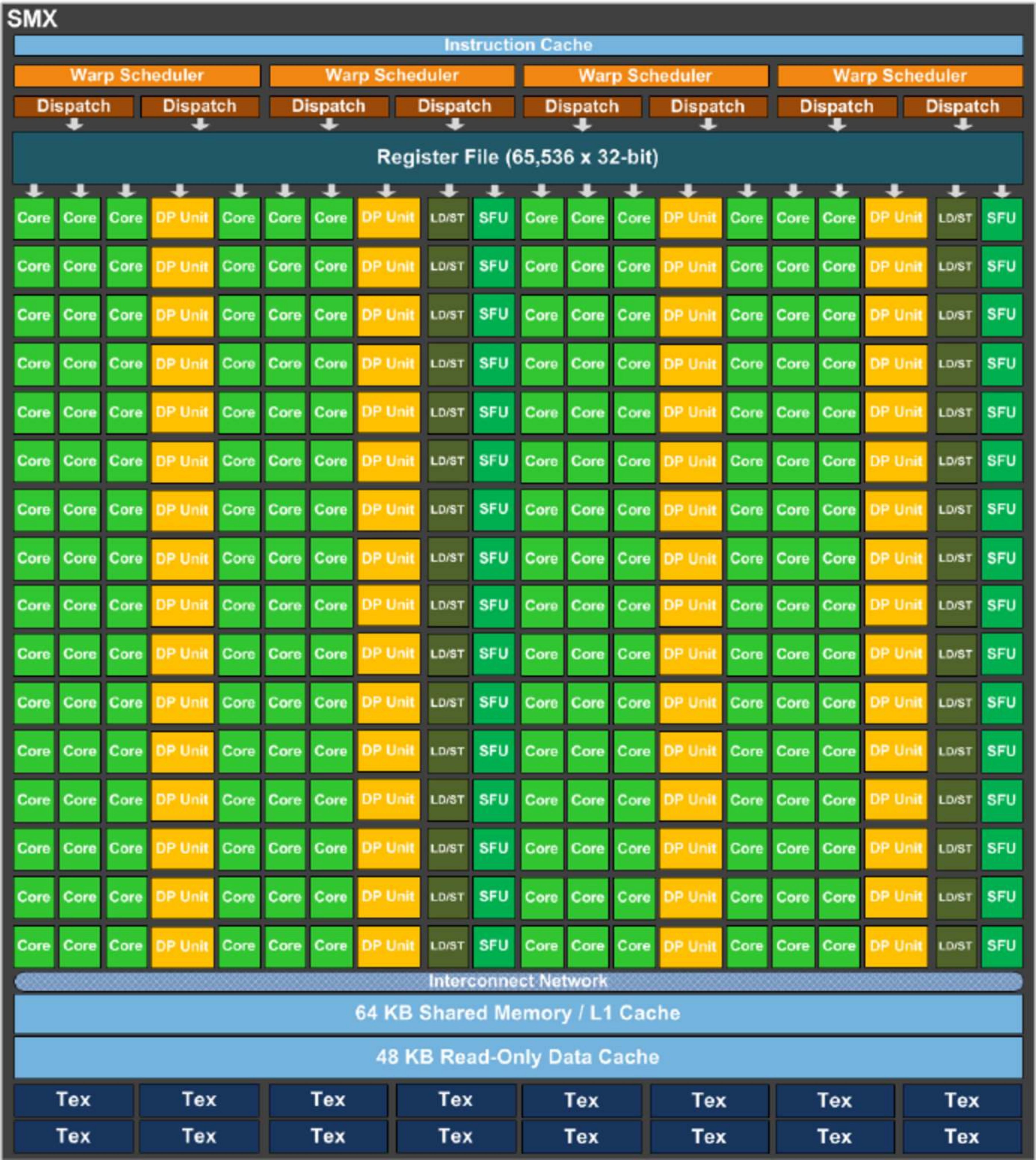}
  \end{center}
  \caption[NVIDIA GK110 SMX block diagram]{NVIDIA GK110 SMX Block Diagram: 192 single precision CUDA cores, 
      64 double precision units, 32 special function units (SFU), and 32 load/store units (LD/ST). Reproduced from \cite{Kepler2012}}
  \label{keplerFigure}
\end{figure}

The main architectural component in GPUs is a scalable array of multi-threaded Streaming Multi-Processors (SMX), each composed of a number of Scalar Processors (SP), special functions units, double precision units, memory load and store units, and a register file, amongst additional components. The decomposition of a single Kepler SMX processor is shown in figure \ref{keplerFigure}. The multiprocessor creates, manages and executes concurrent threads in hardware with minimal scheduling overhead, supporting very fine-grained parallelism via fast barrier synchronisation and fast thread creation. This allows low granularity decomposition of problems, by assigning one thread to each data element.

The multiprocessor employs an architecture which NVIDIA call Single Instruction 
Multiple Thread (SIMT), where it maps each thread to one SP core, and each 
scalar thread executes independently with its own instruction address and 
register state, allowing hundreds of threads to run several different programs 
in parallel. The SIMT unit handles threads in groups of 32 parallel threads 
called warps (a half-warp is either the first or second half of the warp). 
Threads within a SIMT warp start together at the same address, but are free to 
branch and execute independently. Every instruction issue time, the SIMT unit 
selects a ready warp to execute the next instruction. The warp executes one 
common instruction at a time, therefore execution becomes fully efficient when 
all the threads within the warp are on the same instruction address. If threads 
undertake a data-dependent conditional branch, each branch path will be executed 
serially, with threads which are not on the execution path being disabled until 
their path is processed, or the branches converge back to the same execution 
path. Branch divergence occurs only within warps, different warps can execute 
independently. The ratio of the number of active warps to the maximum number of 
processable warps is called the occupancy of the GPU when running the specific 
kernel.

Each thread has access to several memory types: (i) one set of local 32-bit registers, (ii) a global read-write memory space accessible from all the processing units in a GPU, (iii) a parallel data cache which is shared by all SP cores, where shared memory resides, (iv) a fast, read-only constant memory space accessible from all SP cores and (v) a read-only texture cache which is shared by all SP cores. A SMX's register and shared memory are split among all the threads of the batch of blocks, so the more resources the threads use, the less blocks which can be processed at one go. A multiprocessor can execute as many as eight thread blocks concurrently. The Fermi GPU series introduced an L2 cache which caches global memory data, and is global to all the SP cores.

CUDA is a general purpose computing architecture that leverages the parallel compute engine in NVIDIA GPUs to solve complex and demanding computational problems. It comes with a software environment which allows the use of C as a high-level programming language. CUDA has three key abstractions: a hierarchy of thread groups, shared memories and barrier synchronisation, which are exposed to the programmer as simple language extensions. They provide fine-grained data and thread parallelism, guiding the programmer to partition the problem into coarse sub-problems
which can be solved by parallel threads within the block. Each block can be scheduled on any processor core so that a compiled CUDA program can execute on any number of of processor cores. Thread blocks are in turn grouped as a grid, which is a high level abstraction of the problem at hand, with the thread blocks representing a coarse partitioning of the problem, and threads in turn represent fine-grained decomposition.

\subsection{State of the Art Many-core Devices}

\begin{table}[]
  \centering
  \begin{tabular}{| r | c | c| c |}
    \hline
         & {\bf NVIDIA K20}   & {\bf Xeon Phi SE10$^1$} & {\bf Xeon E5-2670$^1$}  \\
    \hline
    Single PTP   & 3.52 TF/s  & 2.15 TF/s       & 332.3 GF/s \\
    Double PTP    & 1.17 TF/s   & 1.07 TF/s       & 161.4 GF/s \\
    Cores         & 2688 (CUDA) & 61              & 8 (16 HT\footnotesize{$^2$}) \\
    Clock Speed   & 732 MHz     & 1.1 GHz         & 3.8 GHz\footnotesize{$^3$}  \\
    Memory        & 5 GB GDDR5  & 8 GB GDDR5      & $\leq$750GB (DDR3)\footnotesize{$^4$}  \\
    Memory BW     & 208 GB/s    & 352 GB/s        & 102.4 GB/s     \\
    TDP\footnotesize{$^5$}            & 235 W       & 300 W           & 135 W     \\
    \hline
  \end{tabular}
  \caption[Comparison of current many-core architectures]{Specification comparison between the NVIDIA K20, Intel Xeon Phi SE10, Intel Xeon E5-2670. Peak Theoretical
       Performance (PTP) is calculated as follows: $F \times N_{\text{cores}}
       \times C$, where $F$ is the number of FLOPS per clock cycle and $C$ is the clock speed in Hz.
        \newline
        \footnotesize{$^1$ Devices developed by Intel}
        \newline
        \footnotesize{$^2$ Hyper-Threaded Cores}
        \newline
        \footnotesize{$^3$ With Intel Turbo Boost enabled}
        \newline
        \footnotesize{$^4$ RAM DIMMS attached to host motherboard which must support similar specifications}
        \newline
        \footnotesize{$^5$ Thermal Design Power}}
  \label{deviceTable}
\end{table}

The latest Kepler-based GPUs, the architecture of which is depicted in figure 
\ref{keplerFigure}, have a single-precision peak theoretical performance of 
around 3.5 TFLOP/s, vastly
exceeding that of current high-end CPUs (order of 330 GFLOP/s). Table \ref{deviceTable} lists
some specifications for the NVIDIA K20, Intel Xeon Phi SE10 and Intel Xeon E5-2670. The Intel Xeon Phi is a new
many-core PCI device developed by Intel having up to 60 cache-coherent Xeon cores connected in a ring buffer around 8 GB of GDDR5 RAM with a high memory bandwidth. The most attractive feature of this device is the ease of deployment
of existing threaded and MPI applications due to their support for the x86-based instruction set. Bandwidth-limited
algorithms will also benefit from the larger cache and higher memory bandwidth. To date there has been no published
investigation into the suitability of this device for radio astronomy signal processing. It should be noted
that the peak theoretical performance of CPUs as well as the Intel Xeon Phi, can only be achieved through the
use of vector registers (256-bit AVX\footnote{\;Advanced Vector Extensions: http://software.intel.com/en-us/avx}
in the case of recent Xeon CPUs and 512-bit AVX for the Xeon Phi). Recent compilers can automatically vectorise data parallel implementations to make use of these features, however in order to achieve optimised high performance low-level intrinsics/assembly code is required.

\subsection{Mapping algorithms to Many-core Architectures}

Many-core architectures exhibit a number of characteristics that can impact the 
performance of an algorithm. To fully utilise these massively-parallel 
architectures, algorithms must exhibit a high level of parallel granularity, 
such that the same operations can be performed on a number of data items in 
parallel. Algorithms which satisfy this requirement are considered to be data 
parallel. Additionally, these devices generally have high internal memory 
bandwidths and high latency (memory transfer time) costs can limit the 
achievable performance of an implementation if the data access pattern does not 
reflect the underlying memory mechanism of the device. These can achieve a high 
memory throughput if data locality (neighbouring threads access memory elements 
which are physically close in memory) is maximised. In CUDA-terms, when threads 
in a warp access contiguous memory locations, each individual request will be 
coalesced into a single request, thus reducing the overall bandwidth required.

The cores in many-core devices should be kept busy as much as possible, so algorithms should have a high arithmetic intensity, the lack of which would result in a low compute to memory access ratio. Such algorithms are referred to as bandwidth-limited, such that their performance is limited by the rate at which memory can be accessed rather than the arithmetic instructions which can be processed. Fermi and Kepler GPUs have an inbuilt block-level cache and latency hiding schemes which can help, however for algorithms with large strided or irregular access patterns it is usually beneficial to implement a custom caching scheme using shared memory as this will avoid loading extra data values within cache lines. These are limited resources, so care should be taken in designing optimal data partitioning schemes. 

For compute-limited algorithms a high level of Instruction-Level Parallelism (ILP) is required to achieve high intensity computations, coupled with the avoidance, or limited use of, high latency calls such as atomic operators, modulus and division, and math library calls. ILP is a measure of how many independent operations can be performed simultaneously by a single thread, and depends highly on the hardware design of the underlying architecture, generally requiring multiple pipelines to overlap instructions. ILP is essential to achieve the peak theoretical performance of Kepler GPUs.

GPUs, and performance accelerators in general, are connected to the host motherboard via interfacing links, with PCI-express currently being the interface of choice. The data on which the GPU operates has to reside in GPU memory\footnote{\;This is not technically necessary with the latest version of CUDA, however in principle data still has to be transferred between the two memory spaces.}, therefore a CPU to GPU transfer has to be issued, and an additional transfer is required to copy the results back to host memory for further processing. The transfer rates are limited by the interface bandwidth, which for a 16x-lane PCIe-3 link is about 15.75 GB/s (bi-directional). This can pose a bottleneck for high throughput systems, or for algorithms with a low arithmetic intensity, and can limit the applicability of GPU to certain scenarios, some of which will be encountered in chapter \ref{skaChapter}.

\section{GPU Framework}

\begin{figure}[t]
  \centering
  \includegraphics[width=150mm]{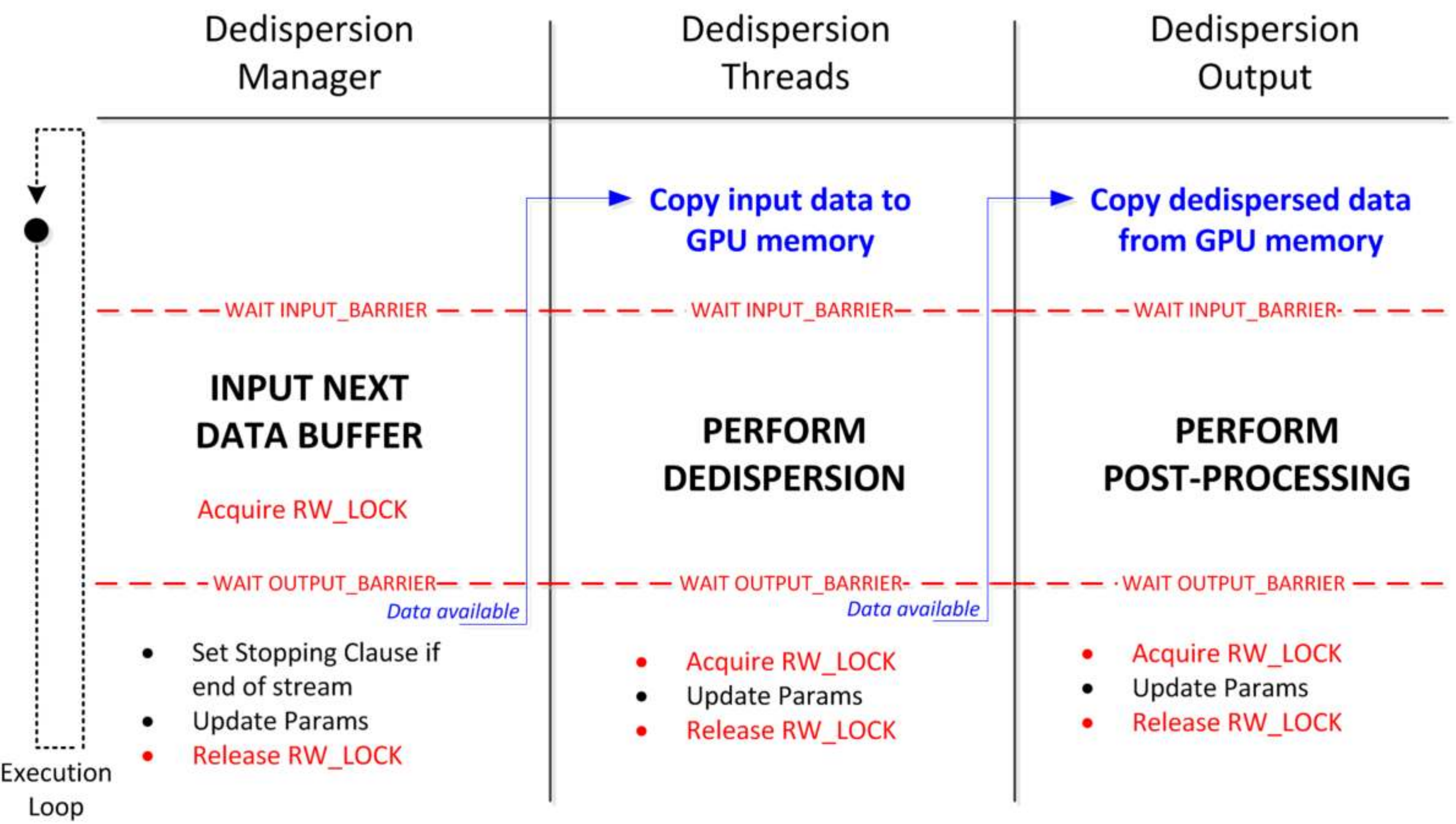}
  \caption[GPU framework workflow]
         {Each thread has three main stages: the input, processing and output stages. Data has to flow from one thread
          type to the next, so synchronisation objects have to be used to make 
sure that no data is overwritten or re-processed. Barriers and RW locks are used 
to control access to critical sections.}
  \label{mdsmFlowFigure}
\end{figure}

GPU kernels require data transfers to and from host memory, as well as some interfacing mechanism to synchronise
GPU execution with data acquisition, data transfer and host execution. This overlapping overhead must be minimised
in order to reduce the run time of the entire application. A pipelining mechanism was developed which overlaps
data acquisition, GPU execution and post-processing of dedispersed data, as the schematic in figure \ref{mdsmFlowFigure}
shows. This pipeline consists of three main stages:
\begin{description}
 \item[Data Acquisition:] GPU kernels require data to process, which can originate from different locations, such
	  as files and real-time network streams. These sources can have different latencies, and so data acquisition needs to be performed in parallel so as not to interfere with kernel execution, especially when the pipeline
	  is run in real-time mode. An interface is provided which accepts large data buffers that are copied directly
	  to GPU memory for processing. These buffers are populated by custom data interpreters. Additional operations
	  performed by this thread include: parsing and initialisation of input parameters, threads and synchronisation objects, as well as error handling.
 \item[Kernel Execution:] Execution parameters are parallelised across multiple GPUs so as to meet realtime requirements.
	  A CPU thread is created for each execution instance and is mapped to a single GPU. Multiple CPU threads
	  can be mapped to the same GPU. Each thread is responsible for copying its section of the input buffer to
	  GPU memory, launching the execution kernel and copying the resultant output to host memory.
 \item[Post Processing:] A single host thread is responsible for post-processing, and where required OpenMP threads are
	  used to speed up compute-intensive operations.
\end{description}

Currently this framework assumes an homogeneous system (with only one type of GPU device attached to it), and does
not perform any load-balancing between the devices. The host threads are split into three conceptual “processing stages”, which are guarded by several thread-synchronisation mechanisms. The three stages are: (i) the input section,
where the thread inputs data to be processed (by itself, or to send to other threads for processing)
(ii) the processing section, which is the main section in the thread and the part which takes the longest to
complete and (iii) the output section, where the processed buffer is output and 
made available to the next thread, and any parameter updates are performed. The 
input and output sections are combined, so no synchronisation is required 
between them.

This simple pipelining framework was extended further to include additional capabilities,
such as ``Transient Buffer Board'' mode, buffer persister, multi-beam mode and beamforming mode, which are discussed
in detail in chapters \ref{pipelineChapter} and \ref{beamformingChapter}. All performance benchmarks presented in this chapter do not include
pipeline overheads, however these are generally minimal and negligible when compared to dedispersion runtime.

\section{Direct Dedispersion GPU Implementation}
\label{incoherentSection}

An analysis for suitability of this algorithm to many-core architecture determines that:
\begin{itemize}
 \item the algorithm is best parallelised over the ``embarrassingly parallel'' DM and time dimensions, with the sum
       of frequencies being performed sequentially
 \item the algorithm has a very high theoretical arithmetic intensity (same order of magnitude
       as the number of DMs computed)
 \item the memory access patterns generally exhibit reasonable arithmetic locality, however
       its non-trivial nature make it difficult to achieve a high arithmetic intensity
\end{itemize}

This technique has been investigated and implemented on GPUs by several authors \cite{Magro2011,Armour2012,Barsdell2012}, all of whom report a speedup of at least an order of magnitude when compared to optimised CPU-based implementations. In the following sections we'll describe our newest implementation of this method and compare it to the implementation by \cite{Armour2012}, the major differences being the use of shared memory by the former, and cache memory by the latter to minimise data read accesses to global memory.

\begin{figure}[t!]
  \begin{center}
  \includegraphics[width=400pt]{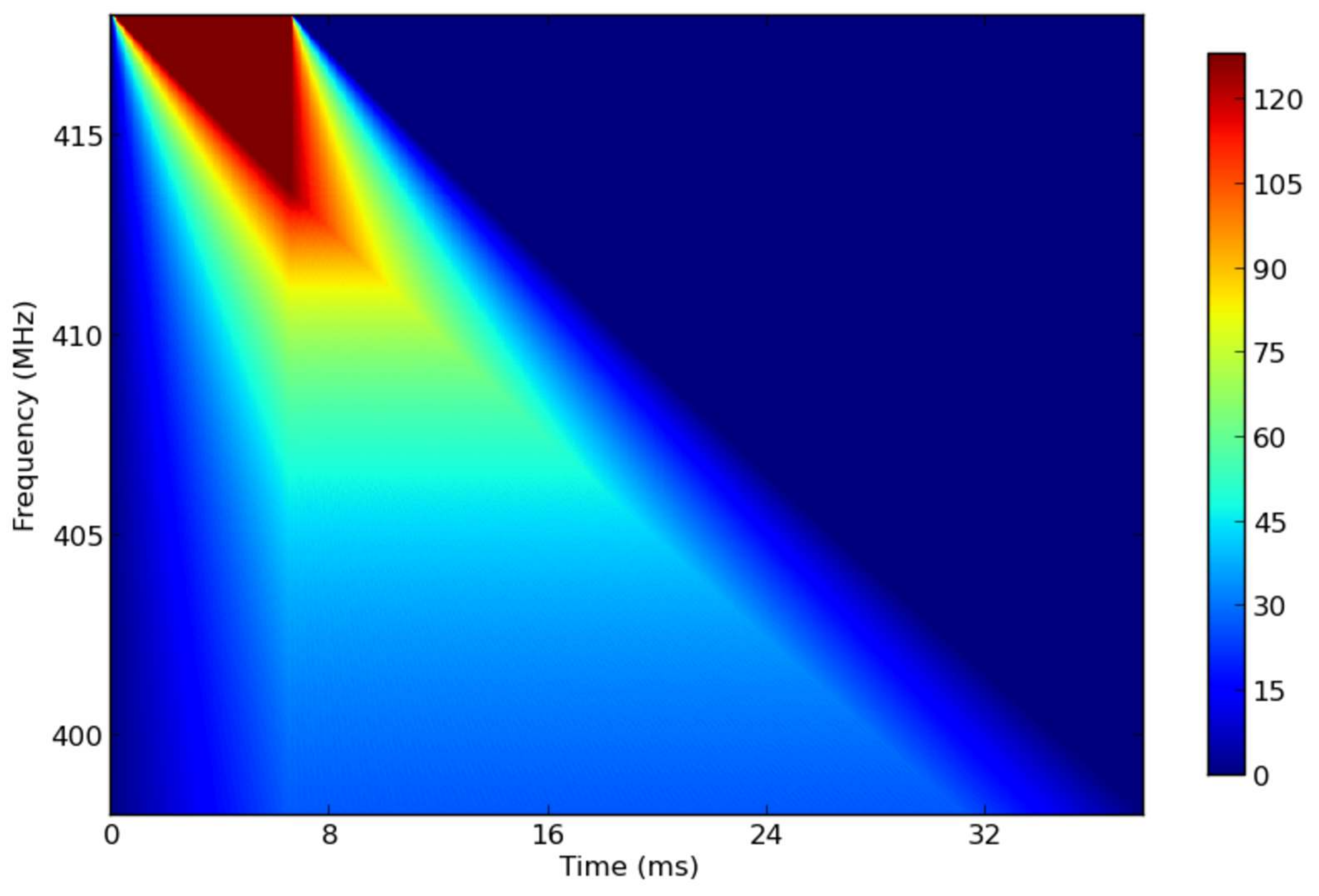}
  \end{center}
  \caption[Dedispersion memory access pattern]{Memory access pattern simulation for a 20 MHz band, centred at
	    408 MHz and channelised into 1024 channels. A group of 128 adjacent time samples were traversed along the
	    dispersion trail for 128 DM values (maximum DM of 12.8 pc$^3$ cm$^{-3}$) and when a cell in ($f$,$t$)
	    space was reached its counter was incremented.}
  \label{dedispMemoryFigure}
\end{figure}

The key factor determining the performance of a GPU implementation of direct dedispersion is the memory access
pattern used to read data from global memory, together with the structures used to keep data local to the thread in order to maximise data reuse. In a na\"ive implementation, when dedispersing a set of adjacent time samples, representing the data elements required to process for a DM range ${N_{\text{DM}}}$, several cells in the input space ($f,t$) need be to read multiple times. This behaviour was simulated by generating an empty $N_t$ by $N_f$ grid and a trail for each DM value was traversed for a number of time samples. The resultant grid is shown in figure \ref{dedispMemoryFigure}. Without any optimisations global memory would need to be accessed $N_{\text{DM}}\times N_f$ times for every dispersion measure. It is clear that this performance penalty can be alleviated by taking advantage of this behaviour.

\begin{figure}[t!]
  \begin{center}
  \includegraphics[width=400pt]{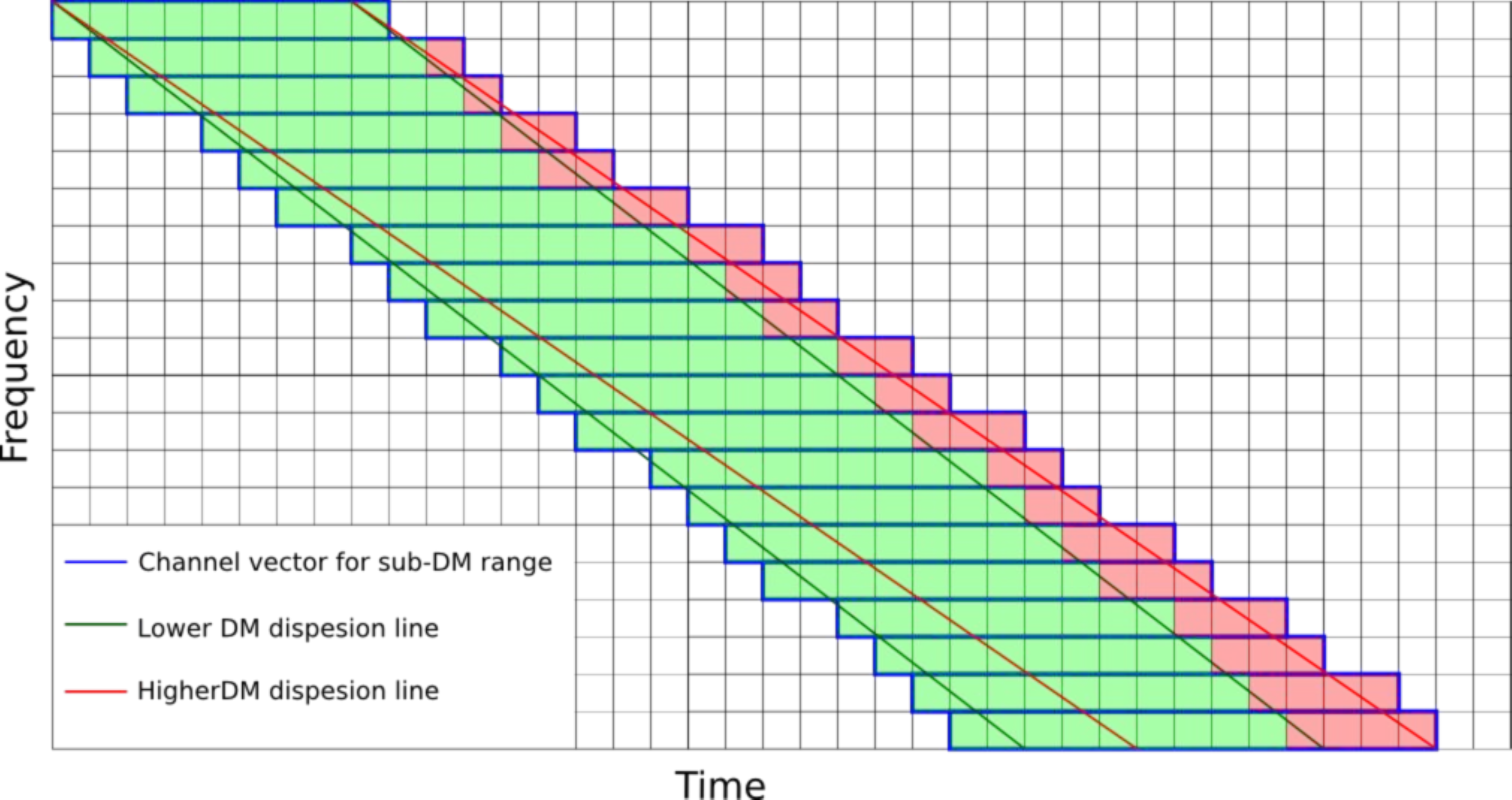}
  \end{center}
  \caption[GPU memory access optimisation]{GPU memory access optimisation for direct dedispersion.
  The channel vectors required to dedisperse a subset of the time samples,
	    for a subset of DM space, is loaded once to fast memory. Cells in green represent
	    data values which are needed by all the time samples being processed, while cells in read are only required
	    by a subset of these.}
  \label{gpuMemoryFigure}
\end{figure}

Consider a set of adjacent time samples $S$ being processed for a subset of the DM range $D$.
When visiting a frequency channel $c$, the required number of data elements $C$ can be calculated
by taking the shift from the highest DM value $D_n$ for the last sample $S_n$ and subtracting the shift required by the lowest DM value
$D_0$ for the first sample $S_0$, as follows:
\begin{equation}
\label{channelDispersionEquation}
 C_n = \Delta t(S_n,D_n, c) - \Delta t(S_0, D_0, c)
\end{equation}
where $C_n \geq S$. $C_n$ can be thought of as the vector of values required to process $S$ for $D$, which we will
refer to as the {\it channel vector}. Figure \ref{gpuMemoryFigure} provides a visual illustration of this behaviour,
where the channel vector for every frequency channel is highlighted. Green cells are required by all of the samples
in $S$, while cells in red are only needed by a subset of the time samples being processed. These vectors can be loaded only once from global memory and stored in fast memory whereby they can then be read for accumulation with
lower latency.

Following this behaviour, a CUDA thread block needs to process a subset of the time samples for a subrange
of DM space, meaning that the CUDA grid will be a direct mapping to the output 
space (DM, $t$). Each thread within
a thread block is associated with one input time sample and a number of DM trials. Local accumulators, stored
in registers, one per DM value, are updated once for every frequency channel. Threads within the same thread block
cooperate to load data from the input buffer into shared memory.

\begin{algorithm}[t!]
  \caption{Direct dedispersion shared-memory implementation}
  \begin{algorithmic}
      \REQUIRE input, output, dm\_delays, nchans, nsamp, tdms, maxshift
      \STATE Declare shared $accumulators[dedisp\_threads]$, $vector$ and $delays[dedisp\_dms]$
      \item[]
      \FOR{$c = 0$ \TO $nchans$}
	\STATE synchronise threads
	\STATE $inshift \gets dm\_delays[block\_shift]$
	\item[]
	\IF{$threadIdx.x < dedisp\_dms$}
	   \STATE $delays[threadIdx.x] \gets dm\_delays[block\_shift + threadIdx.x] - inshift$
	\ENDIF
	\STATE synchronise threads
	\item[]
	\FOR{$s = threadIdx.x$ \TO$ s < blockDim.x + delays[dedisp\_dms - 1]$}
	  \STATE $vector[s] \gets input[shift + inshift + s]$;
	\ENDFOR
	\STATE synchronise threads
	\item[]
	\FOR{$d = 0 \to dedisp\_threads$}
	  \STATE $accumulators[d]\; += vector[threadIdx.x + delays[d]]$
	\ENDFOR
      \ENDFOR
      \item[]
      \FOR{$d = 0$ \TO $dedisp\_dms$}
	\STATE $output[output\_shift] \gets accumulators[d]$
      \ENDFOR
  \end{algorithmic}
  \label{bruteAlgorithm}
\end{algorithm}

Details of our implementation are described in algorithm \ref{bruteAlgorithm}. Two shared memory buffers are declared,
one for storing the channel vector and another for storing the sample shifts for each channel and DM combination.
These shifts are required to be located in fast shared memory due to repeated access in the innermost loop. Accumulators
are stored in registers for fastest possible access (care needs to be taken not to define a large DM subrange for each
thread block, otherwise these accumulators will be spilled to local memory, which is much slower). Then, for
every frequency channel:
\begin{enumerate}
 \item The required shifts are cooperatively loaded by all the threads within a thread block and stored in shared 	      memory. A block-wide barrier is required to synchronise all the threads and ensure that all the shifts are                          
      loaded before any computation is performed.
 \item The channel vector is cooperatively loaded to shared memory. The input buffer is stored in channel order, with
       time changing the fastest (and thus assumes that the data has been transposed beforehand), which means that threads within a warp will access this buffer in a coalesced manner. An extra transfer request might be issued by the GPU due to misaligned accesses caused by the shift induced by the first sample. A block-wide barrier is required to ensure all input data is available to all threads for processing.
 \item For each DM value, the required shifted value from the channel vector is added to the accumulator associated
       with it
\end{enumerate}
Once all the frequency channels have been processed, the dedispersed and summed values are written to the output buffer in global memory.

\begin{figure}[t!]
  \begin{center}
  \includegraphics[width=400pt]{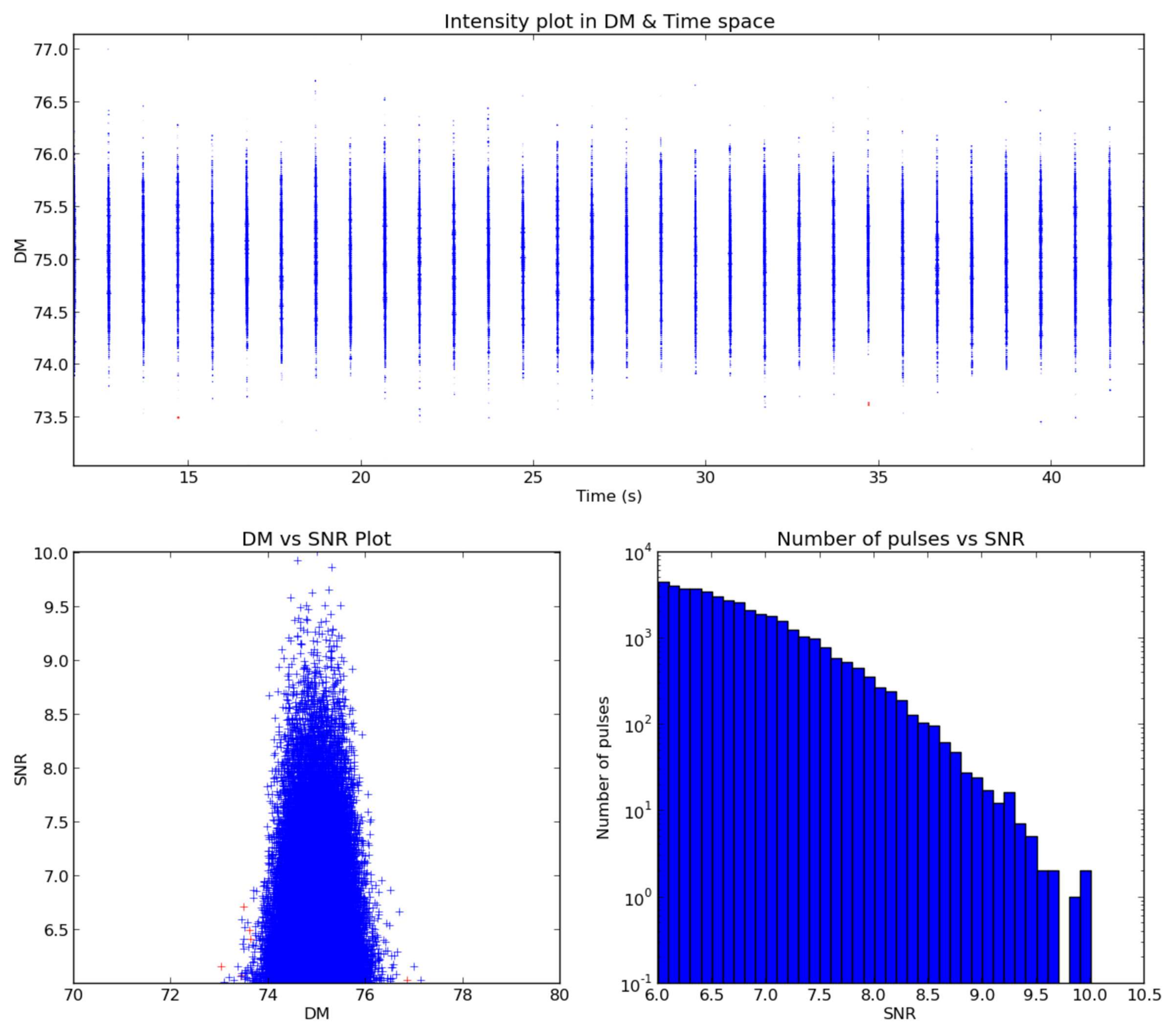}
  \end{center}
  \caption[Direct dedispersion test case]{Direct dedispersion output for an input filterbank file containing a simulated pulsar. Plots show DM vs Time, weighted by the detection S/N (top), S/N vs DM (bottom left) and number of pulses vs S/N (bottom right).}
  \label{lofarTestFigure}
\end{figure}

To test the functionality of the implementation, a simulated data file centred at 153 MHz, with 6 MHz bandwidth, channelised into 1024 frequency channels, containing a pulsed signal with period 1s, duty cycle 1\% and dispersed at a DM of 75 pc cm$^{-3}$ was generated. The pulses were modelled as a top-hat pulse of height 8$/\sqrt{1024}$ = 0.25, which were embedded in random Gaussian noise with mean 0 and standard deviation 1. The S/N of the average simulated pulse, integrated over the frequency band, has a mean value of 8. Direct dedispersion over 4096 DM trials and a DM step of 0.02 pc cm$^{-3}$ was performed. Figure \ref{lofarTestFigure} shows the output of the dedispersion code, which captures all detections with S/N greater than 6.

\subsection{Performance and Benchmarks}

Greater emphasis was put on making the dedispersion kernel optimised for Kepler GPUs, which makes the implementation
more future proof and capable of fully utilising current top-range, high performance GPUs. These GPUs have a higher number
of registers allocated per thread block, however the core clock speed is lower than in Fermi GPUs. This means that
more data can be kept in fast memory, however care must be taken to better hide access latency and
make sure that a high level of instruction-level parallelism is achieved. Our implementation
requires two configuration parameters to be defined, and consequently optimised upon:
\begin{description}
  \item[Number of accumulators] defines how many DM values each thread will process. These are stored in registers,
      and thus are limited by the number of registers available to a thread block. A lower number of accumulators
      will decrease channel vector reuse and
      increase global memory bandwidth requirements, while a high number will reduce the
      occupancy on the GPU, and in the worst case can be spilled to local
      memory, significantly reducing performance.
  \item[Threads per block] determines the number of time samples a single thread block
      will process (one per thread). A low number will reduce the number of active warps in a SMX, leading to
      diminished parallelism, while a higher number will increase shared memory and register requirements,
      reducing occupancy.
\end{description}

\begin{figure}[t!]
\begin{center}
\includegraphics[width=280pt]{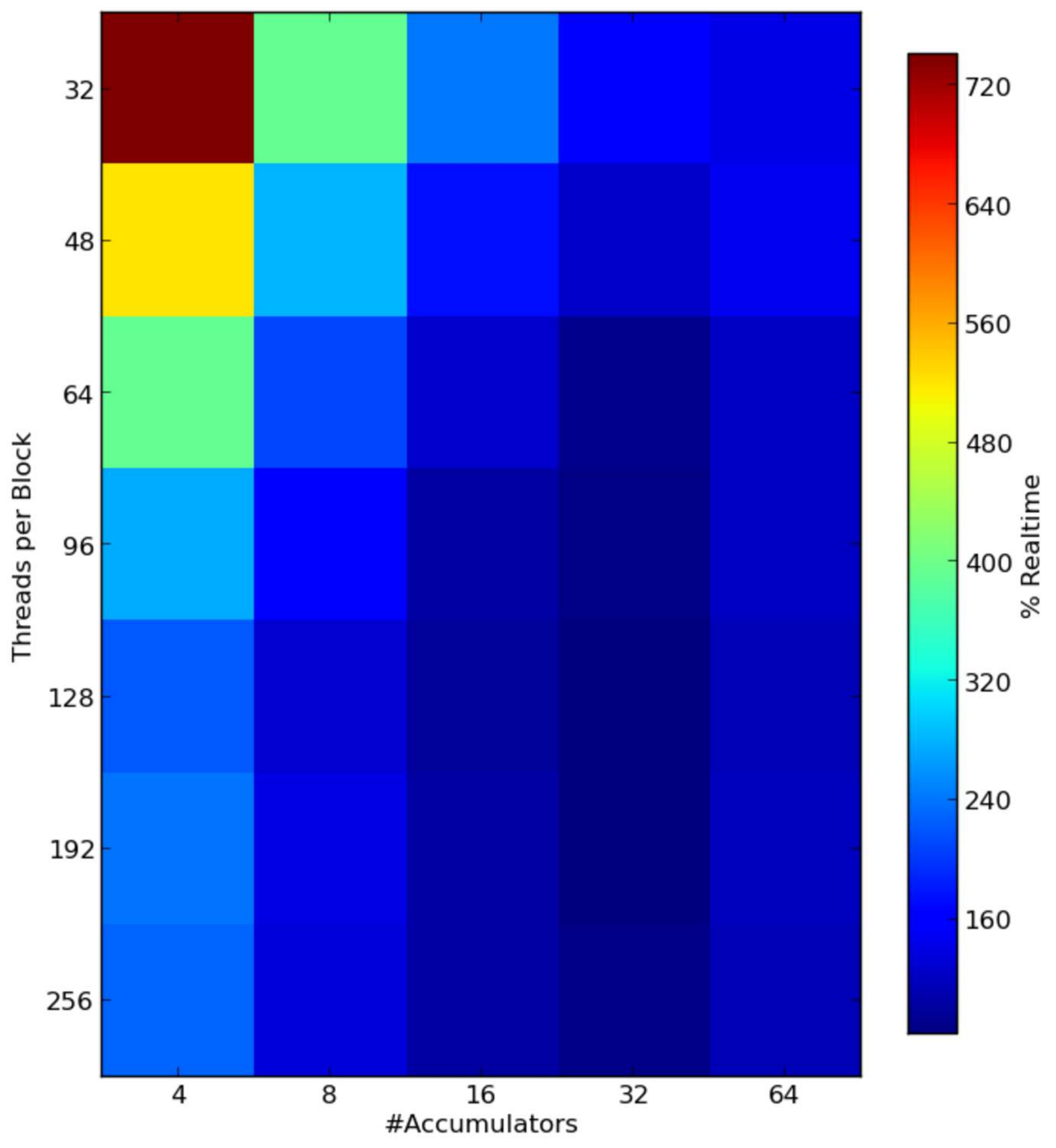}
\end{center}
\caption[Configuration optimisation for direct dedispersion kernel]{Configuration optimisation for direct dedispersion. On the tested GPU, the optimal combination was 32 accumulators and 128 threads per block.}
\label{dedispConfigFigure}
\end{figure}

In order to find the optimal configuration for the above parameters a series of benchmark tests were conducted on an
NVIDIA GTX 670 card for a wide range of parameters combinations. A 6.7 s simulated dataset centred at 408 MHz, with a bandwidth of 20 MHz, split into 1024 frequency channels, was generated and dedispersed over 4096 DM trials with a maximum DM of 409.6 pc cm$^{-3}$. The results of these tests are depicted in figure \ref{dedispConfigFigure}, which shows that increasing the number of threads per block and number of local accumulators yields a performance benefit until a global maximum is reached, after which performance starts to degrade. On the tested GPU, the optimal combination was 32 accumulators and 128 threads per block. These values were then used to benchmark other aspects of the implementation.

\begin{figure}[t!]
  \centering
  \subfloat[Performance for varying number of DM trials]{\includegraphics[width=410pt]{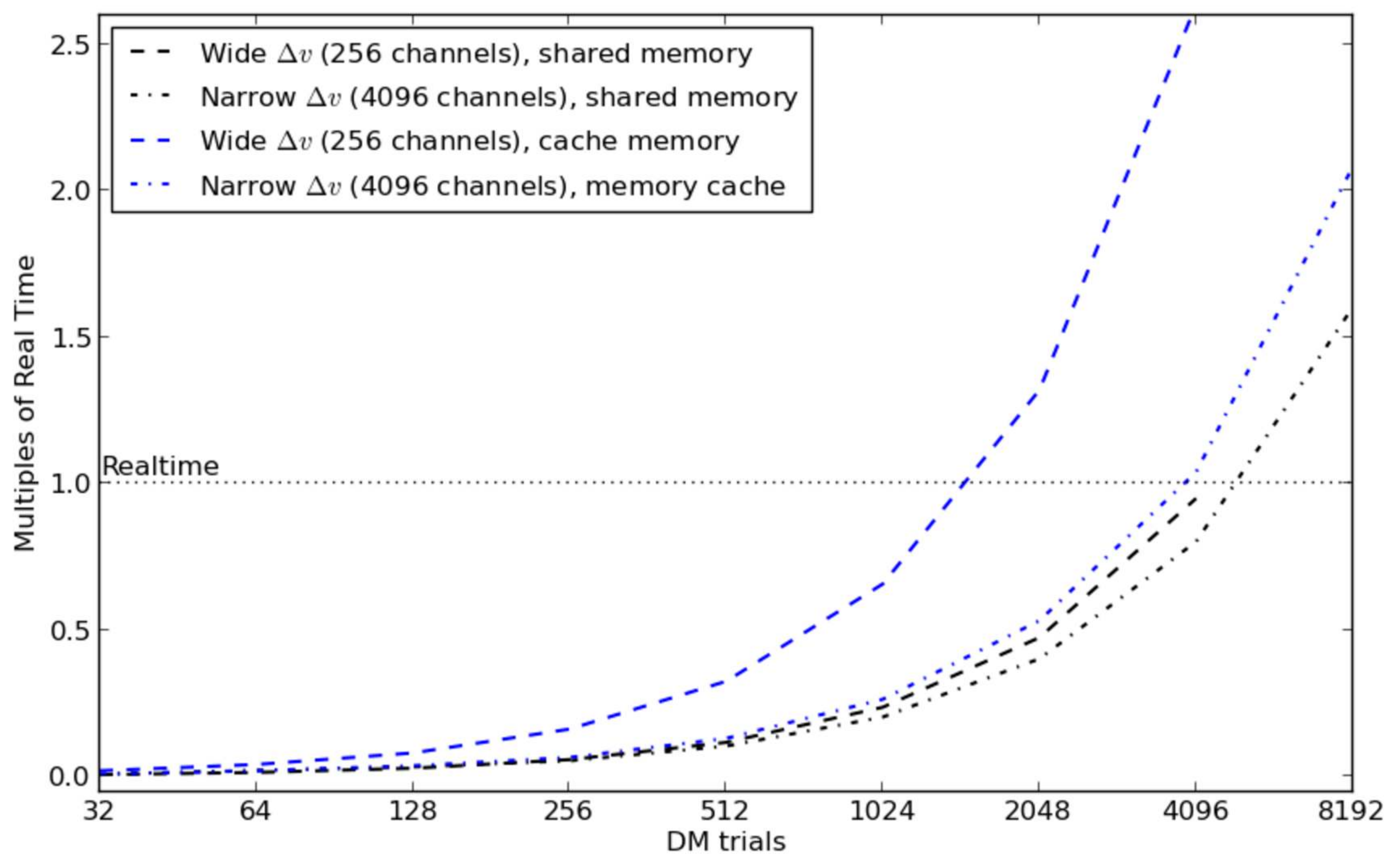}}
  \hspace{8mm}
  \subfloat[Performance for varying number of time samples]{\includegraphics[width=410pt]{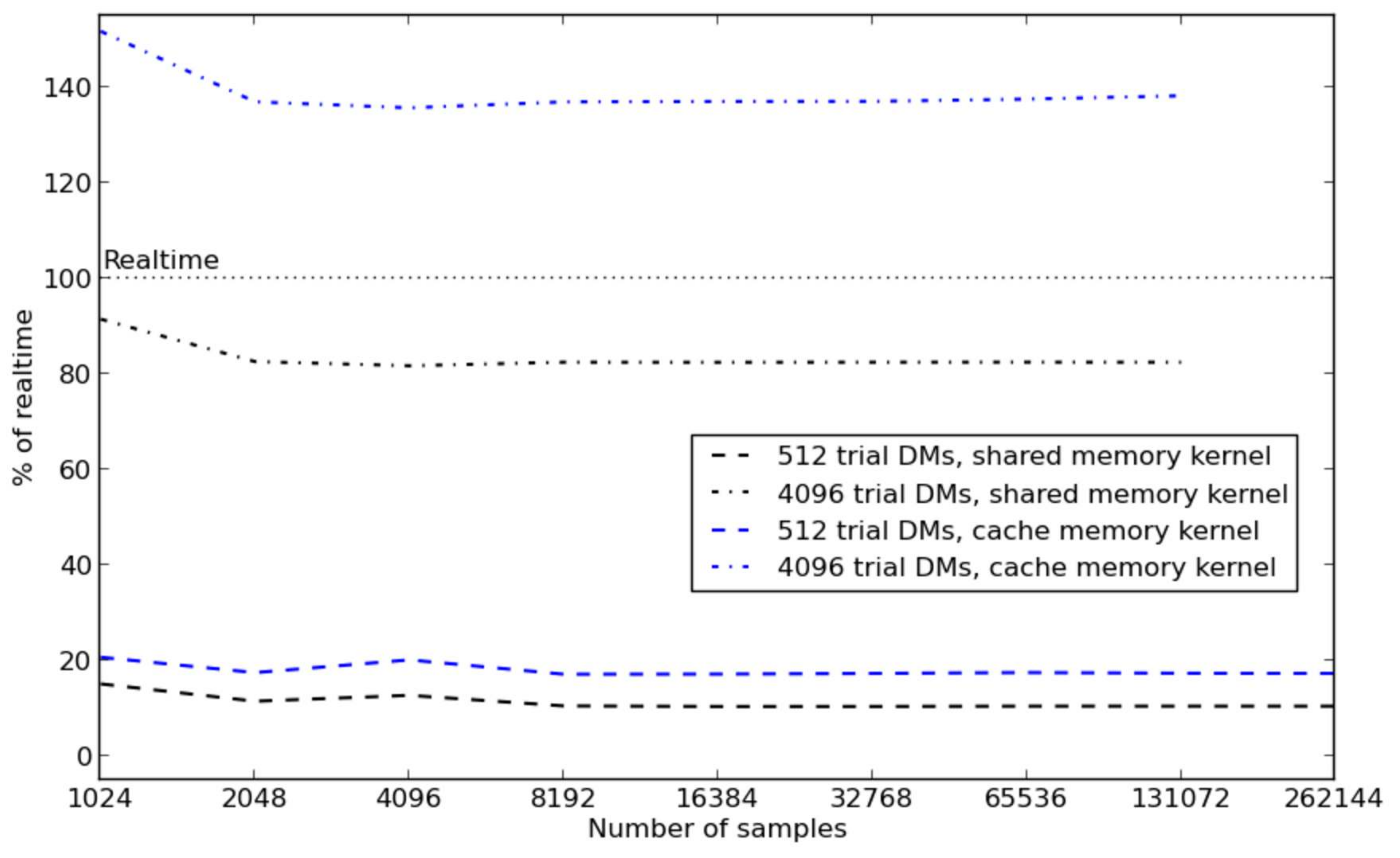}}
  \caption[Direct dedispersion performance] {Direct dedispersion performance benchmarks for varying
	    (a) number of DM trials and (b) number of time samples in buffer. Our implementation was compared
	    with the one by \cite{Armour2012}. Both implementations scale linearly over both varying dimensions,
	    with the cache-optimised version showing a performance degradation for lower number of channels (wider
	    channels bandwidth).}
  \label{dedispBenchmarkFigure}
\end{figure}

\begin{figure}[t!]
\begin{center}
\includegraphics[width=400pt]{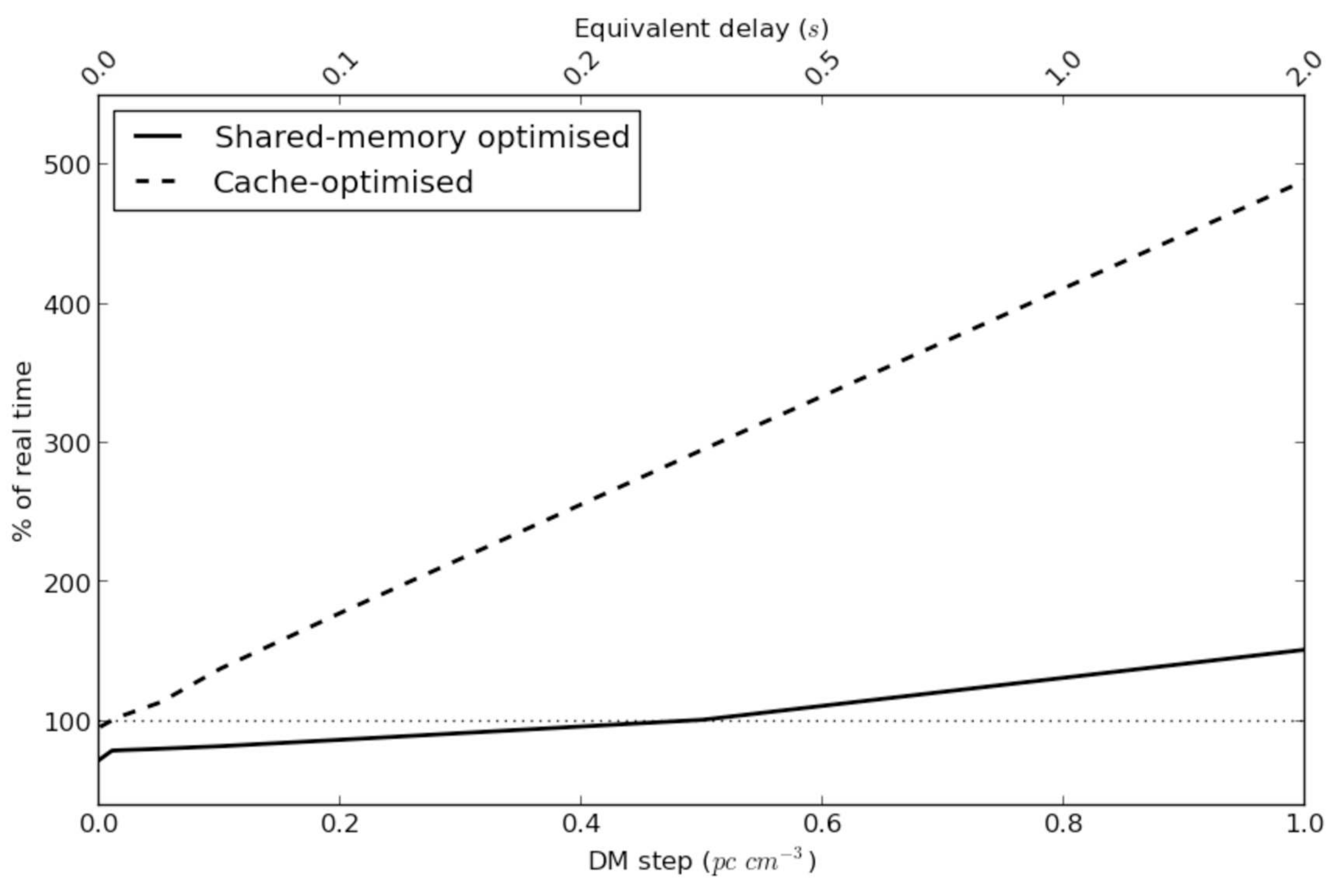}
\end{center}
\caption[Performance degradation relative to time shifts]{This plot shows the performance degradation
     of both implementations when the DM step is increased, leading to higher shifts between
     consecutive frequency channels for higher DM values. This results in a higher rate of cache misses
     for the cache-based implementation and a higher amount of shared memory usage for the shared memory
     based implementation.}
\label{dedispMaxshiftFigure}
\end{figure}

The scalability performance of the algorithm was also analysed and directly compared with the cache-optimised
implementation of \cite{Armour2012}. The cache-optimised implementation uses the same ``channel vector`` concept, however it is assumed that this will be performed by the L1 caching mechanism introduced in Fermi GPUs. This was an improvement over the previous work on which it was based (\cite{Magro2011}), however here we showcase the fact that implementing the caching in the kernel itself, using shared memory, provides a performance benefit. The benchmark
results are presented in figure \ref{dedispBenchmarkFigure}, where scaling 
performance for varying number of (a) DM trials and (b) time samples in the 
buffer are shown. Both implementations scale linearly over both varying 
dimensions, with the cache-optimised version showing a performance degradation 
for lower number of channels (wider channel bandwidth). For plot (a) a simulated 
dataset centred at 408 MHz, with 20 MHz bandwidth, was split into 256 (wide 
$\Delta v$) and 4096 (narrow $\Delta v$) frequency channels, and 1.6 s worth of 
data were buffered to the GPU, where both implementations were executed for a 
varying number of DM trials. For plot (b) the same observation parameters were 
used and a varying number of samples were buffered to the GPU.

The channel width-dependent performance degradation affecting the cache-optimised implementation can be attributed to the shift-dependent performance degradation of both implementations, as depicted in figure \ref{dedispMaxshiftFigure}, where a 6.7 s simulated dataset centred at 408 MHz with a bandwidth of 20 MHz, split into 1024 frequency channels, was generated and dedispersed over 4096 DM trials with varying DM steps. The equivalent delay in $s$ refers to the delay for a DM of 1 pc cm$^{-3}$. Increasing the DM step generates larger shift values between consecutive channels, which could result in ''skipped samples`` within the channel vector. This effect is much more severe for the cache-optimised implementation since the GPU will load an entire cache line for every data value accessed, and wider shifts would result in a higher frequency of cache misses. Our implementation counters this by increasing the shared memory buffer, which results in a much lower degradation gradient. It should be noted that in real-world transient surveys this effect would be minimal since consecutive time samples are usually combined for better performance and high time resolution is not required due to excessive signal smearing within the band.

It is clear from these benchmark tests that the key to performance is data movement within the GPU. Although GPUs
have very high compute capabilities, data transfers are struggling to keep the SMXs busy. On the GTX 670 a performance
of around 600 Gops/s is achieved, roughly evenly split between single precision floating point, integer and shared memory access.
This amounts to about 25 - 30\% of peak compute capability and communication. Whilst this is a noticeable improvement
from previous implementations, higher bandwidth and lower latency features in GPUs would be ideal. The Intel Xeon Phi
might prove to be a more suitable platform, however the direct dedispersion kernel has not yet been benchmarked
on this device.

\section{Coherent Dedispersion}
\label{coherentSection}

Incoherent dedispersion is limited in two ways. First, the goal of recording power vs. time make sense only on a
timescale greater than
\begin{equation}
 dt_1=\frac{1}{dv}
\end{equation}
where $dv$ is the width of a frequency channel. This is due to time-frequency uncertainty. Secondly, it is limited
by the width of the individual frequency channels, which inherently detain a small dispersion delay $dt(v)$:
\begin{equation}
  \label{delayEquation}
  dt(v) = 2(\mathcal{D}\cdot DM)\frac{dv}{v^3}
\end{equation}
Since the signal in each frequency channel has a differential dispersion delay of $dt(v)$, incoherent
dedispersion cannot localize a pulse better than this (for more details see \cite{Astropulse}). Coherent
dedispersion is an alternative technique which allows better time resolution by performing the mathematical
inverse of the ISM's dispersion operation. After measuring
the complex voltage $v\left( t \right)$ coherent dedispersion recovers the intrinsic complex voltage as it originated
from the source $v_{\text{int}}\left( t \right)$. This is then transformed into a real signal which is not affected by the
dispersive effect of the ISM.

Coherent dedispersion relies on the fact that the modification of the signal can be described as the work of a
'phase-only' filter, or transfer function.  In the frequency domain, for a signal centred at $f_0$ and
bandwidth $\Delta f$:
\begin{equation}
 V\left(f_0 + f \right) = V_{\text{int}}\left( f_0 + \Delta f\right)H\left(f_0 
+ \Delta f\right)
\end{equation}
where $V\left(f\right)$ and $V_{\text{int}}\left(f\right)$ are the corresponding Fourier transforms of the raw voltages
$v\left(t\right)$ and $v_{\text{int}}\left( t\right)$, which are non-zero for $|f| > \Delta f/2$. The delay
in the ISM depends on the frequency and path length travelled, $\Delta \phi = -k\left( f_0 + f\right)d$, where
$k\left( f \right)$ is the wavenumber and $d$ is the distance to the pulsar. From this it follows that the
transfer functions becomes
\begin{eqnarray}
 H\left( f_0 + f\right) & = & e^{-ik\left( f_0 + f\right)d} \\
                        & = & e^{+i\frac{2\pi \mathcal{D}}{(f+f_0)f^2_0}\text{DM} f^2}
\end{eqnarray}

This transfer function is determined for a given pulsar and its inverse applied to the measured,
Fourier-transformed voltage. The result is then transformed back into the time domain to obtain the
desired dedispersed signal. See \cite[section 5.3]{LorimerKramer2005} for an in-depth discussion of coherent dedispersion.

A Fourier Transform can be performed using the Fast Fourier Transform (FFT) algorithm, which has a time complexity of
$\mathcal{O}(NlogN)$ where $N$ is the number of input samples. This process has to be applied separately to each DM trial and all frequency channels (in cases where the observing band is very wide and thus must be
channelised into a number of narrower subbands), leading to a time complexity of $\mathcal{O}(N_f N_{\text{DM}}  NlogN)$, suggesting that this is a slower technique when compared to incoherent dedispersion. However, coherent dedispersion can test for more DM trials due to higher time resolution, and thus makes it suitable for such cases. Also, it is the algorithm of choice for pulsar timing observations.

\subsection{GPU Implementation}

\begin{figure}[t!]
\begin{center}
\includegraphics[width=400pt]{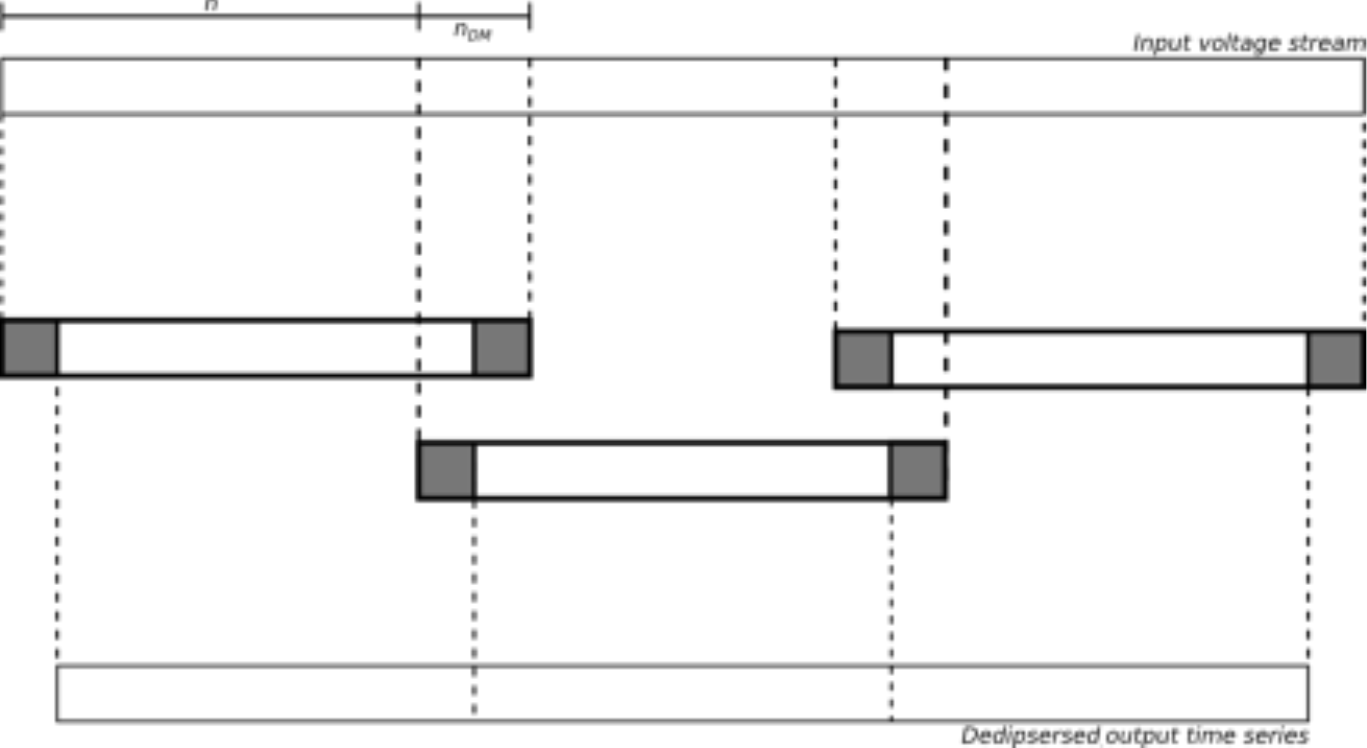}
\end{center}
\caption[Overlap-save method]{Schematic showing the overlap-save method, which is an efficient way of evaluating the
		 discrete convolution between a very long signal, the input complex voltages in this case, and an FIR
		 filter, here represented by the chirp function.}
\label{overlapSaveFigure}
\end{figure}

Several authors have investigated the applicability of the coherent dedispersion process to GPU processing
\cite{Allal2009,VanStraten2011}. Here we describe a simple implementation based on the overlap-save method, which
is an efficient way of evaluating the discrete convolution between a very long signal (the input complex voltages
in this case) and an FIR filter (the chirp function). Figure \ref{overlapSaveFigure} provides a visual schematic
of this technique. \cite{VanStraten2011} also use this scheme for performing coherent dedispersion within their software package, and present several benchmarks of their implementation. Due to the similarity of both implementations, these tests were not replicated. Here we simply outline the algorithmic design of our implementation.

The multiplication of the chirp function with the voltage stream in the Fourier domain corresponds to a convolution
in the time domain. The length of the voltage data for dedispersion needs to have a length of at least the dispersion
delay between the upper and lower edges of the bandpass $dt(v)$ (equation \ref{delayEquation}). For a bandwidth $dv$ which
is Nyquist sampled, this amounts to $n_{\text{DM}}=dt(v)\cdot dv$, where $n_{\text{DM}}$ represents the fact that this value is
DM-dependent. A discrete convolution of each point of a time series of length $n$ depends on $n/2$ points at both
ends of the buffer, requiring the input voltage series to be padded. Thus the shortest data set which can be
coherently dedispersed must be at least $2n_{\text{DM}}$ samples long. Long data sets are generally chosen to increase
efficiency and performance. For GPU implementations this will be generally limited by the amount of memory
available on the device and the performance of the FFT library being used. The 'wings' of length
$n_{\text{DM}}/2$ at the beginning and end of each data set need to be ignored after convolution.

In order to minimise GPU execution time, an optimal FFT size needs to be determined for a given dedispersion length.
This value is then used to fill up GPU memory using multiple overlapping blocks as per the overlap-save method. The
mapping between the input CPU buffer (left) and working GPU buffer (right) is defined below:
\begin{equation}
 N_c \cdot \left( (N_{f} - \frac{n_{\text{DM}}}{2}) \cdot N_B  + \frac{n_{\text{DM}}}{2} \right) \Rightarrow N_c \cdot N_{f} \cdot N_B
\end{equation}
where $N_c$ is the number of frequency channels, $N_f$ is the FFT size and $N_B$ is the number of FFT blocks to be
processed in parallel. Our GPU implementation consists of several processing steps:

\begin{enumerate}
 \item A buffer containing the (possibly coarsely channelised) complex voltages is copied to GPU memory. Each frequency channel is copied separately whilst expanding each $N_f$ set to overlap FFT execution.
 \item $N_c \cdot N_B$ $N_f$-point forward FFTs are performed in parallel, one for each channel-block pair, using the
       cuFFT\footnote{\;https://developer.nvidia.com/cufft} CUDA library. The FFT plan is created once during initialisation.
 \item The coherent dedispersion kernel is launched for execution, which calculates the discretely sampled chirp
       function on the fly and multiplies the coefficients with the Fourier components. The kernel configuration
       consists of a 2D grid, with Fourier components changing in the x-direction and frequency subband in the y-direction. Each thread is responsible for the same Fourier component across all blocks. With this scheme, the chirp coefficient for the frequency bin can be calculated once and stored in registers.
 \item The result is inverse Fourier transformed in place, resulting in $N_c \cdot N_B$ $N_f$-point buffers with        
        $n_{\text{DM}}/2$ extra points at both edges.
 \item The dedispersed buffer is copied back to CPU memory whilst simultaneously collapsing the overlap, thus removing
       the invalid points at the FFT edges.
\end{enumerate}

The computational bottlenecks in this implementation are the Fourier transform steps. The actual computational performance will depend on the library and implementation used. Several cuFFT benchmarks are readily available. These can be used to calculate the optimal FFT size, which depends on the dispersive smearing length within the lowest frequency channel, where this smearing is widest. Data movement to and from GPU memory can also pose a considerable execution bottleneck. For this reason, when deployed on a real-time system, data expansion should be performed on the GPU itself to reduce the input data rate. Depending on the observation requirements, the output data buffer can either be folded, which is generally the case for pulsar timing, or encoded to a lower bit rate.

\section{Conclusion}

Online fast transient detection pipelines need to dedisperse the incoming data 
stream in real time, generally for $\mathcal{O}(10^4)$ DM trials when performing 
a blind search. For this reason, fast implementations of dispersion removal 
techniques have been studied and proposed in recent years. In this chapter we 
demonstrate the applicability of GPUs for such systems. We implement the direct 
dedipsersion method and perform a performance benchmark and scalability study 
for typical usage parameters. This implementation forms part of the GPU-based, 
real-time pipeline which will be described in detail in the following chapter. 
We also describe a typical GPU implementation for coherent dedispersion 
systems.


\chapter{GPU-Based Transient Detection Pipeline}
\label{pipelineChapter}

Fast transient surveys generate a large volume of data, which is not cost-effective to store, and therefore have to be conducted in real-time, where data from a telescope, which is continuously observing the sky, is streamed to a processing backend. In section \ref{fastPipelineSection} we have described a generic fast transient search process, and in this chapter we introduce our GPU-based transient detection pipeline. We start off by examining the current state--of--the--art of similar pipelines, and then discuss in detail the various processing stages of our pipeline, including RFI mitigation, dedispersion, event detection and classification, as well as data quantisation and persistence. These stages are encapsulated as a standalone framework which can be used in offline mode, for processing archival data, as well as within an online application with additional real-time capabilities. It uses the incoherent dedispersion kernel described in the previous chapter. We conclude by analysing the clustering and classification stage of the pipelines, and produce performance benchmarks.

\section{Transient Detection Pipelines}
\label{otherPipelinesSection}

Conventional pulsar surveying machines were essentially baseband recorders, 
where raw voltage data were recorded to disk and then processed offline on 
conventional compute clusters (for example, see \cite{Voute2002,Stairs2002}). 
The need for real-time surveys has resulted in a conceptual shift, where most of 
the processing is now performed online and only data containing interesting 
events is recorded to disk. The new generation of surveying pipelines require a 
considerable amount of processing power, motivating the adoption and 
investigation of alternative compute platforms. Standard commercial, 
off-the-shelf (COTS) compute clusters are still popular. \cite{Roy2009} have 
developed a fully real-time 32 antenna, 32 MHz dual-polarisation backend for the 
GMRT, which acts as a real-time correlator and beamformer, as well as a baseband 
recorder. Extensions for online transient detection are then discussed in 
\cite{Bhat2013}.

The Advanced Radio Transient Event Monitor and Identification System 
(ARTEMIS\footnote{\;http://www.oerc.ox.ac.uk/research/artemis}) 
\cite{Serylak2012} was the first pipeline to explore the use of GPUs for 
transient detection systems, using the dedispersion kernel developed by 
\cite{Magro2011} and eventually optimised further by \cite{Armour2012}. This is 
implemented as a PELICAN pipeline. The Pipeline for Extensible Lightweight 
Imaging and CAlibratioN (PELICAN\footnote{\;https://github.com/pelican/pelican}) 
framework is an efficient, lightweight, C++ library for processing data in 
qausi-realtime that separates data acquisition from data processing and output, 
allowing the scalability and flexibility to fit in different scenarios. ARTEMIS 
was mainly developed for deployment on LOFAR international stations, where the 
coarsely channelised, beamformed data are streamed toward a number of servers, 
each processing a subset of the beams. The processing stages include: data 
reception and interpretation, RFI filtering, polyphase-filter bank 
channelisation, dedispersion and event detection. The dedispersion stage is the 
only one which is performed on GPUs. A similar system is being developed for the 
Parkes radio telescope \cite{Barsdell2011}.

The use of FPGAs for online transient detection is also being investigated by \cite{Addario2013}, who are developing an FPGA-based, real-time backend for ASKAP, to act as the CRAFT processing platform. They use a Pico Computing EX-500\footnote{\;http://picocomputing.com/ex-series/ex-500} backplane and up to 6 M-501\footnote{\;http://picocomputing.com/m-series/m-501} FPGA modules, 4 of which dedisperse and detect 9 of 36 304 MHz dual-polarisation beams, each channelised into 1 MHz frequency channels, over a range of 442 DM trials, with 0.9 ms time resolution. We compare this system with the one presented in this chapter in section \ref{comparisonLabel}

\section{Pipeline Overview}

We consider the case where beamformed data are processed to extract 
astrophysical radio bursts of short duration
and designed and implemented a generic, standalone, scalable, high throughput, GPU-based transient detection
pipeline which can be deployed on telescope backend processing servers for real-time transient surveys.
We extended the pipeline  design adopted in \cite{Magro2011}, described in detail in chapter \ref{dispersionChapter}, to include a fast buffering system, the possibility of writing streamed data to disk at different stages in the pipeline, the ability to process multiple beams concurrently and an online candidate selection mechanism to separate RFI events from astrophysical ones. The main design emphasis was high-performance and scalability across many beams.

\begin{figure}[t!]
\begin{center}
\includegraphics[width=420pt]{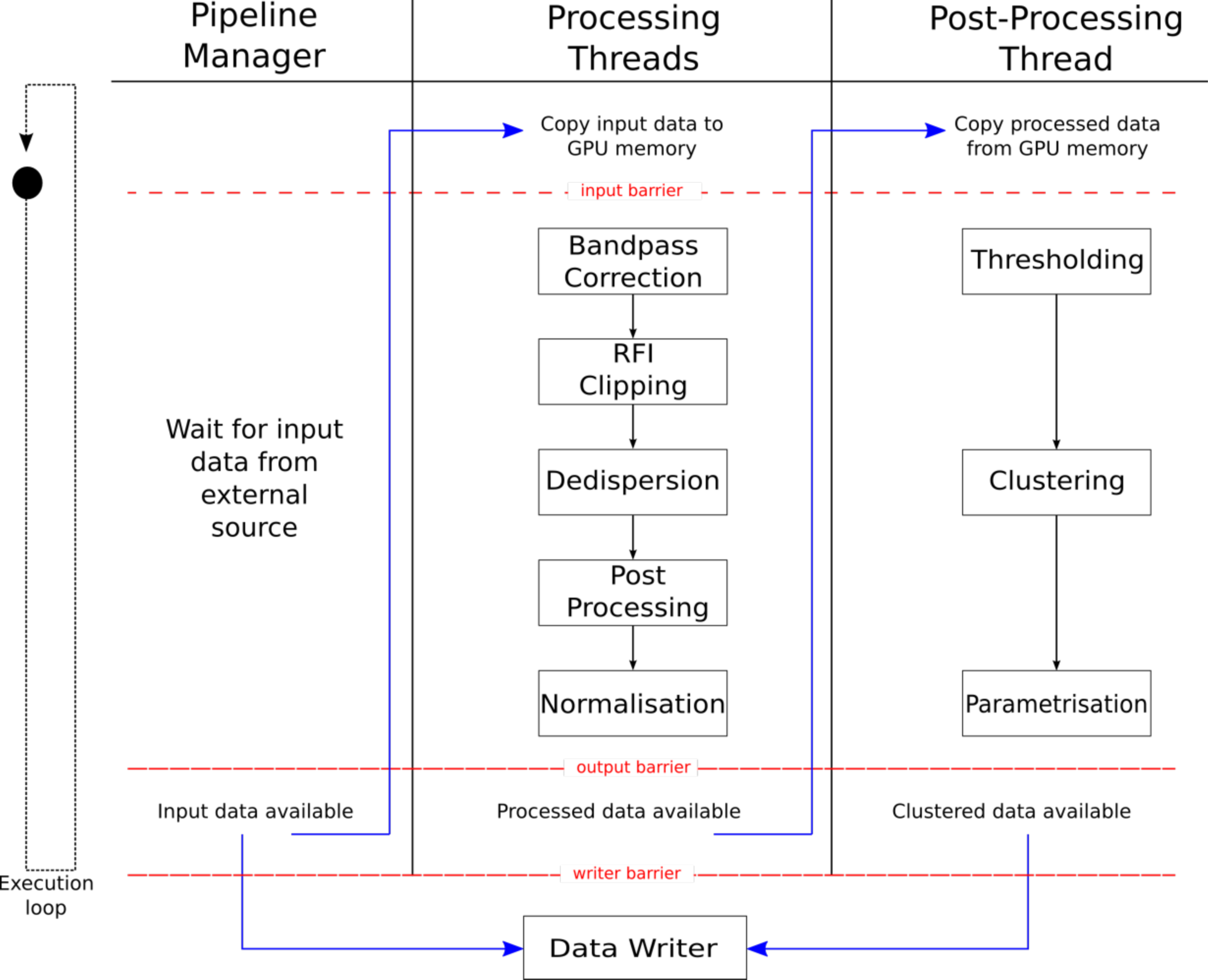}
\end{center}
\caption[Real-time transient pipeline schematic overview]{An extension of the architecture presented in figure
	 \ref{mdsmFlowFigure}. A processing thread is initialised for every input telescope beam which needs to
         be processed. These are processed independently and in parallel. The data writer can be triggered either
         from the pipeline manager in situations where the entire data series needs to be written to disk, or the post-processing
         thread when an interesting event is detected in the processed data. This will quantise, encode and write
         to disk the entire unprocessed input buffer, including data from all the beams being processed.
         Arrows in blue show data buffer movement (copies to/from GPU memory or across host memory buffers)
	 while the dashed red lines are CPU barriers required to synchronise the three processing stages.}
\label{pipelineFigure}
\end{figure}

The high-level architecture of this pipeline is depicted in figure \ref{pipelineFigure}.
Data acquisition is performed by a custom external data interpreter which forwards a data buffer
containing either filterbank data or a voltage series which has been channelised into finer frequency channels. 
This buffer can contain data from multiple telescope beams which are distributed to multiple processing thread instances, one per
thread. Each thread is associated with a single GPU device, however each GPU can have multiple processing threads
assigned to it. These threads execute independently from each other. Once data becomes available each thread is
responsible for copying the input buffer segment associated with it to GPU memory. This scheme maximises PCIe
bandwidth on multi-GPU systems. Once the copies finish the processing threads pass through several
processing stages: bandpass correction, RFI clipping, dedispersion and optional post-processing and normalisation.
The CPU imprint of these threads is almost negligible and mostly consists of CUDA kernel timing and synchronisation,
as well as a one off $p^\text{th}$ order polynomial fit over $N_{c}$ values per iteration, where $N_c$ is the number of
frequency channels. The dedispersed time series are then copied back to GPU memory and passed to the post-processing
thread where they are thresholded, clustered and classified. Any data points belonging to interesting clusters
are written to disk, together with the unprocessed
data buffer after being quantised to 8 or 4 bits. There is also the possibility of writing the entire data stream
to disk, including buffers without interesting events, after passing through an encoding and quantisation stage, provided that the disk drives can manage the data rate. All operating parameters are provided by an XML configuration file.

This architecture has several in-built assumptions and limitations, all of which help to speed up certain aspects
of the workflow. The pipeline works on channelised data which are stored in 
beam/channel/time order in GPU memory, with time changing the fastest, and thus 
assumes that the data have been channelised and transposed during an earlier 
stage. For real-time systems the transpose is performed in chunks by the 
external data interpreter, as described in
section \ref{packetReceiverSection}, whilst channelisation is assumed to have been performed on the digital backend
prior to packetisation and streaming. A GPU implementation of a polyphase filter bank channeliser to generate
finer frequency channels would be a required addition to the pipeline in cases where the backend lacks this functionality, or only coarse-grained frequency channels can be streamed, as this would severely limit the sensitivity of transient searches due to signal smearing (for example, see \cite{Veldt2011}).

Beams are assumed to be identical (same observation and survey parameters), however this can easily be extended to allow the possibility of performing different operations on each beam. Also, having the capability to write data to disk from either the pipeline manager or post-processing thread requires a considerable amount of host memory, since an input buffer needs to remain accessible for three iterations. This feature is implemented as a triple buffering scheme, with buffer pointers updated after each iteration.

All GPU buffers are allocated once during the initialisation phase (per processing thread). The input and
output buffers are recycled across different kernel launches to maximise use of available GPU memory, which is
a limiting resource. Also, even though we could get a potential speedup and possibly a reduction in memory requirements, by combining some of the GPU kernels into monolithic functions, separating functionality this way makes it easier to add intermediary stages as well as disabling certain functionality when not required.

\section{RFI Mitgiation}
\label{rfiMitigarionSection}

Terrestrial RFI, and methods to try and mitigate or correct for it, is one of the major problems in transient surveys,
especially for radio telescopes which are close to urban centres. Several methods have been
investigated to cater for this (see \cite{Hodgen2012} for a comparison of RFI mitigation techniques for dispersed
pulse detection), the major challenge being to try and remove as much RFI as possible without affecting any true astrophysical dispersed pulses which might be present in the data.
For our pipeline we implement a simple RFI-thresholding process which limits the amount of strong, bursty RFI,
whose effectiveness can be tweaked by applying different thresholding factors. Any undetected RFI will result
in incorrect detections after dedispersion, and although these are never welcome, it would be more advantageous to allow some low-power RFI to seep through rather than increasing the probability of clipping astrophysical events. The detection and classification stage should then be able to discern between real and terrestrial signals. Not performing any RFI mitigation would result in a large number of incorrect detections as well as a higher variance in the data, which lowers the probability of detecting weak astrophysical signals. The thresholding process consists of three stages as described below.

\subsection{Bandpass Correction}

A bandpass is a telescope's response function across the frequency band being observed. For an ideal telescope
this would be flat, however several effects, including the behaviour of the channelisation algorithm used,
can result in different bandpass shapes. To properly mitigate narrowband RFI the bandpass shape needs 
to be modelled in order to detect isolated high power frequency channels. This process can either be performed offline
prior to the observation, after which the model parameters are passed to the pipeline, or online where the bandpass
is computed once per data buffer being processed. The former scheme is ideal for telescopes which have a well behaved and time invariant response function (although the power levels will change during the observation with variations in 
sky and telescope noise) whilst the latter is beneficial for experimental setups where the bandpass can change during
the observation. Another approach is to model the bandpass during the first iteration in an online observation and then add weighted incremental updates to the model
during the observation. For our implementation we decided to perform a bandpass fit at every iteration due to the dynamic nature of the setup at the BEST-II array (as described in section \ref{medicinaObservationSection}).

Once a data buffer is available in GPU memory a $p^{\text{th}}$-order polynomial is fitted to an averaged bandpass, 
providing a smoothed, approximate description of the telescope's response to the frequency band being 
processed. Accumulation and averaging is performed on the GPU, mainly consisting of a reduction sum across the 
frequency channels, resulting in $N_c$ values which are copied back to CPU memory. The resulting bandpass is then fitted using least-squares. The root mean square error (RMSE) between the fitted bandpass and the averaged, interpolated bandpass is then computed and used as a thresholding factor for the channel thresholding stage, whilst the mean and standard deviation are used for the spectrum thresholding stage. The fitted bandpass is then subtracted from all the spectra on the GPU, generating corrected spectra, having the effect of equalising the telescope's response, as well as moving the mean to, or near, 0.

Online, iteration-based fitting can be ineffective if significantly powerful narrowband RFI is present,
as this will distort the fit and generate incorrect thresholds. Such channels are usually intermittently
or permanently on, depending on the source of the noise. For this reason these channels need to be masked prior to fitting
and their values interpolated from neighbouring channels when calculating the averaged bandpass.
Also, a smooth telescope response function can generate a very smooth averaged bandpass when accumulated over a
large time frame (order of seconds, depending on the sampling rate) resulting in a very low RMSE value after
fitting. In these situations using the mean and standard deviation of the fit can generate better thresholds.

\subsection{Channel Thresholding}

Each frequency channel should have a uniform power level across short time spans. Sudden changes can generally
be attributed to narrowband RFI (or alternatively, a very strong astrophysical burst). Strong, short duration, 
narrowband signals can result in incorrect detections across a wide DM range after dedispersion. If these signals
are frequent and occur throughout the frequency range they could lead to detections exhibiting similar properties
to dispersed pulses. For this reason such signals need to be removed from the data prior to dedispersion. If the 
RFI environment is properly modelled then persistently noisy channels can be 
masked before an observation. Also, if the duration of short duration signals is 
known a priori then a windowed thresholder can be implemented with window 
length, $W_c$, matching the signal's duration. When applied to the data buffer 
any such window having a mean value greater than the channel's threshold can be 
removed. Such a scheme will also remove longer duration signals in segments of 
length $W_c$. A small $W_c$ can be used to implement a more generalised 
thresholder based on the same principle, where long duration signals will be 
removed one window length at a time. An optimal value for $W_c$, as well as the 
associated threshold, will depend on the RFI environment on-site, the width of 
astrophysical transients being searched for and the observing  parameters, and 
can be empirically set through test observations.

We implemented this functionality using two successive GPU kernels. The first kernel applies a non-overlapping 
sliding window to each frequency channel, thereby partitioning them into chunks of width $W_c$. The mean of each
chunk is calculated and compared to a pre-calculated channel threshold. If this threshold is exceeded then it is 
flagged for removal. Neighbouring blocks are also flagged so as to include the edges of long signals which might not
be aligned with the partitioned grid. The list of flags is then used in the second kernel which replaces the values
of flagged chunks with the frequency channel's fitted bandpass value. This is preferred to setting these values to 
0 since the statistical properties of the data are maintained. For every frequency channel, this implementation takes the form of
\begin{equation}
 \forall_c \in C_v  : M(c) < B_\sigma \Rightarrow (\forall_i \in c: c_i = B_v)
\end{equation}
where $C_v$ is the set of partitions of length $W_c$ for frequency channel $v$, $M(c)$ is a function which returns the mean of the partition, $B_\sigma$ is the threshold value and $B_c$ is the fitted bandpass value for channel $v$.

\begin{figure}[t!]
\begin{center}
\includegraphics[width=400pt]{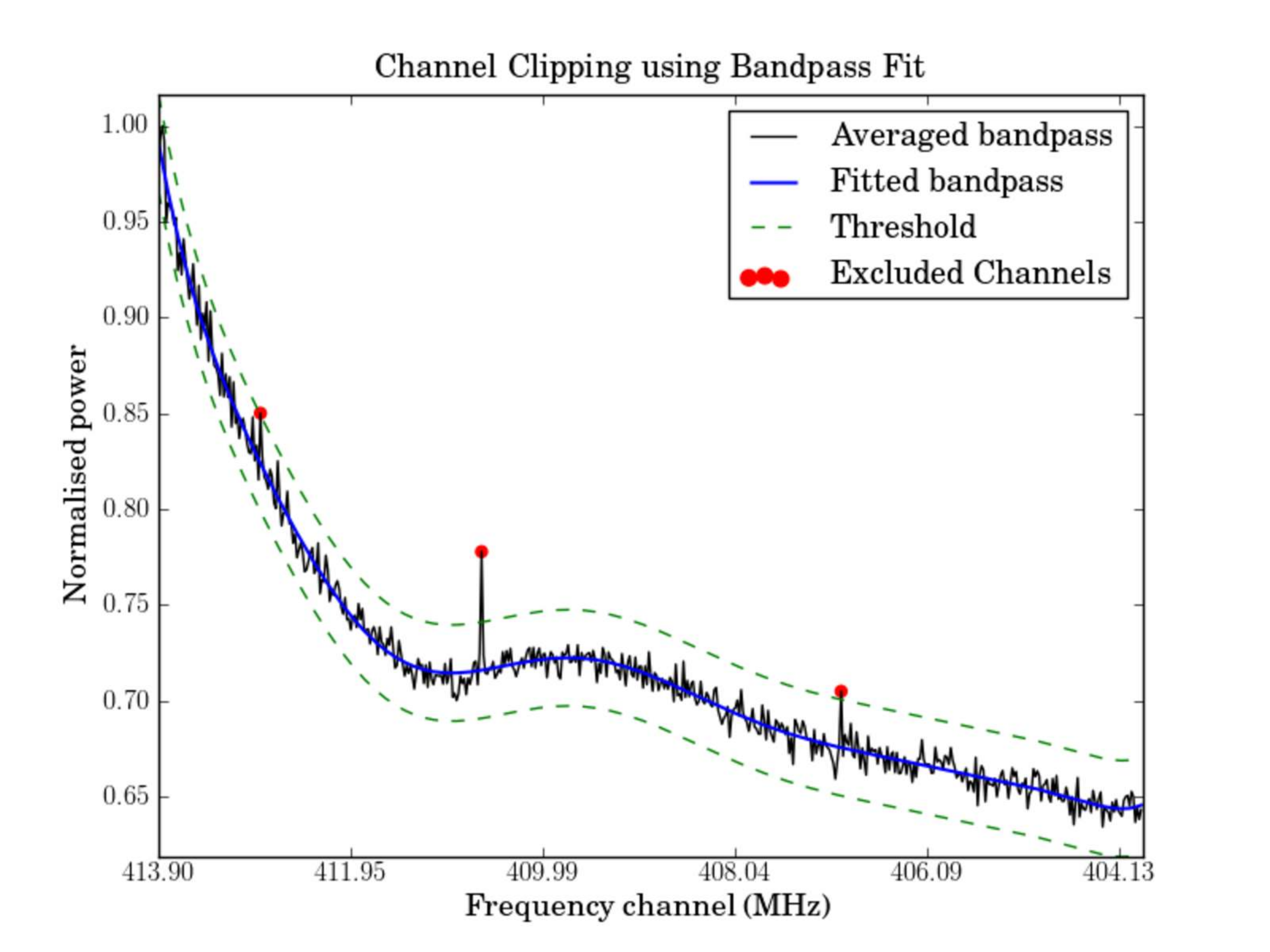}
\end{center}
\caption[Bandpass fitting example]{The averaged bandpass (black) is fitted with a $p^{\text{th}}$ order polynomial (blue) 
          and the RMSE of the two is used to generate channel clipping thresholds (green). This threshold is then 
          applied to subsets of spectra and any chunk exceeding it will be replaced with the channel's 
          fitted bandpass value, as is the case for the channels marked in red in this plot. }
\label{bandpassFitFigure}
\end{figure}

Figure \ref{bandpassFitFigure} shows a one iteration snapshot during bandpass 
fitting, where the averaged bandpass is fitted with a $p^{\text{th}}$ order 
polynomial and the RMSE of the two is used to generate the channel clipping 
thresholds in green. This threshold is then applied to subsets of spectra of 
size $W_c$ and any chunk exceeding it will be replaced with the channel's fitted 
bandpass value, as is the case for the channels marked in red in this figure. 
The lower threshold is required for when input data are lost, such as packet 
loss in real-time mode.

\subsection{Spectrum Thresholding}

Broadband RFI generally has the least impact on the dedispersion process as most of the detections will be
concentrated at a DM of 0 pc cm$^{-3}$, which can however span over a wide range for long duration pulses.
The general mitigation mechanism for broadband pulses is to ignore all detections for DM values less than a lower limit $D_{\text{min}}$ We decided to implement a simple spectrum thresholding mechanism which removes strong broadband pulses to reduce the  number of detections that need to be clustered in the post-processing stage. This process is performed on the GPU  as well. The mean of each full-band spectrum is calculated and compared to a pre-calculated
spectrum threshold. If the mean exceeds this threshold then the entire spectrum is replaced with:
\begin{equation}
  \forall_c \in C : S_c = S_c - (S_c - B_c)
\end{equation}
where $C$ is the set of all channels, $S$ is the entire spectrum and $B$ is the fitted bandpass. This has the effect
of removing the noise induced by RFI whilst maintaining the background properties of the affected signal.
Dispersed pulses should not be affected by this thresholding, unless the threshold is
exceptionally low, since their power is distributed across multiple spectra 
shifted depending on the pulses' DM value. The threshold is empirically set such 
that time spectra are rarely affected by this stage, although
this will depend on the RFI environment.

\subsection{RFI Mitigation Test}

\begin{figure}[t!]
  \begin{center}
    \includegraphics[width=300pt]{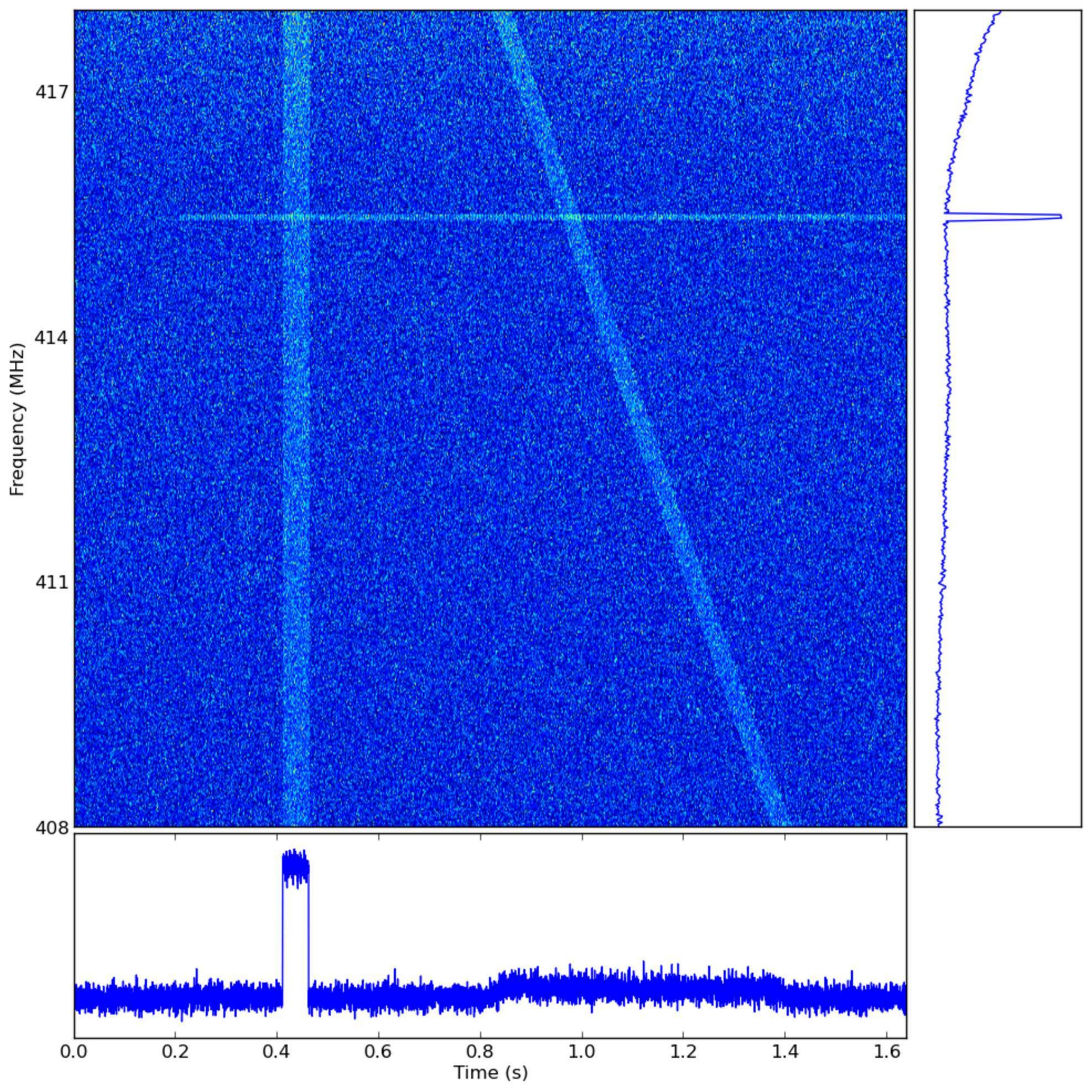}
  \end{center}
  \caption[Test case for RFI mitigation]{Simulated data file containing narrowband, broadband and dispersed signals, used to test the RFI mitigation stage. All these signals are clearly visible in the central waterfall plot, while the accumulated bandpass is shown on the right and summed bandwidth in the bottom plot.}
  \label{fakeRFIFigure}
\end{figure}

\begin{figure}[t!]
  \begin{center}
    \includegraphics[width=300pt]{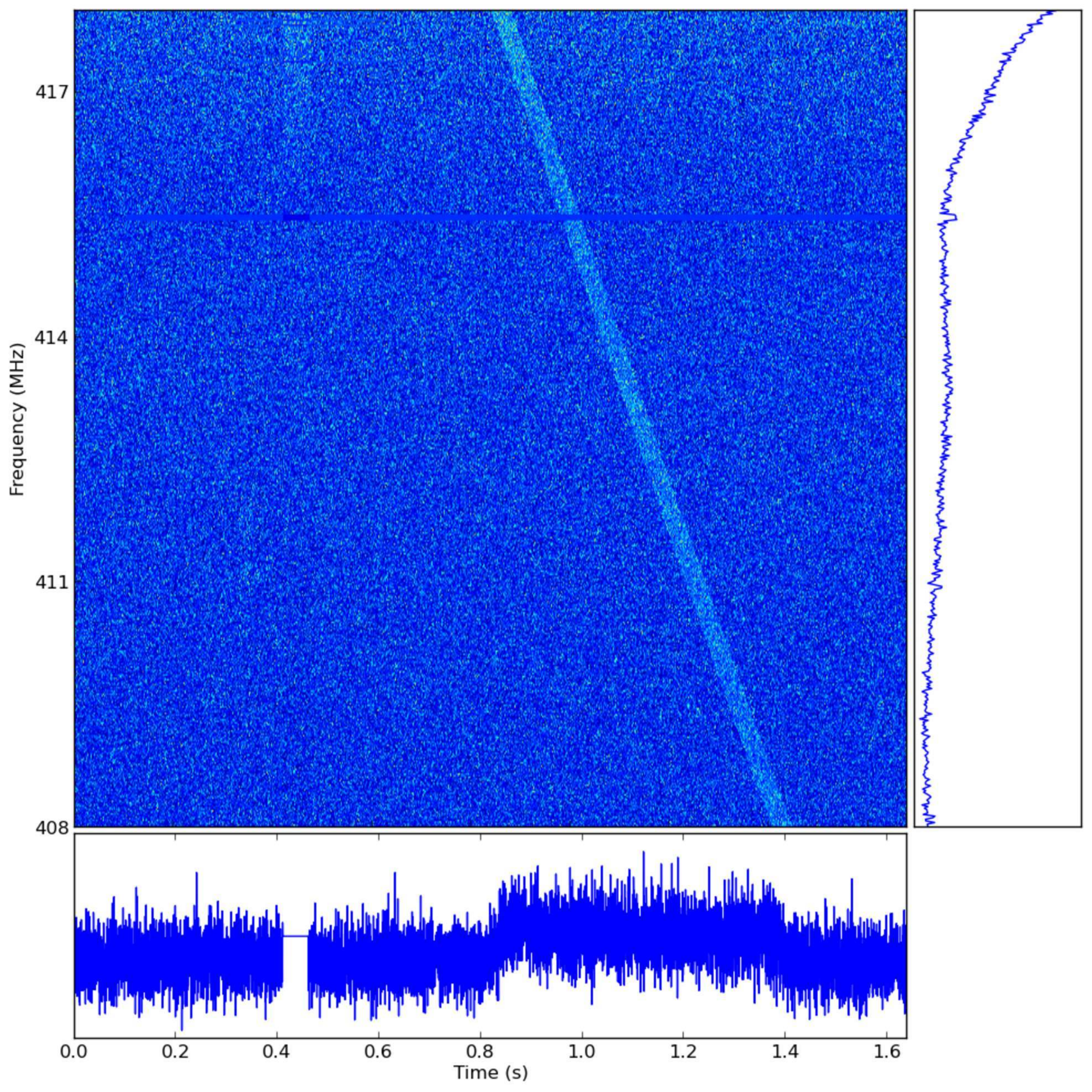}
  \end{center}
  \caption[RFI mitigation test case output]{The simulated data from figure \ref{fakeRFIFigure} was passed through
       the RFI mitigation stage and the resultant output is shown in this plot. Both the broadband and narrowband signals were mitigated by the implemented algorithms, whilst the dispersed pulse was left untouched.}
  \label{cleanFakeRFIFigure}
\end{figure}

To demonstrate the effectiveness of the implemented RFI mitigation algorithms, a 
simulated filterbank dataset containing a dispersed pulse with a DM of 120 pc 
cm$^{-3}$, as well as narrow and broadband RFI, was generated. The simulated 
observation is centred at 413 MHz having a bandwidth of 10 MHz channelised into 
512 frequency channels. 1.6 s of random noise with $\mu=0$ and $\sigma=1$ were 
first generated to which a 10 ms square, dispersed pulse having a S/N of 10 was 
added. Narrowband RFI (78 kHz, periodic signal with a period of 6.5 ms) and a 
broadband 2 ms pulse with an S/N of 10 were also added. Finally the data were 
multiplied by a bandpass shape modelled on the BEST-II beamformer output 
(discussed in the next chapter). 
Figure \ref{fakeRFIFigure} shows the resulting waterfall plot, together with the 
accumulated bandpass and summed bandwidth. The simulated data were then passed 
through the RFI mitigation stages. The dispersed pulse didn't trigger any 
thresholding mechanism, whilst the narrowband RFI was replaced with the value of 
the fitted bandpass at the noisy frequencies. The broadband pulse was replaced 
by the value of the noisy signal, its difference from the bandpass values 
subtracted. A channel block length of 512 was used, while the channel threshold 
was set to 7 and spectrum threshold to 5. The resulting output is shown in 
figure \ref{cleanFakeRFIFigure}

\section{Dedispersion}

Dedispersion is by far the most time consuming process in the entire pipeline. Its implementation underwent a
series of optimisation runs with evolving GPU architecture, as discussed in chapter \ref{dispersionChapter}, where
implementation details, as well as performance benchmarks, were presented. When integrating this kernel with a
real-time pipeline several factors need to be taken into consideration, including how to maximize the use of 
available GPU memory and how to handle inter-buffer overlaps due to DM shifts.

Maximising the number of spectra which can be processed in a single iteration increases the compute to copy time
ratio since high PCIe transfer rates can only be achieved with large transfer sizes. The dedispersion kernel is the
only one which requires a separate output buffer to store the dedispersed time series, all the other stages
process the buffer in place. A simple calculation can be used to compute this value:
\begin{equation}
 N_s = \frac{M \cdot \nicefrac{b}{32} - (\Delta t_{\text{max}} \cdot N_c) - \alpha}{N_{\text{DM}} + N_c}
\end{equation}
where $N_s$ is the number of time samples, $N_c$ is the number of frequency channels, $N_{\text{DM}}$ is the number of DM 
trials, $M$ is the amount of GPU global memory in 32-bit words, $\Delta t_{\text{max}}$ is the dispersion shift in 
time samples between the edges of the observing band for the largest DM value, which will be referred to as maxshift
throughout this chapter, $b$ is the input bitwidth and $\alpha$ represents other smaller buffers which are required 
(such as the shift buffer for dedispersion and fitted bandpass buffer for bandpass correction). 

Maxshift represents the extra number of spectra required to process the data in one iteration. These extra spectra
still need to be dedispersed, so during the following iteration they are first copied to the beginning of the buffer
and the next input buffer is appended to it. Due to the fact that the data have 
been transposed, this actually needs to be performed for each
frequency channel separately, so a series of asynchronous memory transfer calls are issued. Also, in order to
avoid extra computation in the dedispersion kernel the shifts in unit sample intervals per DM and frequency channel are computed
once during thread initialisation and transferred to GPU memory. These shifts are stored in channel/DM order so as 
to assure coalesced memory access.

\section{Signal Post-processing}

The dedispersed time series are then passed through a post-processing stage, 
which smooths and normalises
the data. This is mostly done using two techniques:

\begin{description}
 \item{\bf Median Filtering: } Noisy, isolated outliers in the time series can be suppressed by replacing the
	    value $y_m$ with one which is calculated using neighbouring points:
	    \begin{equation}
	      \label{medianFilterEquation}
               y_m = operation\{x_i, i \in w\}
	    \end{equation}
	    where $w$ represents a neighbourhood centred around the location $m$, and $operation$ is the averaging
            function used. We chose to use the median of the neighbourhood points, since it better preserves the
            edges of the signal than the mean. We have implemented a windowed-median filter on the GPU, where thread
            blocks are split across a 2D grid with each row processing one dedispersed time series, for a single
            DM value, partitioned along the columns of the grid. Each thread block loads a subset of the series to shared memory, including some overlapping values at the edges, and then each thread computes the median of its associated data point neighbourhood and stores this value back to global memory. Finding the median of
            a sequence of values requires a partial sort (of length $w$) so it can be inefficient since thread warps will generally have to divert branch execution during comparisons. A $w$ value of 5 or 7 generally gives a good ratio between performance and filtering efficiency, however this also depends on the time resolution. It should be noted that median filtering will limit the detectability of short duration and unresolved signals, with widths close to the sampling time.
            
 \item{\bf Detrending and Normalisation:} The mean power of the incoming data stream can change gradually in time, the
            rate of which depends on the cause of this change, such as steady temperature changes in telescope electronics, sky temperature variations, or an astrophysical radio source moving toward, or away from, the beam centre. This effect can be alleviated by subtracting a best-fit line to the dedispersed time series, which effectively centres the series to a mean of 0. The detrending process is also performed on the GPU where thread blocks are assigned a single dedispersed time series to process, which essentially compute a best-fit line using linear regression. The kernel requires two passes of the data, one to calculate the regression parameters  and another to subtract the trend-line from the series. This is performed by having the thread block move in a non-overlapping window fashion across the series, with each thread accumulating values locally. The final value is then computed collaboratively using a reduction sum mechanism. During the second pass the standard deviation is also computed, so by adding a third pass the data can be normalised and prepared for thresholding.
	    
\end{description}

After this stage, the post-processed dedispersed time series are copied back to CPU memory and forwarded to the event
detection stage, which is started as soon as all the beams have been processed (and a new input data buffer is available
for processing).

\section{Detection and Candidate Selection}
\label{detectionSelectionSection}

During the first stage of event detection, the dedispersed time series are thresholded using a suitable threshold value ($n\sigma$). This value should be low enough to allow low S/N pulses to pass through, even at the cost of incorrect RFI detections, which will be filtered in the clustering stage. A list of detections is generated for each beam, containing (time, DM, intensity) triplets. Astrophysical transients, as well as RFI signals, will result in a number of entries in this list which should be grouped together and treated as a single candidate. This is the main function of the clustering stage.

We start by applying a density-based clustering technique, DBSCAN \cite{Ester1996}, to group neighbouring data points together. Its definition of a cluster is based on the underlying estimated density distribution of the dataset. The shape of the clusters is determined by the choice of the distance function for two points. The {\it Eps-neighbourhood} of a point $p$, denoted by $N_{Eps}(p)$ is defined by $N_{\text{Eps}}(p)=\{q\in D\; |\; dist(p,q) \leq Eps\}$ where $Eps$ is an argument defining the neighbourhood extent of a point, $D$ is the set of points and $dist(p,q)$ is the distance function for points $p$ and $q$. DBSCAN distinguishes between two types of cluster points, those inside the cluster ({\it core points}) and other at the border of the cluster ({\it border points}) which generally have less points in their neighbourhood. A point $p$ is {\it directly density-reachable} from $q$ if $p \in N_{\text{Eps}}(q)$ and $|N_{\text{Eps}}(q)| \ge$ MinPts, where MinPts is the minimum number of data points. A point $p$ is {\it density reachable} from a point $q$ if there is a chain of points $p_1,...,p_n$, $p_1=q$, $p_n=p$ such that $p_{i+1}$ is directly density-reachable from $p_i$. Border points within the same cluster
might not be density-reachable from each other, but they are {\it density-connected} if there is a point $o$ such that both of them are density-reachable from $o$. Following these definitions, a {\it cluster} is defined as a non-empty subset of $D$ satisfying the following two conditions:

\begin{itemize}
 \item  {\it Maximality}: $\forall p,q : p \in C \wedge (q\mbox{ is density-reachable from } q) \Rightarrow q \in C $
 \item {\it Connectivity}: $\forall p,q \in C : p\mbox{ is density-connected to } q$
\end{itemize}

Any points which do not belong to a cluster are regarded as noise: $N = \{p \in D\;|\; \forall i:p \notin C_i\}$, where $N$ is the set of noise points, $C_1,...,C_k$ are the clusters in $D$ and $i=1,...,k$ with $k$ being the total number of clusters. This technique has several advantages  which makes it a suitable candidate for clustering dedispersed thresholded detections:
\begin{enumerate}
 \item it does not require a seed to specify the number of clusters in the dataset, which is a desirable property since the number of events occurring within a time-frame, be they astrophysical or RFI, is unknown unless a cluster approximation step is performed beforehand
 \item it has a notion of noise
 \item it is also capable of separating overlapping clusters having different density distributions, such as when an RFI signal overlaps a transient event, if the input arguments are sensitive enough
\end{enumerate}

\begin{figure}[t!]
  \centering
  \includegraphics[width=340pt]{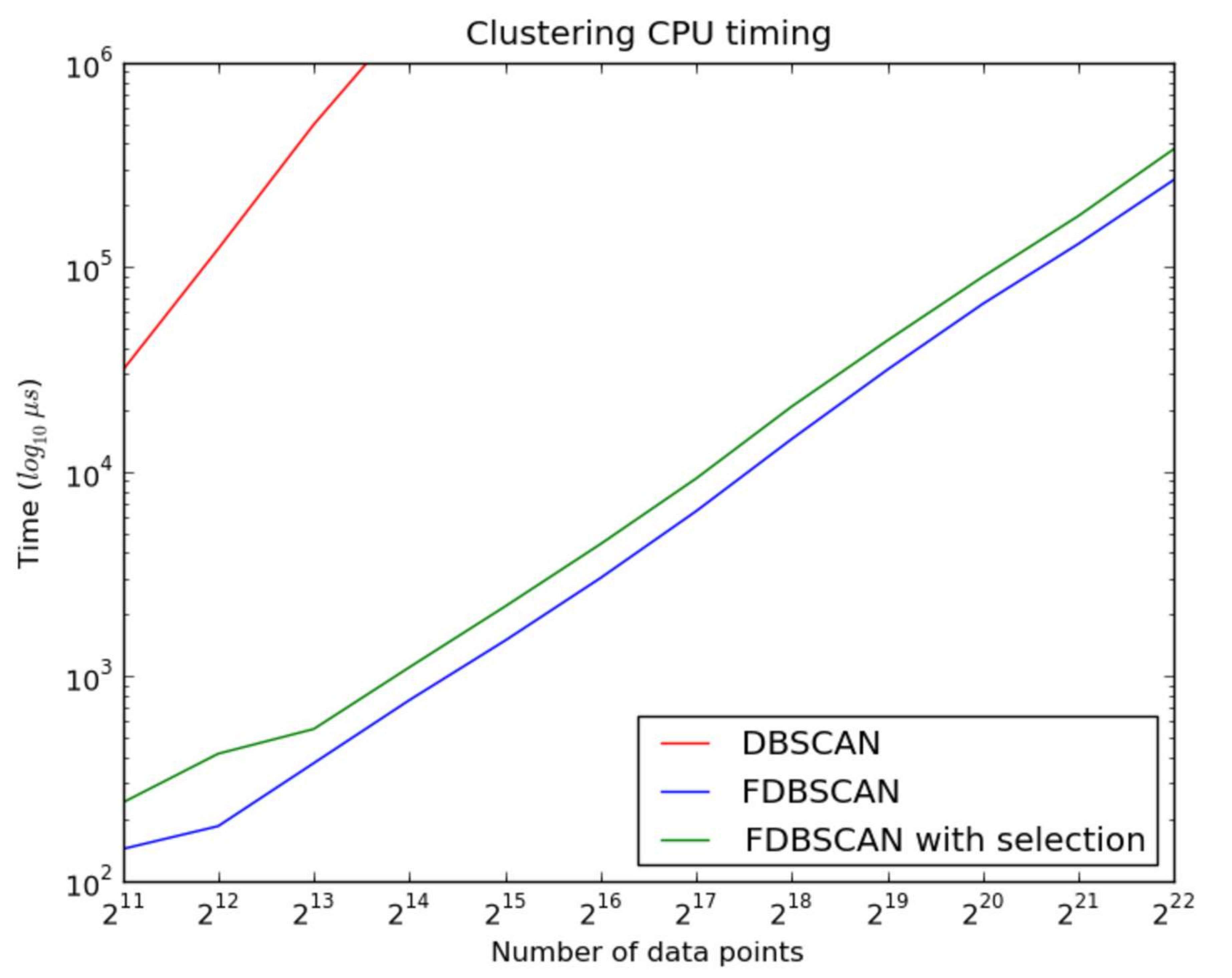}
  \caption[DBSCAN implementation performance]{ The significant difference in runtime between DBSCAN and FDBCAN, with candidate  selection only increasing this by a small factor.}
  \label{dbscanFigure}
\end{figure}

The main drawback of this technique is its runtime complexity, dominated by the neighbourhood calculation for every point, which is of order $\mathcal{O}(N^2)$ unless an indexing structure is used or a distance matrix is computed beforehand, however this needs $\mathcal{O}(N^2)$ memory, which is unfeasible for large datasets. To counter this we implement a faster, albeit less accurate, version of the algorithm, FDBSCAN \cite{Zhou2000}, which only uses a small subset of representative points in a core point's neighbourhood as seeds for cluster expansion, reducing the number of region query calls. The representative points are chosen to be at the border of a core point's neighbourhood, two for each dimension, one for each direction when placing the core point at the origin. Due to this approximation, some points might be lost, and in some cases clusters might be split apart, however the probability of this happening is very low (see \cite{Zhou2000}).

The distance function used in our implementation assigns a different value to each dimension, and thus requires three different values for $Eps$: ($T_\epsilon$, $S/N_\epsilon$, $DM_\epsilon$). Detections after dedispersion will have a specific shape along the time dimension due to incorrect dedispersion, with higher DM values detecting events before lower ones. The width of a cluster will also reflect the true pulse width, being narrower near the true DM. The value of $T_\epsilon$ should depend on the lowest pulse width being searched for. A higher $T_\epsilon$ might result in pulses close in time to be fused together into one cluster. The value of $S/N_\epsilon$ and $DM_\epsilon$ should be large enough to allow clusters to encompass the entire range, whilst allowing DBSCAN to distinguish between core and border points. 

The output of the clustering stage is a list of detected clusters. The main challenge is then to discern between RFI-induced clusters and transient candidates. In the case of transients the highest S/N detections will be centred around the pulse's true DM, diminishing in power when moving away from this value until the threshold level is reached due to dedispersion at an incorrect DM value. Following \cite{Cordes2003}, assuming a rectangular bandpass function, which should resemble the telescope's bandpass shape after bandpass correction, a Gaussian-shaped pulse with a width at FWHM of $W$ in milliseconds, the ratio of the measured peak flux density $S(\delta \text{DM})$ to true peak flux $S$ for a DM error $\delta$DM is

\begin{equation}
 \frac{S(\delta \text{DM})}{S}=\frac{\sqrt{\pi}}{2}\zeta^{-1}\erf\zeta
\label{incorrectDedispersion}
\end{equation}
where
\begin{equation}
 \zeta = 6.91\times10^{-3}\delta \text{DM} \frac{v_{\text{MHz}}}{W_{\text{ms}}v^3_{\text{GHz}}}
\end{equation}

Comparing this model with a cluster's DM-S/N signature provides us with a classification mechanism. Our implementation
performs the following steps for each detected cluster:

\begin{enumerate}
 \item Generate its DM-S/N signature by collapsing the time dimension.
 \item Smooth this signature by running a $k$-element moving average.
 \item Find the DM value containing the highest number of detections and associated maximum S/N. If this DM value is less than 1.0 then it is assumed that it was caused by broadband RFI and the cluster is discarded.
 \item Approximate the pulse's FWHM by computing the difference between the pulse's start and end time at the maximum S/N value.
 \item Normalise DM-S/N signature.
 \item Compute the analytical curve for incorrect dedispersion using equation \ref{incorrectDedispersion}.
 \item Calculate the RMSE between the modelled curve and pulse's DM-S/N signature.
 \item If the RMSE exceeds a preset threshold, then the cluster is discarded (RFI), otherwise it is classified as a potential candidate. This threshold can be set empirically through test observations or by using simulated data.
\end{enumerate}

This procedure is most effective for detecting relatively strong pulses and differentiating them from RFI. The 
classification of clusters having a small number of detections can be incorrect if the width of the pulse cannot be 
determined. To increase the likelihood of detecting lower-S/N pulses, the detection threshold during the first stage 
of event detection should be lower than typically used for similar searches in order to allow more pulse detections to 
be clustered together. This will result in a higher number of background noise detections, however these will be 
filtered out by DBSCAN.

\section{Data Persister}

The pipeline can either write the entire incoming data stream to disk or dump 
data buffers containing interesting detections when triggered from the event 
detection stage, for future off-line processing. If the incoming data rate being 
processed is higher than the attached disk drives' write speed then the data 
have to be quantised first. The quantisation mechanism was developed for the 
pipeline's deployment at the Medicina BEST-II array (discussed in chapter 
\ref{medicinaChapter}) so the accepted data formats are currently limited to two 
types: signed 16-bit complex channelised voltage series which are quantised to 
signed 4-bit complex values, with two complex components packed into one byte, 
and the equivalent detected unsigned 32-bit single-precision floating point 
power series which is quantised to unsigned 8-bit bytes (\cite{Backer1997} have 
thoroughly discussed the effect of quantising a data stream down to 4-bits).

\begin{figure}[t!]
\begin{center}
\includegraphics[width=340pt]{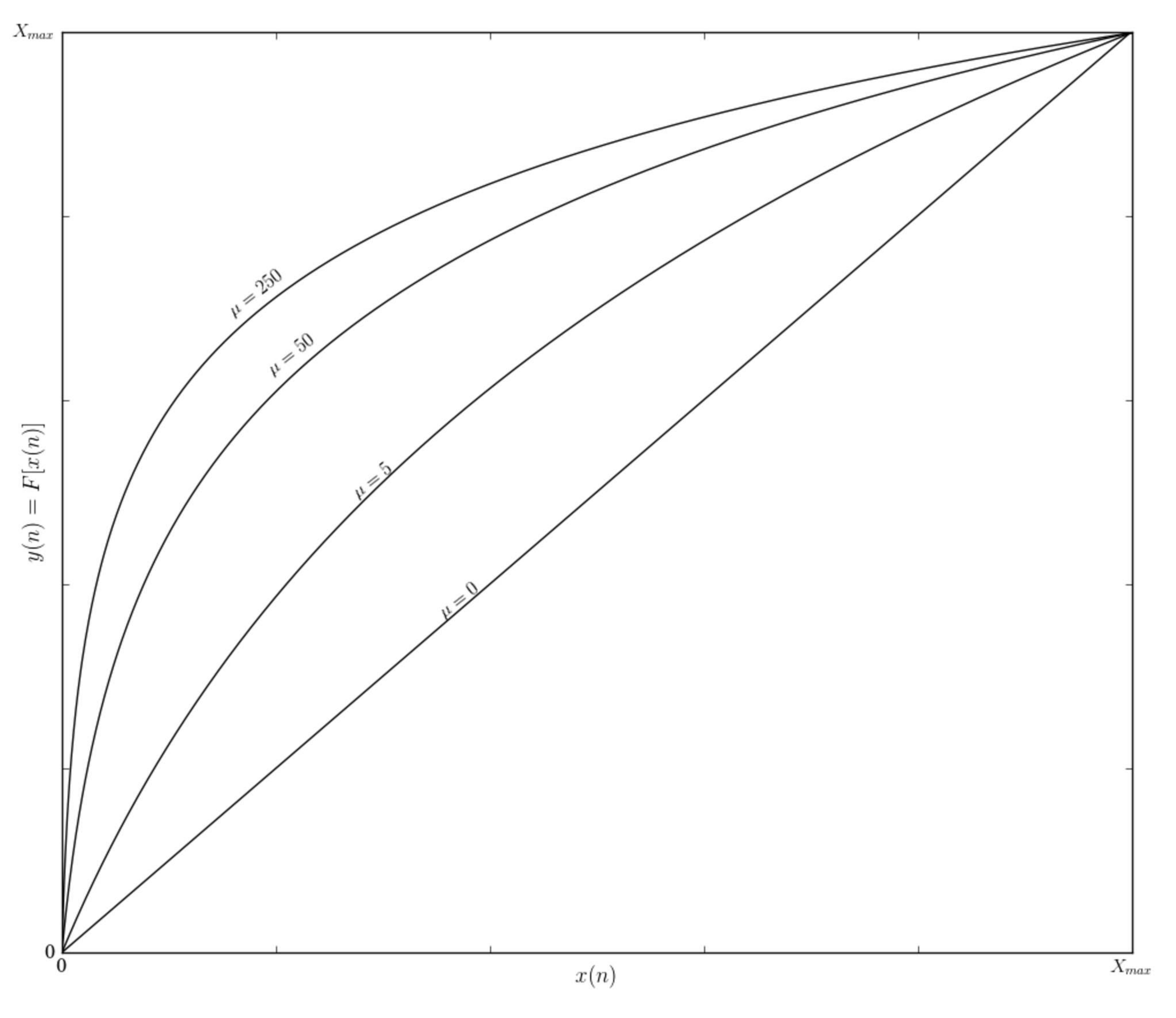}
\end{center}
\caption[Effect of compression factor]{The mapping between input and output values for a given range when applying the
	   $\mu$-law algorithm}
\label{compressionFigure}
\end{figure}

\begin{figure}[t!]
\begin{center}
\includegraphics[width=400pt]{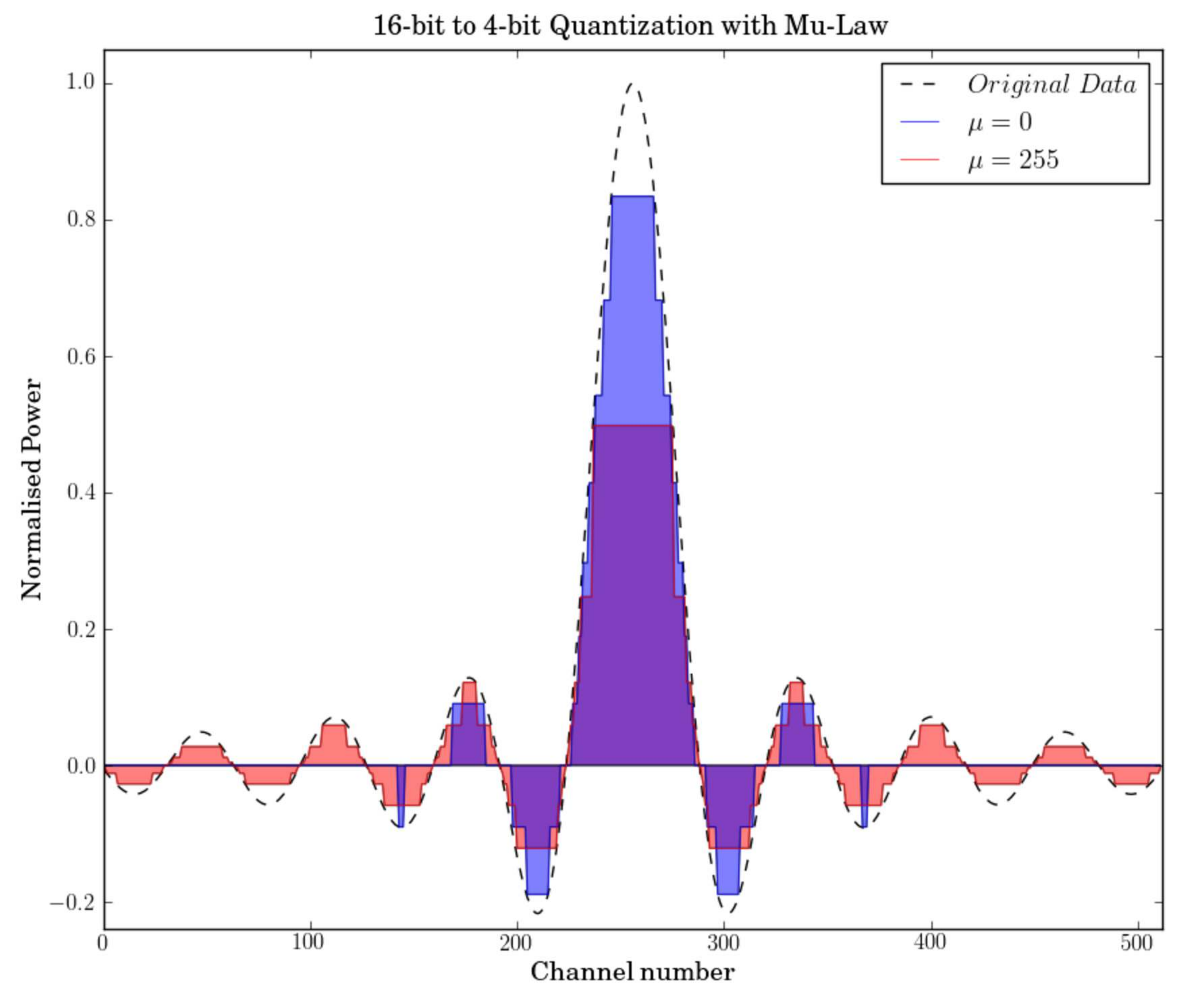}
\end{center}
\caption[$\mu$-law 16-bit to 4-bit quantisation]{The effect of the compression factor has when quantising signed 16-bit         input values down to signed 4 bits. When the compression factor is high most of the detail in the trailing      curve is retained, at the expense of losing information at higher values. Applying this to channelised raw      voltages would have the effect of allocating more bits to the region around the mean and clipping high-valued       outliers.}
\label{muLawQuantFigure}
\end{figure}

\begin{figure}[t!]
\begin{center}
\includegraphics[width=340pt]{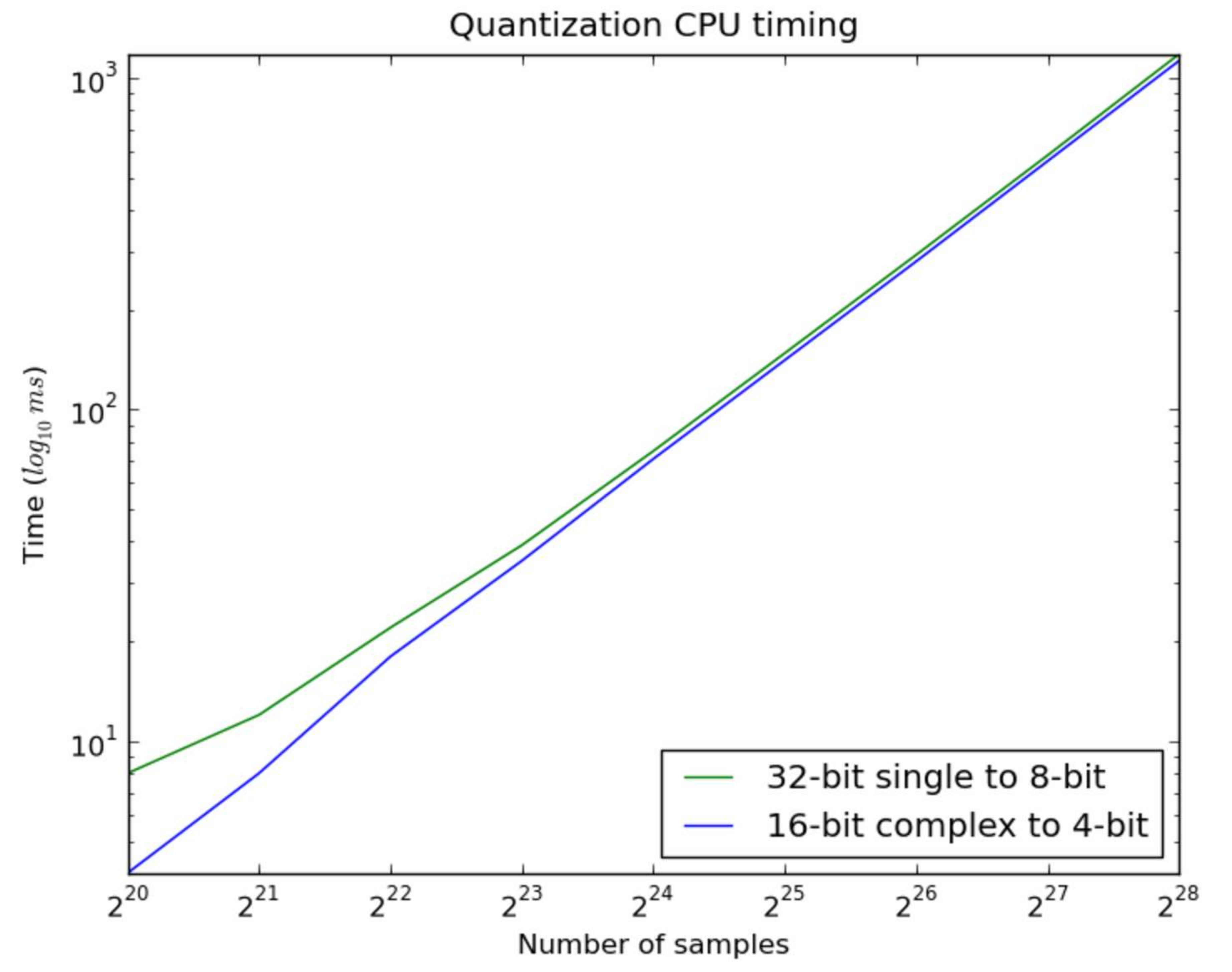}
\end{center}
\caption[Quantisation scaling performance]{Scaling benchmarks of our CPU implementation when converting 32-bit
	   single precision values to 8-bit and two 16-bit components to two 4-bit values packed into one byte.}
\label{quantScalingFigure}
\end{figure}

Incoming voltage series typically (depending on the quantisation mechanism on the backend) follow a normal distribution with mean 0. When this voltage series is detected the distribution changes to a half-normal one (if $X$ follows an ordinary normal distribution $N(0,\sigma^2)$ then $Y=|X|$ follows a half normal distribution). In both cases most of the data values will be concentrated around the mean, around $\pm3\sigma$ for normal and $3\sigma$ for half-normal distributions. Regular quantisation schemes, which represent a signal using $L$ regularly spaced levels in increments of $\epsilon\sigma$, will assign a single bit to each level. For example, if an 8-level, $\epsilon = 1$ scheme is used, the region between $[3\sigma,4\sigma]$ will have a dynamic range equal to the region $[\mu,\sigma]$ even though the latter represents many more data points. Therefore, using a linear quantiser would result in a loss in sensitivity within the region centred around the mean. Also, any strong signals present in the data, be they astrophysical or RFI-induced, will be represented in the bins at the edges, potentially reducing the S/N.

An alternative method would be to use a logarithmic quantiser, where quantisation levels are distributed according to
a logarithm function, resulting in finer resolution (smaller quantisation steps) around the signal mean. In our pipeline we implement a $\mu$-law quantiser, adapted from the G.711.0 standard\footnote{\;See http://www.itu.int/rec/T-REC-G.711}:
\begin{equation}
 y = X_{\text{max}}\frac{log\left[1+\mu\frac{|x|}{X_{\text{max}}}\right]}{log\left[1+\mu\right]}sign\left[x\right]
\end{equation}
where $x$ represents points in the data series, $X_{\text{max}}$ is the maximum value in this series, $\mu$ is the compression
factor and $y$ is the quantised data series. A higher compression factor results in more output bits being allocated
to the high data point concentration range of the input distribution, as shown in figure \ref{compressionFigure}, while figure \ref{muLawQuantFigure} shows the effect the compression factor has when quantising signed 16-bit         input values down to signed 4 bits. Logarithmic quantisers are more computationally intensive since for every input sample a logarithm needs to be calculated. We counter this by pre-computing a log lookup table for the input data range, small enough such that it can be stored in cache.

The data buffer is first encoded using this mapping and then the output values are quantised and written to disk.
This is performed on the CPU, so as not to interfere with the main processing pipeline and induce delays.
The data series is split across multiple OpenMP threads, whose processing is interleaved with file I/O calls,
in order to overlap CPU-processing and disk I/O. Figure \ref{quantScalingFigure} shows the linear performance
scalability of our implementation when processing a single beam.

\section{Pipeline Analysis}
\label{pipelineAnalysisSection}

We have described in detail the various processing stages composing our GPU-based transient detection pipeline. In this section we analyse its performance scalability and define upper bounds on the number of beams and DM trials which can be processed on a standard COTS GPU server. We limit this analysis to BEST-II observing parameters, as described in chapter \ref{medicinaChapter}, while in chapter \ref{skaChapter} we analyse the scalability of this pipeline to a wider parameter range, with special emphasis to SKA$_1$ specifications. We also provide several figures of merit for a number of processing stages by evaluating them against simulated data over a wide parameter range.

\begin{figure}[t!]
  \begin{center}
    \includegraphics[width=350pt]{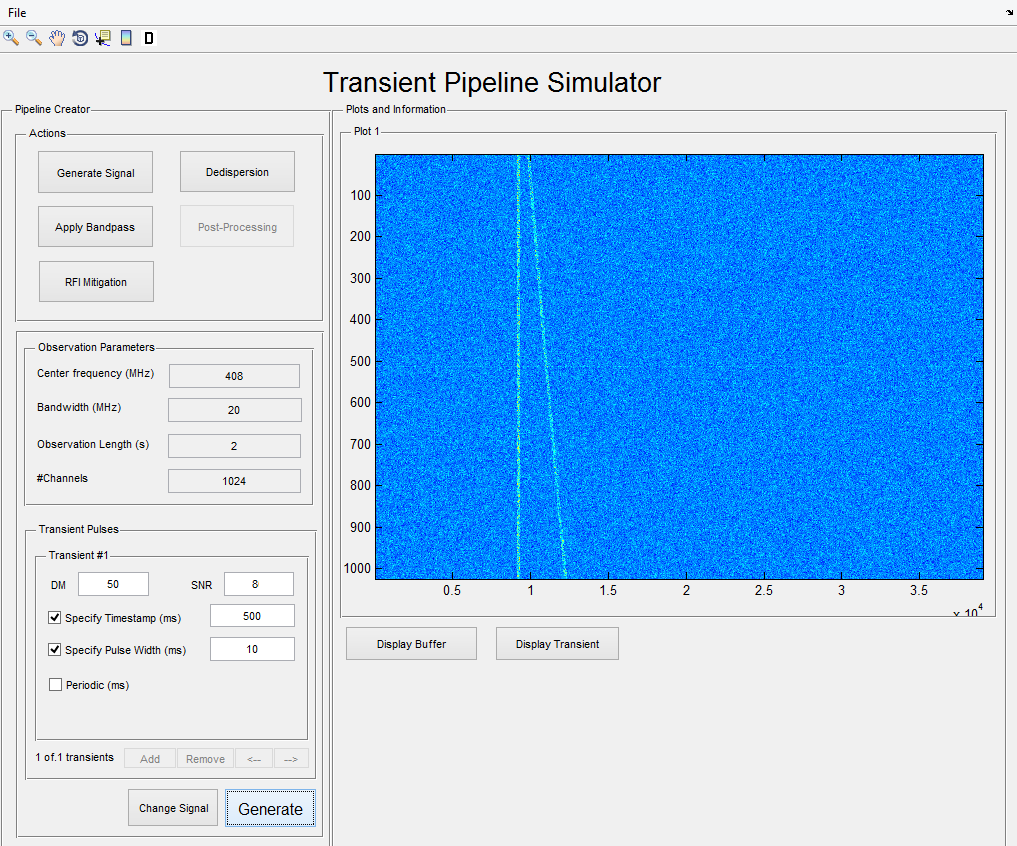}
  \end{center}
  \caption[Matlab transient pipeline simulator]{Screenshot of Matlab transient detection pipeline simulator which was used to analyse the algorithms implemented in the GPU pipeline and to generate simulated data for evaluation. The  data generation stage is shown in this figure.}
  \label{simulatorFigure}
\end{figure}

We have developed a lightweight transient pipeline simulator, written in Matlab, which was used to generate test 
filterbank data containing both dispersed and RFI signals, as well as to act as a prototyping platform to test 
the applicability of different algorithms to all the stages within the pipeline. The generated pulses can be tracked
as they pass through different processing stages and then the performance of the single pulse detection stages, as
well as the RFI mitigation algorithms, can be measured using appropriate figures of merits. The aim of this tool
is to provide a quick and easy way to test novel algorithms, and provide an intuitive Graphical User Interface (GUI)
for quick visual inspection and plotting mechanism. Figure \ref{simulatorFigure} provides a screenshot of the data
generation stage. This tool was used to generate the test data described below.

\subsection{Clustering and Classification Evaluation}

The principal aim of a transient detection pipeline is to detect as many ``interesting'' signals as possible, over a wide parameter range, including DM, pulse width, S/N and pulse shape. The standard technique which is generally used is template matching, where boxcar functions of varying width are convolved with each dedispersed time series, the result of which is then thresholded. Each boxcar template has the effect of downfactoring the time series, thus increasing the S/N of wider pulses. This process is compute intensive, with a time complexity of
\begin{equation}
 \mathcal{O}(N_{\text{DM}} N_t N_s log N_s )
\end{equation}
where $N_{\text{DM}}$ is the number of dispersion measure trials, $N_s$ is the number of time samples and $N_t$ is the number of templates being matched. This process is generally performed in the Fourier domain, meaning that a forward and backward Fourier Transform are required. The output of this process still needs to be clustered in some way such that detections belonging to the same event are grouped together. Resulting groups are then classified as either RFI or not, after which appropriate action is taken.

Matched filtering techniques were briefly discussed in section \ref{singlePulseSection}, where it was stated that narrower pulses are more easily detected than broader ones, however low-amplitude, broad pulses are more easily detectable than sharp, narrow pulses if its area is sufficiently large. There are, however, several downsides to this technique. First of all, as was already stated, it is a compute intensive process. Secondly, boxcar functions are generally used, and thus can be insensitive to pulses which cannot be approximated well by a top hat pulse, such as highly scattered pulses. Template matching also increases the number of detections when compared to standard thresholding, as multiple templates are generally used. 

\begin{table}[b!]
 \centering
 \begin{tabular} {  l  c }
  \hline
  Centre Frequency           & 418 MHz \\
  Bandwidth                  & 20 MHz \\
  Frequency Channels         & 1024 \\
  Sampling Time              & 51.2 $\mu s$ \\
  \hline
  DM                         & 25 pc cm$^{-3}$ \\
  Dispersed pulse S/N        & 0.5 - 5, increments of 0.5 \\ 
  Dispersed pulse widths     & 0.1 - 102.5 ms (10 unique values) \\ 
  Repetitions per event      & 10 \\
  \hline
 \end{tabular}
 \caption[Simulated data parameters for event detection accuracy test]{Simulated data parameters for event detection accuracy test}
 \label{clusteringTestParametersTable}
\end{table}

As we discussed in section \ref{detectionSelectionSection}, we apply a density 
based clustering technique on the thresholded dedispersed time series, which 
groups detections that were caused by the same event together. An unoptimsed 
DBSCAN implementation, with an indexing structure for region queries, has a 
complexity of $\mathcal{O}(N_{\sigma} log N_{\sigma})$, where $N_{\sigma}$ is 
the number of points with standard deviation greater than a preset threshold. 
FDBSCAN reduces the $log N_{\sigma}$ factor to a scalar value defined by the 
number of representative seeds taken from a core point's neighbourhood which are 
used for cluster expansion. Therefore, our implementation has a complexity of 
$\mathcal{O}(N)$, which makes it considerably faster than template matching, and 
an ideal candidate for real-time detection. The cluster parametrisation and 
candidate selection stage have a negligible effect on execution time. 

\begin{figure}[t!]
  \begin{center}
    \includegraphics[width=320pt]{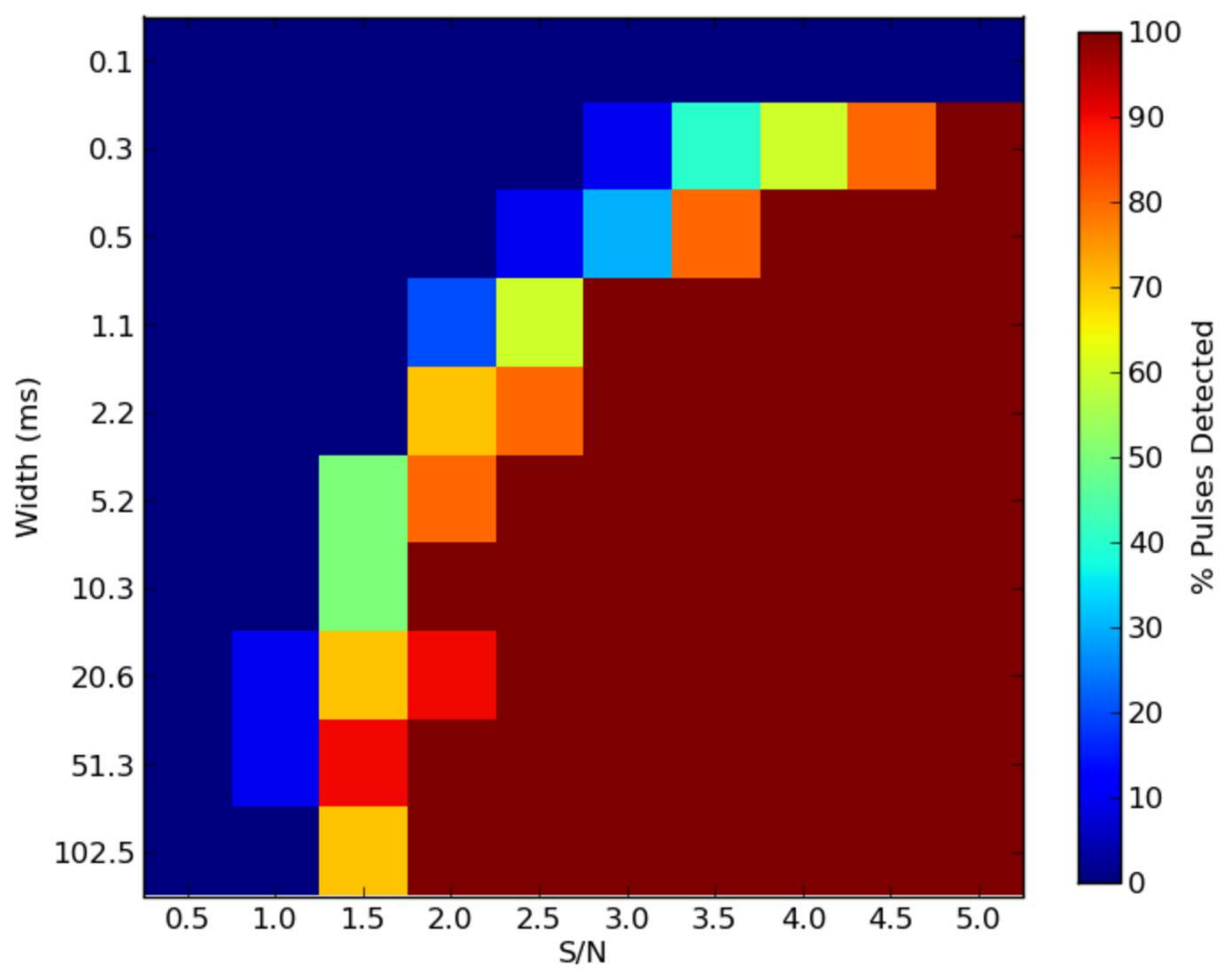}
  \end{center}
  \caption[Clustering and classification accuracy]{Output for clustering and classification accuracy test. 10 dispersed pulses for each combination were present in the dataset, the number of which were detected is show here. A pulse width of 0.1 ms is equivalent to $\sim$2 time samples.}
  \label{pipelineSurfaceFigure}
\end{figure}

A simple metric which can be used to test the accuracy of the event detection 
stage is the percentage of pulses which are correctly detected and classified. 
To measure this, a simulated dataset was generated, the parameters of which are 
listed in table \ref{clusteringTestParametersTable}. A total of 100 different 
dispersed pulses, with a DM of 25 pc cm$^{-3}$, each repeated 10 times, were 
injected into a stream of Gaussian noise with mean 0 and standard deviation 1. 
The pulses are Gaussian shaped, with a value at the mean equal to $\sqrt{S/N}$ 
in each channel, with width at FWHM equal to the width of the pulse. This data 
set was then processed by the transient detection pipeline, dedipsersed over 
2048 DM trials with a DM step of 0.04 pc cm$^{-3}$. The RFI thresholding and 
post-processing stages were disabled. The event detection threshold was set to 
4$\sigma$, resulting in a high number of background noise detections as well as 
better sensitivity to low S/N pulses. The detected pulses are shown in figure 
\ref{pipelineSurfaceFigure}, where the number of pulses which were correctly 
detected and classified with respect to the original dataset are listed. Narrow 
pulses, of the order of 2 to 10 samples wide, were only fully detected for a 
pulse S/N of 4.0 or greater, while wider pulses were detected for pulse S/N as 
low as 2.0. This emphasises the fact that such a classification scheme is 
sensitive to pulses of varying widths and S/N. It should be noted that the S/N 
of wide pulses can be increased by integrating consecutive time samples 
together. Also, applying a median filter to the dedispersed time series will 
reduce the detectability of very narrow pulses, since they might be considered 
as noisy outliers. 

\begin{figure}[t!]
  \begin{center}
    \includegraphics[width=400pt]{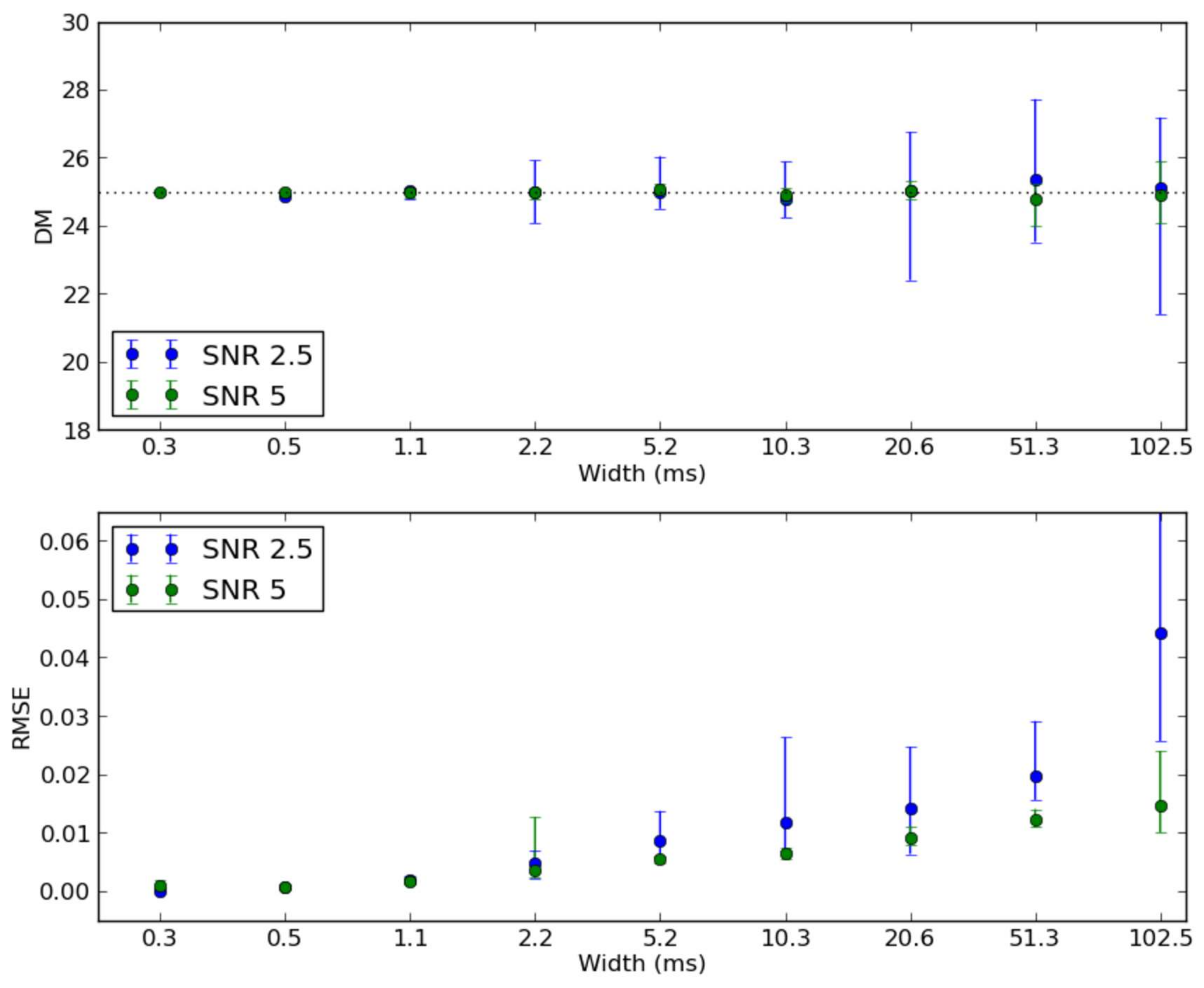}
  \end{center}
  \caption[Pipeline detection accuracy]{DM and RMSE values for the simulated 
pulses as calculated by the classification stage. Uncertainties increase as 
pulses get wider. Error bars represent the range between the maximum and 
minimum values.}
  \label{pipelineDmSnrFigure}
\end{figure}

An additional benefit of the classification stage is that it automatically detects the DM and width of the pulse, which are used to compute the analytical curve for incorrect dedispersion. Figure \ref{pipelineDmSnrFigure} shows the accuracy of this process with respect to accurate (a) DM detection and (b) RMSE. Wider pulses induce a wider DM error margin due to random fluctuations within the pulse peak. For this reason, the DM is approximated by calculating the median of all the values in the DM histogram having a normalised value greater than 0.75. As the pulse gets wider, random fluctuations also induce a higher error in the calculated RMSE value, resulting in a higher error margin.

\subsection{RFI Mitigation}

A simple metric which can be used to test the pipeline's resilience to RFI events is the percentage of RFI events which pass through the event detection stage and are incorrectly classified as astrophysical events. In order to be able to come up with a value for a particular implementation, the actual number of RFI events which occur during an observation needs to be known. There are two ways in which this can be done. The first is to generate a simulated observation where all the RFI events, and other additional signals, are accounted for. The second method is to conduct a real observation of an ``empty'' area of sky with the telescope of interest, such that all events will be RFI-induced. For the latter method an accurate statistical model of the RFI environment is required, and the test observation should be conducted for enough time to obtain a statistically significant value. Here we use the former approach.

\begin{table}[b!]
 \centering
 \begin{tabular} {  l  c }
  \hline
  Centre Frequency           & 418 MHz \\
  Bandwidth                  & 20 MHz \\
  Frequency Channels         & 1024 \\
  Sampling Time              & 51.2 $\mu s$ \\
  \hline
  Broadband RFI width range  & 1 - 30 ms (10 unique values) \\
  Broadband RFI S/N          & 1 to 10, increments of 1 \\
  Repetitions per event      & 10 \\
  Narrowband RFI frequency   & 97.6 kHz \\
  \hline
 \end{tabular}
 \caption[Simulated data parameters for RFI thresholding accuracy test]{Simulated data parameters for RFI thresholding accuracy test}
 \label{rfiTestParametersTable}
\end{table}

A simulated dataset was generated to test the pipeline's resilience to RFI 
events, the parameters of which are listed in table 
\ref{rfiTestParametersTable}. A total of 100 different broadband events, each 
repeated 10 times, were injected into a stream of Gaussian noise with mean 0 and 
standard deviation 1. A 97.6 kHz (5 frequency channels) constant narrowband 
signal was also included, as well as random narrowband events of 1 s duration, 
which should incur some detections since the GPU buffer used to run the test was 
3.3 s long. A 6$^{\text{th}}$-order polynomial bandpass was also applied to the 
frequency band, which decreases the response at the edges of the band. This data 
file was then processed by the transient detection pipeline, where the RFI 
mitigation, dedispersion and clustering stages were enabled. RFI thresholding 
parameters were set such that dispersed pulses would not be affected (see 
previous section). The time series data were dedispersed over 2048 DM trials, 
with a DM step of 0.04 pc cm$^{-3}$. All the detections were then clustered and 
evaluated.

Most of the simulated signals did pass through the RFI mitigation stage, as the 
thresholds were set a large value, however 100\% of the broadband signals were 
correctly classified as RFI, whilst a single cluster caused by a narrowband 
burst was incorrectly classified as an astrophysical event. This emphasises the 
fact that a clustering technique, coupled with a classification mechanism, is a 
viable and accurate candidate for event detection. It should be noted that the 
simulated RFI events were simple signals, in that they are vertical or 
horizontal lines in time and frequency space. More complicated artificial 
signals can induce erroneous positive detections, and this points towards the 
need for a more robust classification mechanism. Pattern matching and machine 
learning techniques can be employed for this, however any supervised learning 
algorithm would require a labelled training set consisting of RFI and non-RFI 
signals. The RFI environment of a radio telescope can be determined by observing 
an empty area of sky, where all detections would be RFI-induced. All the 
resultant detections can then be used as a training set.

\subsection{Performance Benchmarks}

The transient detection pipeline can be thought of as a soft real-time system, where all the parallel stages should keep up with the incoming data stream for maximum quality of service, however if for some reason one of them does not meet a processing deadline the system does not become unstable, but rather the processing of input data will be delayed until the pipeline progresses by one iteration. This has no effect when running in offline mode, however when deployed as a real-time system input data might be lost. These processing hiccups might happen when, for example, a very high-power RFI spike induces a large  number of detections resulting in a large number of data points to cluster. The GPU processing times are fixed regardless of the quality of the data, so these issues are only applicable to CPU threads. For this reason, a pipeline iteration should be limited by the GPU processing time. The scaling performance for the two most time consuming stages on the CPU, the clustering and quantisation stages, are shown in figures \ref{dbscanFigure} and \ref{quantScalingFigure} respectively, each performed by a single thread. The number of samples which need to be quantised for a single pipeline iteration is fixed ($f_{\text{bw}} \times N_{\text{beams}} \times N_{\text{components}} \times t_s$), while the number of data points which need to be clustered depends on the number of events present in the data stream. 

The upper bound on the number of dispersion measure trials which can be processed in real-time depends on several  factors, including: the observing bandwidth and centre frequency, the number of simultaneous beams which have to be processed, the DM step and the setup on which it is deployed. As a general rule, for a given setup, the number of DM 
trials is inversely proportional to the number of beams, which leads to a compromise between field of view and observable distance, assuming an appropriate DM step is used. We choose to evaluate the performance of this pipeline by using the observing parameters of the BEST-II array, discussed in chapter \ref{medicinaChapter}.

Table \ref{timingTable} lists the average timings for 10 pipeline iterations for all the processing stages for one pipeline iteration when processing eight 20 MHz beams, channelised into 1024 frequency channels and centred at 408 MHz. A simulated data file containing a 500 ms periodic pulse with 1\% duty cycle and a S/N of 3 was input to the system, with the data copied to all the beams in 5 s buffers, this length being limited by the amount of GPU memory available. This results in about 18 clusters, with a total of $2^{16}$ points, per iteration. GPU timings are for a single GTX 660 Ti processing 4 beams while CPU timings are for all the beams when processed using a single output thread and quantisation thread. The host system on which the benchmark tests were conducted consists of 2 Intel Xeon E5-2630 2.3 GHz processors, 2 NVIDIA GTX 660 Ti GPUs with 3GB of GDDR5 RAM, 32 GB DDR3-1600 system RAM and a Fujitsu D3118 system board. The table is split into two columns which work in parallel, the GPU-based processing stages and the CPU-based processing stages, with data transfer to and from the two performed at synchronisation barriers. It is clear that the processing bottleneck is the dedispersion kernel, which takes up 93\% of the GPUs’ processing time, while the rest of the kernels have a negligible effect on the overall running time of the pipeline. The number of DM trials which can be processed during an observation is limited by this as well, and currently this value is around 864 DMs.


\begin{table}[t!]
  \centering
  \begin{tabular}{ | p{3.2cm} | r | p{3.2cm} | r | }
    \hline
    \multicolumn{4}{|c|}{Copy to GPU: 295.30 ms} \\
    \hline
    \hline
    \multicolumn{2}{|c}{GPU} \vline & \multicolumn{2}{c|}{CPU} \\
    \hline
    Bandpass Fitting & 36.79 ms   & Thresholding     & 690.96 ms  \\
    RFI Filtering    & 74.97 ms   & Clustering       & 1148 ms \\
    Dedispersion     & 3863 ms    & Classification   & 168.64 ms  \\
    Median Filtering & 135.11 ms  & Clusters to file & 728.73 ms  \\ \cdashline{3-4}
    Detrending       & 59.06 ms   & Quantisation\footnotesize{$^*$}  & 3532 ms     \\
    \hline
    \hline
    \multicolumn{4}{|c|}{Copy from GPU: 155.97 ms} \\
    \hline
    \hline
    \multicolumn{4}{|c|}{Total iteration time: 4619.72 ms} \\
    \hline
  \end{tabular}
  \caption[Transient detection pipeline timings]{GPU and CPU timings, averaged over 10 pipeline iterations, when processing 8 20MHz beams split between 2 NVIDIA GTX 660Ti cards. GPU timings are for a single GTX 660 Ti processing 4 beams while CPU timings are for all the beams when processed using a single output thread and quantisation thread.
	   \newline		
	  \footnotesize{* Quantisation is performed on a different CPU thread from the event detection stages}}
  \label{timingTable}
\end{table} 

\section{Comparison with Other Work}
\label{comparisonLabel}

In section \ref{otherPipelinesSection} we discussed the current state-of-the-art for fast transient pipelines, with implementations ranging from conventional CPU-based systems to ones which use GPU or FPGA accelerators to speed up processing. We will not compare the performance between CPU- and GPU-based systems, as this has been thoroughly performed in other work (for example, \cite{Barsdell2011}). No performance benchmarks have been published for the ARTEMIS system to date, so a direct comparison is not possible. It should be noted, however, that the principal ideology for these two systems is different. ARTEMIS uses the concept of ``data blobs'', where small data buffers are passed through the CPU-based pipeline modules, after which the dedispersion module generates larger buffers to offload to GPU memory for processing. This increases CPU performance for certain algorithms due to higher cache coherency, however it also requires a number of additional memory copies amongst data blobs, and an additional buffering scheme. The CPU-based RFI clipper is also based on a median filter, which requires a partial sort and thus is one of the primary CPU bottlenecks. In our implementation, buffering is performed at the start of the pipeline and all processing stages are then executed on the GPU, resulting in less overhead, and is less coupled with the scheduling policy of the operating systems, which could result in a temporary degradation in performance. 

\begin{figure}[t!]
  \begin{center}
    \includegraphics[width=420pt]{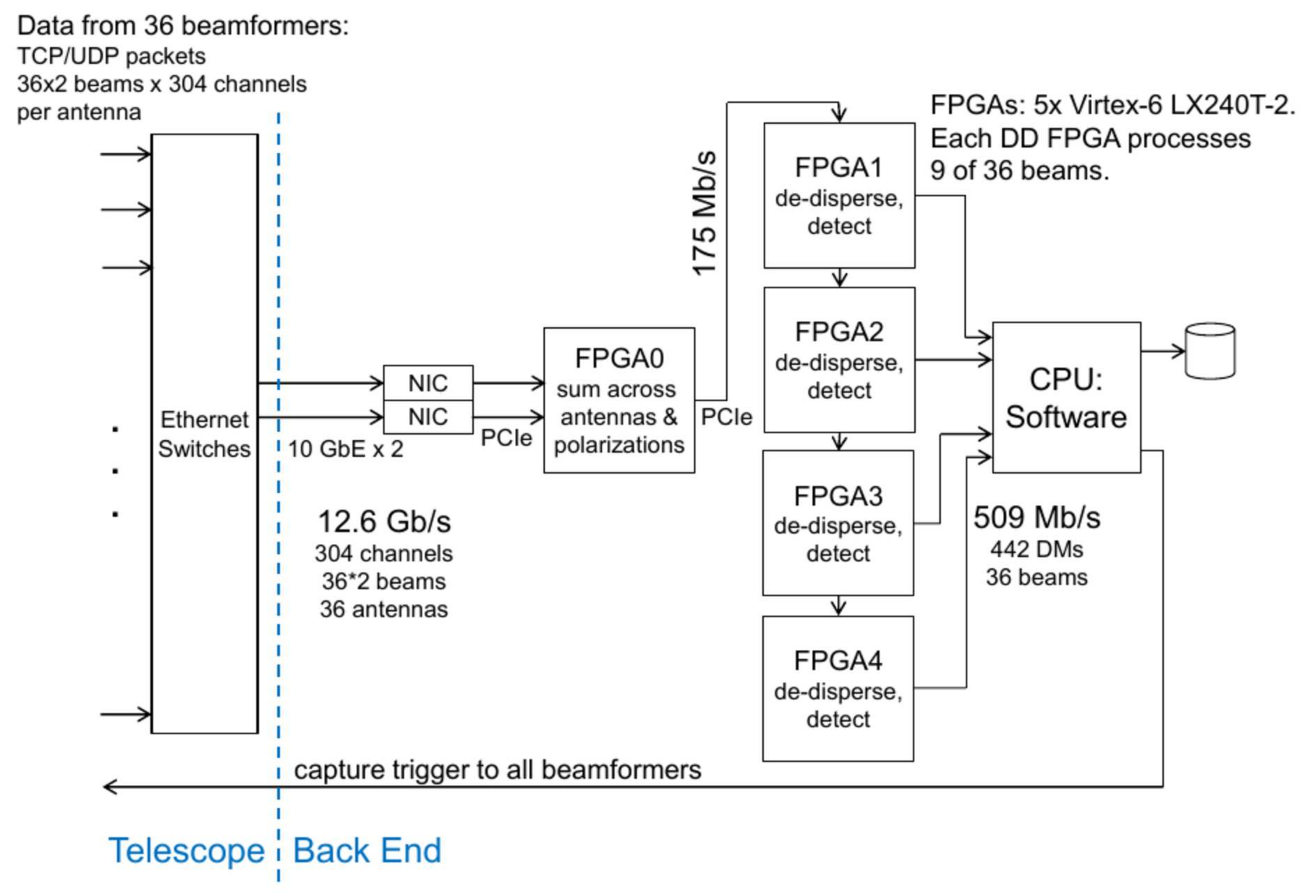}
  \end{center}
  \caption[CRAFT backend implementation]{TARDIS: CRAFT fast transient backend implementation. Source: \cite{Addario2013}}
  \label{tardisFigure}
\end{figure}

\cite{Addario2013} provide some performance benchmarks as well as a detailed architectural description, which is reproduced in figure \ref{tardisFigure}. Dedispersion and detection is split across 4 Virtex 6 LX240T, whilst an additional FPGA performs an incoherent summation across antennas and also sums the two polarisations. Each dedispersion FPGA processes 9 power beams, split into 304 channels, with a time resolution of 0.9 ms, for up to 442 DM trials in real-time. Ignoring the operations required for summation and detection, this results in approximately 5.7$\times$10$^9$ operations per second for dedispersion spread over 4 FPGAs, at peak performance. By comparison, the test benchmark presented in the previous section managed 138$\times$10$^9$ operations per seconds on two NVIDIA GPUs. These approximations are based on the direct-dedispersion method, however \cite{Addario2013} use a different version which joins contiguous samples together if the dispersion lines span multiple time bins, thus increasing the operation cost by a small factor. Even so, these figures suggest that GPUs are more suited for real-time transient pipelines than FPGAs. The latter would also suffer from lack of on-board memory available when faster sampling times and high DM values are required. An additional advantage in favour of GPUs, when compared to FPGAs, is the ease with which new algorithms can be implemented and attached to existing pipelines. On the other hand, although a high end Virtex 6 costs more than a GPU, the running costs of an FPGA-based system are less, due to lower power requirements. This can be countered by appropriate upgrade paths throughout the lifetime of the system, as newer generation GPUs will generally have a higher performance per watt ratio, and upgrading the devices would yield savings in the long run.

\section{Conclusion}

Fast transient detection pipelines require considerable computational resources, 
and several processing platforms are being investigated to suite these 
requirements. In this chapter we propose a GPU-based solution, moulded around a 
custom GPU framework, whereby data input and buffering is performed by the CPU, 
after which this data are copied to GPU memory where compute-intensive tasks 
are performed. It is capable of processing multiple independent beams in 
parallel across as many GPUs as are attached to the host. RFI mitigation through 
bandpass correction and thresholding removes high power spatial and temporal 
terrestrial signals which would result in a high number of false detections. 
The dedispersion kernel presented in chapter \ref{dispersionChapter} is 
integrated within this pipeline as well. After an optional post-processing 
stage, the dedispersed time series from each beam are copied to host memory 
where event detection through thresholding and density-based clustering is 
performed. The DM-S/N curve of each cluster candidate is then compared to a 
fitted model which filters out RFI-induced clusters and sends a trigger to the 
data persistence module, which quantises and writes the input buffer to disk 
containing the event, together with cluster parameters. 

The accuracy of the RFI mitigation and event detection stage was also examined. 
Simulated data sets containing RFI and dispersed pulses were passed through the 
pipeline, the output of which was examined. Narrow pulses, of the order of 2
to 10 samples wide, were only fully detected for a pulse S/N of 4.0 or greater, while wider pulses were detected for pulse S/N as low as 2.0. Most of the simulated RFI signals did pass through the RFI mitigation stage, however 100\% of the broadband signals were correctly classified as RFI, whilst a single cluster caused by a narrowband burst was incorrectly classified as an astrophysical event.


\chapter{Deployment at the Medicina BEST-II Array}
\label{medicinaChapter}

In the previous chapter we have described in detail a generic transient detection pipeline and demonstrated the 
applicability of GPUs for deploying such a system to the backend of radio telescopes. In this chapter we enhance this 
pipeline with real-time streaming capabilities and attach it to the digital backend of the Basic Element for SKA Training II (BEST-II) array in Medicina, Italy. This serves as an excellent test-bed for benchmarking and conducting
test observations. We begin by introducing BEST-II and the ROACH-based digital backend, and then move on to 
describe the deployment setup for our transient detection pipeline and present the results from various test observations, most notably of PSR B0329+54.

\section{The BEST-II SKA pathfinder}
\label{best2Section}

\begin{figure}[b!]
  \centering
  \subfloat{\includegraphics[width=190pt]{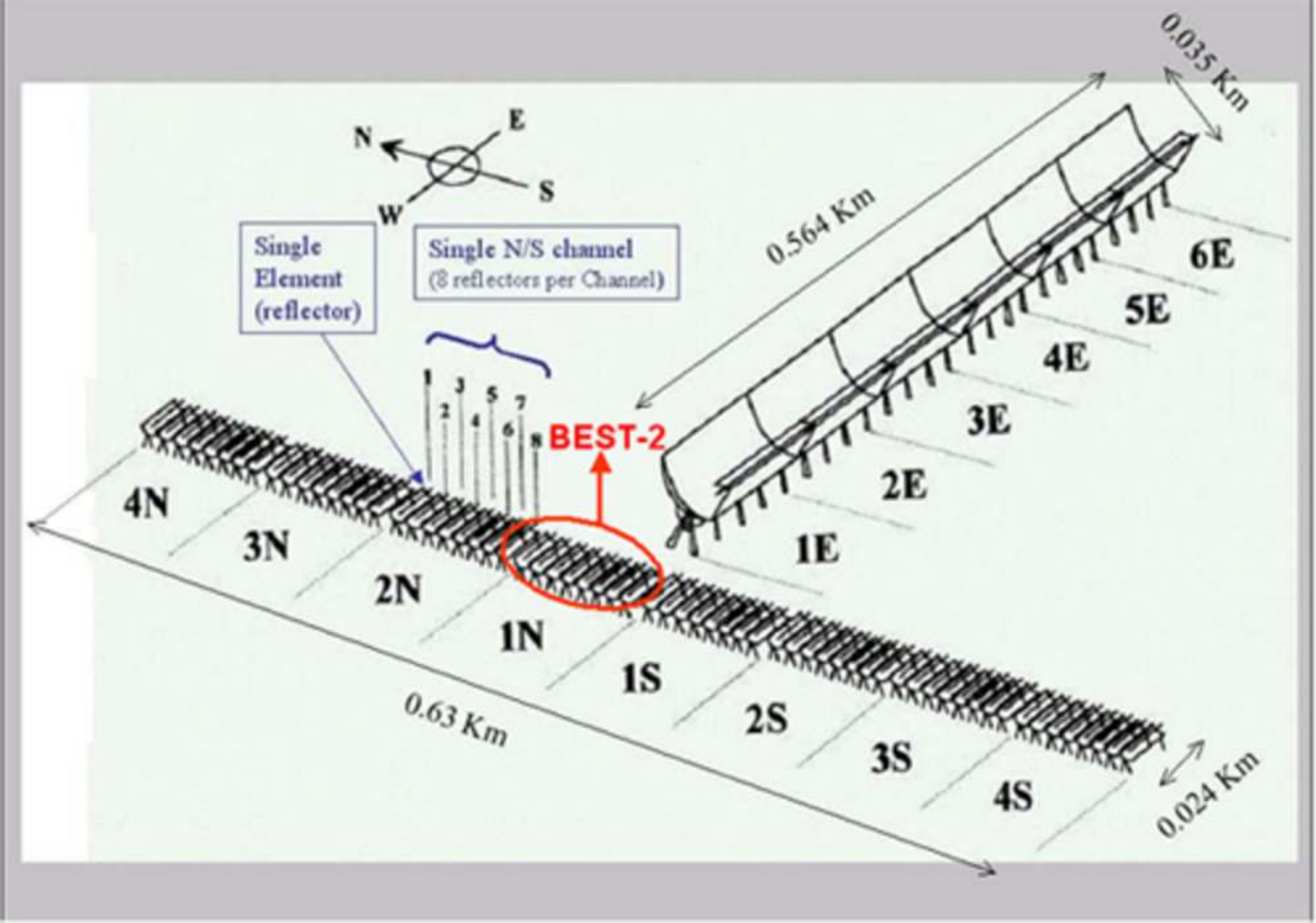}}
  \hspace{8mm}
  \subfloat{\includegraphics[width=190pt]{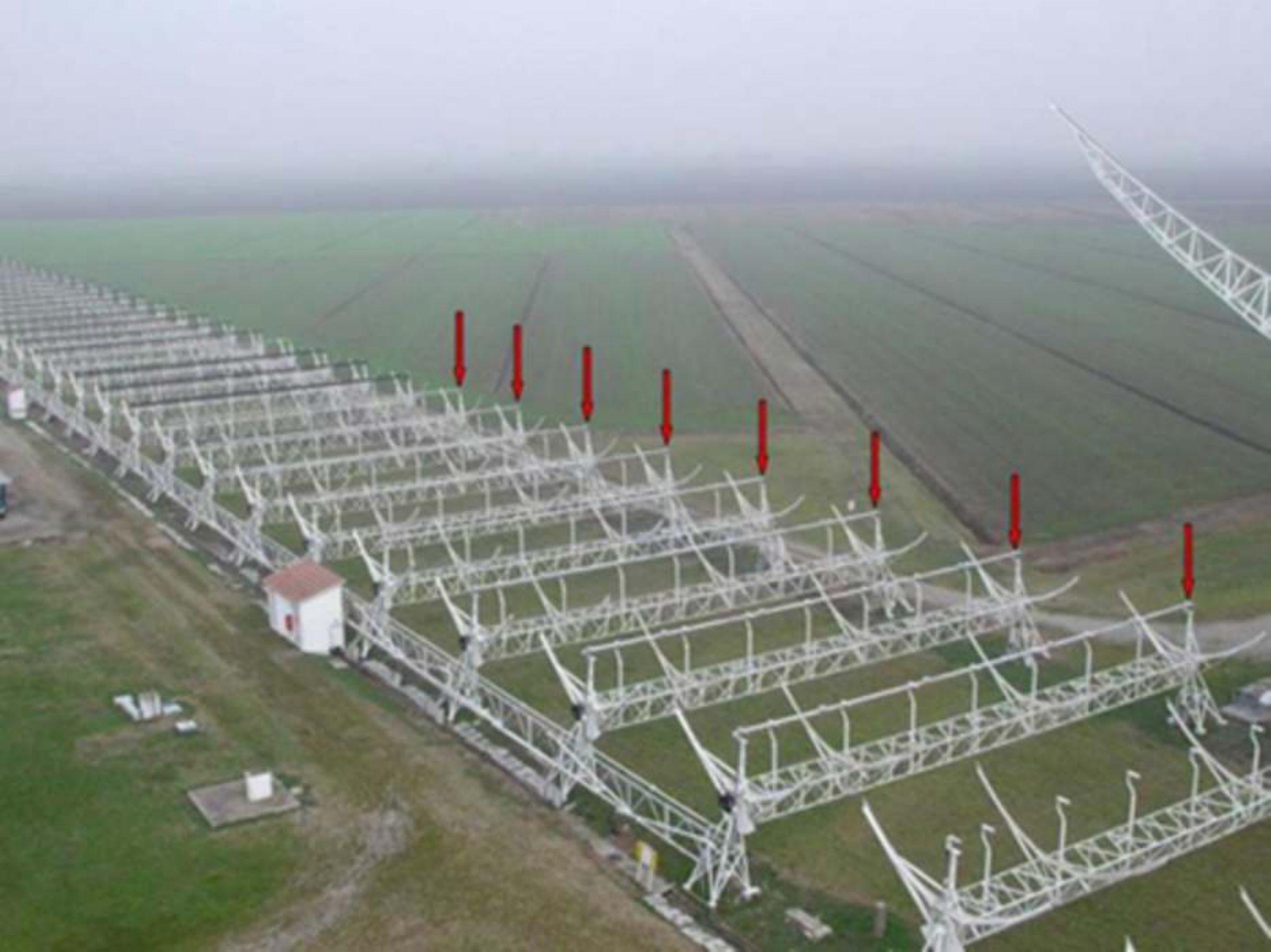}}
  \caption[The BEST-II array]{(a) A schematic of the Northern Cross Telescope, indicating the subset of eight 
	  reflectors of the N-S arm which form the BEST-II array. (b) A photo of the BEST-II array. }
  \label{best2Figure}
\end{figure}

The Basic Element for SKA Training II (BEST-II) \cite{Montebugnoli2009} is a subset of the Northern Cross cylindrical 
array (figure \ref{best2Figure}) at the Medicina observatory near Bologna, Italy. The array is composed of eight
East-West oriented cylindrical concentrators, each with 64 dipole receivers spaced such that the cylinder focal line 
is critically sampled at 408 MHz. Signals from the 64 dipoles are combined in groups of 16 using analogue circuitry,
resulting in four analogue channels per cylinder. This results in a total of 32 effective receiving elements
positioned regularly on a 4 x 8 grid, as shown in figure \ref{best2Figure}a. The top level specifications
of the BEST-II array, which provides $\sim$1$^{\circ}$ resolution over $\sim$38 square degrees FoV, are summarised in table \ref{best2SpecsTable}.

\begin{table}[t!]
 \centering
 \begin{tabular} {  l  c }
  \hline
  \multicolumn{2}{|c}{BEST-II Specifications} \vline  \\
  \hline
  \hline
  Total Collecting Area          & 1411.2 m$^2$ \\
  Total Effective Area           & 1001.95 m$^2$ \\
  Receiver Temperature           & 51 K \\
  System Temperature             & 86 K \\
  $A_{\text{eff}}/T_{\text{sys}}$              & 11.65 m$^2$/K \\
  \hline
  Longest Baseline (N-S)         & 70 m \\
  Longest Baseline (E-W)         & 17 m \\
  RF band                        & 400 - 416 MHz \\
  Total analogue channels        & 32  \\
  \hline
  Primary FoV (Dec, 408 MHz)     & 5.7$^{\circ}$ \\
  Primary FoV (RA, 408 MHz)      & 6.6$^{\circ}$ \\
  Synthesised FoV (Dec, 408 MHz, at zenith)     & 0.52$^{\circ}$ \\
  Synthesised FoV (RA, 408 MHz)  & 1.73$^{\circ}$ \\
  \hline
 \end{tabular}
 \caption[BEST-2 specifications]{The top-level specifications of the BEST-2 array. Source: \cite{Hickish2013}}
 \label{best2SpecsTable}
\end{table}

\begin{figure}[t!]
\begin{center}
\includegraphics[width=420pt]{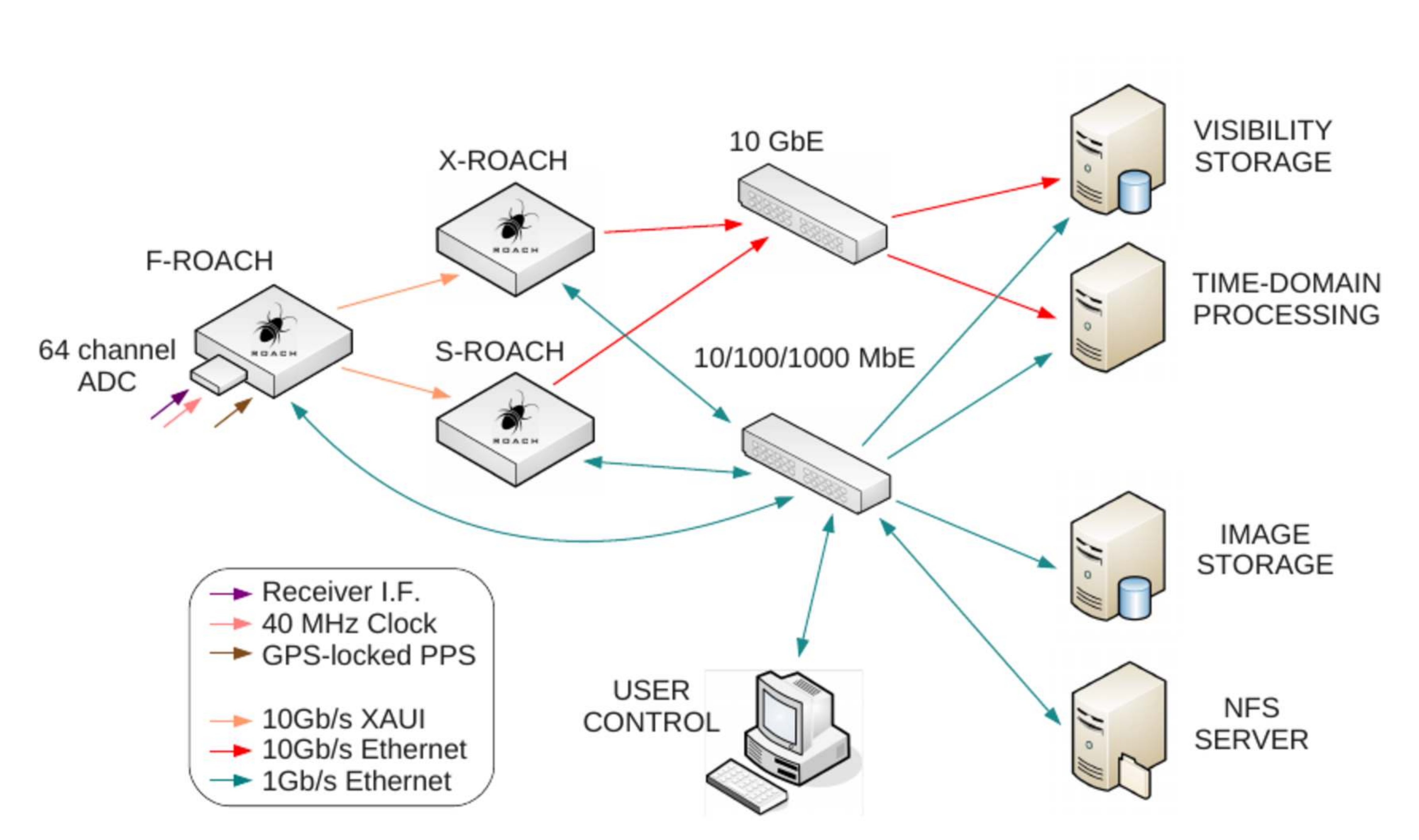}
\end{center}
\caption[BEST-II digital backend]{The top level architecture of the BEST-II backend. 10Gb and 1Gb links are used to transport data from the ROACHes to CPU and GPU nodes. 10Gb/s
	 XAUI is used for ROACH-ROACH connections. ``F'', ``X'' and ``S'' ROACH boards are responsible for 
	 channelization, cross correlation and spatial FFT respectively. The roles performed by any of the 
	 CPU-based nodes need not be split amongst physically distinct hosts. Source: \cite[figure 4.6]{Hickish2013}}
\label{best2BackendFigure}
\end{figure}

Such a receiver configuration leads to a set of highly redundant baselines, which allows efficient spatial 
processing using Fourier techniques. In the case of BEST-II, the array possesses 496 possible antenna pairings (excluding 
pairings of an antenna with itself), forming 52 unique baselines, which is an ideal test-bed for testing and deploying a Direct Imaging Correlator. The digital backed, developed and deployed by Jack Hickish \cite{Hickish2013} and Griffin Foster \cite{Foster2013} from the University of Oxford, is based on three Reconfigurable Open Architecture Computing Hardware (ROACH\footnote{\;https://casper.berkeley.edu/wiki/ROACH}) processing platforms and a number of software-based processing nodes. The overall topology and architecture of this system is shown in figure \ref{best2BackendFigure}. The three ROACH boards are
referred to as the:

\begin{description}
 \item[``F''-engine] for frequency transform, which is responsible for digitisation, channelisation of the processed 
      20 MHz bandwidth into 1024 frequency channels, and 
      transmission of coarsely quantised antenna signals to downstream processing boards. All 32 analogue 
      streams are digitised on a single processing node using a custom 64ADCx64-12 board\footnote{\;https://casper.berkeley.edu/wiki/64ADCx64-12}. A 4-tap Hann-windowed
      polyphase filter bank is used for channelisation.
      
 \item[``S''-engine] for spatial transform, which is responsible for formation of electric-field and total
      power beams on the sky by spatial Fourier transform and SPEAD packetisation for 10GbE streaming to the time 
      domain processing server. The current implementation allows for the arbitrary selection of 8 beams from
      the generated 128-beam grid for output as 16-bit complex-valued words.
      
 \item[``X''-engine] for cross-multiplication, which performs cross multiplication and accumulation of antenna
      signals as well as SPEAD packetisation of the generated visibility matrices and streaming to the visibility
      storage server for offline imaging. This is also used to calibrate the S-engine.
\end{description}

Data is passed between ROACH boards using the Ten Gigabit Attachment Unit Interface (XAUI) over a copper cable connection, which is a lightweight, bidirectional, point-to-point transmission protocol. Downstream processing of time-domain beam data is accomplished by using a Linux-based server, which hosts 2 NVIDIA GPUs running our transient detection pipeline. The 8 beams selected for output from the S-engine are packed as 16-bit complex
values and transmitted over 10GigE using a custom Streaming Protocol for Exchanging Astronomical Data (SPEAD\footnote{\;https://casper.berkeley.edu/wiki/SPEAD}) \cite{Manley2012} packet format. SPEAD is a flexible, self-describing, lightweight application-level datastream format. It is designed to provide a standard output format for radio astronomy instruments outputting UDP streams. The output rate of a single beam from the beamformer is 640 Mbps excluding packet headers, which is calculated using D = C $\times$ T $\times$ W , where C is the number of frequency channels, T is the number of time samples per second and W is the word length. In our case, C = 1024, T = 19531.25 (20 MHz processable bandwidth channelised into 1024 channels) and W = 32 bits (16 bits for each complex component). A total output bandwidth of 5.12 Gbps is required to send out all 8 beams, which is manageable over a single 10GigE
link. 

\section{BEST-II transient detection pipeline}

\begin{figure}[t!]
\begin{center}
\includegraphics[width=420pt]{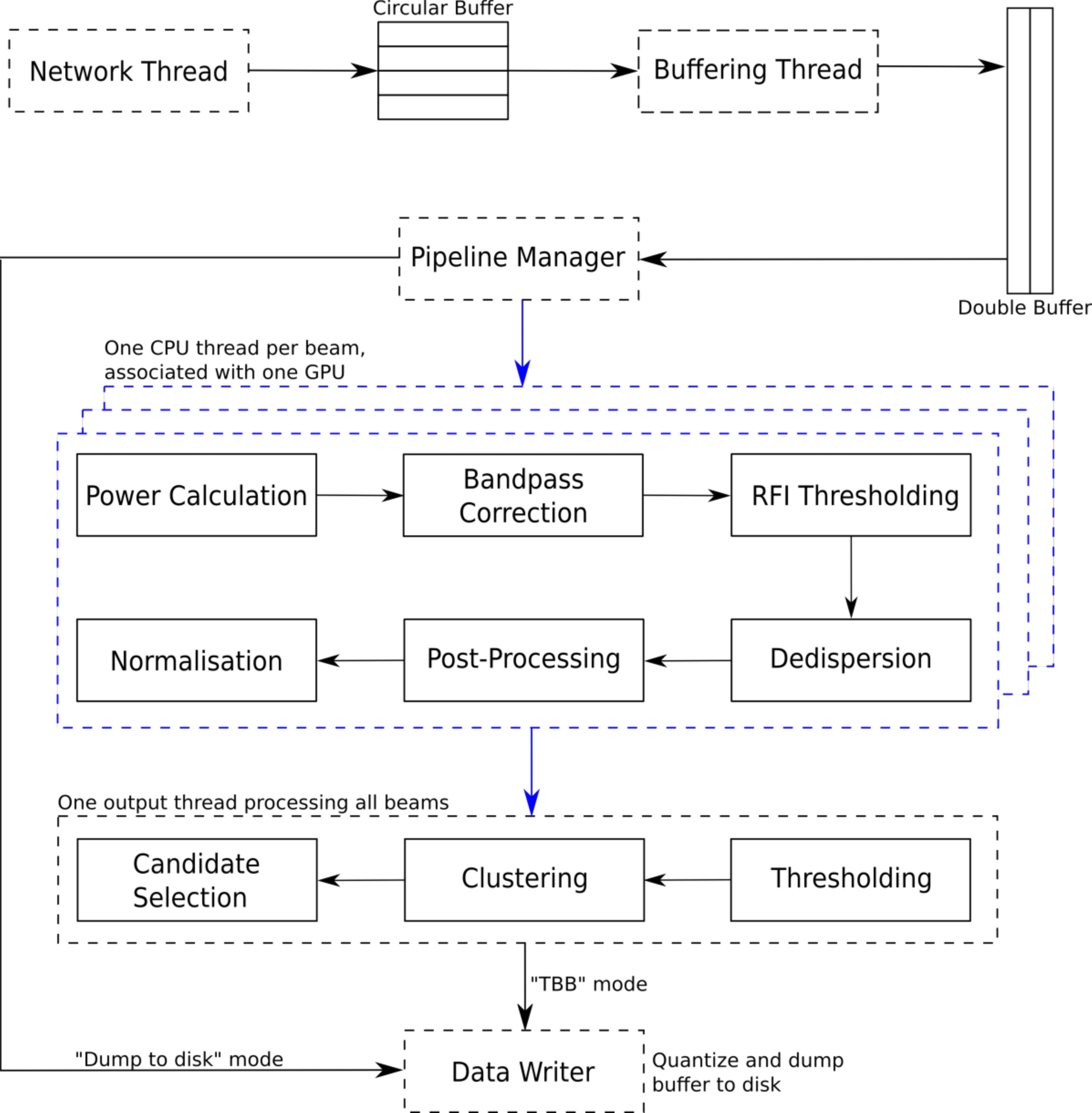}
\end{center}
\caption[BEST-II system architectural overview]{Overview of the system and pipeline architecture. Boxes in dashed lines represent distinct CPU threads, whilst dashed blue overlays represent CPU threads which use GPUs for processing. Blue arrows represent data transfers to/from GPU memory.}
\label{architectureFigure}
\end{figure}

In chapter \ref{pipelineChapter} we have described in detail a generic GPU-based transient detection pipeline. 
Here we enhance this with real-time capabilities, which takes the form of a high-speed, optimised packet
receiver. The high-level architecture of this pipeline is depicted in figure \ref{architectureFigure}, which is split into three main processing stages: the data reception and buffering stage, the GPU-based processing pipeline stage and the post-processing stage. Beams are processed independently across multiple GPUs, and a CPU thread is associated with each beam, where multiple beams can reside on a single GPU. 

Packet reception, interpretation and buffering is performed on the CPU, which then forwards the
data to the attached GPUs where it passes through several processing stages including: power calculation, bandpass correction, RFI
clipping, dedispersion and optional post-processing and normalisation, after which the dedispersed time-series 
are copied back to CPU memory and passed through a detection stage where it is thresholded,
clustered and classified. Any data-points belonging to interesting clusters are written to disk, together with the unprocessed data buffer after being
quantised to 8 or 4 bits depending on whether the data has been converted to a channelised power-series
in the receiver thread. There is also the possibility of writing the entire data stream to disk after passing
through an encoding and quantisation stage, provided that the disk drives can manage the reduced data

\subsection{Packet Receiver}
\label{packetReceiverSection}

\begin{figure}[t!]
\begin{center}
\includegraphics[width=450pt]{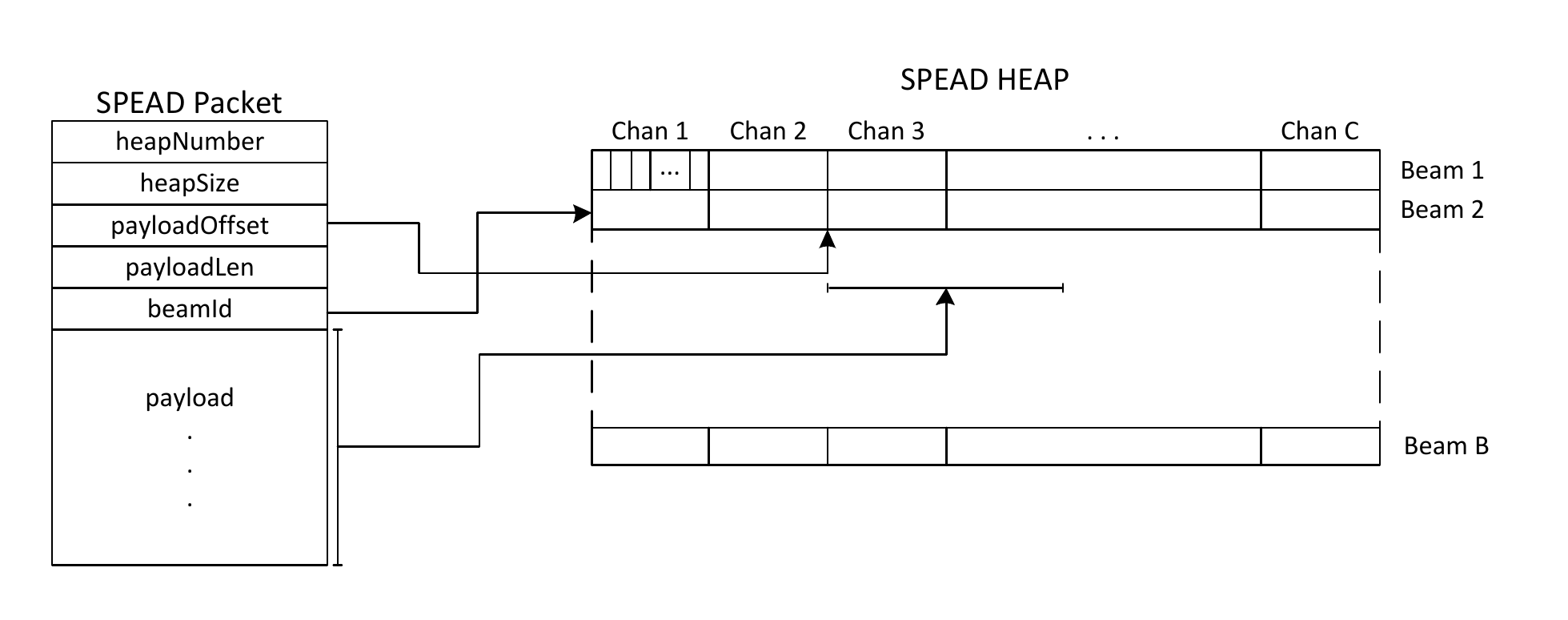}
\end{center}
\caption[BEST-II S-engine output packet format]{SPEAD packet format and heap data organisation used to send data between the S-engine and GPU server. A heap contains a number of spectra for all the beams being processed, for all the frequency channels. This is split up across multiple packets which contain time samples from a subset of the frequency channels, for a single beam. The packet header defines where the data should be located within the heap, as well as information to be able to identify heaps and extract timing information.}
\label{packetFormatFigure}
\end{figure}

The S-engine packetises the channelised beamformed data using a custom SPEAD packet format designed to 
reduce the overhead of heap generation and data movement on the receiver side. A heap consists of a time-slice 
containing several spectra (composed of 16-bit complex values sampled for all channels) from multiple beams, organised 
in beam/channel/time order, which maps directly to the data organisation in GPU memory, thus considerably reducing the 
overhead for memory re-arrangement. Figure \ref{packetFormatFigure} describes in further detail the data organisation 
of a heap and how this maps to the packet format. The UDP transmission protocol is used to send the packet stream, 
since it is lightweight and the probability of losing packets, or receiving them out of order, is very small and can 
be easily handled using simple error checking routines on the receiver side, where heap buffers provide a small time 
window during which an out-of-order packet can still be processed correctly. The heap size is set by the digital backend.

Care needs to be taken when designing receiver codes for these types of data streams. A saturated 10GigE link can 
achieve a total data rate of $\sim$1.25 GB/s, which needs to be processed in real-time, usually by a single CPU 
thread, with minimal resource consumption on the host system. Some of the major factors which affect receiver 
performance include:
\begin{description}
 \item[Packet Size] Every received packet needs to pass through the protocol stack before reaching its destination and 
       is ready for processing, where the respective headers are stripped, checked and processed. This induces a 
       processing overhead which can become severe when the payload size to frame header ratio is small. Ideally the 
       packet size would be the maximum possible allowed by the transmitting backend and network devices between the 
       two end points. The S-engine transmits packets containing a payload of 4096 bytes, resulting in 156,250 
       packets per second.
 \item[Data Movement] Even if the incoming packets are read once from the network socket and then discarded there's an 
       automatic penalty of data movement between kernel space and user space, together with the associated context 
       switch required to access kernel memory space. Once in user space, care needs to be taken with the buffering 
       scheme used, as large memory access strides will result in a large number of cache misses and performance 
       degradation.
 \item[System Calls] System calls require context switches, and therefore must be reduced as much as possible. At 
       least one is usually required to read a packet for the kernel UDP buffer, and sometimes polling mechanisms are 
       employed to wait for incoming data, resulting in at least two system calls per packet, which is 
       not the ideal case.
\end{description}

\begin{figure}[t!]
\begin{center}
\includegraphics[width=420pt]{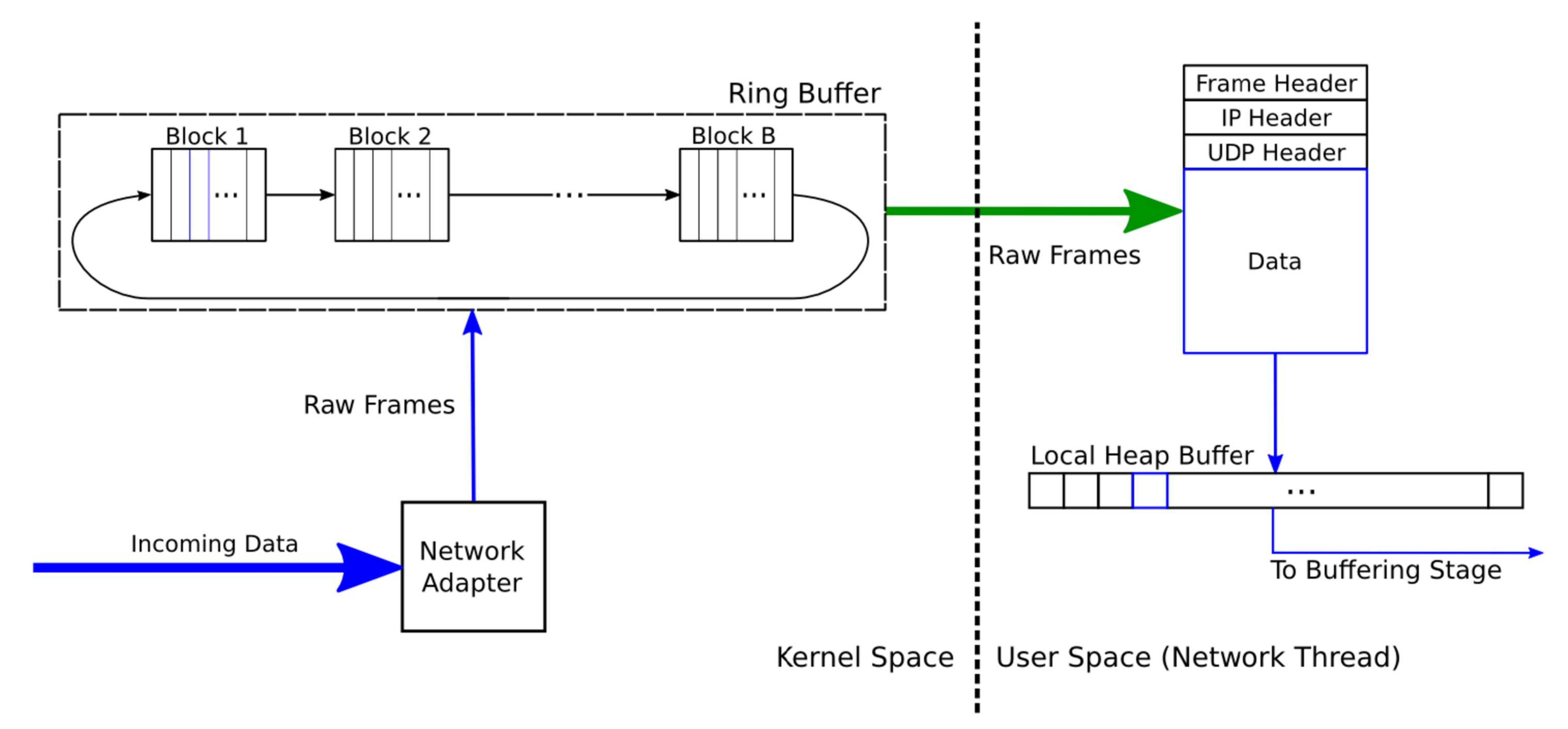}
\end{center}
\caption[Packet receiver schematic]{Packet receiver schematic, which uses PACKET sockets and memory mapping between kernel and user space for high speed packet reception. Incoming packets follow the arrows in blue, while the green arrow depicts the zero-copy mechanism between the two memory spaces.}
\label{packetReceiverFigure}
\end{figure}

The latter two factors can be greatly alleviated by careful design of the receive code. In our implementation we make 
use of the PACKET\footnote{https://www.kernel.org/doc/Documentation/networking/packet\_mmap.txt} socket interface, which provides a size configurable circular buffer, mapped to user space that can 
be used to either send or receive packets. This buffer is partitioned into 
blocks, each containing number of frames, where each frame is a placeholder for 
a network packet. With this mechanism, the user code simply waits for a frame to 
become available in this ring buffer (essentially waiting for a DMA transfer 
from the network adapter's internal memory to the circular buffer) and most of 
the time there is no need to issue system calls since the frames are read 
directly from user memory. Using a shared buffer between the kernel and the user 
also minimises packet copies, and can be thought of as zero-copy memory access. 
The major downside of the scheme is higher implementation complexity in the user 
code, which has to strip down the protocol headers itself. The internal workings 
of this scheme is illustrated in figure \ref{packetReceiverFigure}

The incoming UDP stream is received and buffered for processing using two CPU threads, the network
thread and the buffering thread. When a new UDP packet is received, the network thread reads and
interprets the SPEAD header, which defines the heap it belongs to and where its payload should be placed
within the heap, after which it is copied to the specified offset. This requires a heap buffer which is kept
local to the network thread until the entire heap has been received. A circular heap buffer is shared between
the network and buffering threads, allowing their operations to overlap and minimise locking overheads.
When a heap is fully read, the
buffering thread is notified and a new heap buffer is returned to the network thread, which starts receiving
the next heap. If a packet from the next heap is received before the current heap is fully populated, then
all missing packets are considered dropped, thus resulting in zeroed-out ``gaps`` in the heap, which are then
handled by the RFI thresholding stage. Increasing the number of slots in the circular buffer reduces the
probability that the network thread is kept waiting for the buffering thread to free up slots, which might
happen when CPU-intensive tasks are scheduled on the same core as the buffering thread.

The buffering thread's main function is to create larger data buffers to be copied directly to GPU
memory, composed of multiple heaps. To overlap the creation of these buffers with GPU execution, a
double-buffering system is used. Heaps from the circular buffer are separated into chunks containing a
time-slice from a single channel per beam, and these are copied to their respective locations within the
GPU buffer. When this is fully populated, the main pipeline thread is notified and the pipeline is advanced
by one iteration after all the GPU-based processing has finished for the previous one.

Mapping the network and buffering threads to specific CPUs and cores can have an effect on overall
system performance. On a multi-CPU system, specific Interrupt Requests (IRQs) are generally handled by a subset of
the available cores, the mapping of which is determined by a hexadecimal bit mask (on Linux systems). 
In cases where PCIe slots are controlled by multiple I/O chips, assigning the 
IRQ related to the 10GigE card
to a core within the physically closest CPU will reduce latency and overhead. The network thread's affinity
can then be assigned to the same core. An additional optimisation option is to set the buffering thread's 
affinity to a core residing on the same CPU in order to avoid false sharing situations among cache lines residing
on different CPUs, which can result in performance degradation, and thus assures that cache lines containing circular
buffer values are not invalidated. The actual performance gain achievable with these fine tunings is system dependent.

\subsection{Data Interaction Tool}

\begin{figure}[t!]
  \begin{center}
    \includegraphics[width=420pt]{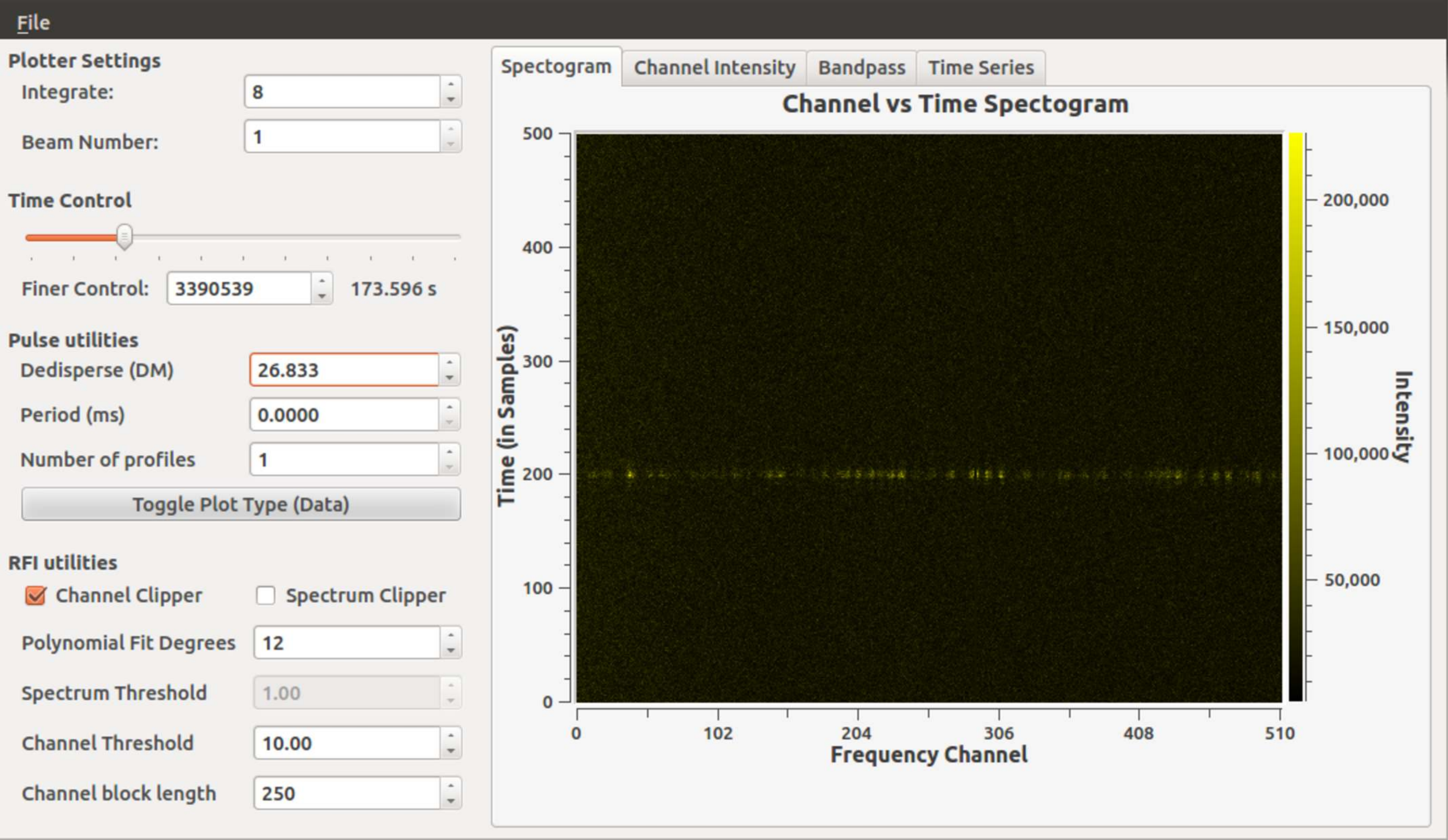}
  \end{center}
  \caption[Data interaction tool]{Data interaction tool written in C++ and QT
	  which can process data files written to disk by the transient detection pipeline. The waterfall plot
	  shows a dedispersed pulse from PSR B0329+54, integrated by a factor of 8.}
  \label{plotterFigure}
\end{figure}

The transient detection pipeline can persist data to disk in several formats, with varying word lengths, depending
on the stage at which the triggering occurred. A custom tool for data interaction was required to check the
correctness of the pipeline, as well as the data streamed by the S-engine in order to make sure the digital
backend was properly initialised. This tool would also serve the purpose of ``empirical parameter setting'', in 
which several threshold factors needed to configure the pipeline, including the RFI thresholds, can be tested and
determined. Due to the potentially large data files which would need to be processed, and the processing steps 
required for this visualisation, including decoding the $\mu$-law encoded data, this tool was written in C++, whilst
the user interface was designed and developed with QT\footnote{\;http://qt-project.org/}. OpenMP was used to accelerate
compute-intensive features. This front-end of this tool is shown in figure \ref{plotterFigure}.

A typical use case for this tool involves loading a persisted data buffer, or a SIGPROC-style \cite{sigproc}
filterbank data file, and decoding it, separating the beams and generating a different filterbank file per
beam. Successive data buffers can also be combined into single observation files in this stage as well. This 
operation is parallelised across multiple beams. Once an observation file has been generated, the visualisation
parameters are reset and the first group of spectra are plotted. The user can then use any of the available 
features to interact with the data:
\begin{itemize}
 \item Analyse the data using various plot types, including: waterfall plot, frequency channel intensity plot, 
       logarithmic bandpass plot and the time series plot (with frequency channels summed for each spectrum).
       These plots can also be saved to disk.
 \item Use the time control features to instantly analyse any part of the file, even if this involves seeking
       through gigabytes of data. Consecutive time spectra can also be combined, or integrated.
 \item Use transient and pulsar-related features such as dedispersion and folding.
 \item Apply the RFI thresholding techniques described in section \ref{rfiMitigarionSection}, with an optional
       user-defined channel mask.
 \item Export segments of the buffer, including folded profiles, as separate filterbank files to disk.
\end{itemize}

\section{BEST-II Observation Results}
\label{medicinaObservationSection}

\begin{figure}[t!]
  \begin{center}
    \includegraphics[width=400pt]{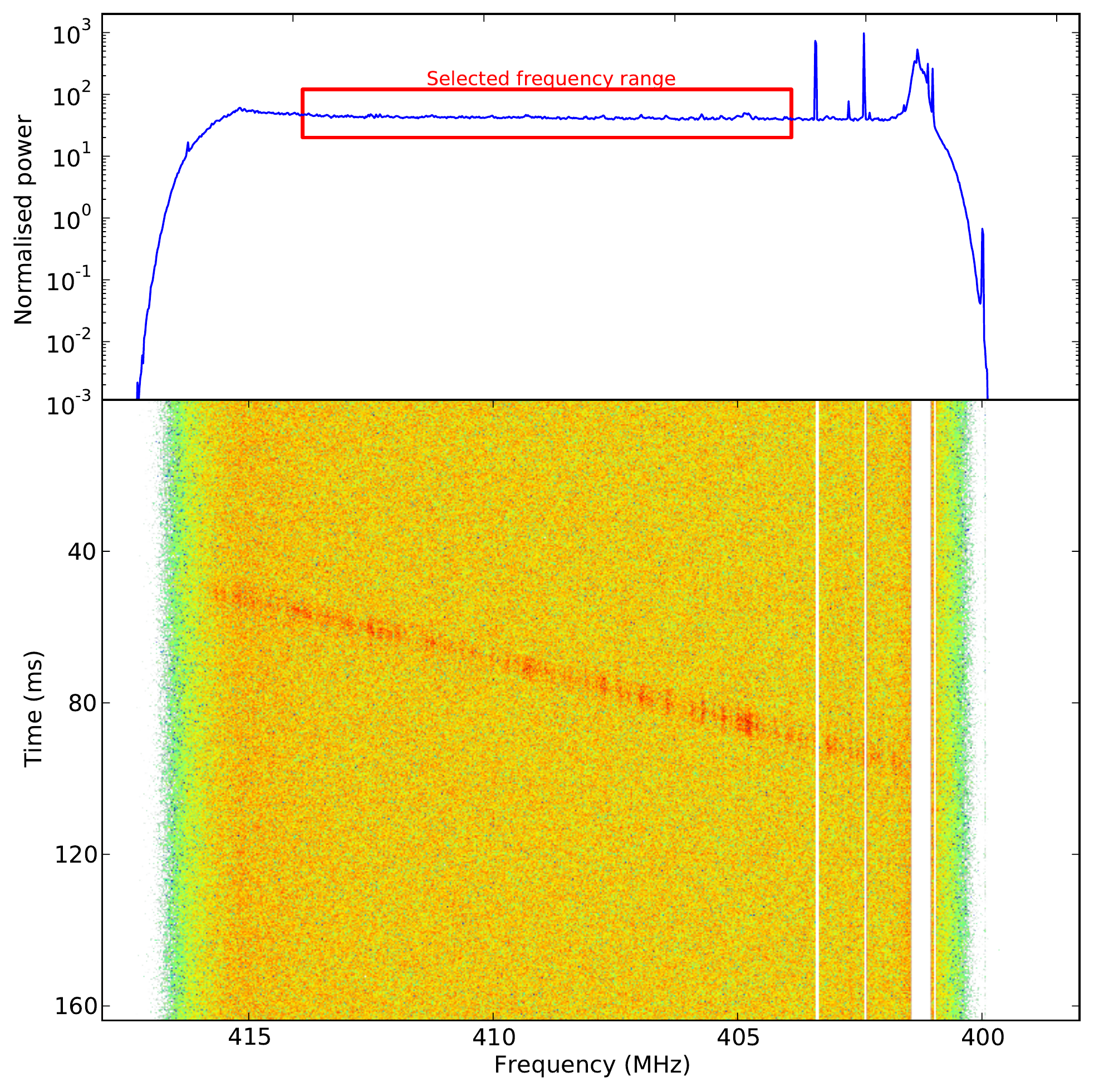}
    \end{center}
    \caption[BEST-II bandpass]{Top plot shows the full-band BEST-II bandpass, 10 MHz of which were selected
	  for further observations, depicted by the red border. Bottom plot is a full-band waterfall plot of a
	  single dispersed pulse from PSR B0329+54. The colour map represents normalised, logarithmic power.
	  Noisy channels were masked.}
  \label{fullBandpassFigure}
\end{figure}

The transient detection pipeline was deployed at the BEST-II array, attached via a 10GbE link to the S-engine.
Initial tests indicated that the full 20 MHz band contained significant 
narrowband RFI, with 2 MHz at both edges of the band lacking any detectable flux 
since they lie outside of the BEST-II RF band. For this reason, the S-engine was 
configured to discard half of the band, using only 10 MHz for the rest of the 
observations, from 413.9 MHz to 403.9 MHz. Figure \ref{fullBandpassFigure} shows 
the full bandpass, with the selected subband highlighted, as well as a waterfall 
plot of a single dispersed pulse from PSR B0329+54, for which the noisy channels 
were masked. An on-pulse, frequency-dependent, instrumental effect is also 
clearly visible in this plot, the source of which is still unknown. This results 
in a loss in S/N of any detected signals.

\begin{figure}[t!]
  \centering 
  \subfloat[Broadband RFI spike, no clipping]{\includegraphics[width=190pt]{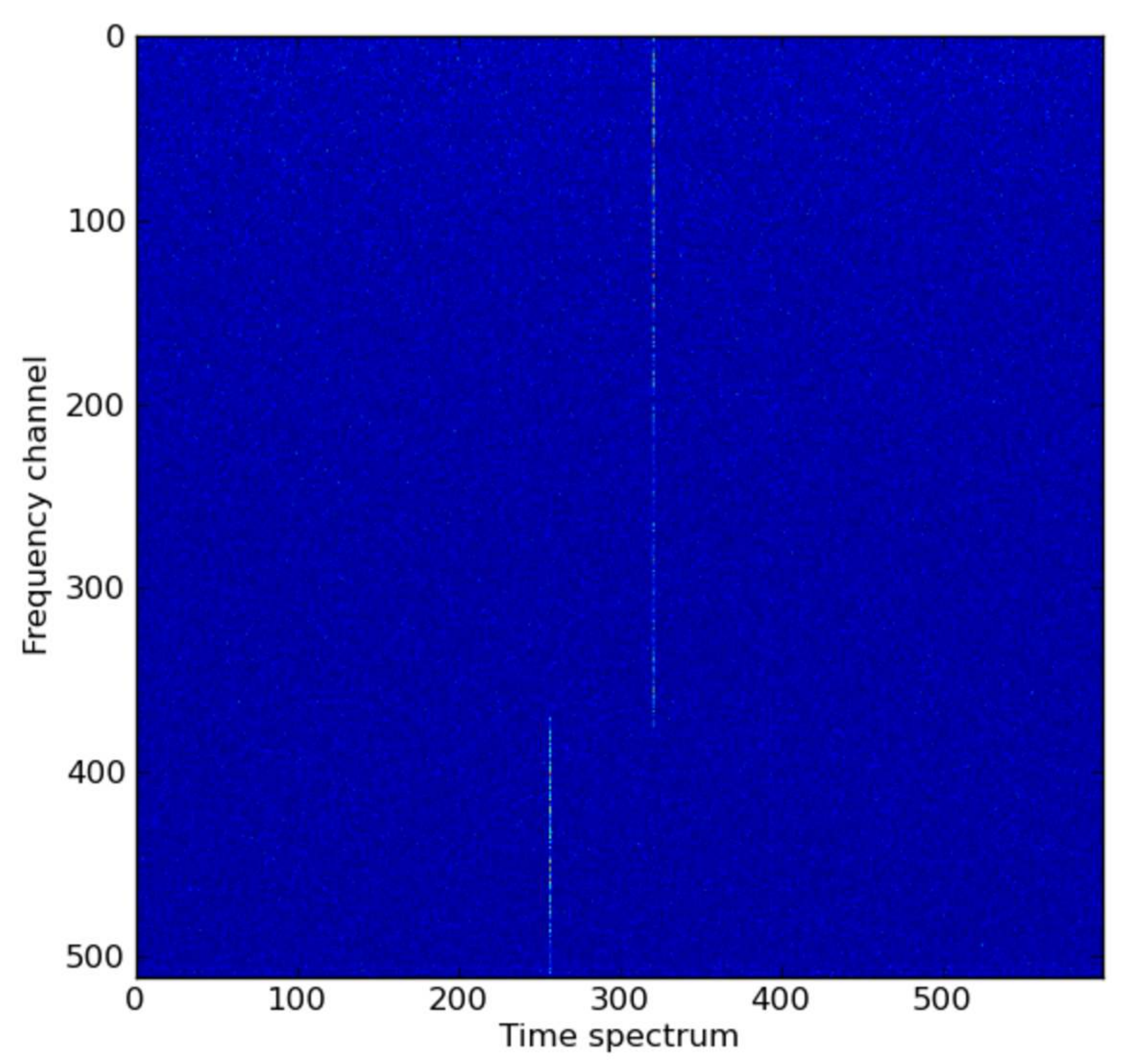}}
  \hspace{8mm}
  \subfloat[Clipped broadband RFI spike]{\includegraphics[width=190pt]{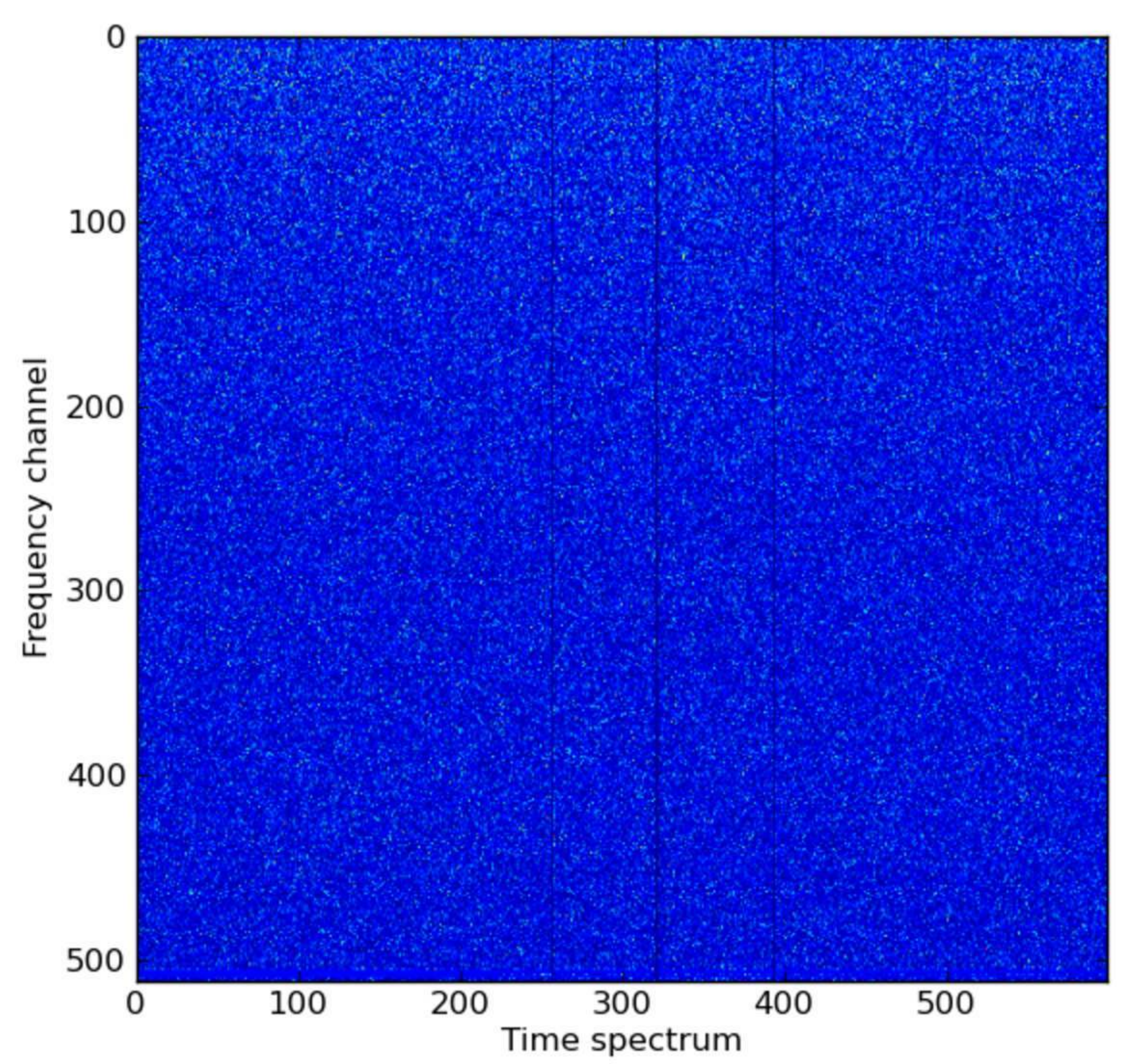}}
  \hspace{8mm}
  \subfloat[Pulse with narrowband RFI]{\includegraphics[width=190pt]{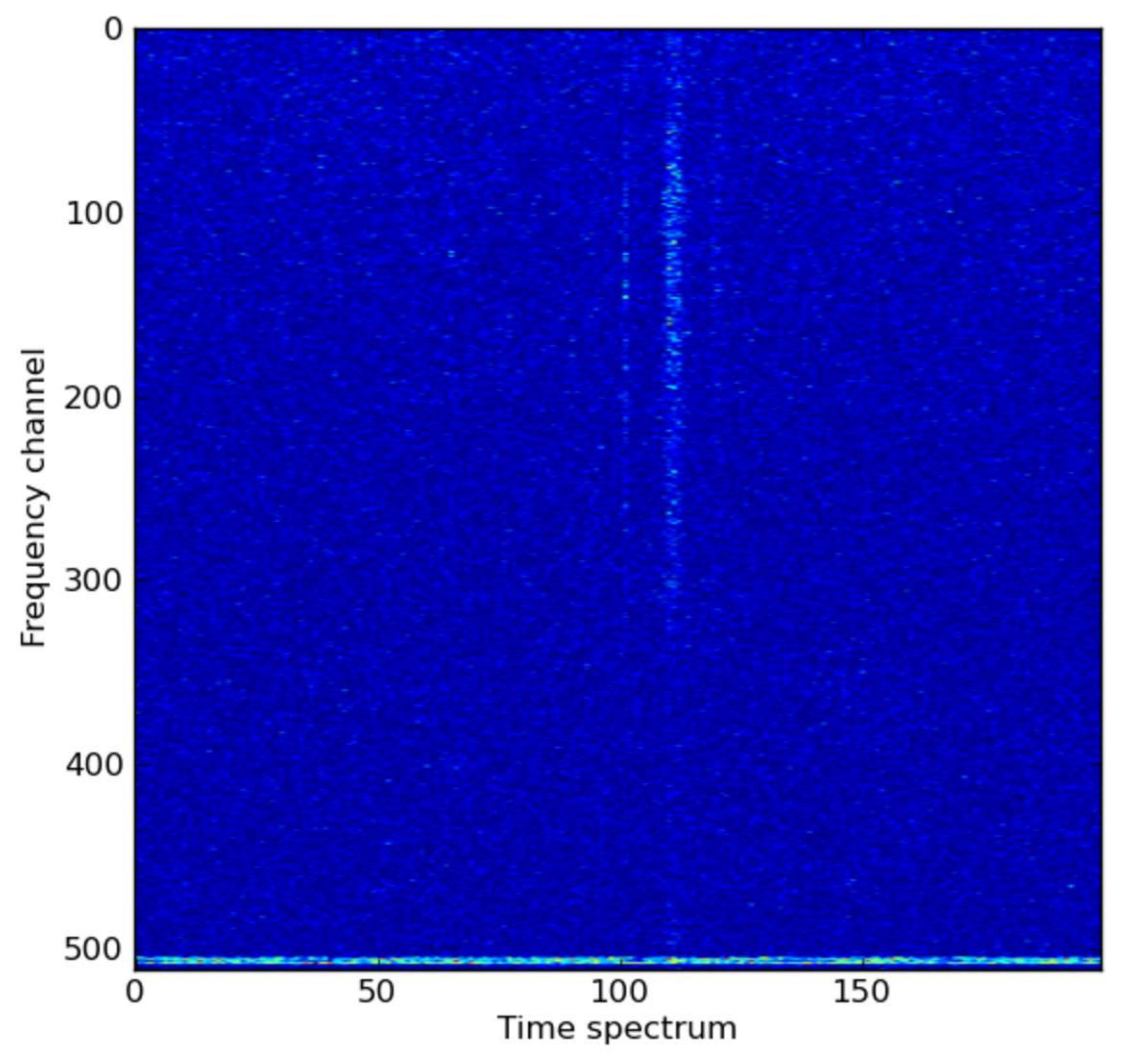}}
  \hspace{8mm}
  \subfloat[Narrowband RFI clipped]{\includegraphics[width=190pt]{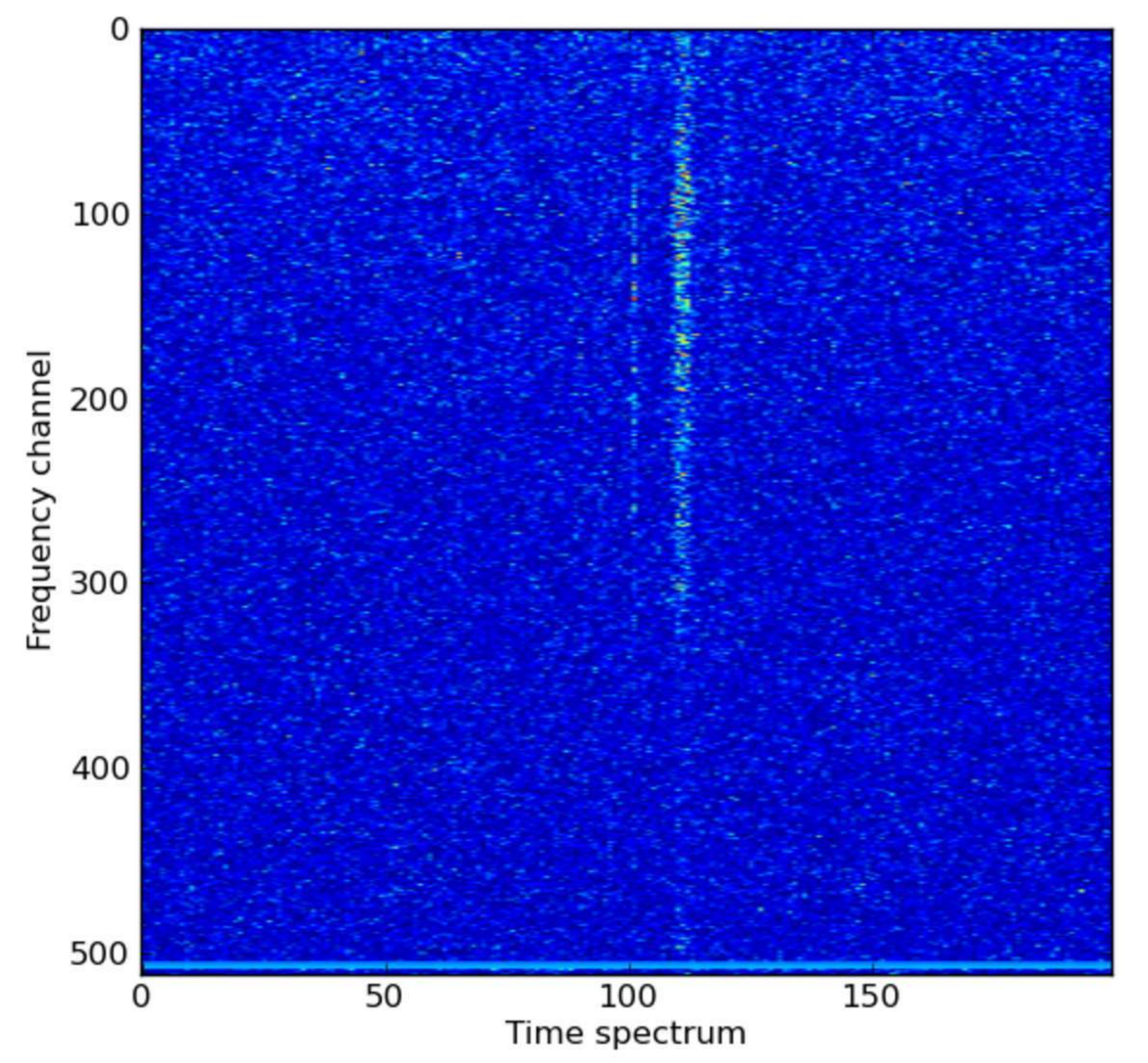}}
  \caption[Examples of BEST-II RFI events and their removal]{RFI events during a test observation using a 10 MHz 
	   bandwidth between 413 and 403 MHz. Plot (a) shows three broadband 
spikes affecting spectra at 260, 330 and 395, while (c) shows narrowband RFI 
affecting frequency channels 504 to 508. Plots (c, d) also contain a transient 
signal originating from pulsar PSR B0329+56 which was being observed during the 
event. For both cases all thresholding stages were enabled and had the same 
threshold parameters. The power intensity values are arbitrary. }
  \label{medicinaRFIFigure}
\end{figure}

\begin{figure}[t!]
\begin{center}
\includegraphics[width=425pt]{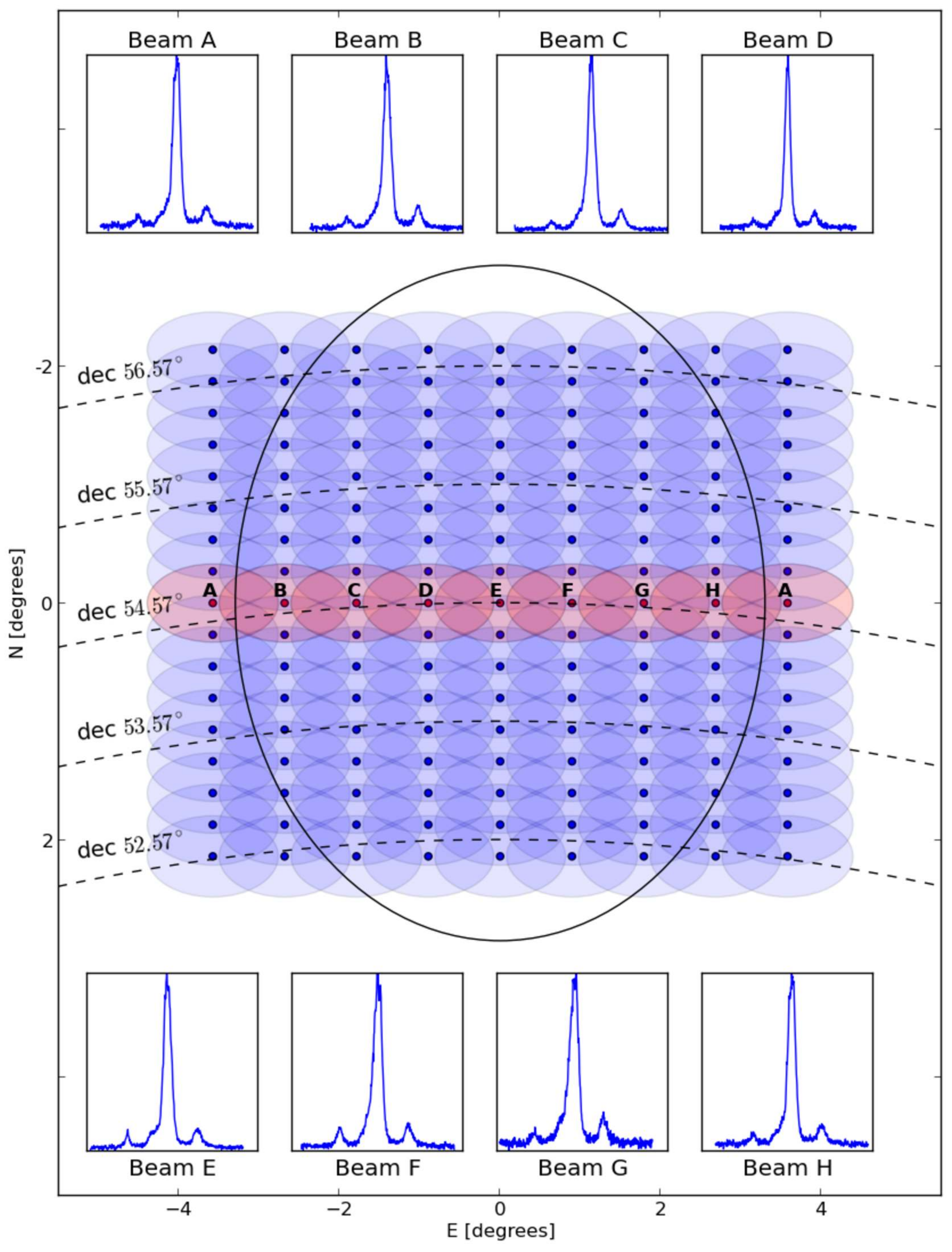}
\end{center}
\caption[PSR B0329+54 transiting through 8 synthesised beams]{S-engine beam configuration, also showing a transiting
	test with PSR B0329+54 during a 1800 s observation. The integrated pulse 
profile for each beam is also shown, generated by folding 50 profiles.}
\label{beamConfigurationFigure}
\end{figure}

Since the BEST-II array is located next to an urban area, RFI is frequent and strong. Figure \ref{medicinaRFIFigure}
shows two examples of both narrowband and broadband RFI events which occurred during our test observations. The
efficacy of the RFI thresholding stage is also demonstrated, where both RFI events were detected and mitigated.
Plots (c, d) also contain a transient signal originating from PSR B0329+54 which 
was being observed during the event.
The transient signal was not affected by the RFI thresholding stage.

The S-engine creates a 2D grid of synthesised beams within the primary beam, out 
of which eight can be selected for output. The primary beams and centres of each 
synthesised beam, together with their FWHM, are shown in figure 
\ref{beamConfigurationFigure}. Beams get wider further from the zenith, which 
for BEST-II is at 44.52$^\circ$. These beams were chosen to create a ``strip'' 
along the E-W direction (along RA) so that pointing towards a transient source 
would result in it transiting across multiple beams, which is useful for testing 
the digital beamformer, the pipeline setup as well as the data transmission 
between the two. It should be noted that neighbouring beams overlap at 
$\sim$81\% of peak power, therefore S/N is reduced when a source lies between 
beam centres. 

 \begin{figure}[t!]
  \centering 
  \includegraphics[width=400pt]{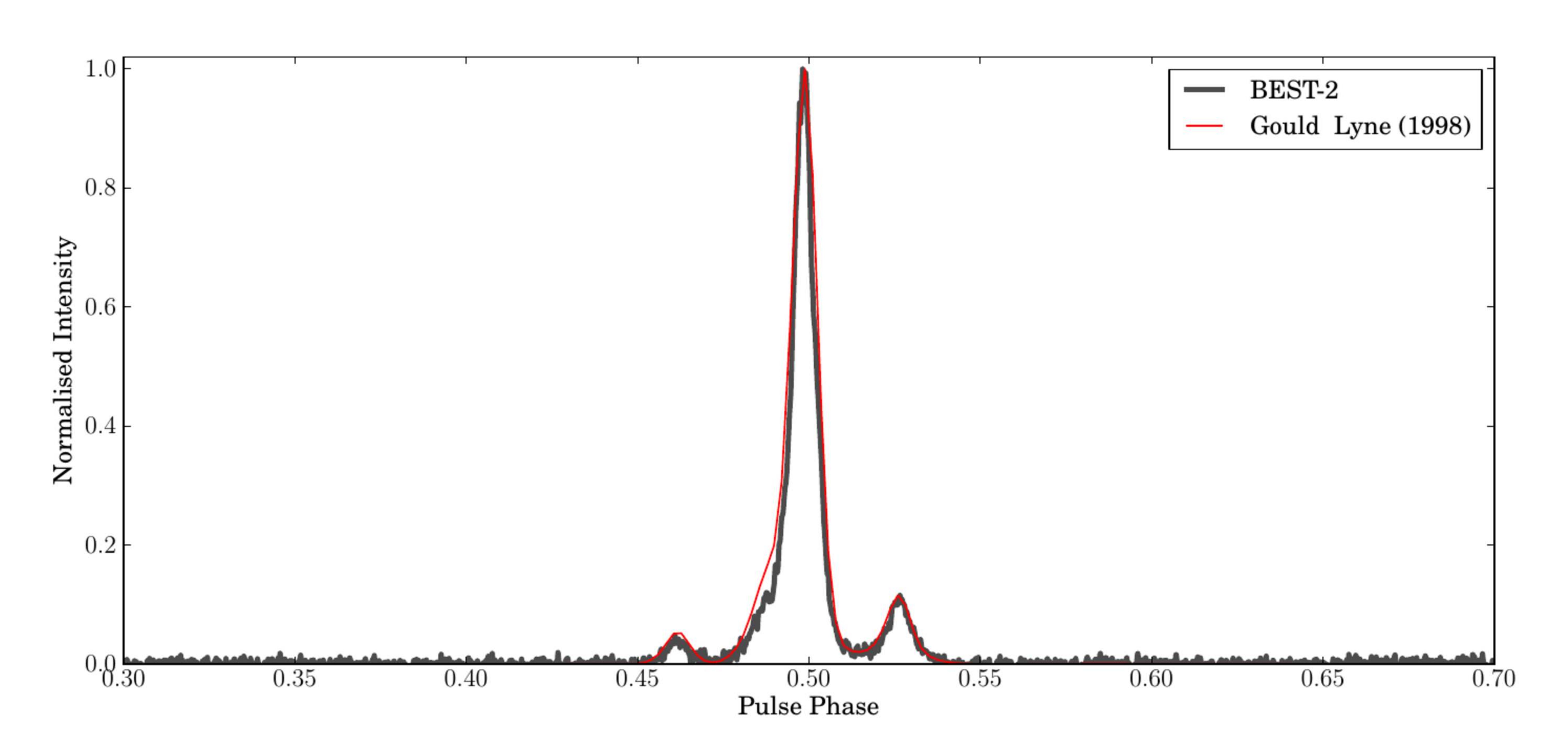}
  \caption[Integrated pulse profile for PSR B0329+54]{Integrated pulse profiles of PSR B0329+54 and consisting of 200 profiles overlaid over a profile obtained using the Lovell telescope at 408 MHz \cite{Gould1998}. (Credit: Jack Hickish)}
  \label{profilesFigure}
\end{figure}

Several test observations were performed on known bright pulsars, especially PSR B0329+54, which is the brightest 
transient source that can be observed with BEST-II, located at RA 03:32:59.37 
and DEC +54:34:43.57. The beam 
configuration for these observations is shown in figure \ref{beamConfigurationFigure}, where the central 
row of beams is selected for output (shaded in red). The pipeline was used in ``persistence mode'', where 
all the channelised complex-voltages from all the beams are quantised and persisted to disk. The integrated pulse
profiles were then generated for each beam using 50 profiles. The differences between the profiles can be attributed 
to RFI events which occurred during the observation, sometimes resulting in a recalculation of the quantisation 
factors (this only happens when an extremely powerful RFI event occurs, as was the case during this observation).

\begin{figure}[t!]
\begin{center}
\includegraphics[width=440pt]{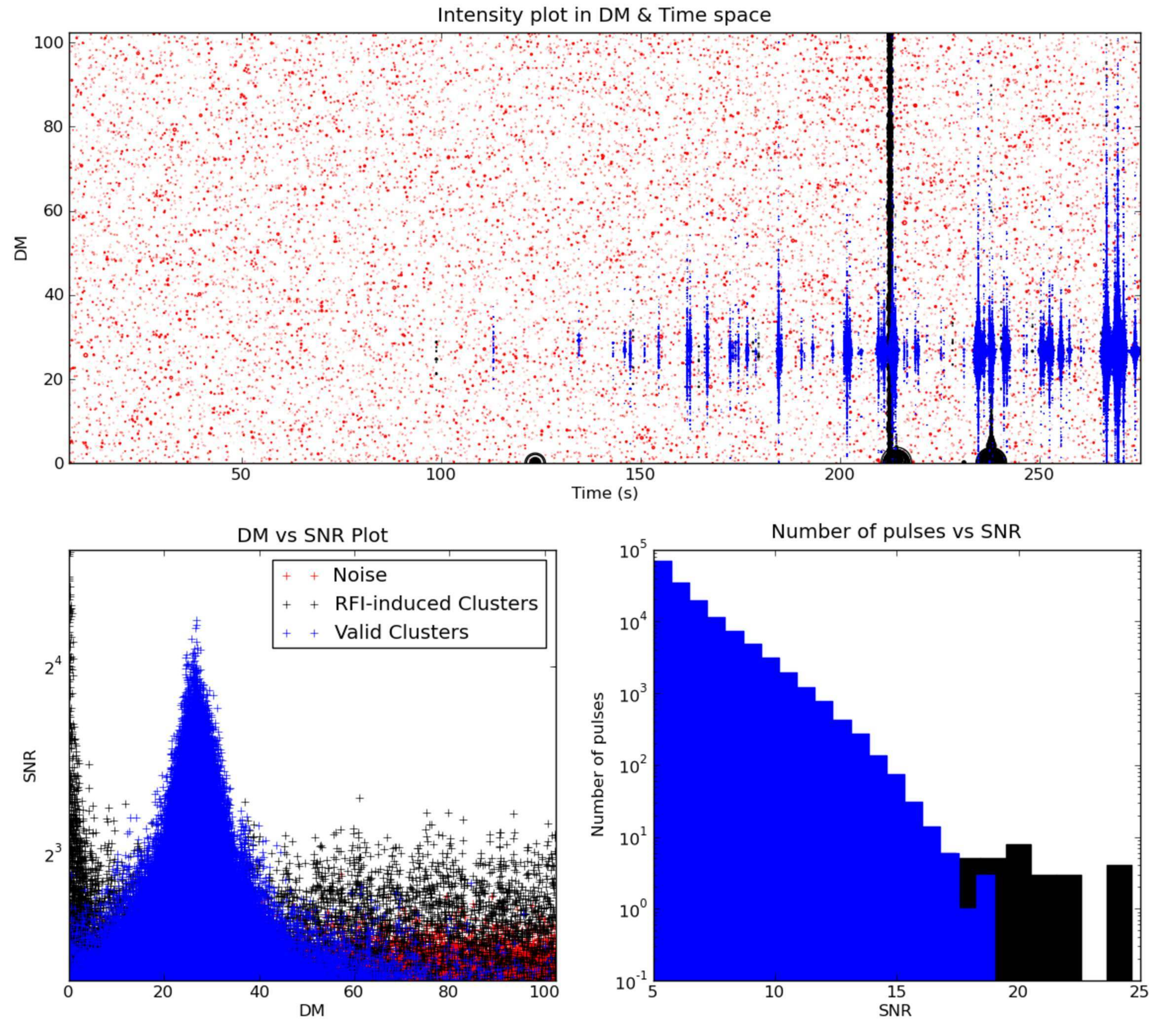}
\end{center}
\caption[Pipeline output for a test observation of PSR B0329+54]{A $\sim$280 s observation of pulsar PSR B0329+54, 
         which enters the beam at around 100 s with the pulse S/N increasing as the pulsar moves towards the centre of 
         the beam. This plot shows the output generated by the clustering and candidate selection stages, partitioning 
         the data into noise (red), RFI-induced clusters (black) and selected candidates (blue). Some of these 
         clusters are also shown in figure \ref{classificationFigure}.}
\label{b0329Figure}
\end{figure}

The integrated pulse profiles for PSR B0329+54 is shown in 
figure \ref{profilesFigure}, generated by folding 200 profiles offline with raw observation data persisted 
to disk from a single beam and overlaid over a profile obtained using the Lovell telescope at 408 MHz \cite{Gould1998}.

\begin{sidewaysfigure}[t!] 
  \centering 
  \subfloat[High S/N pulse from B0329+54]{\includegraphics[width=215pt]{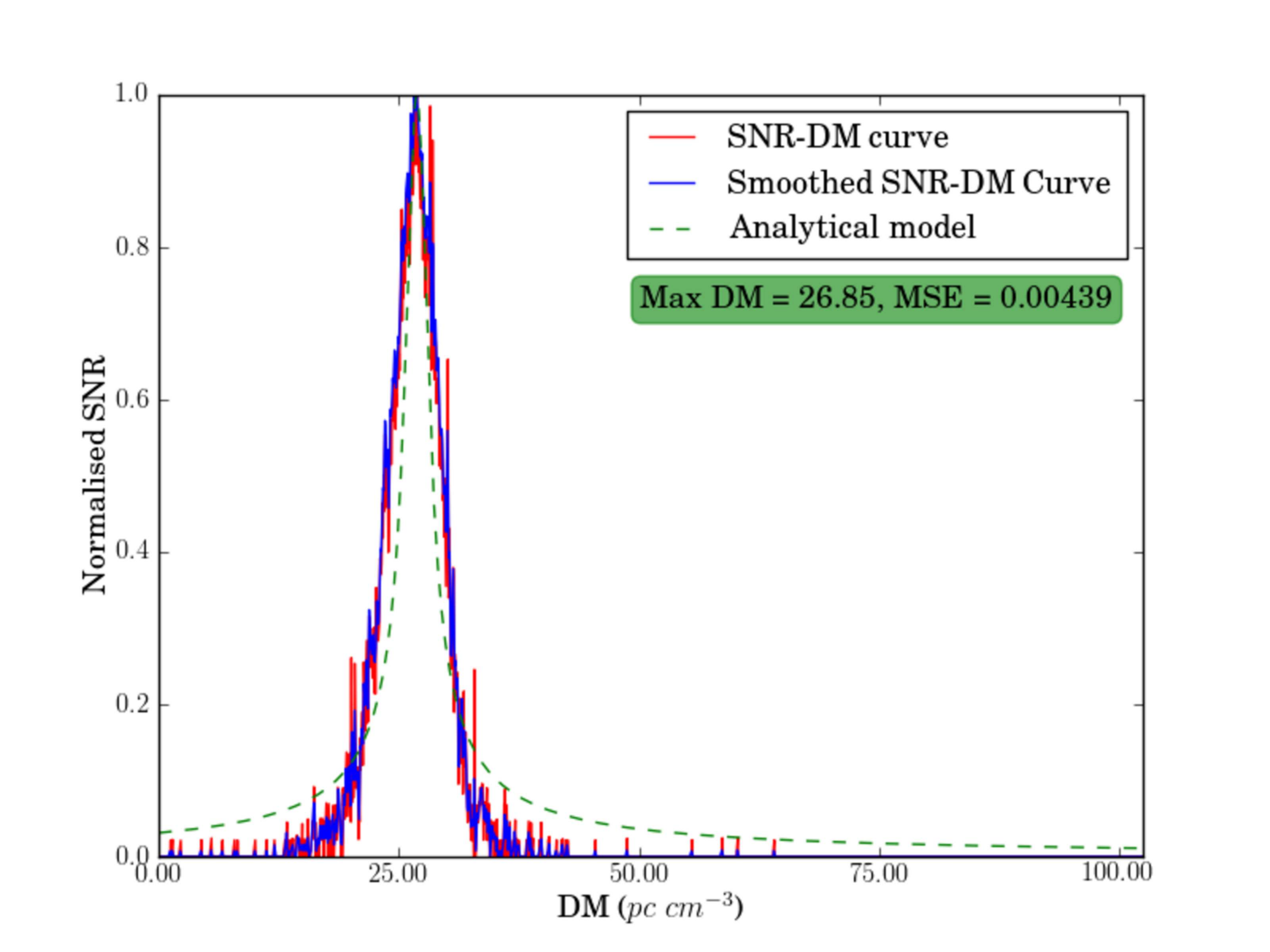}}
  \hspace{8mm}
  \subfloat[Low S/N pulse from B0329+54]{\includegraphics[width=215pt]{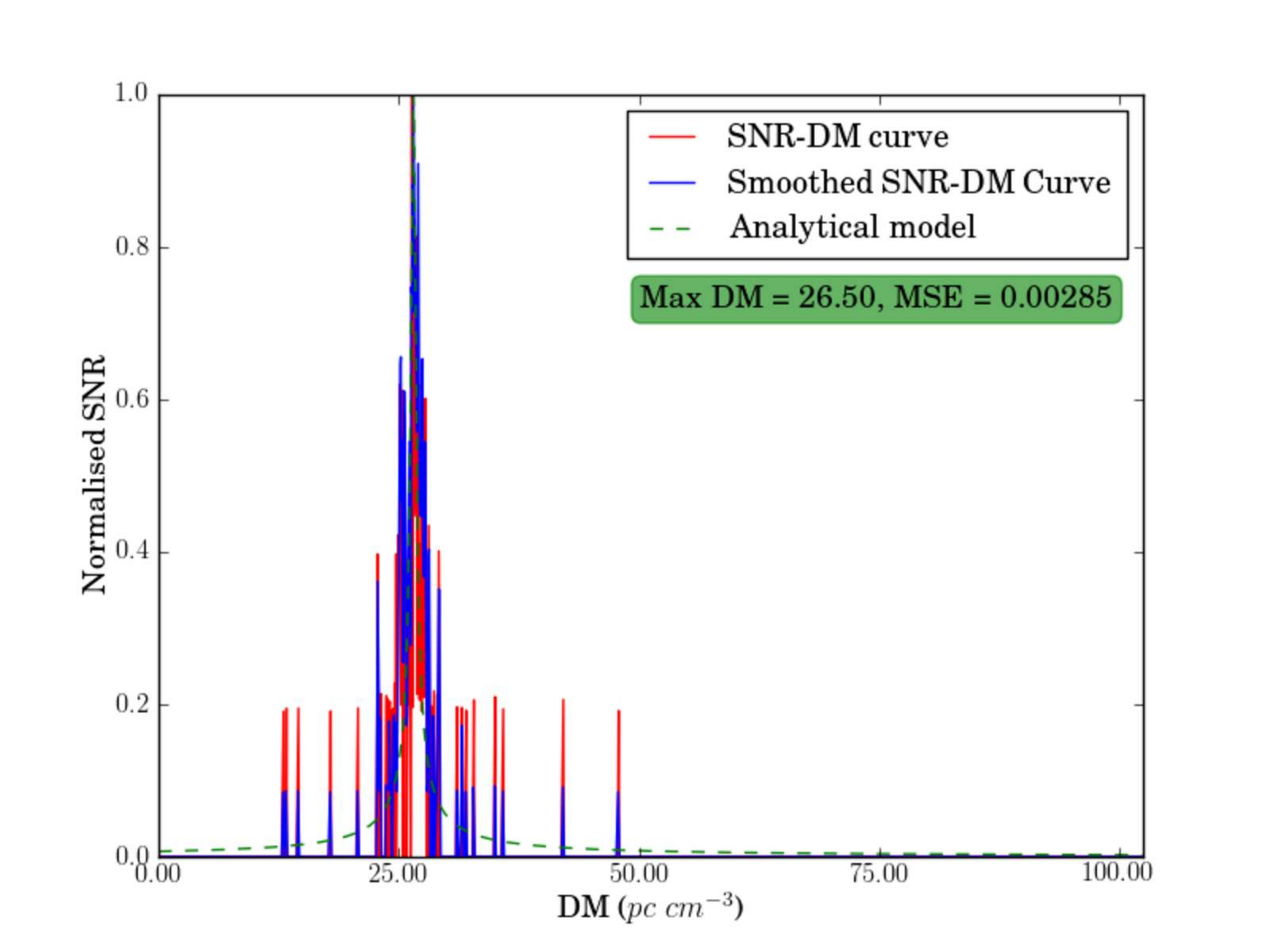}}
  \hspace{8mm}
  \subfloat[High-intensity narrowband RFI]{\includegraphics[width=215pt]{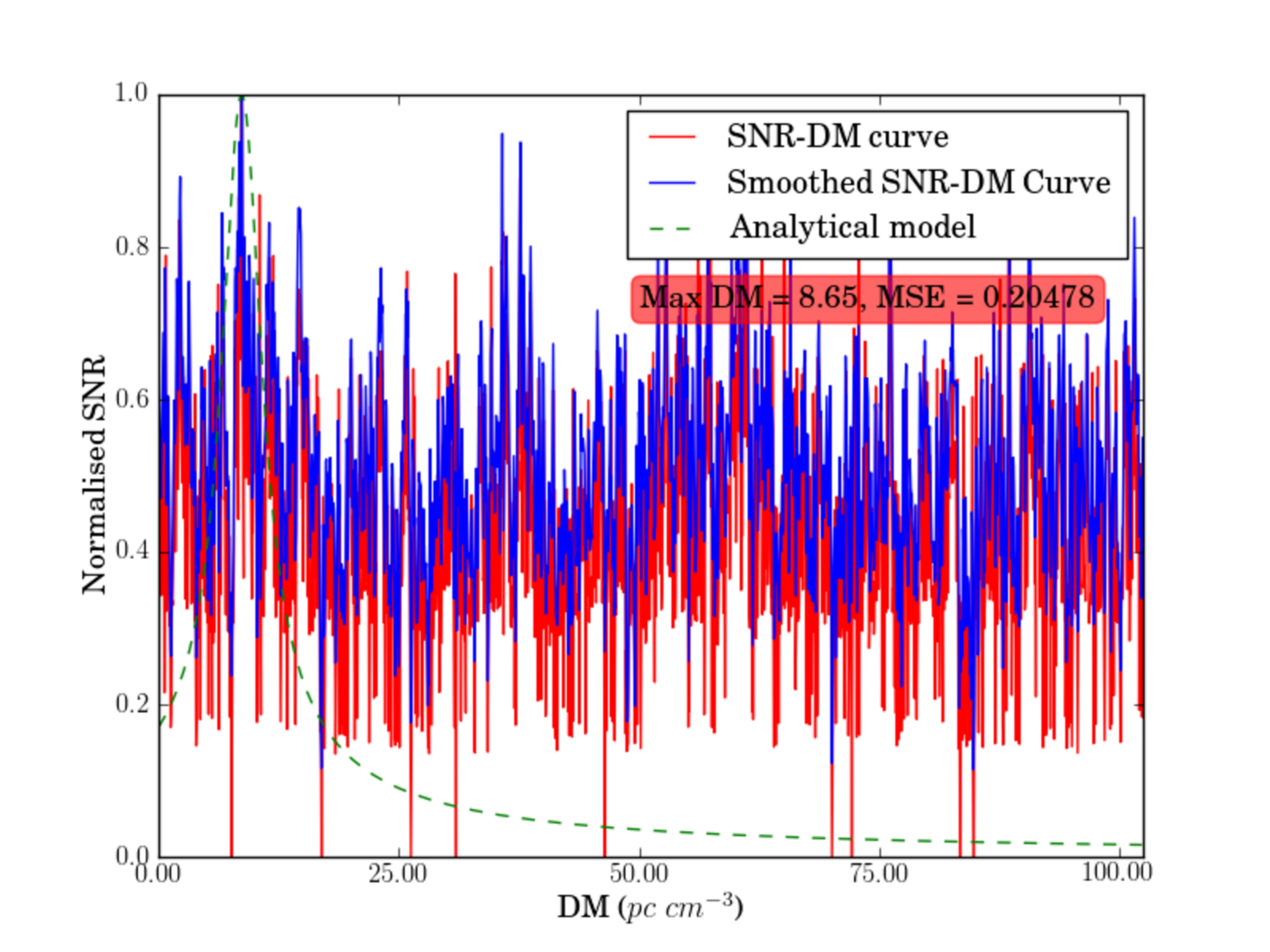}}
  \hspace{8mm}
  \subfloat[Broadband RFI]{\includegraphics[width=215pt]{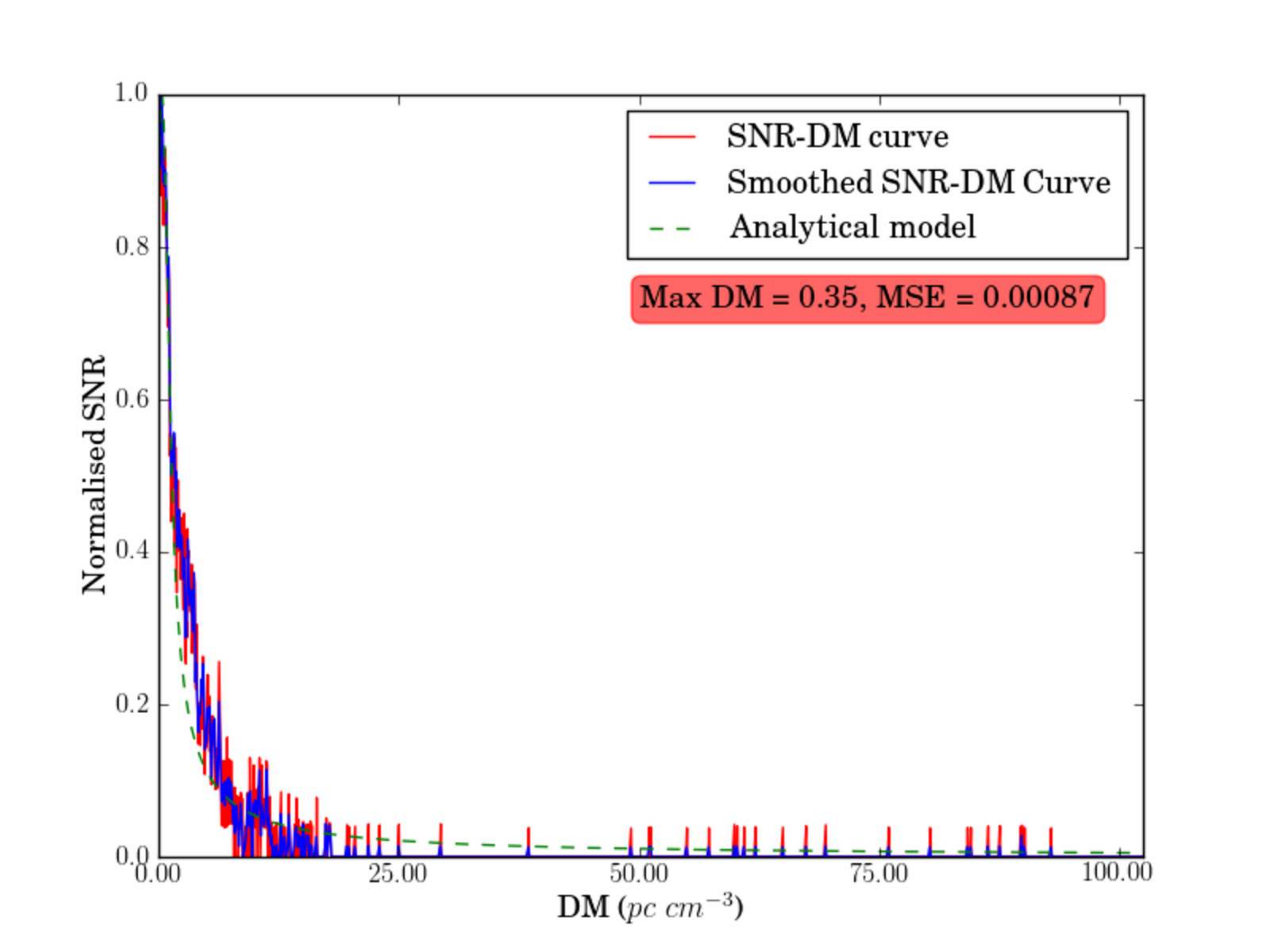}}
  \caption[Positive and negative examples of cluster classification]{Examples of clusters during the candidate 
          selection stage. The first two plots belong to pulses from B0329+54 whilst the rest are due to RFI. Plot c represents the black vertical cluster at around 215 s in figure \ref{b0329Figure}, whilst plot d is the low DM, high-S/N detection at 240 s.}
  \label{classificationFigure}
\end{sidewaysfigure}

Figure \ref{b0329Figure} represents the output of the pipeline during an observation of PSR B0329+54
for a single beam. The pulsar enters the beam at $\sim$100 s, with pulses getting stronger as it moves towards the beam's centre. Several RFI events were also detected, the most noticeable of them being three broadband events (the large detections at DM $\approx$ 0) and a bright narrowband event at around 215 s which was detected across the entire DM range. Figure \ref{classificationFigure} provides visual snapshot of four clusters during the classification stage,
with two clusters originating from pulse detections from B0329+54 and two additional clusters attributed to RFI.
Both these RFI clusters were correctly filtered.

\section{Conclusion}

We have enhanced the GPU-based transient detection pipeline described in the 
previous chapter with real-time capabilities, consisting primarily of high-speed 
stream processing functionality capable of processing a 5.12 Gb/s SPEAD stream. 
We have also described the deployment of this system on the Medicina BEST-II 
array, where a ROACH-based digital backend performs all the required 
pre-processing prior to transmitting beamformed data to the transient detection 
system. Several test observations were conducted, especially on PSR B0329+54, 
which is the brightest pulsar observable by BEST-II. Through these observations 
appropriate threshold parameters were set to various stages in the pipeline, 
leading to the automatic classification of pulses originating from B0329+54 from 
RFI-induced events, serving as an online test case for the pipeline.


\chapter{GPU-Based Beamforming for Transient Discovery}
\label{beamformingChapter}

In the previous chapters we have described the design, implementation and deployment of a real-time, GPU-based 
transient detection pipeline, where the host servers receive beamformed antenna signals for processing. The design
is linearly scalable with increasing number of beams, which can be parallelised across multiple GPUs and servers. However,
without appropriate feedback mechanisms to the beamformer, observations will be limited to the observational
parameters set during initialisation. The beams' shape, distribution across the array's FoV, sensitivity and
sky coverage depend on the beamformer's implementation and supported parameter range. This is especially limiting
for FPGA-based designs, where simply changing the beamformer's mode (for example, from incoherent to coherent)
would generally require a separate design to be loaded and configured. An alternative approach is to have the 
digital backend stream the raw antenna voltages to the host system, where beamforming is performed within the 
transient detection pipeline itself.

This setup offers several advantages. Switching between beamforming modes can be performed in real-time and with
minimal overhead, with the possibility of generating a mixture of beam types. Such a system would be useful
for on-the-fly localisation and tracking of transient candidates, whereby coherent beams can be generated 
whilst still using the default beam setup for full FoV scanning. Having the raw antenna voltages available
on the host system also enables the possibility of generating an image of any detected candidates, where sources
can be better localised. Performing this in real time allows immediate follow-up observations, however this 
requires accurate source extraction techniques in images, which are still being investigated, especially for 
slow transient surveys. Several authors also investigated the benefits of sub-arrays for transient detection (for example, \cite{Cordes2009,Macquart2011,Colegate2011,Bhat2013}), where an array is partitioned into sub-arrays, each of which is incoherently beamformed and processed, with the benefit of being more resilient to localised RFI when using coincidence filtering techniques.

In this chapter we investigate the applicability of GPUs for beamforming purposes. We design and implement 
a standard, GPU-based, coherent beamformer which can generate multiple beams with arbitrary directionality.
We also integrate this kernel with the transient detection pipeline described in chapter \ref{pipelineChapter}
and demonstrate the performance and flexibility of this system.

\section{Beamforming}
\begin{figure}[t!]
  \begin{center}
  \includegraphics[width=400pt]{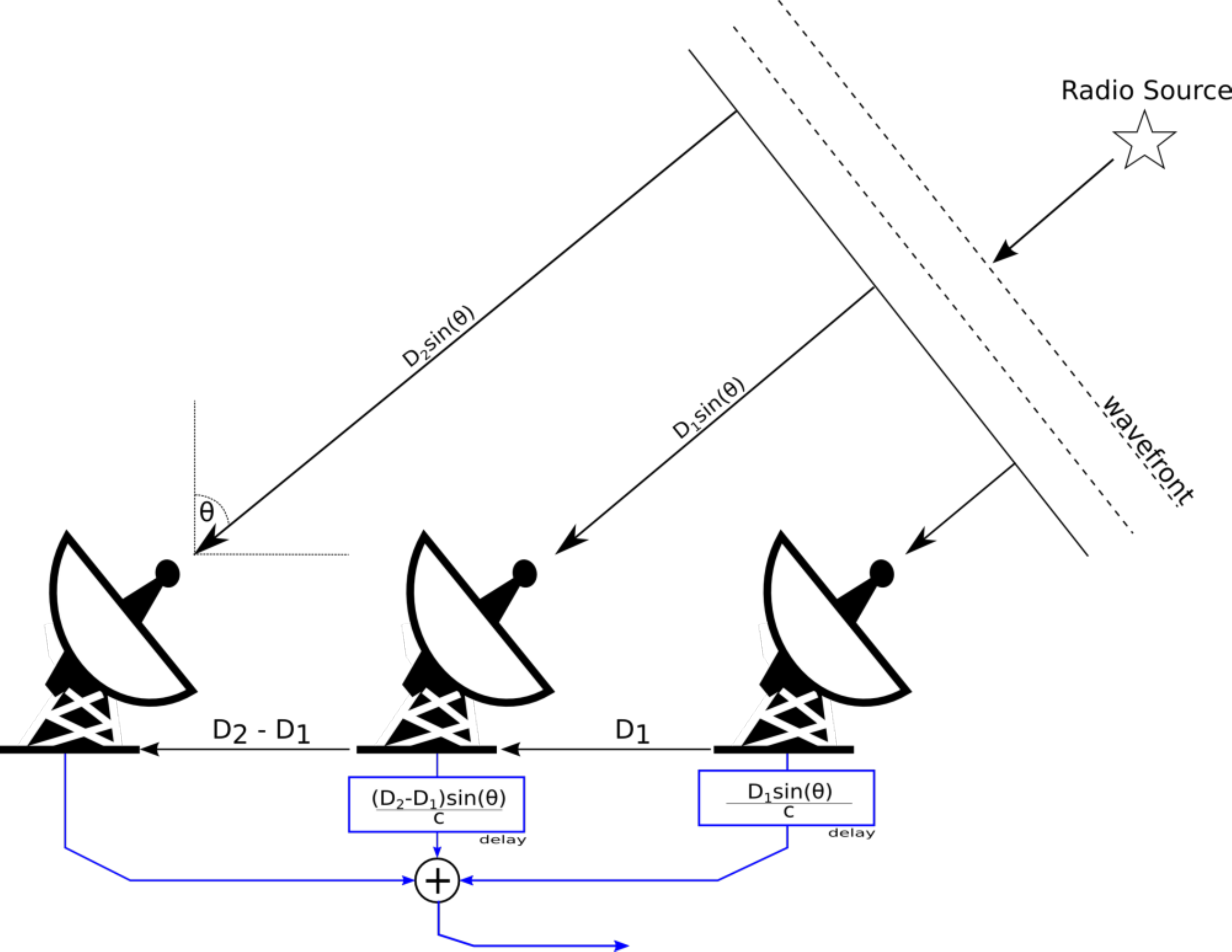}
  \end{center}
  \caption[Beamformer schematic]{Schematic of a simple 1D, 3 antenna beamformer forming a beam in direction $\theta$. Signals 
	  from each spatially separated antenna are delayed and summed such that the response is greatest in the 
	  desired direction. Figure adapted from \cite{Hickish2013}.}
  \label{beamformerFigure}
\end{figure}

In an array with $N$ elements a radiation wavefront originating from a particular direction will reach element
$n$ at time $n_t$. Adding signals from all these elements together without 
compensating for 
element-dependent delay is the process of incoherent beamforming. The maximum amplitude is achieved when the 
signals originate from a source perpendicular to the array, where they are highly correlated and add 
constructively. Alternatively, if the signals originate from a non-perpendicular direction they will arrive
at different times, will be less correlated and result in a lower output 
amplitude. Incoherently combined arrays are referred to as incoherent arrays 
(IA). Coherent beamforming relies on the fact that for a 
given array configuration the relative delays between arrival
times $t_{0..N-1}$ are a function of the direction of propagation of the incident wave. Artificial delays 
or phase shifts, in the time and frequency domain respectively, can be applied to the received signals from each 
antenna, causing them to add constructively when element signals are summed. This process is illustrated in figure
\ref{beamformerFigure} for a simple 3-element, one-directional array. Coherently combined arrays are referred to as phased arrays (PA).

After summing to form a beam towards the incident radiation with wavevector $k_0$ the response $B(k)$ of the beam 
to radiation with other wavevector directions $k(\theta, \phi)$ is given by summing the signals contributions 
from individual antennas. If each individual antenna in an array has a response to radiation given by $A_n(k)$, 
this is given by:
\begin{equation}
\label{beamResponseEquation}
 B(k)=\sum\limits_{n=0}^{N-1}A_n(k)e^{i(k-k_0)\cdot r_n}
\end{equation}
where $r_n$ is the position vector of antenna $n$. This is equivalent to the Fourier transform of the contributions from receiving elements weighted by their individual antenna responses. In cases where the antennas have very similar response functions equation \ref{beamResponseEquation} reduces to the Fourier transform of the antenna distribution modulated by the response pattern of each antenna. In this case, the FoV of the synthesised beam is approximately given by $\nicefrac{\lambda}{r_{\text{max}}}$, where $r_{\text{max}}$ is the maximum separation of the antennas used to form the beam.

Since prior to beamforming each individual antenna in an array sees a portion of the sky determined by its
response function $A_n(k)$, it is possible to effectively form beams in any direction where $A_n$ is
significantly non-zero. It is also possible to generate multiple beams by making copies of the input antenna
signals and adding them with different phase weightings, thereby increasing the filling factor of an array's
full FoV. Choosing the number of beams to form amounts to balancing the FoV observed by an array
(which increases linearly with each beam added) with the number of signals required to be processed.
Each beam can be treated as if it is a signal from a single antenna with a response equal to the beam 
response, and so can be fed directly to backend detectors and used for time-domain observation, such a
cosmic transient event observations and transient surveys. The latter generally require a wide FoV 
and high sensitivity, however achieving this for larger arrays can be prohibitively expensive in signal 
processing costs since the number of required beams goes as $(D/d)^2$, where $D$ is the physical extent of the 
array and $d$ is the size of an individual element. Alternative strategies include sacrificing sensitivity
to achieve maximal FoV by using IA mode, or partitioning the array into multiple PAs and processing them
independently.

\section{Beamforming implementation on GPUs}

In this section we describe a standalone GPU implementation of a coherent beamformer which is capable of generating
multiple beams concurrently. Performing this in real-time poses several challenges. First of all the digitised 
antenna voltage streams need to be transported to the backend server, which has to cope with the generally high data rates. 
Network links and devices can achieve very high transfer rates, however in order to offload processing to a 
GPU all this data needs to be transferred through PCIe connections to GPU memory, which have limited peak bandwidth. 
This can create an upper bound on the number of beams which can be generated. Also, depending on the representation 
scheme used for the generated beams, the amount of available GPU memory might limit the level of parallelism 
achievable. These factors will need to be considered when a beamforming kernel is deployed within a real-time 
pipeline, and will be discussed in greater detail shortly. 

Ignoring the weight calculation scheme used to point each beam, the computational complexity for a generic 
coherent multi-beam beamformer is $\mathcal{O}(N_b  N_f  N_a N_t)$
where $N_b$ is the number of synthesised beams, $N_t$ is the number of time samples, $N_f$ is the number of 
frequency channels and $N_a$ is the number of antennas. The beamforming process is essentially a matrix-vector
multiplication, which involves multiplying a vector of $N_a$ antennas by an $N_b \times N_a$ matrix of
coefficients to form $N_b$ beams, per frequency channel. This incurs a computational cost of $N_a$ complex multiply
accumulates per beam and frequency channel. 

This algorithm is trivially paralellisable across the frequency, time and beam dimensions. A simple and na\"{i}ve 
implementation would have a single thread combine all the antennas, applying appropriate weights, for a single
$(f,t,b)$ triplet. This would require $2N_a$ global memory requests for every thread, since $N_a$ antenna values
and $N_a$ complex weights are needed, with an instruction count of $8N_a - 1$, resulting in a flop to 
data request ratio of 4. Due to the large global memory latencies, such a ratio would make this implementation
bandwidth limited within the GPU, and thus data reuse schemes need to be employed to increase performance.
By making the assumption that beam coefficients do not need to change within small time frames, it is possible
to reuse the same weights for all the time bins residing in a GPU buffer. This is especially true for wide beams, however
does not hold when very narrow beams are required, or when tracking non-astrophysical objects such as satellites.
This drastically reduces the number of coefficients which need to be generated and read from global memory within
a thread block. We use this assumption to partition beam generation, where a 3D CUDA grid is mapped to the input 
space, with time bins varying in the x-dimension, frequency channel along the y-dimension and beam subset
along the z-dimension. Thus each thread block generates a subset of the beams for a number of time bins, for a
single frequency channel. Within each thread block, local beam accumulators are declared and stored in registers,
one per beam, to which weighted antenna values are added. The antenna coefficients are loaded once per antenna group
and stored in shared memory. This antenna partitioning is required as otherwise too much shared memory would be 
utilised, especially for large arrays, resulting in a decrease in occupancy and GPU compute resource utilisation.

Our kernel was originally implemented to match BEST-II backend specifications, most notably the output data format
of the F-engine, which sends out antenna signals as 4-bit, two's complement, floating point, complex voltages. This 
format enables the packaging of each antenna value as a single byte, and groups of 4 antennas can in turn be
packaged as 32-bit words. This is beneficial for the GPU kernel as each group can be loaded with a single
memory request, which can be coalesced if they are accessed contiguously by a thread warp. For this reason,
antennas are processed in groups of 4 in the beamformer, and complex coefficients are loaded in groups of the 
same size as well. In the inner-most loop, where these antennas are accumulated to form beams, this also has the
effect of increasing instruction level parallelism, thus decreasing execution time. Breaking down antennas into 
groups of 4 has the additional benefit of making the kernel extensible to larger arrays without affecting 
performance. Therefore, we have effectively implemented a generic multi-beam beamformer which scales linearly with
increasing bandwidth, number of antennas, number of frequency channels and number of output beams.

\begin{algorithm}[t!]
 \caption{Coherent beamforming GPU implementation}
 \begin{algorithmic}
    \REQUIRE input, output, coefficients, nsamp, nchans, nants, nbeams
    \STATE Declare $coeffs$ shared memory
        \item[]
    \FOR{time bins to process}
      \STATE Declare local beam accelerators $beams[beams\_per\_tb]$
              \item[]
      \FOR{$ant=1$ \TO $nants/4$}
        \item[]
        \STATE Load antenna group from global memory
        \STATE Cooperatively load required coefficients and store in $coeffs$ 
        \STATE Synchronise threads
            \item[]
        \FOR{$b=1$ \TO $beams\_per\_tb$}
	    \STATE Update local beam accumulator
	\ENDFOR
	        \item[]
      \ENDFOR
        \item[]
      \FOR{$b=1$ \TO $beams\_per\_tb$}
	\STATE Store generated beam to global memory
      \ENDFOR 
              \item[]
    \ENDFOR
 \end{algorithmic}
 \label{beamformingAlgorithm}
\end{algorithm}

\begin{figure}[t!]
\begin{center}
\includegraphics[width=250pt]{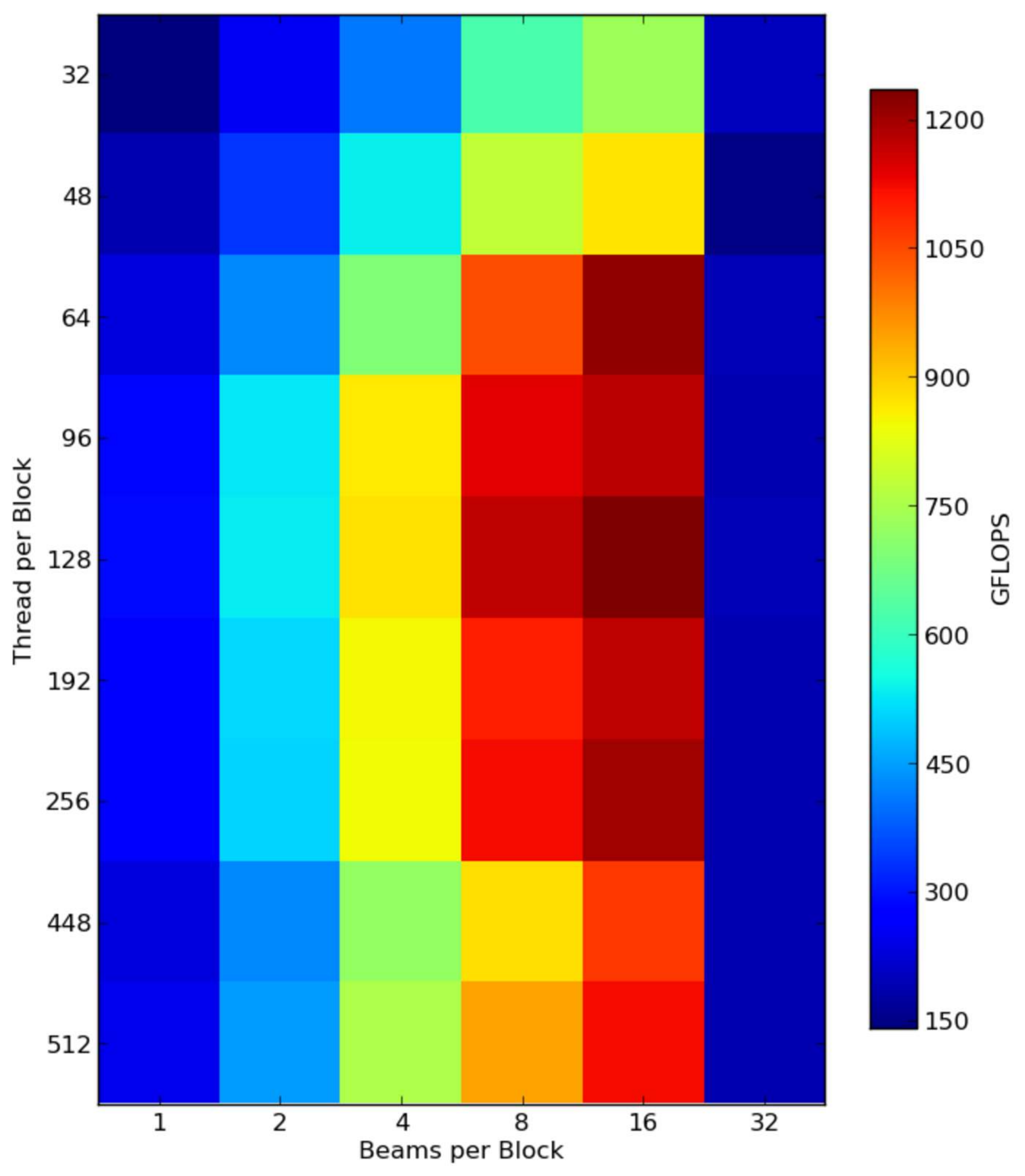}
\end{center}
\caption[Configuration optimisation for beamforming kernel]{Configuration optimisation for the beamforming kernel,
    where 128 threads per block, each generating 16 beams, provides the best parameter combination on the test device.}
\label{beamConfigFigure}
\end{figure}

\subsection{Performance and Benchmarks}
\label{beamformerBenchmarkSection}

Algorithm \ref{beamformingAlgorithm} provides a detailed breakdown of our implementation. The algorithmic design is 
similar in concept to our direct dedispersion kernel, with two configuration parameters determining the number
of threads per blocks and the number of local accumulators where antenna values are combined to form beams.
Performance benchmarks demonstrate that increasing both their values yields a performance benefit until a global maximum is reached, after which performance starts to degrade. Figure \ref{beamConfigFigure} determines the optimal values for these parameters on the test device, an NVIDIA GTX 670, where 16 accumulators and 128 threads per block provide the best performance. For these tests a 420 ms simulated buffer containing 32 4-bit complex sampled antenna voltage streams having a 20 MHz bandwidth channelised into 1024 frequency channels, was generated and processed 10 times for each combination. The resulting mean was used as a measure of the configuration efficiency.
The optimal configuration is capable of sustaining more than 1.2 TFLOPs when 16 or more beams are generated, which amounts to approximately 50\% of the device's peak theoretical performance. This kernel's limiting factor is the high number of shared memory requests needed to access phase weightings for each beam, antenna and channel triplet.

\begin{figure}[t!]
\begin{center}
\includegraphics[width=400pt]{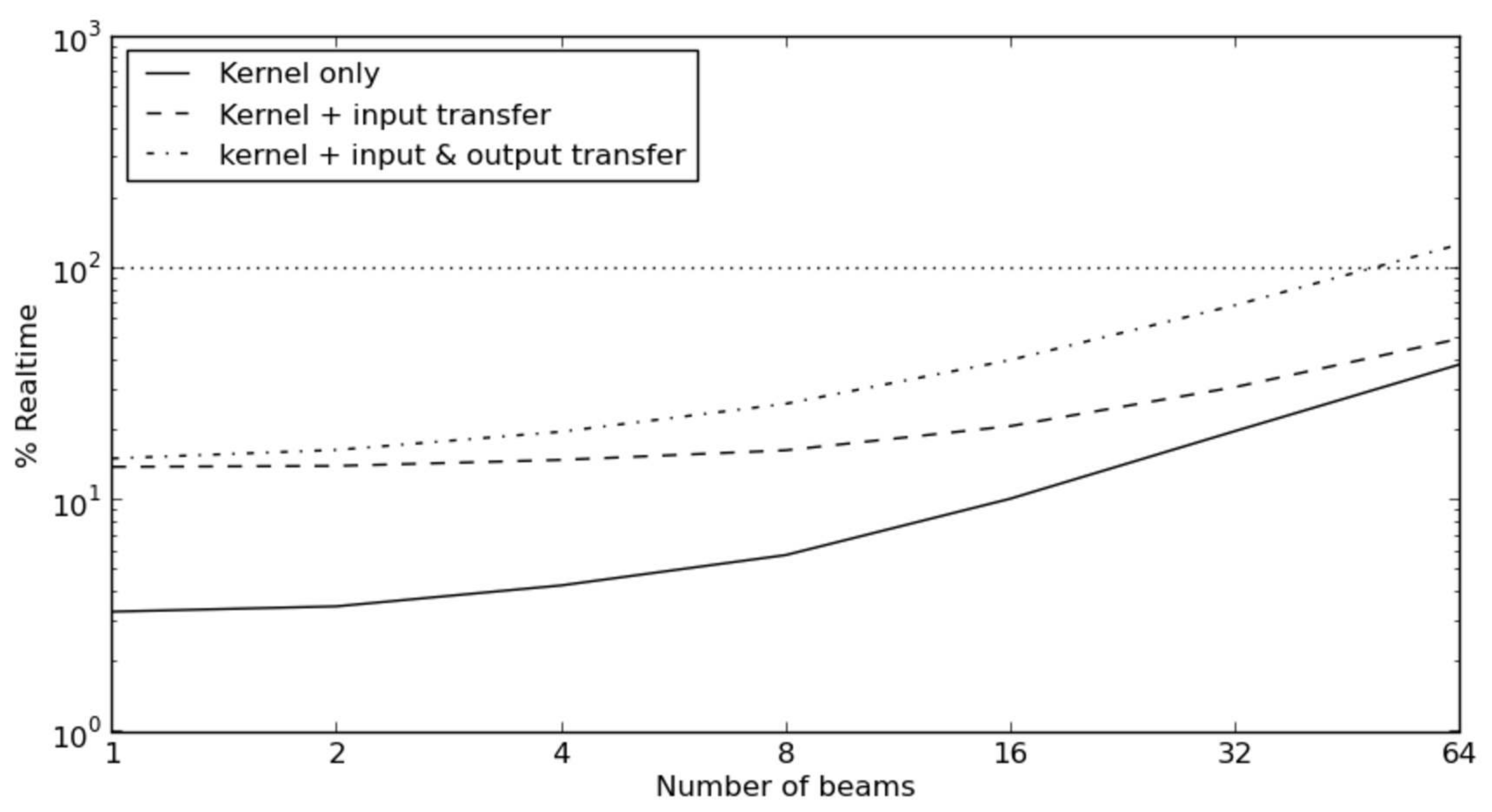}
\end{center}
\caption[Beamformer performance for varying number of beams]{Coherent beamforming performance benchmarks for varying
	 number of synthesized beams, for BEST-II parameters.}
\label{beamParamsFigure}
\end{figure}

A scalability analysis of the algorithm was also performed, where a 420 ms data buffer containing voltage data from 32 single polarisation antennas with a bandwidth of 20 MHz, channelised into 1024 frequency channels, was used to generate an increasing number of synthesised beams in one GPU iteration. Figure \ref{beamParamsFigure} shows that performance scales linearly with increasing number of beams and the execution time is a fraction of realtime for
BEST-II parameters. The maximum number of output beams which can be optimally synthesised is determined by the amount
of global memory available on the GPU. The GPU spends 50\% of its time waiting for raw voltage data to be copied 
from host to GPU memory, thus indicating that the implementation is bandwidth limited over the PCIe link. The kernel is not affected by the number of frequency channels for a given bandwidth. 

\begin{figure}[t!]
\begin{center}
\includegraphics[width=380pt]{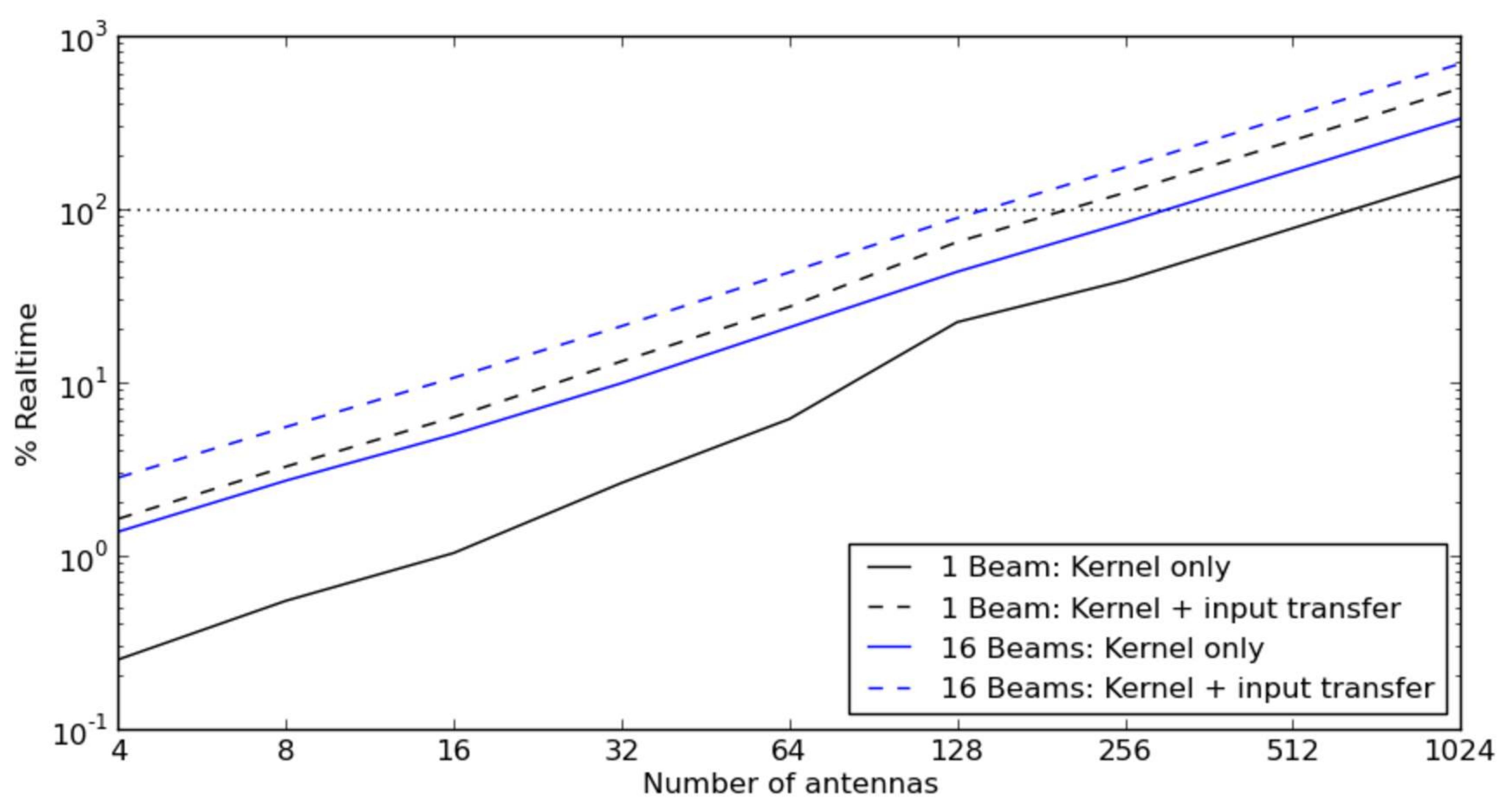}
\end{center}
\caption[Beamformer performance for varying number of antennas]{Coherent beamforming performance benchmarks for varying number of antennas when synthesising 1 (blue) or 16 (black) beams. }	 
\label{beamAntennaFigure} 
\end{figure}

We have also 
benchmarked the scalability of the beamformer for increasing number of antennas, the results of which are shown in 
figure \ref{beamAntennaFigure}. Several 150 ms complex voltage buffers, each containing inputs from an increasing 
number of single polarization antennas having a bandwidth of 20 MHz, channelised into 1024 frequency channels, were 
generated and beamformed into 1 and 16 beams in one GPU iteration. Performance scales linearly with increasing number of antennas. 16 beams can be synthesised in realtime for 256 antennas when excluding data transfer time. The transfer-to-compute ratio is approximately 1 across the entire range. 

The above two benchmarks clearly demonstrate the PCIe bandwidth bottleneck for beamforming kernels. When excluding
the time required to copy the generated beams out of GPU memory, the total execution time is evenly split
between kernel execution and transferring antenna voltages to GPU memory. This data movement overhead can essentially 
be masked by using two CUDA streams if the pipeline is capable of overlapping 
kernel execution and data 
transfer for the following iteration, however this would also require additional global memory, which is a scarce 
resource. The situation is worse when the generated beams are post-processed externally to the GPU on which they were
generated, and the pipeline will essentially be dominated by PCIe transfer overheads. This poses several challenges 
for GPU-based beamforming pipelines and raises doubts on whether this is a viable solution for 
large-N aperture array telescopes. This problem can be somewhat alleviated by keeping the generated beams in 
GPU memory and performing further processing, thus increasing the compute-to-copy ratio. This can take the form 
of correlation, accumulation for generating images, or transient detection.

GPU-based beamforming is a relatively unexplored area in radio astronomy, possibly due to the assumption that 
any such system will be severely limited by the PCIe bandwidth required to transfer data to the GPU, as previously 
discussed. \cite{Sclocco2012} have implemented a GPU-based version of their Blue Gene/P beamformer within the LOFAR 
pipeline, and state that they achieve a 45-50 time speed up, at 642 GFLOPs when using CUDA, on an NVIDIA GTX 580 
(using 40\% of the peak theoretical performance), with a power efficiency improvement of 2-8 times. In this 
section we have shown that our implementation achieves a higher GPU utilisation rate, being 25\% more optimal than 
their quoted benchmarks. This utilisation rate is also sustained across a much wider parameter range 
(see \cite[figure 12]{Sclocco2012}). Nevertheless, both implementations suffer from bandwidth limitations.
In chapter \ref{skaChapter} we examine the scalability of this implementation to SKA$_1$low and SKA$_1$-mid.

\section{Pipeline Integration}

We have integrated our beamforming kernel into the transient detection pipeline described in chapter 
\ref{pipelineChapter}. The beamforming kernel has to be executed before the RFI excision and dedispersion stages, 
after which each beam is then processed by a separate GPU instance. The main challenge lies in determining the 
most efficient way to generate these beams, minimising as much as possible data movement between the host and GPUs, 
as well as amongst GPUs. Three basic schemes can be employed:

\begin{itemize} 
 \item The simplest scheme, implementation wise, would be to have each CPU processing thread generate its own 
       coherent beam. This would require the input data buffer to be to copied to each GPU instance, thus 
       replicating this buffer multiple times within a GPU, greatly reducing the number of time spectra which 
       would fit in global memory. As clearly indicated in figure \ref{beamParamsFigure}, this would result in a severe
       degradation in performance due to the time spent transferring data over the PCIe links. 
       Also, the beamforming kernel is optimised for generating multiple beams in parallel, and this scheme would
       not be fully utilising the kernel's performance capabilities.
       
 \item At the other extreme, a single GPU can generate all the beams required by all the processing threads and 
       then copy each beam to its destination, which could be on a separate GPU. Only one host to GPU transfer
       is required, and kernel execution time is minimised when compared to the scheme above where multiple kernel
       launches are required, each generating one beam, which is clearly inefficient. However, a copy per generated 
       beam is required, some of which can be across GPUs for multi-GPU systems. This scheme also introduces 
       heterogeneity across GPUs in the pipeline, where one GPU acts as a master which provides clients with data to work on.
       
 \item An alternative approach is to combine both schemes, where the beams required by all the processing threads 
       allocated to the same GPU are generated in one kernel launch on the GPU itself. The input antenna
       voltages need to be copied once to every GPU, where a ``master'' thread launches the beamforming kernel,
       generating all the required beams. After completion each processing thread is provided with a pointer
       to its input beam, and processing advances as per standard transient detection.  
\end{itemize}

\begin{figure}[t!]
\begin{center}
\includegraphics[width=420pt]{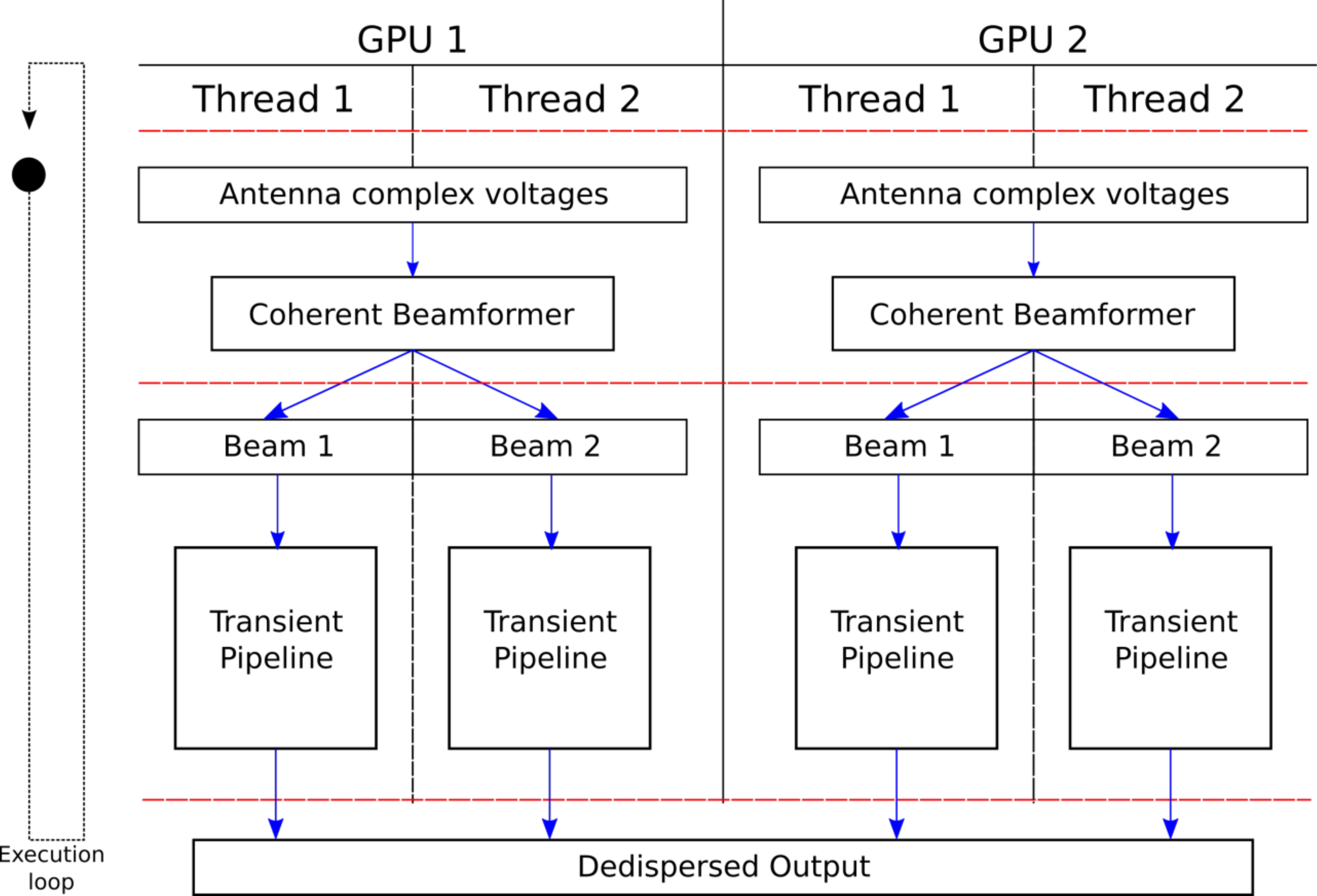}
\end{center}
\caption[Integration of coherent beamformer in transient detection pipeline]{A schematic of the transient detection
         pipeline described in chapter \ref{pipelineChapter} with integrated coherent beamforming kernel.}
\label{beamPipelineFigure}
\end{figure}

The last scheme above provides a good balance between input data replication, data transfers over PCIe links and beamforming kernel efficiency, and is the scheme we implemented for our pipeline. The resulting design is illustrated in figure \ref{beamPipelineFigure}. The buffered antenna voltages are copied to an input buffer allocated on every attached GPU. These transfers occur in parallel over different hardware PCIe links, thus maximising the available motherboard PCIe bandwidth with minimal overhead. A master thread per GPU (belonging to the first allocated beam) is responsible for generating all the GPU buffers and launching the beamforming kernel. The kernel will generate all the beams in a shared output buffer on the GPU, which can be accessed directly from other processing threads allocated on the same GPU, thus not requiring a copy per generated beam. The synthesised beams are then processed in place by 
independent CPU threads.

The main shortcoming of this design is that the ``slave'' processing threads have to wait for the beamforming
kernel in the master thread to finish before they can start processing data, thus wasting valuable clock cycles. 
This duration is relatively short when compared to the time required
for dedispersion. In the current implementation, the slave threads are blocked in a barrier synchroniser. An alternative
implementation would have these threads perform somethings useful in the meantime, such as phase/gain calibration.
The complex coefficients are computed by the pipeline manager whilst waiting for the processing threads to finish
working on the previous input buffer. This allows beam pointings to change dynamically during pipeline execution, 
which is useful for tracking observations or online follow-up of interesting transient events, provided appropriate 
feedback mechanism exist within the pipeline itself. The weight computation overhead is negligible relative to the 
execution time of a single pipeline iteration. The array configuration is stored in an XML file containing relative antenna 
locations. 

\section{Processing analysis of beamforming pipeline}

In order to test the correctness of the beamforming kernel, and in order to benchmark the performance of the 
entire pipeline, we set up a mock BEST-II backend, where the pipeline was deployed on a GPU server 
connected to a ROACH running the BEST-II F-engine. Due to FPGA resource limitations, a SPEAD 
packetisation block could not be included in the F-engine design, so the XAUI output stream was simply branched
to a 10GigE block, with UDP headers encapsulating groups of 128 time samples from 32 antennas, for each frequency 
channel, thus resulting in 4 KB packets with an additional 64-bit header containing the timestamp of the first 
time bin in the packet, as well as the frequency channel index. The total output data rate can be calculated using
$D=C\times T \times A \times W$, where $C$ is the number of frequency channels, $T$ is the number of samples
per second, $A$ is the number of array elements and $W$ is the word length. In our case, $C = 1024$, 
$T = 19531.25$, $A = 32$ and $W = 8$ bits (4-bits for each complex component), resulting in a total output
bandwidth of 5.12 Gbps, excluding packet headers. Therefore a single 10GigE link is sufficient for data transfer
between the F-engine and GPU server. At the receiving end, the packet receiver described in section 
\ref{packetReceiverSection} was updated to be compatible with this format. 
Groups of 128 time samples are considered 
as heaps, and thus heap functionality is still applicable. An additional lookup table is required to match the 
antenna order sent by the F-engine to the element ordering within the array and XML file, and is performed in 
the buffering thread. 

\begin{figure}[t!]
\begin{center}
\includegraphics[width=360pt]{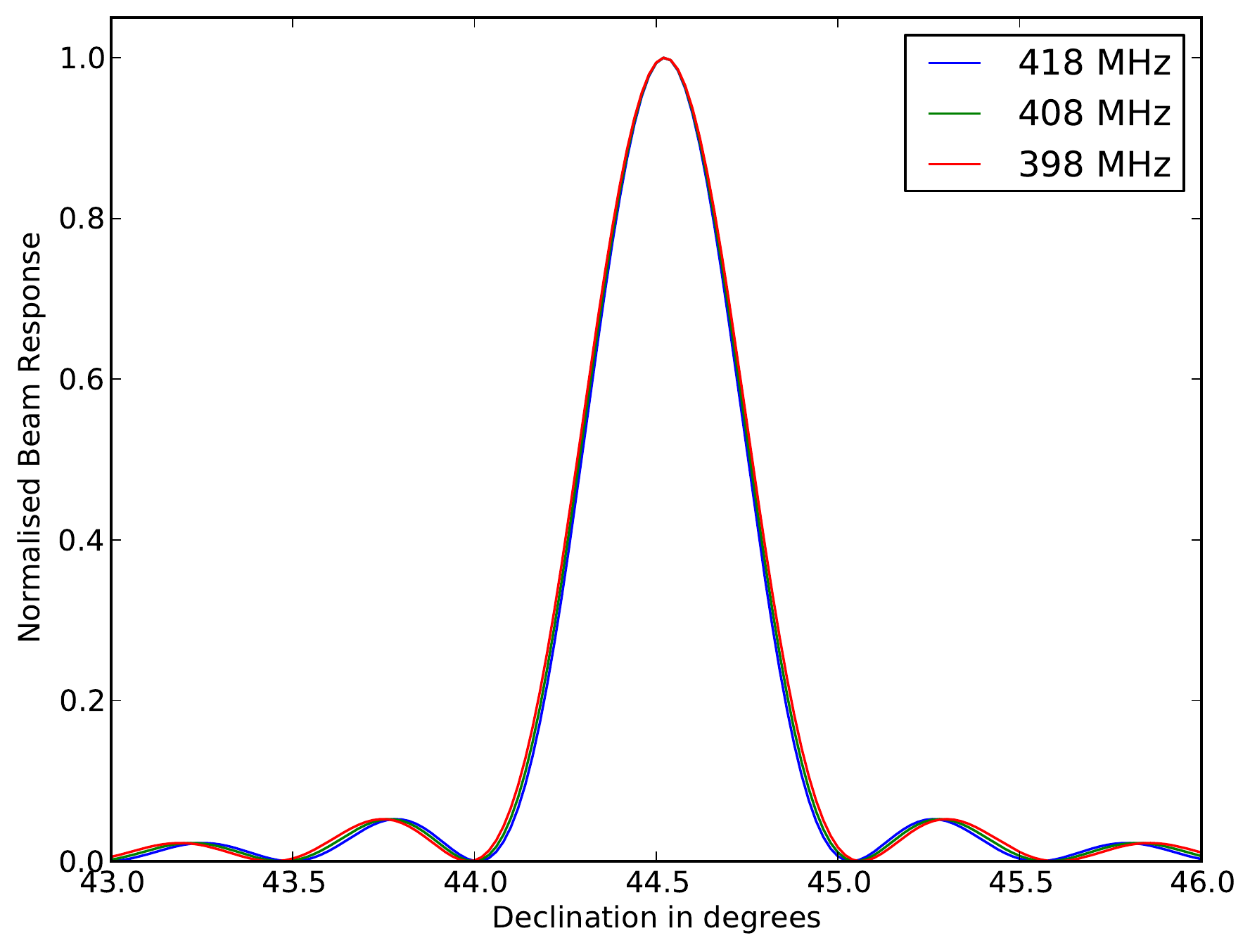}
\end{center}
\caption[Beam pattern for test setup]{Beam pattern for the central and edge frequency channels for 6$^{\circ}$ around the zenith.} 
\label{beamPatternFigure}
\end{figure}

A beam response test was conducted in order to test the correctness of the beamforming kernel. The 32 ADC inputs were
left unconnected such that they only measure background and instrument noise. The signal from one of the inputs 
was mirrored across all connections, resulting in identical antenna signals. Signal equilisation and gain calibration
in the F-engine scales 
up this noise to use the entire bit width dynamic range. These signals were then phase calibrated to point to 
the zenith and transmitted continuously to the GPU server. The beamforming pipeline was set to generate 4 beams per
iteration, each having a pointing difference of 0.02$^{\circ}$, where the synthesised beams were written to disk
for offline processing. BEST-II antenna positions were used to generate the 
beam coefficients, with the zenith at a declination of 44.524$^{\circ}$. The current implementation can only 
generate pointings along the y-dimension, so for this test
the interference pattern should resemble an 8-slit system with a central beam width of approximately 1$^{\circ}$ 
and 7 nulls between the central beam and first grating lobe. The resultant beam pattern, showing the central
beams and two nulls, is shown in figure \ref{beamPatternFigure}.

The entire pipeline was also benchmarked, using BEST-II parameters as a reference telescope, with observational 
and survey parameters based on 
those used to generate timings for table \ref{timingTable}. A 2.52 s buffer containing voltage data from 32 20 MHz
single polarisation antennas centred at 408 MHz, split into 1024 frequency channels, was generated and processed
in a single pipeline iteration. One NVIDIA GTX 660 Ti card was used for this test. Four synthesised beams were 
generated from this buffer, each of which was then dedispersed over a range of 864 DM trials with a maximum DM 
of 86.4 pc\ cm$^{-3}$. Table \ref{beamPipelineTable} lists the accumulated execution time of each stage  (since no kernel parallelisation is possible on the test device). The GPU to CPU transfer of antenna voltages and the beamforming kernel launch is performed by the ``master'' processing thread, and the input data buffer is limited by the amount of global memory available. This table shows that, for a 32 element array, the beamforming cost is negligible when compared to the total execution time of the entire pipeline, and amounts to approximately 3\%. This hints to the possibility of deploying a fully GPU-based beamforming and transient detection system for small, low bandwidth arrays. Analogue signal reception, digitisation, channelisation and equilisation still need to be performed prior to beamforming, and these operations are more suitable to an FPGA-based system, such as the F-Engine deployed at the BEST-II array.

\begin{table}[t!]
 \centering
 \begin{tabular} { | c | c |}
  \hline
  \multicolumn{2}{|c}{Antenna Voltage Copy Time{\footnotesize{$^*$}}: 291.92 ms} \vline  \\
  \hline
  \hline
  Beamforming\footnotesize{$^*$} & 74.98 ms \\
  Bandpass Fitting    & 80.16 ms \\
  RFI Thresholding    & 36.44 ms \\
  Dedispersion        & 1780.6 ms \\
  Median Filtering    & 70.96 ms \\
  Detrending          & 44.16 ms \\
  \hline
  \hline
  \multicolumn{2}{|c}{Copy from GPU: 101.64 ms} \vline \\
  \hline 
  \hline
  \multicolumn{2}{|c}{Total iteration time: 2379.27 ms} \vline \\
  \hline
 \end{tabular}
 \caption[Beamforming pipeline timings]{GPU timings for one pipeline iteration performing beamforming, bandpass 
	  fitting, RFI thresholding, dedispersion, median filtering and detrending. 
	  \newline		
	  \footnotesize{$^*$ Performed by the master thread on every GPU}}
 \label{beamPipelineTable}
\end{table}

\section{Conclusion}

We have presented a GPU implementation of a coherent multi-beam beamformer which can synthesise a number of 
coherent beams for arrays with an arbitrary number of elements and frequency channels.
The kernel utilises 50\% of the peak theoretical performance on the test device, at 1.2 TFLOPs, and is limited 
by shared memory bandwidth, which is required to load complex beam coefficients. The overall implementation is 
limited by PCIe bandwidth, where for a serial execution pipeline, 50\% of the processing time is spent transferring 
data from CPU to GPU memory. If the beams need to be post-processed outside of the GPU on which they were generated, 
the GPU remains mostly idle, waiting for data transfers request to terminate.

The latter can be alleviated by performing additional operations after beam generation, and to test this assertion
we have integrated this kernel with the transient detection pipeline described in chapter \ref{pipelineChapter}. 
A mock BEST-II backend running just the F-Engine was set up and used to benchmark this system, and we show that
for small, low bandwidth arrays the computational cost for beamforming is negligible when compared to dedispersion. 
This gives rise to potential systems where these operations are integrated, allowing for dynamic observations where
beam pointings can be updated online, in real time, triggered by interesting events during the event detection
stage of the transient detection pipeline. Having the raw voltage data available in memory can potentially
enable on-the-fly correlation and image generation for more accurate transient source localisation.

In the next chapter we determine the hardware resources required to scale up such a pipeline to larger
arrays, specifically to single station, as well as central processing station, specifications for SKA$_1$-low and SKA$_1$-mid.


\chapter{Applicability of GPUs to SKA$_1$}
\label{skaChapter}

The Square Kilometre Array (SKA) provides an excellent opportunity for searching for fast radio transients. The increased sensitivity, field of view and survey speed of the SKA, compared with other radio telescopes, will make it an ideal instrument to search for impulsive emission from high energy density events. The three telescopes making up the SKA will, however, pose considerable processing and data transport challenges. In this chapter we examine some computational aspects of the SKA, concentrating mostly on beamforming and dedispersion.

Recently, \cite{Dewdney2013} revised the specifications for phase 1 of the SKA, 
and we use the configuration and parameters defined here as a basis for the 
analysis presented in this chapter. We start off by introducing the SKA, discuss 
possible advances in technology up till construction of the SKA, and describe 
methods for computing transient search parameters. We then analyse the 
applicability of GPUs for SKA$_1$-low station beamforming and channelisation, as 
well as present a possible use case, together with estimates for computational 
resources required, for SKA$_1$-low  station-level transient searches. GPU-based 
implementation for several stages of the SKA$_1$-mid non-imaging pipeline are 
also investigated, focusing on beamforming, channelisation and dedispersion.

\section{The Square Kilometre Array - Phase 1}

The scientific motivations for building the SKA\footnote{\;http://www.skatelescope.org/} are highlighted by Key Science Projects (KSPs), which represent unanswered questions in fundamental physics, astrophysics and astrobiology that the SKA will play a key role in addressing \cite{Carilli2004,Gaensler2004}. Of particular significance to this thesis is the potential use of the SKA for probing the fast transient phase space. The SKA will provide continuous frequency coverage from 50 MHz up to 10 GHz in the first two phases of its construction, with phase 1 providing 
$\sim$20\% of the total collecting area at low and mid frequencies, and phase 2 seeing the completion of the full array. The major observatory entities will be located in Australia (SKA$_1$-low and SKA$_1$-survey) and South Africa (SKA$_1$-mid), each hosting a supercomputing facility termed the Central Signal Processing (CSP), which will combine and process the signals from all the stations of their respective continent. The SKA$_1$ system baseline design \cite{Dewdney2013} provides a baseline technical starting point and system decomposition for eventual construction of the SKA, and all listed technical specifications and calculations are based on values from this document, unless otherwise stated. We now proceed to give an overview of two of the telescopes comprising SKA$_1$. There are no plans to include a non-imaging infrastructure for SKA$_1$-survey, so it will not be included in the following discussions.

\subsection{SKA$_1$-low}

The main science goal of the low frequency SKA is to study the highly redshifted 21 cm HI line, and key design decisions have largely followed the recommendation of the SKA-EoR science working group. It will also be well suited for conducting low radio frequency observation of pulsars and, possibly, extrasolar planets. This will consist of an aperture array of $\sim$250,000 log-periodic, dual-polarised antenna elements, most of which will be arranged in a very compact configuration (the {\it core}) with a diameter of $\sim$1 km, the rest being arranged in 35 m diameter stations, configured as three equally spaced spiral arms, with a maximum radius of $\sim$45 km.  The antenna array will operate from 50 MHz to $\sim$350 MHz, with peak sensitivity at 108 MHz, and a sensitivity of $\sim$1000 m$^2$/K at the zenith, assuming an instantaneous bandwidth of 250 MHz, and a core brightness temperature sensitivity of $\sim$1 mK. The elements will be grouped into 911 35-m stations (866 in the core, 45 in the spiral arms). Table \ref{ska1LowSpecTable} lists the key specifications of SKA$_1$-low, as defined by \cite{Dewdney2013}. The station beamformer will generate a single, smooth beam with a FoV of 20 deg$^2$. Signals from these beamformers will be transported to the CSP, where they will be channelised and cross-correlated. In this chapter we'll examine the applicability of GPUs for station-level beamforming, as well as examine the possibility of conducting station-level fast transient searching.

\begin{table}[t!]
  \centering
  \begin{tabular}{ l c c c }
    \hline
    Lower Frequency &  50 MHz \\
    Upper Frequency & 350 MHz \\
    Bandwidth       &  300 MHz \\
    \hline
    Antennas per Station & 289 \\
    Number of Stations & 911 \\
    Total Antennas &  263,279 \\
    Antenna Area &  2.25 m$^2$ \\
    Station Diameter &  35 m \\
    Station filling factor & 0.7 \\
    Total Physical Aperture & 8.0 $\times$ 10$^5$ m$^2$ \\
    \hline
    Channel Resolution & 1 kHz \\
    Maximum Number of channels & 250,000 \\
    \hline
    Number of Beams  & 1 \\
    Inst. BW per beam & 250 MHz \\
    \hline
  \end{tabular}
  \caption[Key Specifications of SKA$_1$-low]{Key specifications of SKA$_1$-low, as defined by \cite{Dewdney2013}}
  \label{ska1LowSpecTable}
\end{table}

\subsection{SKA$_1$-mid}

This telescope will primarily address observations of the 21-cm hyperfine line 
of neutral hydrogen, many classes of radio transients, and a survey of the 
entire visible sky for pulsars with a pseudo-luminosity depth of 0.1 mJy kp$^2$ 
at 1400 MHz out to a distance of 10 kpc. For this survey, regions defined by 
dispersion measure will determine the optimum frequency ranges for the Galactic 
plane, the Galactic centre and high Galactic latitudes. The array will be a mix 
of 64 13.5 m diameter dishes from the 
MeerKAT\footnote{\;http://www.ska.ac.za/meerkat/index.php} array and 190 15 m 
SKA$_1$ dishes, arranged in a moderately compact core with a diameter of $\sim$1 
km (to support pulsar searching) and a further 2D array of randomly placed 
dishes out to $\sim$3 km radius. Three spiral arms will extend to a radius of 
$\sim$100 km from the centre. SKA$_1$-mid will cover a frequency range of 350 
MHz to at least 3050 MHz in three receiver bands, with lower frequency receivers 
having a bandwidth of $\sim$1 GHz and higher frequency receivers $\sim$2.5 GHz 
in each polarisation. SKA$_1$ dish sensitivity is expected at 6.9 m$^2$/K, with 
a combined array SEFD of 1.7 Jy. Table \ref{ska1MidSpecTable} lists the key 
specifications of SKA$_1$-mid, as defined by \cite{Dewdney2013}. Only SKA dishes 
were included for these calculations, and all processing requirement 
calculations in the rest of this chapter will focus on these as well. Also, only 
the lower three bands are listed, as these frequency regions will be used for 
transient-related science, except for Galactic centre observations which require 
higher frequencies. For transient searching, signals from the dishes will be 
transported to the CSP, where they will be beamformed and passed through 
specialised transient search equipment.

\begin{table}[t!]
  \centering
  \begin{tabular}{ l c c c }
    \hline
     & Band 1 & Band 2 & Band 3 \\
    \hline
    Frequency Range (MHz) & 350 - 1050 & 950 - 1760 & 1650 - 3050 \\
    Max. Available BW (MHz) &  700 & 808 & 1403 \\
    Aperture Efficiency & $\sim$60\% & $\sim$65\% & $\sim$78\% \\
    \hline
    Average $T_{\text{sys}}$ (K) & 28 & 20 & 20 \\
    SEFD$^*$ (Jy) & 4.4 & 2.1 & 2.1 \\
    Minimum detectable flux ($\mu$Jy s$^{-1/2}$) & 105 & 58 & 54 \\
    $A_{\text{eff}}/T_{\text{sys}}$ (m$^2$/K) &  779 & 1309 & 1309 \\
    FoV @ centre frequency (deg$^2$) & 2.8 & 0.75 & 0.25 \\
    SSFoM$^*$ (10$^9$ m$^4$ K$^{-2}$ deg$^2$ MHz) & 1.2 & 1.0 & 0.08  \\
    \hline
  \end{tabular}
  \caption[Key Specifications of SKA$_1$-mid]{Key specifications of SKA$_1$-mid, including only SKA dishes, as defined by \cite{Dewdney2013}
        \newline
        \footnotesize{$^*$ For all antennas}}
  \label{ska1MidSpecTable}
\end{table}

\section{Technology Configuration}

In section \ref{manycoreSection} we have discussed the current state--of--the--art in processor and accelerator technology, focusing on three devices: the Intel Xeon E5-267, NVIDIA K20 and Intel Xeon Phi SE10. Construction of SKA$_1$ will start in 2016, and a technology freeze date of 2016 is assumed for commercially available computer equipment which will be used for initial science data processing. Due to the dynamic nature of technology, it's difficult to predict the processing specifications of future devices. Here we'll list several assumptions which will be used in the rest of this chapter, with a primary focus on GPU technology.

GPU performance has roughly followed Moore's law in the past few years. Current GPUs can sustain a peak theoretical performance of $\sim$3.5 TFLOPs with an internal memory bandwidth of $\sim$280 GB/s. We therefore project that by 2016, GPUs will have a peak theoretical performance of $\sim$10 TFLOPs, retaining a GPU dissipation of $\sim$250 W. The new upcoming NVIDIA architecture, Maxwell, will introduce Unified Virtual Memory, which will allow the CPUs and GPUs on a host system to see all of system memory, simplifying data transfer between the two. The following architecture, Volta, will provide an internal memory bandwidth of 1 TB/s, greatly improving the performance of bandwidth limited implementations. We will also assume that PCIe 3 will be the interfacing platform for these devices, which supports 15.75 GB/s bi-directional data rate over 16 lanes, as the specifications for PCIe 4 are not due until 2015.

We will also assume that the ratio between CPU and GPU power remains relatively constant, with a proportional increase 
in compute power. Host memory will probably be composed of DDR4 SDRAM, which 
will feature faster clock frequencies, lower voltages and 230 Gbps of maximum 
bandwidth for a 72-bit wide data bus. No doubt, this will greatly decrease the 
latency for internal memory transfers, and prove beneficial when transferring 
high throughput  data from network adapters. These will probably consist of 40 
Gbps Ethernet or 56 Gbps Infiniband adapters. We follow \cite{Dewdney2013} and 
choose the latter, with an additional benefit being that dual-port 56 Gbps 
adapters will match the maximum PCIe 3.0 transfer rate.

We now apply the specifications listed above to the algorithm implementations discussed in the previous chapters, particularly to direct dedispersion and coherent beamforming, to create a ``scalability measure'' for each. The current implementation of the dedispersion kernel can achieve $\sim$25\% of the peak theoretical performance on Kepler devices, as opposed to the 10\% listed in \cite{Dewdney2013}. On Volta devices, this will get an additional boost due to the increased memory bandwidth, which will probably lead to a re-design of the algorithmic behaviour so as to reduce the reliance on shared memory and take advantage of increased global memory bandwidth. We'll therefore assume that $\sim$30\% of Volta GPUs' peak theoretical performance can be utilised, and a 16-lane PCIe 3 interface for transfer between the CPU and GPU. The beamforming kernel achieves $\sim$50\% of the peak theoretical performance, and the kernel will probably gain a further boost due to the increased internal memory bandwidth, so we'll assume that 60\% of Volta GPUs' peak performance can be utilised. 
These scalability measures are listed in table \ref{voltaTable}. Additionally, empirical testing on current GPUs show that GPU PCIe bandwidth utilisation is approximately 90\% of specified peak.

\begin{table}[t!]
  \centering
  \begin{tabular}{ l c c c }
    \hline
       & \% Peak Performance & Arithmetic Intensity & Bandwidth \\
    \hline
    Dedispersion & 30\%  & 3.6 TFLOPs & 13.4 GB/s \\
    Beamforming  & 60\%  & 7.2 TFLOPs & 13.4 GB/s \\
    \hline
  \end{tabular}
  \caption[NVIDIA Volta scalability measure]{Scalability measure for direct dedispersion and coherent beamforming on NVIDIA Volta devices.}
  \label{voltaTable}
\end{table} 

\section{Transient Search Parameters}
\label{surveyParametersSection}

In the following sections we'll be discussing the computational and hardware requirement for conducting transient surveys with SKA$_1$. Here we develop a general framework for calculating the surveying parameters which will be used throughout this chapter. Blind surveys of large total solid angles require full-FOV sampling, which is feasible only for a subarray comprising the innermost antennas, the core array. This is characterised as a circular distribution of $n_a$ antennas with diameter $b_c$. The core array contains a fraction $f_c \equiv n_a/N_a$ of the total number of antennas. The maximum number of independently pointed, coherently combined beams, $N_b$, which can be formed within the FoV of a single element beam is given by:
\begin{equation}
 N_b = \frac{\Omega_0}{\Omega_{\text{arr}}} =  \left( \frac{D_{\text{arr}}}{D_0} \right)^2 
\end{equation}
where $\Omega_0$ is the single element FoV, $\Omega_{\text{arr}}$ is the array beam FoV, $D_{\text{arr}}$ is the diameter of the entire array, $D_0$ is the element diameter.

In this chapter we will only consider the computational requirements for 
coherent signal combination. Incoherent signal combination and subarraying will 
yield different event detection rates and have different processing 
requirements, as discussed by \cite{Colegate2011}.  Irrespective of the signal 
combination mode used, each synthesised beams needs to be searched independently 
for transient events. Here we will focus on post-detection analysis for fast 
transient, single-pulse blind surveys. Dispersion and scattering effects can be 
used to define the bounds of the search.

The maximum DM value, DM$_{\text{max}}$, and consequently the distance to which the search can be conducted, is limited by 
\begin{itemize}
 \item the degree of expected scatter broadening, which can be computed by using the empirical relationship defined in equation \ref{scatteringRelationship}
 \item The amount of signal smearing within one frequency channel, where restricting this to one time sample results in a $\text{DM}_{\text{max}}$ of
 \begin{equation}
  \text{DM}_{\text{max}} = \frac{\Delta t\cdot f_0^3}{8300 \cdot \Delta f}
 \end{equation}
where $\Delta t$ is the sampling time, $f_0$ is the frequency at the lower edge 
of the band and $\Delta f$ is the channel width. This is a rearranged version 
of equation \ref{dispRelationshipEquation}.
\end{itemize}

A model of the galactic distribution of free electrons, such as the NE2001 model (\cite{Cordes2002}) can also be used to restrain this parameter, and is the scheme adopted by \cite{Dewdney2013}. For extragalactic surveys, the updated fit by \cite{Lorimer2013} will be used. Higher DM values can be searched for by altering the beamformed time series, such as downsampling by combining adjacent time samples.

The required spectral resolution depends on the amount of DM smearing within a 
single channel for DM$_{\text{max}}$. By setting a limiting factor on this 
smearing amount, $F_{\text{smear}}$, the total number of frequency channels can 
be computed. Unless explicitly stated, we will not include the computational 
requirements for channelisation in our calculations, however this scales as 
$\mathcal{O}(N_tN_p(\text{log}_2(N_c)+N_{\text{taps}}))$ whereas dedispersion 
scales as $\mathcal{O}(N_tN_cN_pN_{\text{DM}})$ (when direct dedispersion is 
used), therefore the total computational requirements are dominated by 
dedispersion. Here, $N_t$ is the number of time samples, $N_p$ is the number of 
polarisations, $N_c$ is the number of channels, $N_{\text{DM}}$ is the number of 
dispersion measure trials and $N_{\text{taps}}$ represents the number of taps on 
a presumed polyphase filterbank channelisation implementation. $N_c$ is 
therefore
\begin{equation}
\label{channelsEquation}
 N_c = \frac{N_{\text{sub}} \times \tau_c \times \text{DM}_{\text{max}}}{F_{\text{smear}} \times W}
\end{equation}
where $\tau_c$ is the pulse smear within a single channel prior to finer channelisation and $W$ is the expected pulse width.

The DM step, $\Delta \text{DM}$, depends on the expected signal width, and can be restricted by limiting the loss in S/N of a pulse due to dedispersion at an incorrect DM value. The interval should not be large such that a transient with a true DM lying between two trial DMs is significantly broadened, however a value which is too small will greatly increase the required computational resources. This value can be calculated by selecting the maximum $\Delta \text{DM}$ at which equation \ref{incorrectDedispersion} yields a S/N loss less than a limiting factor $\text{DM}_{\text{smear}}$, where $W_{\text{ms}}$ is replaced with the minimum expected intrinsic signal width. The number of DM trials is then simply the DM range divided by $\Delta$DM. 

Further restraints can be applied by relying on the fact that the amount of smearing due to scattering increases with DM. As soon as the scattering timescale of a pulse of width $\Delta t$ becomes $2\Delta t$, the time series can be downsampled by a factor of 2, as the original high temporal resolution in no longer required. This process can be repeated until DM$_{\text{max}}$ is reached. This is the process which PRESTO uses to generate dedispersion plans, and has the effect of reducing the computational requirement with minimal effect on the sensitivity of the survey. For the purpose of this chapter, no downsampling will be applied to the time series, unless otherwise stated.

\section{SKA$_1$-low station beamforming}

\begin{figure}[t!]
  \centering
  \includegraphics[width=380pt]{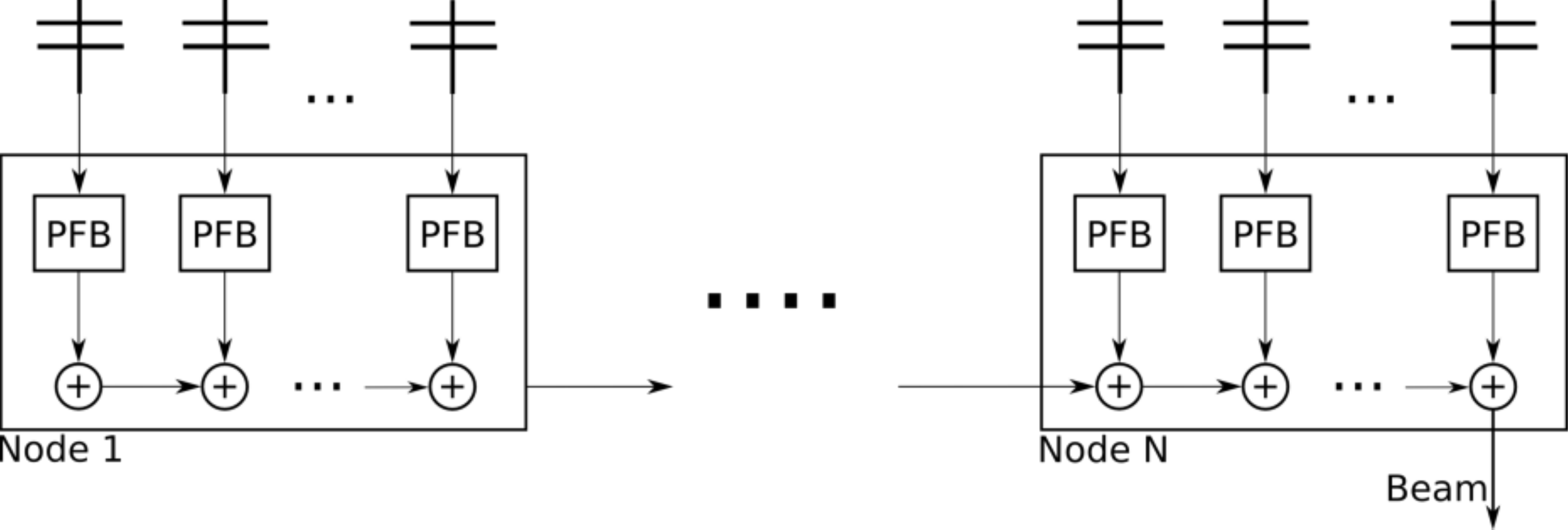}
  \caption[Ring-based beamforming]{Ring-based beamforming schematic}
  \label{ringBeamformingFigure}
\end{figure}

Signals from each antenna within an SKA$_1$-low station need to be combined to generate station beams. Here's we'll concentrate on the applicability of GPU-based systems for station-level beamforming, limiting ourselves to the requirements defined in \cite{Dewdney2013}, who state that one output beam over the entire band is required per station. We'll investigate a setup where GPUs, together with their host systems, will perform all the required processing, including digitisation, channelisation and beamforming. We follow \cite{Hickish2013} and assume a ring-based beamforming architecture, which essentially is a serial chain of addition stages whereby each processing node contributes to the formation of the entire beam. This setup is used by LOFAR \cite{Haarlem2012} to generate station beams. Figure \ref{ringBeamformingFigure} provides a visualisation of this scheme. We also assume Nyquist sampling with a 1.25 coding factor, resulting in 625 MSA/s. Assuming digitisation of antenna signals with $n_s$ bit samplers, the input data rate to each processing node is 
\begin{equation}
 S_{\text{in}} = 2B\eta N_aN_pn_s
\end{equation}
where $B$ is the bandwidth, $N_a$ is the number of antennas $N_p$ is the number of polarisations and $\eta$ is the coding factor. The output data rate is defined by the number of beams required, and is
\begin{equation}
 S_{\text{out}} = 2B\eta N_bN_p n_b
\end{equation}
where ${N_b}$ is the number of beams and $n_b$ is the bitwidth representing each beam sample. Since $N_b$ = 1,  $S_{\text{out}}$ is significantly lower than $S_{\text{in}}$ and therefore I/O bandwidth will be entirely dominated by the input data rate. The number of operations required for beamforming and channelisation are
\begin{eqnarray}
B_{\text{ops}} & = &  8 BN_bN_aN_p \\
C_{\text{ops}} & = &  BN_bN_aN_p \cdot (\text{log}_2(N_c) + 8N_{\text{taps}})
\end{eqnarray}
where $N_c$ is the number of frequency channels and $N_{\text{taps}}$ represents 
the number of taps in a presumed polyphase filterbank channelisation 
implementation. The factor of 8 for $N_{\text{taps}}$ represents the number of 
operations required for multiplying complex coefficients. A PFB implementation 
consists of two processing stages: complex multiplication with a pre-calculated 
function, and a FFT. The first stage can generally be optimally implemented, so 
we assign a 50\% compute efficiency to it, whilst 30\% is assigned to the FFT, 
based on performance benchmarks of the cuFFT library for a 1024-point single 
precision FFT. For the following calculations, we'll assume a 1024-channel PFB 
with 8 taps. Table \ref{skaLowBeamTable} lists the values of the above 
parameters for SKA$_1$-low stations. Note that for ring-based beamforming, the 
input data rate to each node includes the output beamformed data from the 
previous node in the chain, and therefore the total node input data rate is 
$S_{\text{in}}$ + $S_{\text{out}}$, where the latter term is constant for any 
number of input antennas.

\begin{table}[t!]
  \centering
  \begin{tabular}{ l c c l}
    \hline
    Bits per Value             & $n_s$             & 8                \\
    Coding Factor              & $\eta$            & 1.25             \\
    Input Data Rate            & $S_{\text{in}}$   & $\sim$2.89 Tb/s   \\
    Output Data Rate           & $S_{\text{out}}$  & $\sim$10 Gb/s    \\
    Beamforming Operations     & $S_{\text{ops}}$  & $\sim$2.89 TFLOPs \\
    Channelisation Operations  & $C_{\text{ops}}$  & $\sim$26.7 TFLOPs \\
    \hline
  \end{tabular}
  \caption[SKA$_1$-low station beamforming parameters]{SKA$_1$-low station beamforming parameters}
  \label{skaLowBeamTable}
\end{table}

\begin{figure}[t!]
  \centering
  \includegraphics[width=380pt]{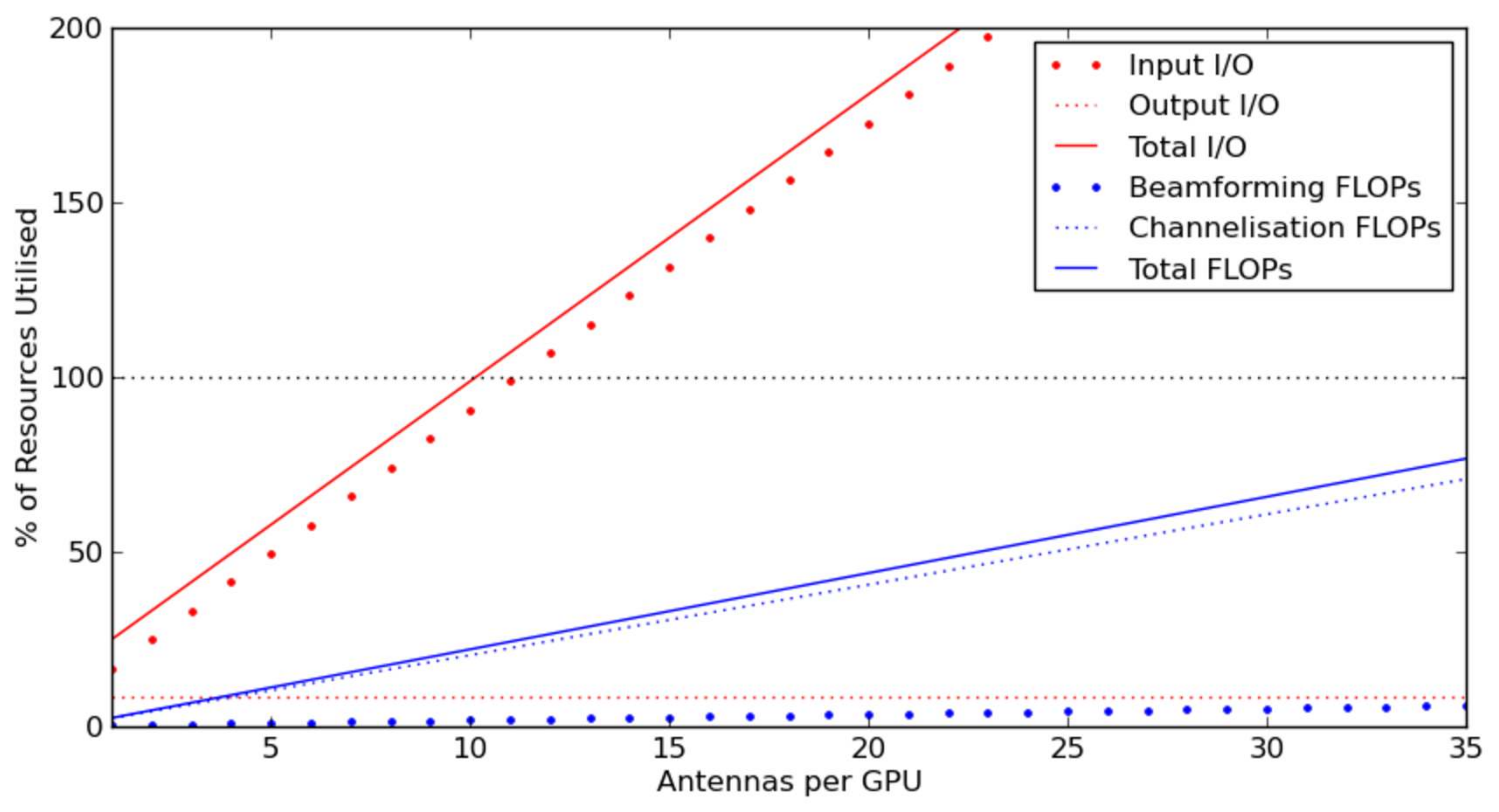}
  \caption[Resource utilisation for SKA$_1$-low station GPU-based processing]{Resource utilisation for SKA$_1$-low station GPU-based processing}
  \label{beamOnlySkaLowFigure}
\end{figure}

Using the hardware specifications listed in table \ref{voltaTable}, figure \ref{beamOnlySkaLowFigure} shows how many antenna signals can be processed by a single GPU (assuming signal digitisation is performed on the host CPU). These plots take into consideration implementation efficiency for each algorithm as well data transfer efficiency over the PCIe link. It is clearly evident that this setup is entirely limited by the data transfer rate onto GPU memory, with the GPU SP cores sitting mostly idle, even if coarse channelisation is performed at the station. Approximately 10 dual-polarisation antennas can be processed by a single GPU, thus requiring at least 29 GPUs per station, or almost 27,000 for all SKA$_1$-low stations. This should be contrasted by the number of Virtex 7 FPGAs which would be required, which would be 12 per station, or $\sim$11,000 in total (see \cite{Hickish2013}). With the latter option the computational resources on the chip are almost fully utilised. 

\begin{table}
 \centering
 \begin{tabular}{l c}
 \hline
  Server Setup                  & 2 GPUs + Host CPU \\
  GPU Cost                      & 1500 per unit \\
  Host Cost                     & 700 \\
  Cost per Server               & 3700 \\
  \hline
  GPU Power Consumption         & 250 W       \\
  Host Power Consumption        & 250 W       \\
  Power Cost (Euro/Watt/Year)   & 2.5         \\
  Replacement parts             & 5\% / year  \\
  \hline
  Station Deployment Cost       & 55,500      \\
  Station Replacement Cost      & 562,500       \\
  Station Power Cost            & 150,000     \\ 
  \hline 
  SKA$_1$-low Deployment Cost   & 50,560,500      \\
  SKA$_1$-low Replacement Cost  & 25,280,250       \\
  SKA$_1$-low Power Cost        & 136,650,000     \\
  \hline 
 \end{tabular}
 \caption[SKA$_1$-low station processing costs]{SKA$_1$-low station processing costs, based approximately on numbers given by \cite{ZarbAdami2013}. Power and replacement costs are for a 10-year operational lifespan. All costs are in 2013 Euro.}
 \label{ska1LowCostsTable}
\end{table}

We now present an approximate cost analysis for this setup, using pricing 
calculations provided by \cite{ZarbAdami2013}. Table \ref{ska1LowCostsTable} 
summarises the cost per station, as well as an aggregate cost for all the 
SKA$_1$-low stations, including deployment and 10-year operational costs. These 
figures exclude communication fabric and hardware, racking, cabling, labour and 
additional infrastructural costs. Two GPUs per server are assumed, since a 
dual-port 56 Gbps network adapter is required per GPU in order to fully saturate 
the PCIe links, requiring  a total of 4 PCIe x16 slots. Clearly, this is a 
costly setup, with very high running costs, primarily due to high GPU power 
consumption. FPGA-based solutions might result in comparable deployment and 
replacement costs, however they provide a much higher performance per Watt, with 
power consumption almost one order of magnitude lower than GPUs. A more detailed 
cost-benefit analysis should be conducted in order to determine the best 
approach for station beamforming and channelisation.

\section{SKA$_1$-low Station-level transient searching}
\label{skaLowTransient}

In this section we examine the possibility of conducting station-level fast transient surveys in ``piggy-back mode'', where the complex voltages from all the antennas are first coarsely channelised by an FPGA or GPU-based processing backend and sent to a network switch, where GPU-based processing nodes then combine the antenna signals and run the transient detection pipeline.  This would be very similar to the ARTEMIS use case for international LOFAR stations, where each station becomes an independent surveying instrument. The digital backend will also form the station beam as required by the central SKA$_1$-low observation, whilst the beam generated by the GPU system can be pointed anywhere within the antennas' FoV, allowing separate manoeuvrability.

In an ideal scenario, data is transferred only once to GPU memory, as additional transfers would increase the number of GPUs required. However, simply partitioning the set of antennas into isolated clusters would drastically reduce the sensitivity of the instrument. A better alternative would be to split the observing band, where each GPU receives signals from all antennas for specific subbands. This would reduce sensitivity by $\sqrt{(N_{\text{subs}} - 1)N_c'}$, where $N_c'$ is the number of channels per subband, which can be countered somewhat by lowering the detection threshold during post-processing, as discussed in section \ref{detectionSelectionSection}. In the previous section we stated that 29 GPUs would be required to process all the raw antenna signals. By assuming that after signal digitisation, channelisation and equalisation the total output data rate becomes $BN_aN_pn_s$, the number of required GPUs is then reduced to 12. Rounding this value up to the nearest power of 2, 16, provides us with the minimum channelisation requirements for splitting up the band. 

For this section we'll assume the observing band is split into 16 frequency subbands, the same number as the minimum number of GPUs required to keep up with the input data rate, however the process below can be repeated for any number of subbands, as will be required by the final SKA$_1$-low design. The number of DM trials which can be processed, when using direct dedispersion, can be computed by calculating the ``left over'' computational resources after GPU beamforming and finer channelisation
\begin{eqnarray}
 \mathcal{F}_{\text{ops}} & = & \mathcal{F}_b + \mathcal{F}_c + \mathcal{F}_d \\
 \mathcal{F}_b & = & 8e_bBN_aN_pN_b \\
 \mathcal{F}_c & = & 8e_cBN_bN_p(log_s(N_c) + N_{\text{taps}}) \\
 \mathcal{F}_d & = & 4e_dBN_bN_{\text{DM}} \\
 N_{\text{DM}} & = & \frac{\mathcal{F}_{\text{ops}} - \mathcal{F}_b - \mathcal{F}_c}{e_dBN_b} 
\end{eqnarray}
where $\mathcal{F}_b$, $\mathcal{F}_c$ and $\mathcal{F}_d$ are the beamforming, channelisation and dedispersion operation counts in FLOPs/s respectively, and $e_b$, $e_c$, $e_d$ are the beamforming, channelisation and dedispersion efficiency factors respectively. For SKA$_1$-low stations, $B$ = 15.625 MHz (250 Mhz split across 16 subbands), $N_b$ = 1, $N_p$ = 2 and $N_a$ = 289. The cost for summing across both polarisations prior to dedispersion is assumed to be negligible. The total channelisation cost is minimal due to the beamforming step, where all $N_a$ antennas signals are collapsed into 1 beam. This would result in $N_{\text{DM}}$ having a value of order $\mathcal{O}(10^5)$, suggesting that this scheme would either be memory or PCIe-bandwidth limited on the GPU.

\begin{figure}[t!]
  \centering
  \includegraphics[width=380pt]{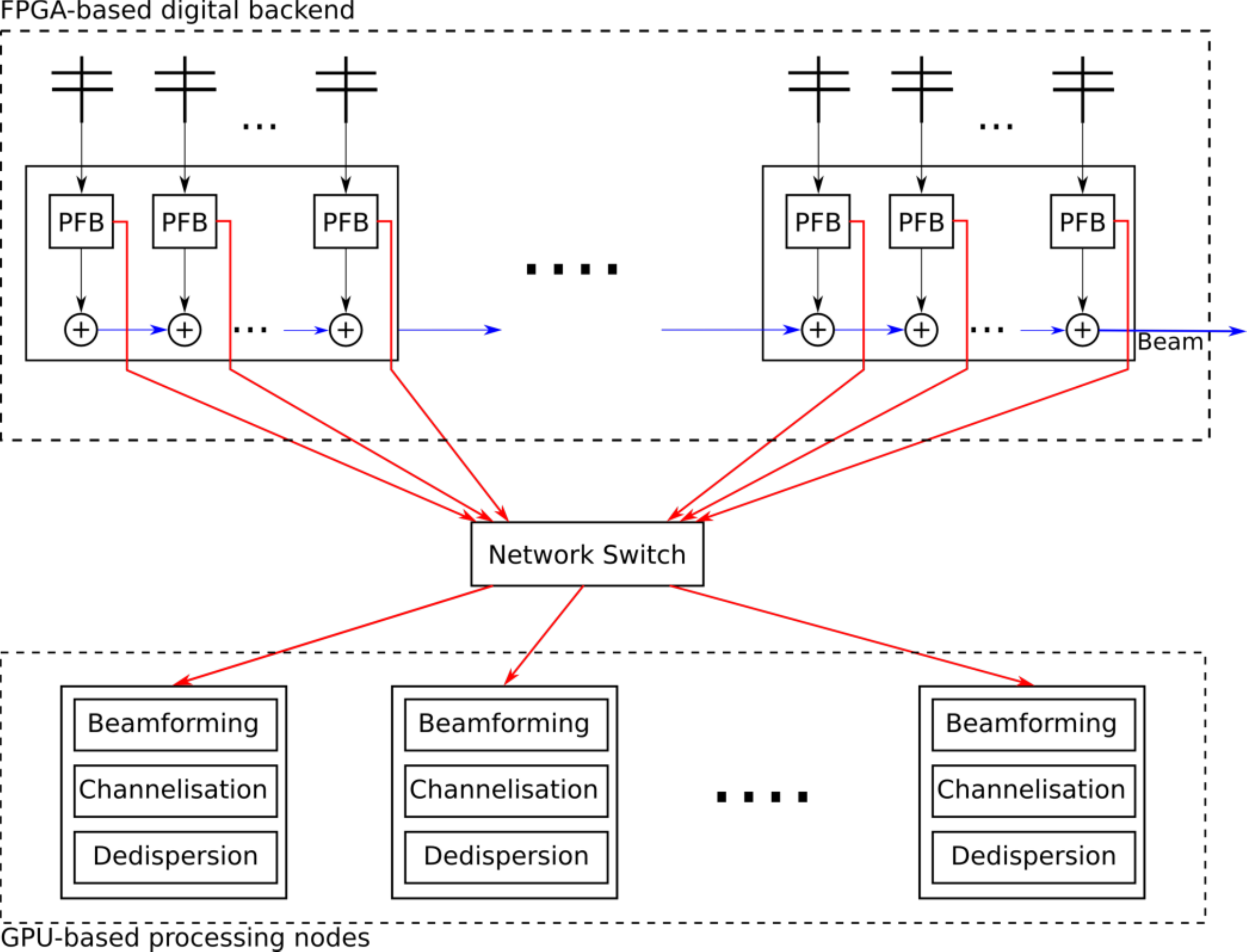}
  \caption[SKA$_1$-low station transient searching]{Ring-based beamforming schematic with transient detection pipeline. Red arrows represent coarsely-channelised data movement, whilst blue arrows}
  \label{skaLowStationTransientFigure}
\end{figure}

The maximum buffering time, such that all of GPU memory is filled up in one transfer, with enough additional memory to store temporary output, is:
\begin{equation}
 \tau_b  =  \frac{8 \cdot {M_{\text{tot}}} \cdot \alpha}{BN_p(n_sN_a + n_dN_b)}
\end{equation}
where $M_{\text{tot}}$ is the amount of GPU memory, $\alpha$ represents the percentage of available GPU memory (other buffers might be required by additional processing stages, or to be used as temporary buffers), $n_d$ is the internal bit-representation of the beamformed time series and $\tau_b$ is the buffering time. This would amount to a total input buffer size of
\begin{equation}
  M_{\text{in}} = BN_aN_p \tau_b
\end{equation}
The input buffer can then be copied to GPU memory in 
\begin{equation}
 \tau_i = \frac{e_tM_{\text{in}}}{B_{\text{PCIe}}}
\end{equation}
where $e_t$ is the PCIe bandwidth efficiency factor and $B_{\text{PCIe}}$ is the available PCIe bandwidth. Using the relationships above, the number of DM values which can be processed in real-time can be calculated. The DM trials which can be processed in parallel is limited by the amount of available GPU memory, and this is equal to
\begin{equation}
\label{dmMemEquation}
 N_{\text{DM-mem}} = \frac{N_cM_{\text{in}}}{B_0n_d}
\end{equation}
where $B_0$ is the number of samples for buffering time $\tau_b$ ($B = \tau_0B)$. The main assumption here is that the output dedispersed time series will be stored in the memory buffer used to store the input antenna voltages. Equation \ref{channelsEquation} can be used to calculate a value for $N_c$, which primarily depends on the maximum DM value. The total number of processable DM trials within the available time frame is:
\begin{equation}
\label{dmProcEquation}
 N_{\text{DM-proc}} = \frac{\tau_b - (\tau_i + \mathcal{T}_o\mathcal{F}_b + \mathcal{T}_o\mathcal{F}_c)}{4e_d\mathcal{T}_oB_0N_b + \tau_d}
\end{equation}
where $\mathcal{T}_o$ is the inverse of the available  GPU floating point resources for execution time $\tau_o$ and $\tau_d$ is the transfer time for a single dedispersed time series, equal to 
\begin{equation}
\tau_d = \frac{e_tB_0n_d}{B_{\text{PCIe}}N_c}.
\end{equation}
 The total number of DM trials processable in real-time is then
\begin{equation}
 N_{\text{DM-tot}} = \text{min}(N_{\text{DM-mem}},N_{\text{DM-proc}})
\end{equation}

The above relationship will be dominated by $\tau_i$, where assuming $\alpha$ = 0.95 and $n_d$ = 8-bits, for SKA$_1$ station parameters, $\tau_b$ = 0.675 s and $\tau_i$ = 0.306 s.  Based on these assumptions, a parameter table for each frequency subband was generated, a subset of which is shown in table \ref{skaLowTransientsTable}. The full table is reproduced in appendix A, where each subband is processed by a single GPU. Treating each station as an independent entity would yield a very small DM range, especially at lower frequencies. However, these would generally operate in unison, pointed to the same coordinates in the sky. This allows the distribution of the DM range over the entire SKA$_1$-low telescope, with each station processing a subset of the full DM range. Alternatively, a more efficient dedispersion algorithm can be used, however we will concentrate on the direct dedispersion as a use case for station dedispersion.

The DM step and maximum DM value are computed as per section \ref{surveyParametersSection}, assuming a minimum expected pulse width of 1 ms, 200\% accepted scattering (such that a 1 ms pulse is scattered across 2 ms) and 150\% pulse smearing. The DM fraction which can be processed on a single GPU is simply
\begin{equation}
 \text{DM}_{\text{frac}} = \frac{\Delta\text{DM} \cdot N_{\text{DM}}}{\text{DM}_{\text{max}}}
\end{equation}

Additionally, since the synthesised beam is generated on the GPU itself, it can be pointed anywhere within the antenna primary beam, thus enabling a wider surveying sky area. The last row in table \ref{skaLowTransientsTable} lists how many stations would be required to survey the entire station FoV, for the entire DM range, with the number of required beams changing per frequency subband. Following \cite{Dewdney2013}, the beam FoV, $\Omega_b$, is
\begin{equation}
 \Omega_b = \frac{\pi}{4} \left( \frac{1.3 \lambda}{D_{\text{station}}} \right) ^ 2
\end{equation}
where $D_{\text{station}}$ = 35 m is the station diamater. The number of beams 
required is then simply $\lceil{\Omega_a/\Omega_b}\rceil$, where $\Omega_a$ = 20 
deg$^2$ is the antenna FoV. A subset of the core stations is enough for a full 
fast transient survey over the entire station FoV, for all the frequency 
subbands.

\begin{table}
 \centering
 \begin{tabular}{ r c c c c }
   \hline
   Low Frequency (MHz)             & 50       & 112.5     & 159.375   & 284.375 \\
   High Frequency (MHz)            & 65.625   & 128.125   & 175       & 300 \\
   Maximum DM (pc cm$^{-3}$)       & 510.185  & 887.71    & 1154.645  & 1699.194 \\
   DM Step (pc cm$^{-3}$)          & 0.000125 & 0.000915  & 0.004095  & 0.02328 \\
   Processable DMs                 & 12,016   & 12,016    & 12,016    & 11,950 \\
   Number of Channels              & 15625    & 15625     & 15,625    & 8,192 \\ 
   Processable Max (pc cm$^{-3}$)  & 1.5023   & 10.996    & 36.114    & 278.186 \\
   DM Range Fraction (\%)          & 0.003    & 0.013     & 0.043     & 0.164 \\
   \#Stations Required             & 340       & 53       & 48       & 42 \\
   \hline
 \end{tabular}
 \caption[SKA$_1$-low station transient parameters subset]{Parameter subsets for 
SKA$_1$-low station blind, fast transient survey. Full table in appendix A.}
 \label{skaLowTransientsTable} 
\end{table}

\begin{figure}[t!]
  \centering
  \includegraphics[width=440pt]{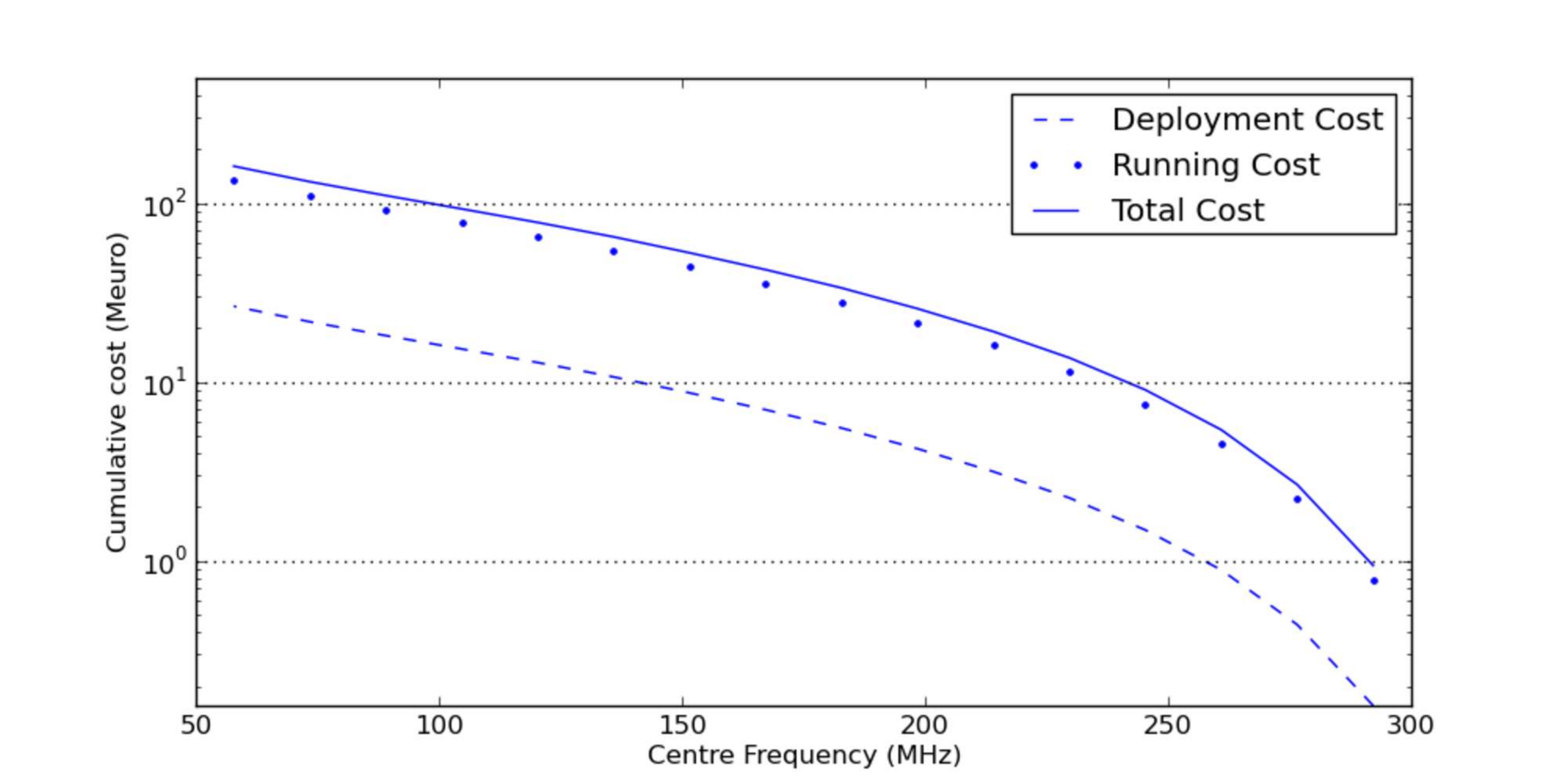}
  \caption[SKA$_1$-low station processing costs]{SKA$_1$-low station processing 
costs, based approximately on numbers given by \cite{ZarbAdami2013}, showing 
cumulative cost across frequency subbands, from higher to lower frequencies. 
These are server-based, and the server number per station changes across the 
observing band.}
  \label{skaLowStationTransientCostFigure}
\end{figure}

Using the values from table \ref{ska1LowCostsTable}, figure \ref{skaLowStationTransientCostFigure} shows the cumulative deployment and running cost across frequency subbands, from higher to lower frequencies, for surveying the entire antenna FoV. These figures are server based, resulting in some stations processing a subset of the subbands. Clearly, this is a costly setup, and should be compared with the hardware and running requirements for performing the same processing at the CSP.

\section{SKA$_1$-mid CSP Beamforming}

Pulsar searching is one of the main science cases for SKA$_1$-mid, and various studies have been conducted on the search parameters and hardware requirements for this telescope (for examples, \cite{Smits2008,Cordes2009}). \cite{Dewdney2013} define the optimal configuration for pulsar searches, a summary of which is listed in table \ref{skaMidParametersTable}, where the worst-case scenario in terms of computational requirements is selected. In this section we focus on the beamforming requirements, whilst in the next we turn to dedispersion.

The dish output streams need to be channelised and beamformed prior to passing through the transient detection pipeline. Since $N_a < N_b$, it is more efficient to channelise the observing band prior to beamforming, so channelisation can be easily integrated within the beamforming architecture. The input data rate to the SKA$_1$-mid beamformer, $B_{\text{in}}$, assuming 8-bits per complex value, is approximately 1.4 Tb/s, with an operation count, $B_{\text{ops}}$, of $\sim$1.5 PFLOP/s and output data rate, $B_{\text{out}}$, of $\sim$10.67 Tb/s, assuming a channelisation output sampling rate of $1/B$. The channelisation operation count, $C_{\text{ops}}$, is a small fraction of $B_{\text{ops}}$, amounting to about 15 TFLOP/s, or 26 GFLOP/s per dish, assuming an 8-tap polyphase filterbank channeliser and 20 kHz spectral resolution. Integrating the channelisation operations into the beamformer induces a minimal cost, even if dishes are channelised multiple times, depending on the beamforming architecture adopted. Simply matching these aggregate values to GPU specifications, taking into account the PCIe transfer and beamforming efficiency, $e_t$ and $e_b$ respectively, at least 350 GPUs would be required. However this is a highly conservative number, as it assumes the antenna signals are transferred once to a single GPU, which is an unlikely scenario. 

\begin{table}[t!]
  \centering
  \begin{tabular}{ l c }
     \hline
     \multicolumn{2}{|c|}{Telescope Parameters} \\
     \hline
     Optimal number of antennas & 141   \\
     Associated array diameter  & 950 m \\
     Centre frequency           & 1400 MHz \\
     Bandwidth                  & 300 MHz \\
     Beam size at 950 MHz       & 1.5 $\times$ 10$^{-4}$ deg$^2$ \\
     Number of beams            & 2222 \\
     \hline
     \multicolumn{2}{|c|}{Search Parameters} \\
     \hline
     Maximum DM                 & 3000 pc cm$^{-3}$ \\
     Achievable maximum DM      & 827 pc cm$^{-3}$ \\
     Number of channels         & 15000 \\
     Sampling time              & 50 $\mu$s \\
     Frequency resolution       & 20 kHz \\
     DM step                    & 0.054 pc cm$^{-3}$ \\
     Number of DM trials        & 15350 \\
     Number of subbands         & 64 \\
  \end{tabular}
  \caption[Key Specifications of SKA$_1$-mid]{Key specifications of SKA$_1$-mid, as defined by \cite{Dewdney2013}, representing the worst case scenario in term of computational requirements, corresponding to searches at galactic latitude $|b| < 5$.}
  \label{skaMidParametersTable}
\end{table} 

\begin{figure}[t!]
  \centering
  \includegraphics[width=400pt]{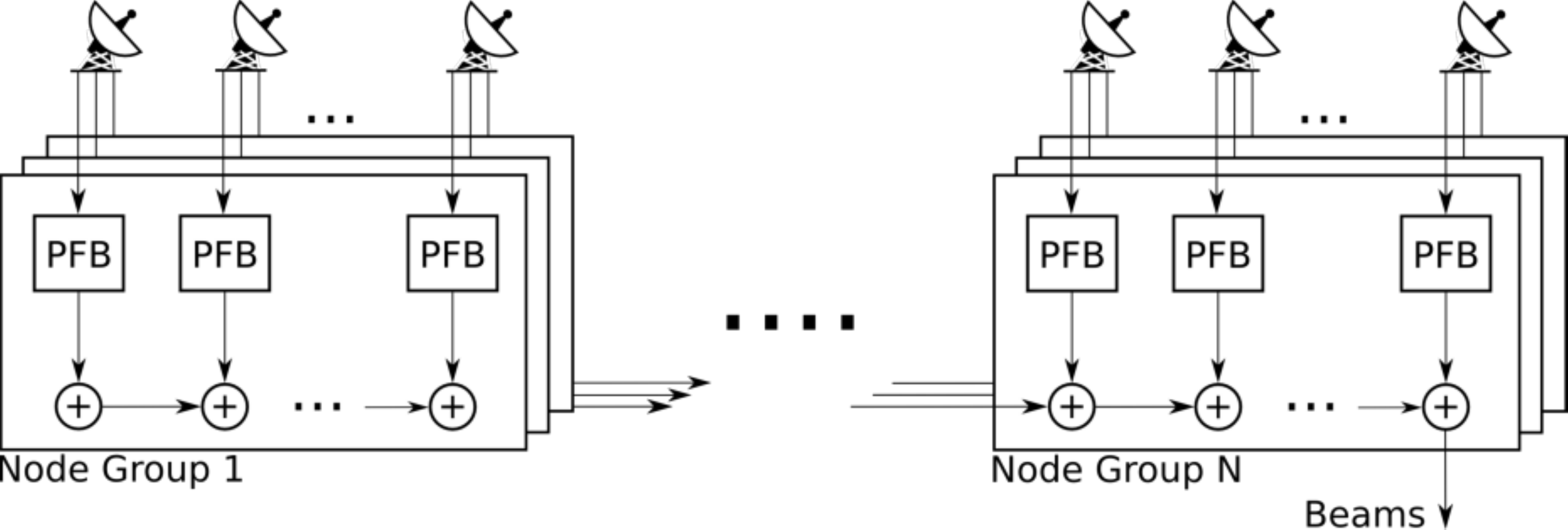}
  \caption[SKA$_1$-mid CSP ring-based beamforming]{SKA$_1$-mid CSP ring-based beamforming schematic. Each antenna signal is streamed to multiple node groups, where each node within the group partially synthesise a subset of the output beams.}
  \label{skaMidCspBeamFigure}
\end{figure}

We now provide a more realistic method for determining the number of GPUs required, assuming a series of ring-based beamforming chains, a schematic of which is depicted in figure \ref{skaMidCspBeamFigure}. Dish signals are streamed to multiple nodes, each of which synthesise a subset of the beams. Node groups are chained together to form the beamforming ring. The beamformer output data rate to the rest of the pipeline can be reduced by computing the power of the voltage series and summing across polarisations directly after beamforming on the last GPU in each chain, resulting in an aggregate output data rate of $\sim$5.34 Tb/s. Assuming no memory constraints ($\tau_b$ = 1) and no execution overlap, the computational and I/O load of each node is
\begin{equation}
\label{beamNodeEquation}
  \tau_b = \tau_i + \mathcal{F}_c + \mathcal{F}_b + \mathcal{F}_p + \tau_o 
 \end{equation}
where $\mathcal{F}_p = 4BN_p$ is the operation count for magnitude calculation 
and summing across polarisations, which is only performed by the last GPU in 
every beamforming chain, prior to transferring the beamformed series to the rest 
of the processing pipeline. It should be noted that input, processing and output
operations can be overlapped on GPUs with dual copy engines, however since 
$\tau_b$ is dominated by $\tau_i$ the benefit will not be large, especially 
when considering the larger memory requirements and higher implementation 
complexity. $\tau_o = BN_b$ is the output data rate. $\tau_i$ 
includes dish inputs as well as the output from the previous node in the chain
\begin{equation}
 \tau_i = 2BN_p(M_a + M_b)   
\end{equation}
where $M_a \subseteq N_a$ and $M_b \subseteq N_b$. Ignoring algorithmic efficiency, arithmetic timing and I/O transfer efficiency, equation \ref{beamNodeEquation} becomes
\begin{equation}
 2BN_p\left(M_an_d + 4N_a(\text{log}_2(N_c) + N_{\text{taps}}) + 4M_bM_a + M_bn_b\right) \leq 1
\end{equation}
Appropriates values for $M_a$ and $M_b$ need to be selected in order to minimise the difference between the sides of this equation, such as to maximise GPU resource utilisation. The total number of GPUs is then simply
\begin{equation}
 N_{\text{GPUs}} = \left\lceil \frac{N_a\cdot N_b}{M_a\cdot M_b} \right\rceil
\end{equation}

\begin{figure}[t!]
  \centering
  \includegraphics[width=420pt]{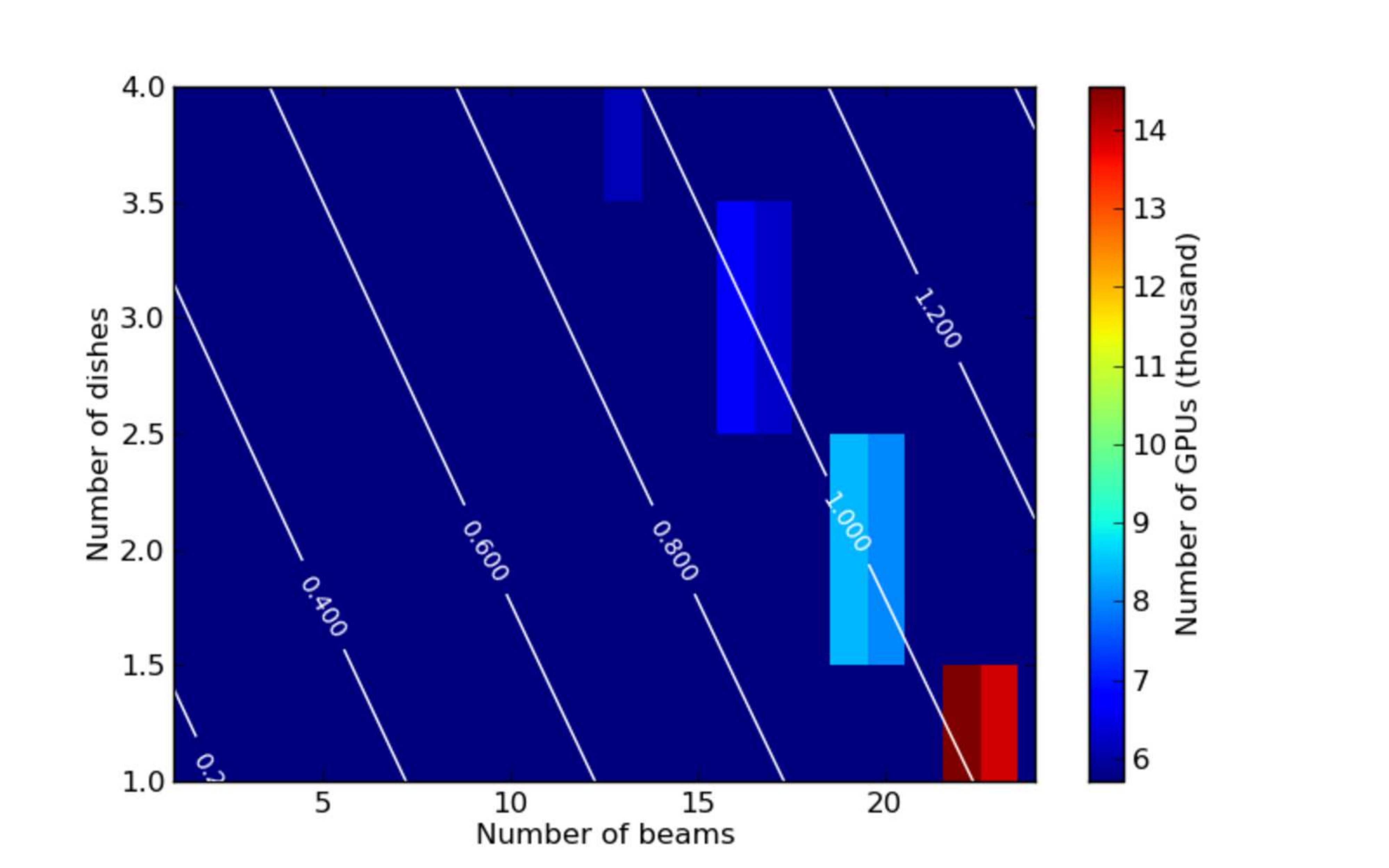}
  \caption[SKA$_1$-mid beamformer configuration]{Assuming a series of ring-based beamforming chains, where every GPU processes a subset $M_a$ of the dishes, partially synthesising a subset $M_b$ of the output beams, the set of optimal values for $M_a$ and $M_b$ lies along the unit contour line. The color mapping represents values for $N_{\text{GPUs}}$ where the GPU utilisation rate is within 1\% of the peak.}
  \label{skaMidBeamOptimalFigure}
\end{figure}

Figure \ref{skaMidBeamOptimalFigure} shows runtime values for various combinations, where the unit contour line represents the set of values for which GPU utilisation is optimised. Appropriate efficiency factors were included in these calculation, and a value for $n_b$ of 8-bits was assumed. The colour mapping in this plot represents values for $N_{\text{GPUs}}$ where the GPU utilisation rate is within 1\% of the peak. The optimal values for $M_a$ and $M_b$ are 4 and 14 respectively, amounting to a GPU utilisation rate of 99.5\%, 86.7\% of which is spent transferring data over the PCIe bus. This is equivalent to about 5700 GPUs. By setting $n_b$ to 4-bits, $N_{\text{GPUs}}$ can be reduced to about 2700 ($M_a$ = 5 and $M_b$ = 23, with similar utilisation rates).

\section{SKA$_1$-mid CSP Dedispersion}

Each synthesised beam has to be independently dedispersed and searched for any interesting signals. Based on the search parameters listed in table \ref{skaMidParametersTable}, a total of 15,350 dedispersed time series have to be generated for 2222 beams, resulting in an operation count of $\sim$10 PFLOP/s when using the direct method, equivalent to about 3,500 GPUs when accounting for $e_d$, which is clearly unfeasible, indicating that alternative dedispersion methods need to be investigated. Some of these techniques were discussed in section \ref{dispRemovalSection}. \cite{Dewdney2013} base their analysis on the Taylor-tree method with piecewise linear approximation, where the frequency band is partitioned into 64 subbands. Following \cite{Barsdell2012}, the maximum DM value which can be processed, referred to as the 'diagonal' DM, with this method is
\begin{eqnarray}
 \text{DM}_{\text{diag}} & = & \frac{N_c' - \frac{1}{2}}{\Delta T(N_c) - \Delta T(N_c-N_c')} \\
 \Delta T(c)&  = & \frac{k_{\text{DM}}}{\Delta t} \left( (f_0 + c\Delta f)^{-2} - f_0^{-2} \right)
\end{eqnarray}
where $\Delta T(c)$ is defined in section \ref{directDispSecion} (equation \ref{deltaTEq}) and $N_c'$ is the length of each subband. The Taylor-tree method requires the number of channels to be a power of 2, therefore each subband must be composed of 
\begin{equation}
 N_c' = 2^{\left\lceil \log_2\left(\frac{N_c}{N_{\text{sub}}} \right) \right\rceil}
\end{equation}
where $N_{\text{sub}}$ is the number of frequency partitions. For SKA$_1$-mid, $N_c'$ = 256, therefore an extra 1384 zero-padded frequency channels have to be injected across the band. The total number of frequency channels becomes 16384, while still retaining the original spectral resolution. Together with the temporal resolution specified in table \ref{skaMidParametersTable}, $\text{DM}_{\text{diag}}$ = 73.37 pc cm$^{-3}$, with a DM step equivalent to $\sim$0.008 pc cm$^{-3}$.

In order to reach the maximum DM required additional processing would have to be applied, such as time binning, or applying time delays to the beamformed time series such that the diagonal DM is at a DM of zero at each iteration. The latter option still requires a band-wide, frequency-dependent delay to be applied for band alignment, however it leads to an overall reduction in computational requirements. This will also result in a loss in pulse S/N for each binning iteration. We base our analysis on this technique. The optimal subband size is a balance between computational speedup and additional signal smearing induced by the linear approximation to a quadratic dispersion curve, however for the scope of this section, we will assume 64 subbands, which also corresponds to the peak performance achieved by \cite{Barsdell2012}. In order to process all DMs up to $\text{DM}_{\text{max}}$, the time series has to pass through 
\begin{equation}
N_{\text{bins}} = \left\lceil \text{log}_2\left(\frac{\text{DM}_{\text{max}}}{\text{DM}_{\text{diag}}}\right) \right\rceil + 1
\end{equation}
binning stages. A total of $N_c'$ DM trials have to be processed in the first stage of each iteration. During the seconds stage, $N_{\text{DM},i}'$ trials need to be processed, where $i$ is the binning stage. This value depends on the maximum DM value associated with each binning stage, $\text{DM}_{\text{max},i}$ and the associated DM step, $\Delta\text{DM}_i$
\begin{equation}
N_{\text{DM},i}' = \left\{ 
  \begin{array}{l l}
    \text{DM}_{\text{max},i}/\Delta\text{DM}_{\text{max}} & \quad i == 1 \\ 
    (\text{DM}_{\text{max},i} - \text{DM}_{\text{max},i-1})/\Delta\text{DM}_{\text{max}} & \quad i > 1
  \end{array} \right.
\end{equation}
where $\Delta\text{DM}_{\text{max}} = \max(\Delta\text{DM}_i, \Delta\text{DM})$. Using these relationships and the values from table \ref{skaMidParametersTable}, table \ref{treeDedispTable} lists the required dedispersion parameters when using the piecewise linear tree algorithm. The operation count for this method is
\begin{eqnarray}
 \mathcal{F}_{\text{tree}} & = & N_b \cdot (\text{stage}_1 + \text{stage}_2) \\
 \text{stage}_1 & = & \sum_{i=0}^{N_{\text{bin}-1}}\left(8\frac{N_t}{2^i}N_{\text{sub}}N_c'\text{log}_2(N_c')\right) \\
 \text{stage}_2 & = & \sum_{i=0}^{N_{\text{bin}-1}}\left(8\frac{N_t}{2^i}N_{\text{sub}}N_{\text{DM},i}'\right) 
\end{eqnarray}

The computational cost for time binning and delaying is assumed to be negligible 
relative to the dedispersion cost, and additionally they could also be performed 
by the CPU. Factors of 8 are included to compensate for delay computation, 
rounding and other operations. Due to the high algorithmic complexity of this 
technique, the lack of available optimised implementations and the use of 
modulus and division operations required during the first stage, an efficiency 
factor, $e_{\text{tree}}$, of 10\% is attributed to the first stage, while the 
same factor as direct dedispersion is applied to second stage. This yields a 
value for $\mathcal{F}_{\text{tree}}$ of $\sim$78 TFLOP/s, which is equivalent 
to to approximately 26 GPUs when taking arithmetic efficiency into account. 
\begin{table}[t!]
  \centering
  \begin{tabular}{ l | c c c c c }
     \hline
     Binning Stage & 1 & 2 & 3 & 4 & 5 \\
     \hline
     Binning Factor & 1 & 2 & 4 & 8 & 16 \\
     Sampling Time ($\mu s$) & 50 & 100 & 200 & 400 & 800 \\
     Maximum DM  (pc cm$^{-3}$)        & 73.37 & 146.74 & 293.48 & 586.95 & 1173.90 \\
     Equivalent DM Step (pc cm$^{-3}$)  & 0.008 & 0.016 & 0.033 & 0.65 &4 0.131 \\
     Number of DMs       & 1358 & 1358 & 2717 & 4486& 1834 \\
     \hline
   \end{tabular}
  \caption[SKA$_1$-mid survey parameters]{SKA$_1$-mid survey parameters when using piecewise linear tree dedispersion.}
  \label{treeDedispTable}
\end{table} 

Next we analyse the memory requirements for this method. If the beamformed time series are binned in place, the piecewise linear algorithm requires
\begin{equation}
\label{treeMemoryEquation}
 M_{\text{tree}} = \text{maxshift} + N_tN_cn_b + \sum_{i=0}^{N_{\text{bins}-1}} \left[ \frac{N_t}{2^i} (N_cn_{\text{tree}} + N_{\text{DM},i}n_{\text{DM}}) \right]
\end{equation}
per beam, where maxshift represents the amount of extra memory required to process up to $\text{DM}_{\text{max}}$.
If these samples are kept in their binned form, then maxshift has a value of $N_c^2N_{\text{bins}}n_b$. $n_{\text{tree}}$ is the bit representation of the output series from the first stage and $n_{\text{DM}}$ is the bit representation of the fully dedispersed time series. $n_{\text{tree}}$ should accommodate the temporary summations during the first stage, and, assuming bit representations of powers of 2, should have a value of at least
\begin{equation}
 n_{\text{DM}} \geq 2^{\left\lceil \text{log}_2(\text{log}_2(N_c'\cdot 2^{n_b})) \right\rceil}
\end{equation}

Equation \ref{treeMemoryEquation} assumes that the intermediary stages for all binning stages, as well as the output dedispersed time series, will be kept in GPU memory. This requirement can be alleviated by overwriting the intermediary stages, and transferring the processed data out of GPU memory after every stage. An additional optimisation is to overwrite the input time series with the output from the second stage, given that they occupy the same amount of memory, otherwise the input buffer would have to be enlarged. Internal GPU bandwidth is of the order of 300 GB/s, so the effect on overall execution time would be negligible. The memory requirement then becomes
\begin{eqnarray}
 M_{\text{tree}} & = & \text{maxshift} + \left\{ 
  \begin{array}{l l}
     2N_tN_cn_{\text{tree}} & \quad N_tN_cn_{\text{tree}} > \beta \\ 
    N_tN_cn_{\text{tree}} + \beta & \quad  N_tN_cn_{\text{tree}} < \beta
  \end{array} \right. \\
 \beta & = & \max({N_tN_{\text{DM},0}},\; \cdots \; ,\frac{N_t}{2^{N_{\text{bins}}-1}}N_{\text{DM},N_{\text{bins}}-1})
\end{eqnarray}
The new size for maxshift will depend on whether the binning process will be performed entirely on the GPU, or on the CPU during the GPU dedispersion stages, with I/O transfer overlapping GPU processing. Here we choose the latter approach, since it requires less memory. Maxshift then simply becomes $N_c^2n_b$. These optimisations reduce the memory requirements to $\sim$1.9 TB, equivalent to approximately 320 GPUs. This value is close to the memory requirement for direct dedispersion, which amounts to approximately 1.2 TB, excluding maxshift.

It is evident that even with the optimisations mentioned above, GPU-based dedispersion for SKA$_1$-mid will be memory limited. The total input and output data rate is approximately 670 GB/s, which can be accommodated by approximately 50 GPUs. Table \ref{dedispReqsTable} summarises the GPU requirements described in this section. A rough estimate suggests that almost 400\% more GPUs are required to keep the data in memory than those required for I/O and computation. 

\begin{table}[t!]
  \centering
  \begin{tabular}{ r | l l }
     Processing    & 78 TFLOP & 26 GPUs  \\
     I/O transfers & 670 GB/s & 50 GPUs \\
     Memory        & 1.7 TB   & 285 GPUs \\
   \end{tabular}
  \caption[Dedispersion requirements for SKA$_1$-mid]{Dedispersion requirements for SKA$_1$-mid.}
  \label{dedispReqsTable}
\end{table} 

\section{Conclusion}

The Square Kilometre Array will pose considerable computational and data transport challenges, some which we have investigated in this chapter, concentrating mostly on station-level processing for SKA$_1$-low and the Central Signal Processing non-imaging pipeline for SKA$_1$-mid. The applicability of GPUs for both station-level and CSP beamforming was investigated, implementations of which will be I/O limited, with PCIe transfers being the major bottleneck for GPU-based systems. In the case for CSP beamforming for SKA$_1$-mid 87\% of total execution time would be spent performing I/O, driving the number of GPUs required for beamforming upward of 5,000. The compute-to-transfer time ratio for SKA$_1$-low station beamforming is higher, however still I/O limited. When considering the running cost for GPU-based systems, especially power consumption costs, it might be more beneficial to deploy an FPGA-based system. GPUs are not well suited for very high I/O scenarios.

We have also analysed the hardware requirements for SKA$_1$-mid CSP dedispersion, in connection with the non-imaging pipeline, and highlight the fact that this process will be memory-limited, where 400\% more GPUs are required to keep the data being processed in memory. This requirement will be even more severe when binary searches are taken into consideration, however the compute requirements will also be much higher, which might tip over the balance between memory requirements and compute requirements. 

The analysis performed in this chapter are based on numbers and parameters listed in the SKA Baseline Design \cite{Dewdney2013}, however the methods described are applicable to any telescope configuration.


\chapter{Conclusion}
\label{conclusionChapter}

The primary aim of the work presented in this thesis was to investigate the use 
and applicability of Graphical Processing (GPUs) for real-time, non-imaging 
pipelines, with significant emphasis on dedispersion, automatic classification 
of detected events and beamforming. This has led to the development of a 
scalable, flexible and high performance software pipeline which has been used to 
process online data from radio telescopes, such as BEST-II and LOFAR, as well as 
post-processing of several data products, including GBT observations of Kepler's 
field of view, which is still underway.

\subsection*{Dedispersion}

The direct dedispersion algorithm was optimised for GPU hardware, based on the work by \cite{Magro2011} and \cite{Armour2012}, registering a speed-up, compared to an optimised CPU implementation, of more than one order of magnitude, comparable to that reported by \cite{Barsdell2012}. This leads to a significant reduction in hardware requirements for online systems. Performance benchmarks presented in chapter \ref{pipelineChapter} demonstrate that 8 beams with 20 MHz bandwidth (or alternatively a single beam of 160 MHz, since the number of beams and processable bandwidth are inversely proportional) can be processed with a single server hosting two mid-range GPUs, for up to almost 1000 DM trials. This was compared to similar state--of--the--art systems currently in development, including an FPGA-based approach by \cite{Addario2013}, and show that GPUs are a favourable architecture for transient searching.

Direct dedispersion scales linearly with increasing number of beams and DM trials, and we show that for SKA$_1$-mid approximately 10 PFLOPs would be required to process all the beams, equivalent to $\sim$3,500 GPUs, this excluding memory requirements and I/O transfer time. Alternative dedispersion algorithms have to be investigated. Following \cite{Dewdney2013}, where the Taylor tree algorithm with piecewise linear approximation was selected as a candidate for dedispersion, we provide a detailed analysis of the computational, memory and I/O requirements. We note that with this technique the system will be memory limited, with 400\% more GPUs required to keep the data in memory than those required for computation and I/O transfer. For SKA$_1$-mid non-imaging requirements, approximately 320 GPUs would be required for dedispersion.

\subsection*{Online Transient Detection Pipeline}

Fast transient detection pipelines require considerable computational resources, and several processing platforms are being investigated to suite these requirements. In this thesis we propose a GPU-based solution, moulded around a custom GPU framework, whereby data input and buffering is performed by the CPU, with compute intensive tasks offloaded to the GPU. Several processing stages were implemented in this pipeline, including RFI mitigation, dedispersion, event detection and classification, as well as data quantisation and persistence. RFI mitigation through bandpass correction and thresholding removes high power spatial and temporal terrestrial signals which would result in a high number of false detections. The set of dedispersed time series are thresholded and then clustered using a density-based clustering algorithm. The DM-S/N curve of each cluster candidate is then compared to a fitted model which filters out RFI-induced clusters. The detection of an astrophysical event triggers the data persistence module, which quantises and writes the input buffer to disk, containing the event together with cluster parameters. These stages are encapsulated as a standalone framework which can be used in offline mode, for processing archival data, as well as within an online application with additional real-time capabilities. The prototype presented here is capable of processing multiple independent beams in parallel across as many GPUs as are attached to the host. 

The accuracy of the RFI mitigation and event detection stages was examined with 
simulated data sets containing RFI and dispersed pulses which were passed 
through the pipeline. Narrow pulses, of the order of 2 to 10 samples wide, were 
only fully detected for a pulse S/N of 4.0 or greater, while wider pulses were 
detected for pulse S/N as low as 2.0. Most of the simulated RFI signals did pass 
through the RFI mitigation stage, however 100\% of the broadband signals were 
correctly classified as RFI, whilst a single cluster caused by a narrowband 
burst was incorrectly classified as an astrophysical event. This leads to two 
conclusions:
\begin{itemize}
 \item Clustering algorithms, coupled with appropriate cluster selection techniques, are a viable candidate for event detection, however more robust candidate selection mechanisms are required to increase the detectability of short duration, low power signals. These include pattern matching and machine learning based techniques. Any supervised learning algorithm will likely require input data sets from several telescope locations, as the RFI environment will change. Ways of generating labelled data sets also need to be investigated.
 \item Provided that the event detection scheme above is available, any RFI mitigation stage need only filter high-power signals so as to reduce the number of detected data points after dedispersion. Any unmitigated RFI signals will then be filtered out by the candidate classification stage.
\end{itemize}

This pipeline was also enhanced with real-time capabilities, consisting primarily of high-speed stream processing functionality capable of processing a 5.12 Gb/s SPEAD stream. This system was then deployed on the Medicina BEST-II array, where a ROACH-based digital backend performs all the required pre-processing prior to transmitting beamformed data to the transient detection system. Several test observations were conducted, especially on PSR B0329+54, which is the brightest pulsar observable by BEST-II. Through these observation appropriate threshold parameters were set to various stages in the pipeline, leading to the automatic classification of pulses originating from B0329+54 from RFI-induced events, serving as a online test case for the pipeline.

\subsection*{Beamforming}

Signals from multiple antennas have to be combined prior to transient searching (except for cases where an aperture array is used in Fly's Eye mode). Beamforming is a computationally expensive process, with considerable I/O requirements for large N aperture arrays where all the antennas cannot be combined using one processing node. Beamforming implementation tend to be I/O limited. In chapter \ref{beamformingChapter} we presented a GPU implementation of a coherent multi-beam beamformer which can synthesise a number of coherent beams for arrays with an arbitrary number of elements and frequency channels. The kernel utilises 50\% of the peak theoretical performance on the test device, at 1.2 TFLOPs, and is limited by shared memory bandwidth. The overall implementation is limited by PCIe bandwidth, where for a serial execution pipeline, 50\% of the processing time is spent transferring data from CPU to GPU memory. If the beams need to be post-processed outside of the GPU on which they were generated, the GPU remains mostly idle, waiting for data transfer requests to terminate.

This discrepancy between computational and I/O requirements can be alleviated by performing additional operations after beam generation, and to test this assertion we have integrated this kernel with the transient detection pipeline described in chapter \ref{pipelineChapter}. A mock BEST-II backend running just the F-Engine was set up and used to benchmark this system, and we show that for small, low bandwidth arrays the computational cost for beamforming is negligible when compared to dedispersion. This gives rise to potential systems where these operations are integrated, allowing for dynamic observations where beam pointings can be updated online, in real time, triggered by interesting events during the event detection stage of the transient detection pipeline. Having the raw voltage data available in memory can potentially enable on-the-fly correlation and image generation for more accurate transient source localisation.

We have also presented a test beamforming setup for SKA$_1$-low station 
beamforming as well as SKA$_1$-mid CSP beamforming, based on a ring setup. We 
demonstrate that GPUs are ill-suited for this task, and the number of processing 
nodes required depends entirely on the input, output and inter-node bandwidth 
requirements. For SKA$_1$-low up to 10 antennas can be combined on a single GPU, 
and even when channelisation is performed in tandem, the devices spend most of 
the time waiting for PCIe transfers. This situation is worse for SKA$_1$-mid 
CPS, where more than 5,000 GPUs are required, with 87\% of execution time spent 
waiting for I/O transfers. GPUs are also power hungry devices, with a TDP of 
approximately 250 W. A GPU-based setup would result in very high operational 
costs, especially when compared to high-end FPGA-based devices with comparable 
processing capabilities, as demonstrated by \cite{Hickish2013}.

\subsection*{Extensions and Future Work}

The work presented in this thesis led to the development of a pipeline prototype for online fast transient detection which can be deployed on any radio telescope and used to process archival data, and could potentially be used as a reference design for future pipelines, and make possible the creation of a dynamic, high performance, transient search software pipeline through collaborative work. By making the data interfacing mechanisms more generic and providing a cleaner model by which new processing stages can be implemented and attached to the system, this software can serve as a starting point for creating specialised backends for current and future radio instruments. We believe that GPUs are an ideal candidate for transient detection, and more work needs to be done in this field. Pulsar search software require additional processing stages, such as harmonic summing and binary searches, both of which are computationally expensive. To date, investigation into porting acceleration search algorithms to GPUs has been limited, and although these algorithms would likely be memory limited, the advances in GPU technology will make it possible for the entire pulsar search pipeline to be ported to GPUs. In order to meet the full non-imaging requirements for SKA$_1$-mid, within a reasonable cost, algorithmic and implementation developments in the coming years will prove to be critical.

\bibliomatter
\bibliographystyle{cell}
\bibliography{ref}

\chapter{SKA$_1$-low Station Transient Parameters}
\label{appendixA}

\onehalfspacing

The table below lists the parameters for SKA$_1$-low station-level transient searching, discussed in section \ref{skaLowTransient}, where the station bandwidth is split into 16 frequency subbands, each processed separately on a GPU processing node.
\begin{sidewaystable}[b!]
  \centering
  \begin{tabular}{ l c c c c c c c c c }
   \hline
   Subband & 1 & 2 & 3 & 4 & 5 & 6 & 7 & 8 \\
   \hline
   \hline
   Low Frequency (MHz)       & 50       & 65.625 & 81.25 & 96.87 & 112.5 & 128.125 & 143.75 & 159.375 \\
   High Frequency (MHz)      & 65.625  & 81.25 & 96.87 & 112.5 &  128.125 & 143.75 & 159.375 & 175  \\
   Maximum DM (pc cm$^{-3}$) & 510.18   & 620.18 & 722.16 & 817.74 & 908.05 & 993.81 & 1075.95 & 1154.64  \\
   DM Step (pc cm$^{-3}$)    & 0.000125 & 0.000285 & 0.000535 & 0.000915 & 0.001435 &  & 0.003005 &  0.004095 \\
   Processable DMs           & 12016     & 12016 & 12016 & 12016 & 12016 & 12016 & 12016 & 12016  \\
   Number of Channels        & 15625   & 15625 & 15625 & 15625 & 15625 & 15625 & 15625 & 15625 \\ 
   Processable Max (pc cm$^{-3}$)  & 1.50 & 3.42 & 6.43 & 11.00 & 17.24 & 25.54 & 36.11 & 49.21  \\
   DM Range Fraction (\%)    & 0.003 & 0.006 & 0.009 & 0.013 & 0.019 & 0.026 & 0.034 & 0.043  \\
   \#Stations Required       & 340 & 182 & 113 & 75 & 53& 39 & 30 & 24\\
   \hline
   \\
   \hline
   Subband & 9 & 10 & 11 & 12 & 13 & 14 & 15 & 16 \\ 
   \hline 
   Low Frequency (MHz)       & 175 & 190.625 & 206.25 & 221.875 & 237.5 & 253.125 & 268.75 &  284.375  \\
   High Frequency (MHz)      & 190.625 & 206.25 & 221.875 &237.5 & 253.125 & 268.75 & 284.375 & 300  \\
   Maximum DM (pc cm$^{-3}$) &1230.39 & 1303.51 & 1374.26 & 1442.86 & 1509.50 & 1574.35 & 1637.54 & 1699.19   \\
   DM Step (pc cm$^{-3}$)    & 0.005416 & 0.007006 & 0.008876  & 0.011056 & 0.013556  & 0.016417 & 0.019647 & 0.023277  \\ 
   Processable DMs           & 12016 & 12016 & 12016 & 12016 & 12016 & 11950 & 11950 & 11950  \\
   Number of Channels        & 15625 & 15625 & 15625 & 15625 & 15625 & 8192 & 8192 & 8192  \\ 
   Processable Max (pc cm$^{-3}$) & 65.08 & 84.19 & 106.66 & 132.86 & 162.90 & 196.19 & 234.80 & 278.19   \\
   DM Range Fraction (\%)    & 0.053 &  0.065 & 0.078 & 0.092 & 0.108 & 0.125 & 0.143 & 0.164 \\
   \#Stations Required       & 19 & 16 & 13 & 11 & 10 & 9 & 7 & 7  \\
  \hline
  \end{tabular}
  \caption[SKA$_1$-low station fast transient survey parameters]{SKA$_1$-low station fast transient survey parameters}
\end{sidewaystable}

\end{document}